\def\MPL #1 #2 #3 {Mod.~Phys.~Lett.~{\bf#1},\  #2 (#3)}
\def\NPB #1 #2 #3 {Nucl.~Phys.~{\bf#1},\  #2 (#3)}
\def\PLB #1 #2 #3 {Phys.~Lett.~{\bf#1},\  #2 (#3)}
\def\PR #1 #2 #3 {Phys.~Rep.~{\bf#1},\ #2 (#3)}
\def\PRD #1 #2 #3 {Phys.~Rev.~{\bf#1},\  #2 (#3)}
\def\PRL #1 #2 #3 {Phys.~Rev.~Lett.~{\bf#1},\  #2 (#3)}
\def\RMP #1 #2 #3 {Rev.~Mod.~Phys.~{\bf#1},\  #2 (#3)}
\def\ZP #1 #2 #3 {Z.~Phys.~{\bf#1},\  #2 (#3)}
\def\IJMP #1 #2 #3 {Int.~J.~Mod.~Phys.~{\bf#1},\  #2 (#3)}

\def\twoloop{two-loop/RGE-improved}
\def\Twoloop{Two-loop/RGE-improved}
\def\dmsq{\delta m^2}
\def\wtilde{\widetilde}
\def\mstopi{m_{\tilde t_1}}
\def\mstopii{m_{\tilde t_2}}
\def\lepii{LEP2}
\def\lammsbar{\Lambda_{\overline{MS}}}
\def\msusy{M_{SUSY}}
\def\emem{e^-e^-}
\def\tevstar{Tev$^\star$}
\def\sighbar{\overline \sigma_{\h}}

\def\rts{\sqrt s}

\def\h{h}
\def\mh{m_{\h}}
\def\gamh{\Gamma_{\h}}
\def\elep{E_{LEP}} 
\def\lam{\lambda}
\def\cale{{\cal E}}
\def\calo{{\cal O}}
\def\eg{{\it e.g.}}
\def\etal{{\it et al.}}
\def\epem{e^+e^-}
\def\mupmum{\mu^+\mu^-}
\def\tauptaum{\tau^+\tau^-}
\def\mm{\mu^+\mu^-}
\def\ee{e^+e^-}
\def\lplm{\ell^+\ell^-}
\def\taup{\tau^+}
\def\taum{\tau^-}
\def\hn{h}
\def\lsim{\mathrel{\raise.3ex\hbox{$<$\kern-.75em\lower1ex\hbox{$\sim$}}}}
\def\gsim{\mathrel{\raise.3ex\hbox{$>$\kern-.75em\lower1ex\hbox{$\sim$}}}}
\def\@versim#1#2{\vcenter{\offinterlineskip
        \ialign{$\m@th#1\hfil##\hfil$\crcr#2\crcr\sim\crcr } }}
\def\alt{\lsim}
\def\agt{\gsim}
\def\zstar{Z^\star}
\def\wstar{W^\star}
\def\slash#1{#1\hskip-6pt/\hskip2pt}

\def\etmiss{\slash E_T}

\def\ie{{\it i.e.}}
\def\ebtag{e_{b-tag}}
\def\emistag{e_{mis-tag}}
\def\gam{\gamma}
\def\egamgam{E_{\gam\gam}}

\def\anti{\overline}
\def\pbi{~{\rm pb}^{-1}}
\def\fbi{~{\rm fb}^{-1}}
\def\fb{~{\rm fb}}
\def\pb{~{\rm pb}}

\def\mev{\,{\rm MeV}}
\def\gev{\,{\rm GeV}}
\def\tev{\,{\rm TeV}}

\def\wt{\widetilde}

\def\rta{\rightarrow}

\def\stop{\wt t}

\def\mstop{m_{\stop}}
\def\msquark{m_{\wt q}}

\def\sq{\wt q}

\def\slep{\wt \ell}

\def\hsm{\phi^0}
\def\mhsm{m_{\hsm}}
\def\hl{h^0}
\def\hh{H^0}
\def\ha{A^0}
\def\hp{H^+}
\def\hm{H^-}
\def\hpm{H^{\pm}}
\def\mhl{m_{\hl}}
\def\mhh{m_{\hh}}
\def\mha{m_{\ha}}

\def\mhpm{m_{\hpm}}
\def\tanb{\tan\beta}

\def\mt{m_t}
\def\mb{m_b}
\def\mc{m_c}
\def\mz{m_Z}
\def\mw{m_W}

\def\wp{W^+}
\def\wm{W^-}
\def\wpm{W^{\pm}}

\def\chitil{\wt\chi}
\def\cnone{\wt\chi^0_1}
\def\cntwo{\wt\chi^0_2}

\def\h{h}
\def\mh{m_{\h}}
\def\cpone{\wt \chi^+_1}
\def\cmone{\wt \chi^-_1}

\documentstyle[12pt,equations]{article}
\def\ie{{\it i.e.}}
\def\etal{{\it et al.}}
\def\9{\phantom 0}      
\renewcommand\linebreak{\unskip\break} 
\begin{document}
\input psfig.sty
\newlength{\captsize} \let\captsize=\small 
\newlength{\captwidth}                     

%
\font\fortssbx=cmssbx10 scaled \magstep2
\hbox to \hsize{
%
%
$\vcenter{
\hbox{\fortssbx University of California - Davis}
}$
\hfill
$\vcenter{
\hbox{\bf UCD-95-28} 
\hbox{\bf ILL-(TH)-95-28} 
\hbox{December 1995}
}$
}

%
\medskip
\begin{center}
\bf
WEAKLY-COUPLED HIGGS BOSONS\\
\rm
\vskip1pc
{\bf J.F. Gunion}\\
\medskip
{\em Davis Institute for High Energy Physics}\\
{\em University of California, Davis, CA 95616}\\
\vskip1pc
{\bf A. Stange}\\
\medskip
{\em Brookhaven National Laboratory}\\
{\em Upton, Long Island, NY 11973}\\
\vskip1pc
{\bf S. Willenbrock}\\
\medskip
{\em Department of Physics, University of Illinois}\\
{\em 1110 W. Green St., Urbana, IL 61801}\\
\end{center}
\vskip2pc

\begin{abstract}
We review the search for the standard Higgs boson, the Higgs bosons of the
supersymmetric standard model, and Higgs bosons from a variety of other models
at present and future colliders.
\end{abstract}

\section{Introduction}

\subsection{Standard Higgs model}

\indent\indent The evidence is overwhelming that the electroweak 
interaction is described by an SU(2)$_L\times$U(1)$_Y$ gauge theory, 
spontaneously broken to electromagnetism, U(1)$_{EM}$.  The symmetry 
breaking provides the masses of the $W$ and $Z$ bosons, as well as the 
masses of the fermions and the Cabibbo-Kobayashi-Maskawa (CKM) mixing between 
the quarks (including CP violation). Electroweak symmetry breaking is 
therefore associated with most of the aspects of the electroweak 
interaction which remain beyond our grasp.  Uncovering the 
electroweak-symmetry-breaking mechanism is essential for progress in our 
quest to describe nature at a deeper level.

Little is known about the mechanism which breaks the electroweak symmetry.
The fact that the relation $\mw = \mz \cos\theta_W$ is nearly satisfied
suggests that the symmetry-breaking sector possesses a global SU(2) symmetry,
often called a ``custodial'' symmetry \cite{S,W,SSVZ}.  
Precision electroweak experiments \cite{V,PT} and
flavor-changing-neutral-current processes \cite{GW} indirectly constrain 
the electroweak-symmetry-breaking sector.

\begin{figure}[htbp]
\let\normalsize=\captsize   
\begin{center}
\centerline{\psfig{file=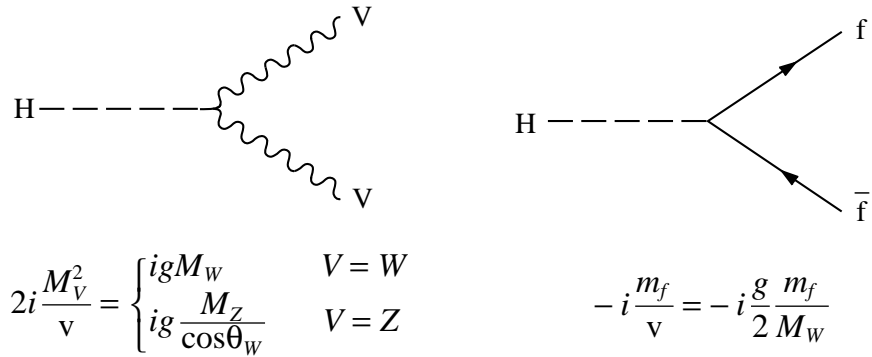,width=12.2cm}}
\begin{minipage}{12.5cm}       
\caption{Couplings of the standard Higgs boson to weak vector bosons and 
fermions. The coupling to weak vector bosons is multiplied by the metric 
tensor $g^{\mu\nu}$.}
\label{dpf1}
\end{minipage}
\end{center}
\end{figure}

The simplest model which is consistent with all constraints is the standard 
Higgs model, in which the symmetry is broken by a fundamental scalar field 
which acquires a non-zero vacuum-expectation value \cite{WS}.  The scalar 
field, $\phi$, is an SU(2)$_L$ doublet with hypercharge $+1/2$, and a potential
\begin{equation}
V(\phi)=-\mu^2|\phi|^2+\lambda|\phi|^4\;.
\label{POT}
\end{equation}
The minimum of the potential is at $|\phi|^2=\mu^2/2\lambda\equiv v^2/2$,
where $v=(\sqrt 2 G_F)^{-1/2}$. The sole
prediction of this model is the existence of a neutral scalar particle, the 
Higgs boson, $\hsm$, of unknown mass ($\mhsm^2=2\lambda v^2$), but with fixed 
couplings to other particles.  The couplings of the Higgs boson to the weak 
vector bosons and the fermions are shown in Fig.~\ref{dpf1}.

Despite the simplicity of the standard Higgs model, it does not appear to 
be a candidate for a fundamental theory.
The introduction of a fundamental scalar field is {\it ad hoc}; the 
other fields in the theory are spin-one gauge fields and spin-half fermion
fields.  Furthermore, the model does not explain why the scalar field
acquires a vacuum-expectation value, nor why it produces the curious
pattern of fermion masses and the CKM matrix. Thus the standard Higgs 
model accommodates, but does not explain, those features of the electroweak 
theory for which it is responsible.  

One response to these criticisms is to regard the 
standard Higgs model as an effective field theory, valid up to some energy
scale above the Higgs mass.  In fact, an ordinary scalar field theory 
{\it cannot} be 
valid up to arbitrarily high energies; it must eventually 
be subsumed by a more fundamental theory.  This is due to the fact that 
the running scalar-field self coupling, $\lambda$, increases with increasing 
energy, which is related to the 
``triviality'' of scalar field theory.  To a good approximation, the 
relation between the Higgs mass and the cutoff $\Lambda$ above which the 
theory becomes incomplete is \cite{MPP}
\begin{equation}
\mhsm^2 < \frac{4\pi^2v^2}{3\ln(\Lambda/\mhsm)}\;.
\end{equation}
If $\Lambda$ is taken to infinity, $\mhsm$ approaches zero, as does the 
Higgs-field self-interaction, $\lambda = \mhsm^2/2v^2$; the theory is 
``trivial''. However, we know for 
certain that new physics enters at the Planck scale, 
$(8\pi G_N)^{-1/2} \sim 10^{18}$ GeV; if we demand that the scalar field 
theory is valid up to this energy, the upper bound on the Higgs mass is about
200 GeV \cite{MPP}.  At the other extreme, if $\Lambda$ is only a few times 
$\mhsm$, the maximum value of the Higgs mass is about 700 GeV \cite{DN,HKNV}.  

A different argument leads to a {\it lower} bound on the Higgs mass for a given 
value of the top-quark mass \cite{MPP,AI}.  The top quark induces a $|\phi|^4$ 
interaction at one loop, which is added to the tree-level scalar potential,
Eq.~\ref{POT}.  This induced interaction enters with the opposite sign from
the tree-level $\lambda |\phi|^4$ interaction, and can destabilize
the potential at large values of $|\phi|$ if it becomes dominant.  The 
value of $|\phi|$ at which this occurs is interpreted as a scale 
$\Lambda$ at which the theory must be replaced by a more 
fundamental theory. CDF has observed the top quark at a mass 
$m_t=176 \pm 8 \pm 10$ GeV \cite{CDF}, and D0 has observed the top quark
at a mass $m_t=199 \pm 20 \pm 22$ GeV \cite{D0}.  These observations are 
consistent with indirect bounds from precision electroweak experiments 
\cite{MNPP}. 
We assume a (pole) mass of $m_t = 175$ GeV throughout this report.  
After including re-summed next-to-leading log corrections and imposing
the somewhat less stringent demand that the vacuum must be metastable,
it is found \cite{quirosi} that for 
$\Lambda \sim 10^{18}$ GeV the lower bound on the Higgs mass is about 
130 GeV; this bound is reduced to roughly 100 GeV for $m_t=160$ GeV.  
For $\Lambda =$ 1 TeV, the lower bound is about 70 GeV, 
close to the existing experimental lower bound.  

Taken together, the above two 
arguments imply (roughly) $100 \gev \le \mhsm \le 200 
\gev$, if the theory is valid up to the Planck scale.  However, another 
argument suggests that the scale, $\Lambda$, at which the 
standard Higgs model must be replaced by a more fundamental theory cannot 
be much larger than the weak scale itself.  This is because the 
vacuum-expectation value of the Higgs field, $v$, depends sensitively 
on the scale $\Lambda$ \cite{S}.  The relation between the bare 
vacuum-expectation 
value, $v_0$, and the renormalized vacuum-expectation value, $v=(\sqrt 2 
G_F)^{-1/2}=246$ GeV, at one loop is
\begin{equation}
v^2=v_0^2+{\cal O}(\Lambda^2)\,.
\label{QUAD}
\end{equation}
If $\Lambda$ is much larger than $v$, then $v_0$ must be adjusted 
precisely to yield the desired value of $v$.  Such a fine tuning is 
unnatural. 
We are led to conclude that the standard Higgs model is an acceptable 
effective field theory as long as it is subsumed by a more fundamental 
theory at an energy not much greater than $v=246$ GeV.  The vacuum-stability 
lower bound is then essentially eliminated, and the triviality bound is 
$\mhsm \lsim 700 \gev$.

It is possible that this fine-tuning problem is solved by a mechanism which
we have not yet imagined.  This point of view is supported by our failure to
find any realistic mechanism to yield a small value of the cosmological
constant, as required by observation.  This is an even more severe fine-tuning
problem than the vacuum-expectation value of the Higgs field.

\subsection{Remarks on general Higgs sector extensions}

\indent\indent The standard model (SM) Higgs sector is particularly
simple, but many possible extensions have been considered \cite{hhg}.
The most attractive possibility is to allow for additional Higgs 
doublet fields; for such an extension $\rho=1$ remains automatic at tree
level. A popular test case is the general two-Higgs-doublet
model (2HDM) \cite{HKS}.
If CP is conserved, then the two-doublet model yields five physical Higgs
CP-eigenstates: two CP-even neutral Higgs bosons, $\hl$ and $\hh$;
a CP-odd Higgs, $\ha$; and a charged Higgs boson pair,
$\hpm$. If the Higgs potential violates CP, then in general all the
neutral states mix, and there are simply three neutral
mass eigenstates of mixed CP character. One drawback of
the 2HDM extension (also shared by all other extensions) should be noted:
it is not {\it a priori} guaranteed that U(1)$_{EM}$ remains unbroken ---
parameters in the Higgs potential can be chosen such that $\mhpm^2<0$.
But requiring $\mhpm^2>0$ is also perfectly natural, and does not
require fine tuning of the parameters. However, the fine-tuning
and naturalness problems associated with one-loop quadratically
divergent contributions to Higgs boson squared masses remain.

An important parameter of a 2HDM is $\tanb=v_2/v_1$,
where $v_2$ ($v_1$) is the vacuum expectation value of the neutral member
of the Higgs doublet that couples to up-type (down-type) quarks. (We restrict
our discussion to the type-II 2HDM models in which up and down
quarks couple to different Higgs doublets; see \cite{hhg} for
a discussion of alternative coupling patterns.)
Assuming that CP is conserved, a second phenomenologically crucial
angle emerges --- $\alpha$, the mixing angle arising from the diagonalization
of the $2\times 2$ mass matrix for the CP-even Higgs sector.
In a general two-doublet model the masses of the Higgs bosons are all 
additional independent parameters, completely
independent of one another. Constraints on the 2HDM Higgs masses 
arise from the requirement that the Higgs sector contribution to $\Delta \rho$
be small (which implies limited mass splitting between any two
Higgs bosons that couple significantly to either the $Z$ or $W$ \cite{hhg})
and from the $b\rta s\gamma$ branching ratio (which places a
significant lower bound on $\mhpm$ \cite{hewett}).

The most direct indication of a two-Higgs-doublet sector for
the standard model would be the observation of more than a single
neutral Higgs boson or detection of the charged Higgs bosons. 
However, even if only a single neutral Higgs boson of the 2HDM is 
observed, in general it will have properties that are very distinct from
the $\hsm$. In the 2HDM, the couplings of any one of the three neutral Higgs 
bosons to  the various channels can be very different from the couplings of the
$\hsm$. Significant deviations from predictions for the $\hsm$ production
and decay rates could easily occur. If such deviations 
are seen, then a general 2HDM becomes a rather attractive first possibility
for the appropriate Higgs sector extension. If the
single neutral Higgs boson is found to have couplings to $b\anti b$,
$\taup\taum$ and $W\wstar$ that are within 10-20\% of the values expected for
the $\hsm$, this will strongly suggest that the only two-doublet extensions
that should be considered are ones in which the other Higgs bosons
of the two-doublet model have decoupled, leaving behind 
one neutral Higgs boson that will then
automatically have couplings that are close to SM values \cite{haberdecouple}. 
The minimal supersymmetric two-doublet extension, to be discussed shortly,
is the most attractive theory of this type.

Higher representations of SU(2)$_L$ can also be considered
in the context of the standard model electroweak gauge group, the next most
complicated being a Higgs triplet. If $Y=0$, the triplet field can be real;
if $Y\neq 0$, it is complex. As is well-known \cite{hhg}
$\rho=1$ is not automatic in such a case. Various possibilities for
obtaining $\rho=1$ at tree level in the presence of a Higgs triplet
can be entertained.  One possibility is that 
the neutral member of the Higgs triplet
has very small or zero vacuum-expectation value \cite{grifolsmendez}.
In this case, the triplet simply decouples from electroweak-symmetry-breaking
physics. If the triplet vacuum-expectation value is significant
there are still a number of ways to obtain $\rho=1$.
For example, it could be that there is fourth generation
with a large $t'-b'$ mass splitting. 
In this case the $t'-b'$ doublet yields a large
positive contribution to $\Delta\rho$ which could by cancelled by the
negative $\Delta \rho$ that would arise from a $T=1$, $|Y|=2$ complex
triplet representation. Obviously, this would require fine tuning the
triplet vacuum expectation value. A second possibility is to combine one
doublet Higgs field with one real $(T=1,Y=0)$ and one complex $(T=1,|Y|=2)$
field \cite{GM,CG,gvw,hhg}.
If the neutral members of the two triplet
representations have the same vacuum expectation value, then $\rho=1$ is
maintained at tree level.  In either case, $\rho=1$ is not maintained at
1-loop. Indeed, unlike the case of doublet models, $\rho$ is infinitely
renormalized in triplet models (due to the fact that the interactions of
the $B$ gauge field with the Higgs bosons violate custodial SU(2))
\cite{CG,gvw}.
Thus, fine-tuning would be required to maintain $\rho=1$ after
renormalization.  

Another issue of concern for triplet models is
grand unification of the gauge couplings.  This will be discussed
in more detail in the next section.  Here we simply make two observations.
First, in the non-supersymmetric
context gauge coupling unification (without intermediate scale physics)
occurs for a single doublet and a single $|Y|=2$ triplet. However,
a $Y=0$ triplet cannot also be present, implying that the $|Y|=2$ triplet must
have a very small vacuum-expectation value 
to achieve $\rho\sim 1$ at tree-level.  
Second, in the supersymmetric context
the presence of either type of triplet completely destroys 
gauge coupling unification.

If the gauge group is expanded beyond SU(2)$_L\times$U(1), 
many new possibilities for Higgs triplets arise. Left-right
symmetric SU(2)$_L\times$SU(2)$_R\times$U(1)$_{B-L}$ models
are an attractive possibility \cite{lrmodels}.  The Higgs
sector of the left-right symmetric models has received
much attention \cite{leftright,hhg}. Typically, the Higgs sector
contains two Higgs doublet fields (in a bi-doublet) and two
separate sets of Higgs triplet fields, one connecting to the
usual $W_L$ sector and the other responsible for giving
mass to the right-handed $W_R$. Despite the presence of Higgs triplets,
gauge coupling unification
is possible in this context for an appropriate choice of the Higgs sector
and intermediate mass scale matter fields \cite{lrunification}.
However, the naturalness problem for $\rho$ is not solved. 
The phenomenology of the left-right Higgs sector is quite intricate,
and will not be detailed here.

An especially interesting feature of Higgs triplet models
is the fact that a triplet field can have lepton-number
violating couplings. In left-right symmetric models, such
couplings can lead in a natural way to the see-saw mechanism for
generating neutrino masses; certain aspects of the phenomenology of the
left-right Higgs sector become closely tied to neutrino physics.

Because of the difficulties with $\rho$ and the
need for intermediate scale physics in order to have grand unification,
models in which Higgs triplets play a substantial
role in electroweak symmetry breaking are generally not in
favor with theorists.  
However, this should not deter the experimental
community from searching for the many new signatures that would arise.
A simple and spectacular signal for a triplet model is direct
detection of the doubly charged Higgs boson(s)
contained in complex Higgs triplet representation(s). 
At an $\epem$ collider, $\epem\rta H^{++}H^{--}$ production 
via virtual $Z$ or photon exchange only
requires adequate machine energy, $\sqrt s\gsim 2m_{H^{++}}$.
$H^{++}H^{--}$ pair production is also possible at a hadron collider,
and yields substantial rates at the LHC for $m_{H^{++}}\lsim 500\gev$.
Assuming a significant vacuum expectation value in the 
triplet sector, a hadron collider of adequate energy
can produce the $H^{--}$ and $H^{++}$ at an observable rate
via $\wm\wm$ and $\wp\wp$ fusion  \cite{gvw}, and
at an $\emem$ collider $H^{--}$ production via $\wm\wm$
fusion is an exciting prospect \cite{ememgunion}.
In addition, there is the possibility of direct $s$-channel
production, $e^-e^-\to H^{--}$ \cite{ememtodelmm} --- 
this process is a direct probe of the very interesting
lepton-number-violating couplings to the Higgs triplet field and would
be observable for an $\emem\to H^{--}$ coupling strength squared
that is some nine order of magnitude smaller than current limits.
(See Ref.~\cite{mendezrecent} for a recent survey of limits.)
Billions of $H^{--}$'s would be produced each year if the
coupling strength were close to current limits.
Since limits on the $\mu^-\mu^-\to H^{--}$
coupling are even weaker, a $\mu^-\mu^-$ collider might yield
even larger numbers of $H^{--}$'s.

\subsection{Minimal supersymmetric standard model}

\indent\indent The partial success of SU(5) grand unification \cite{GG}
encourages one to consider a
scenario in which there is no new physics between the weak scale and the
grand-unified (GUT) scale, $M_U \sim 10^{16}$ GeV.  This is based on the 
observation that each generation of fermions fits into the $\anti 5 + 10$
representation of SU(5), and that the SU(3)$_c$, SU(2)$_L$, and U(1)$_Y$
couplings nearly converge (when normalized according to SU(5))
when extrapolated up to $M_U$.  The fact that the grand-unified scale and
the Planck scale are relatively close provides further encouragement.

One is thus led to search for new physics at the weak scale which improves
upon the partial success of the SU(5) model, and which allows the weak 
scale to be much less than the grand-unified scale. Supersymmetry (SUSY) 
is the unique weakly-coupled theory which eliminates 
the quadratic renormalization of the Higgs-field vacuum-expectation 
value, Eq.~\ref{QUAD}, to all orders in perturbation theory,
and thus allows $v=246$ GeV to occur naturally, if the scale of supersymmetry 
breaking is comparable to the weak scale \cite{QUAD}.  Further,
in the minimal supersymmetric model,
the supersymmetric partners of the gauge bosons and Higgs fields
affect the relative evolution of the gauge couplings \cite{DRW} in
a favorable way. (The supersymmetric partners of the fermions
do not influence the relative evolution at one loop.)
There are {\it exactly} two Higgs doublet fields 
in the minimal supersymmetric model.
At least two Higgs doublet fields are necessary to
provide masses for both down-type and up-type quarks; in addition, 
only for an even number of Higgs doublets do the
gauge anomalies cancel amongst the Higgs superpartners.  
The relative evolution of the gauge couplings in the presence
of multiple doublets and their supersymmetric partners \cite{TWOHIGGS}
is such that for the minimal two-Higgs-doublet sector
the net effect is precisely that desired; if all
particles and their superpartners have masses
in the TeV range, the gauge couplings merge rather precisely at a scale 
of about $10^{16}$ GeV \cite{GUT}. This is a non-trivial result, and is not
easy to reproduce in models of dynamical electroweak symmetry breaking 
\cite{LW}. 

Although the addition of still more pairs of Higgs doublets
and associated superpartners preserves the anomaly cancellation,
gauge coupling unification is destroyed. As noted earlier,
Higgs bosons (and superpartners) in triplet (and higher) representations
also destroy the gauge unification. Only Higgs singlet 
representations can be appended to the minimal two doublets while
preserving coupling constant unification. Thus, the minimal supersymmetric
model with exactly two Higgs doublets is an especially attractive
theory.

The imposition of supersymmetry and the addition of a second Higgs doublet
have far-reaching consequences for the electroweak-symmetry-breaking mechanism. 
The potential for the Higgs doublets, $\phi_1$ and $\phi_2$, including soft
supersymmetry-breaking terms, is 
\begin{eqnarray}
V(\phi_1,\phi_2) &=& m_1^2 |\phi_1|^2 +m_2^2 |\phi_2|^2 - 
m_{12}^2(\epsilon_{ij}\phi_1^i\phi_2^j + h.c.) \nonumber \\
& & + \frac{1}{8}(g^2+g^{\prime 
2})(|\phi_1|^2-|\phi_2|^2)^2+\frac{1}{2}g^2|\phi_1^*\phi_2|^2 
\label{SUSYPOT}
\end{eqnarray}
where $g$ and $g^{\prime}$ are the SU(2)$_L$ and U(1)$_Y$ couplings. 
The most important feature of this potential is that,
unlike in the standard model,
the magnitudes of the quartic potential terms are not arbitrary;
they are constrained by the supersymmetry 
to be of magnitude $g^2,g^{\prime\,2}$.
This has the consequence that triviality and vacuum stability are not issues;
the mass of any physical Higgs boson that is SM-like,
in that it has significant $WW$ and $ZZ$ couplings, is strictly limited, 
as are the radiative corrections to the quartic potential terms.
Electroweak symmetry breaking occurs if the origin in Higgs-field space is a 
saddle point, which requires $S\equiv m_1^2m_2^2-m_{12}^4 < 0$.  In the limit of
exact supersymmetry, one has $m_1 = m_2 = \mu$ (where $\mu$ is the coefficient 
of the term $-\mu \epsilon_{ij}\hat H_1^i \hat H_2^j$ in the superpotential; 
$\hat H_1$ and $\hat H_2$ are the Higgs superfields)
and $m_{12}=0$, so electroweak
symmetry breaking does not occur.  Thus the Higgs-field vacuum-expectation
values, $v_1,v_2$, are naturally of the same size as the supersymmetry-breaking 
scale.  Supersymmetry also makes the presence of fundamental scalar fields 
natural, although the Higgs fields are not the superpartners of any of 
the known fermions.

In supergravity grand-unified models, electroweak symmetry breaking 
arises ``radiatively'', that is by evolving the parameters of the Higgs sector 
(including the soft SUSY-breaking parameters) from the GUT
scale down to the weak scale via the renormalization group equations 
\cite{RAD}.  
Due to the large mass of the top quark, the parameter $m_2^2$ is driven 
more rapidly towards zero (often to negative values)
than are $m_1^2$ and $m_{12}^2$, so that $S$
is driven negative at the weak scale, triggering electroweak symmetry breaking.
Thus, in these models, electroweak symmetry breaking is explained by the 
large top-quark mass (more precisely, by the large top-quark Yukawa coupling). 
Supergravity models can also explain
why the scale of supersymmetry breaking ({\it i.e.}, the scale of the 
soft-supersymmetry-breaking terms) is 
comparable to the weak scale, rather than the GUT scale; this is a natural
result if supersymmetry breaking occurs via a ``hidden'' sector.  

As in the general 2HDM,
the presence of two Higgs-doublet fields implies that the spectrum of Higgs 
particles is much richer in the minimal supersymmetric model than in the 
standard model. However, the two-doublet MSSM is a highly constrained
version of the general 2HDM. First,  the two-doublet
MSSM Higgs couplings are automatically type-II, with one Higgs doublet ($H_1$)
coupling only to down-type quarks and leptons,
and the other ($H_2$) coupling only to up-type quarks.
Secondly, in the MSSM, CP conservation in the Higgs sector is
automatic (for a review and references, see \cite{hhg}), and we find
the previously-mentioned five physical Higgs 
particles: two $CP$-even neutral scalars, $\hl$ and $\hh$; a $CP$-odd neutral 
scalar (often called a pseudoscalar), $\ha$; and a pair of charged Higgs 
bosons, $\hpm$. 
Further, due to the special form of the potential, Eq.~\ref{SUSYPOT}, 
dictated by softly-broken supersymmetry, the
Higgs sector is described (at tree level) by just two free parameters, 
only one more than the standard Higgs model. It is conventional to choose 
the mass of the pseudoscalar Higgs boson, $\mha$, and 
the ratio of the Higgs-field vacuum-expectation values, $\tanb\equiv 
v_2/v_1$, as the free parameters. Other parameters of 
the model affect the Higgs sector after including loop
corrections \cite{haberperspectives}; the most important of 
these are the top-quark and the stop-squark masses, with parameters
that influence stop-squark mixing also playing a significant role.

The dominant radiative corrections arise from an incomplete cancellation
of the virtual top-quark and stop-squark loops.  The two stop-squark
masses ($\mstopi$ and $\mstopii$) are obtained by diagonalizing
a $2\times 2$ stop-squark squared-mass matrix; the off-diagonal
elements of this matrix involve the parameters $\mu$ and $A_t$,
where $\mu$ is the coefficient 
of the term $-\mu \epsilon_{ij}\hat H_1^i \hat H_2^j$ in the superpotential
($\hat H_1$ and $\hat H_2$ are the Higgs superfields),
and $A_t$ is the coefficient appearing in the soft-supersymmetry-breaking
potential term $A_th_t \epsilon_{ij} H_2^j \wtilde Q^i \wtilde U$ where,
in the case
of the stop squarks, $\wtilde Q=\pmatrix{\wtilde t_L \cr \wtilde b_L}$
and $\wtilde U= \wtilde t_R^*$. Here,
$h_t$ is the top-quark Yukawa coupling, $h_t=g\mt/(\sqrt 2\mw)$.

We first give a summary of the corrections to the Higgs sector
at one-loop, and then discuss the most important two-loop effects
and renormalization group improvement of the one-loop terms.
The leading effects (\ie\ those proportional
to $\mt^4$, neglecting terms\footnote{The subdominant 
terms of ${\cal O}(\mw^2\mt^2)$, 
not shown explicitly in the following equations, can
be found in Ref.~\cite{habertwoloop}.} 
of order $\mw^2\mt^2$
and $\mb^4\tan^2\beta$)
at one-loop on the Higgs sector can be expressed \cite{erz} in terms of 
\begin{eqnarray}
\dmsq_{11} & \equiv & 
\frac{3g^2}{16\pi^2\mw^2}m_t^4 {\mu^2 X_t^2}g(\mstopi^2,\mstopii^2) \nonumber \\
\dmsq_{12} & \equiv & 
\frac{3g^2}{16\pi^2\mw^2}m_t^4 \mu X_t \left\{h(\mstopi^2,\mstopii^2)
+A_tX_t g(\mstopi^2,\mstopii^2) \right\} \nonumber \\
\dmsq_{22} & \equiv & \frac{3g^2}{16\pi^2\mw^2}m_t^4
\left\{2\ln\left(\frac{\mstopi\mstopii}{m_t^2}\right)
    +A_tX_t\left[2h(\mstopi^2,\mstopii^2)+
            A_tX_tg(\mstopi^2,\mstopii^2)\right]
     \right\} \;,\nonumber \\
 & & 
\label{dmsqdefs}
\end{eqnarray}
where $X_t\equiv A_t-\mu\cot\beta$.
The functions $h(a,b)$ and $g(a,b)$ are defined by
\begin{equation}
h(a,b)\equiv {1\over a-b}\ln\left({a\over b}\right)\,,\qquad
g(a,b)\equiv {1\over (a-b)^2} \left[ 2-{a+b\over a-b}\ln\left({a\over
b}\right)\right]\,.
\end{equation}
Note that in the $X_t=0$ limit, where squark mixing is negligible,
only $\dmsq_{22}$ is non-zero, with its magnitude being
determined by $\ln(\mstopi\mstopii/\mt^2)$, which vanishes in
the supersymmetric limit of $\mstopi=\mstopii=\mt$.

For $\mha\sim {\cal O}(\mz)$, the squared-masses of the CP-even 
neutral scalar Higgs bosons are obtained by diagonalizing the matrix:
\begin{equation}
{\cal M}^2\equiv
\pmatrix{\mha^2\sin^2\beta+\mz^2\cos^2\beta+{\dmsq_{11}\over\sin^2\beta} &
         -(\mha^2+\mz^2)\sin\beta\cos\beta-{\dmsq_{12}\over\sin^2\beta} \cr
         -(\mha^2+\mz^2)\sin\beta\cos\beta-{\dmsq_{12}\over\sin^2\beta} &
         \mha^2\cos^2\beta+\mz^2\sin^2\beta+{\dmsq_{22}\over\sin^2\beta} \cr} 
\label{msqmatrix}
\end{equation}
yielding
\begin{eqnarray}
m_{\hl,\hh}^2 & = & \frac{1}{2}[\mha^2+\mz^2+(\dmsq_{11}+\dmsq_{22})/\sin^2\beta] 
\nonumber \\
& & \pm\frac{1}{2}\biggl\{[(\mha^2-\mz^2)\cos 
2\beta+(\dmsq_{22}-\dmsq_{11})/\sin^2\beta]^2       \nonumber \\
& & \phantom{\pm\frac{1}{2}\biggl\{[}
+[(\mha^2+\mz^2)\sin 2\beta+\dmsq_{12}/\sin^2\beta]^2\biggr\}^{1/2}
\label{SCALARMASS}
\end{eqnarray}
where $\mhl<\mhh$ by definition. Note that the neutral Higgs bosons satisfy 
the mass sum rule
\begin{equation}
\mhl^2+\mhh^2=\mha^2+\mz^2+(\dmsq_{11}+\dmsq_{22})/\sin^2\beta\;.
\end{equation} 

For $\mha\gg\mz$, one of the two Higgs doublets decouples, leaving
behind one light physical Higgs state with couplings identical to
those of the SM Higgs boson.
In this limit, $\mhl$ is obtained simply as the large-$\mha$
limit of Eq.~\ref{SCALARMASS}, 
\begin{eqnarray}
\mhl^2 & \simeq  & \mz^2\cos^22\beta+
[\dmsq_{22}+\dmsq_{11}\cot^2\beta-2\dmsq_{12}\cot\beta] \nonumber \\
        & = & \mz^2\cos^22\beta+ {3g^2\mt^4\over 8\pi^2\mw^2}\ln \left(
{\mstopi\mstopii\over\mt^2}\right) \nonumber\\
 & & +{3 g^2\mt^4 X_t^2\over 16\pi^2\mw^2} \left[ 2h(\mstopi^2,\mstopii^2)
+X_t^2g(\mstopi^2,\mstopii^2)\right]\,.
\label{HBOUND}
\end{eqnarray}
Eq.~\ref{HBOUND} sets an upper bound on $\mhl$
which is critical to ensuing discussions.
Without the inclusion of the one-loop correction, the bound would be
$\mhl < \mz$, and the lightest scalar Higgs boson would have been unable to 
escape detection at \lepii. After including the one-loop correction,
$\mhl$ can be substantially larger than $\mz$.
For $\mha\gg\mz$, the mass of the heavier $\hh$
cannot be obtained as the large $\mha$ limit of Eq.~\ref{SCALARMASS}
since radiative corrections must be computed at the large
scale ${\cal O}(\mha)$. Instead, one obtains the result
\begin{equation}
\mhh^2  \simeq  \mha^2+\mz^2\sin^22\beta+
{3 g^2\mt^4 X_tY_t \cot^2\beta\over 16 \pi^2\mw^2} 
\left[ 2h(\mstopi^2,\mstopii^2)+X_tY_tg(\mstopi^2,\mstopii^2)\right]\,,
\end{equation}
where $Y_t=A_t+\mu\tanb$.
The corrections proportional to $\mt^4$ are
absent for $X_t=0$, in which case the leading corrections
are ${\cal O}(\mw^2\mt^2)$.

The lower bound on $\mhh$ for any given $\tanb$ occurs for $\mha=0$.
An approximate form for the lower bound on $\mhh$, obtained
by evaluating Eq.~\ref{SCALARMASS} at $\mha=0$
and neglecting terms proportional
to $[\dmsq_{ij}]^2$ relative to $\mz^2\dmsq_{ij}$, is
\begin{equation}
\mhh^2\geq \mz^2+[\dmsq_{22}+\dmsq_{11}\cot^2\beta]\,.
\label{mhhlowerbound}
\end{equation}

One-loop corrections proportional to $\mt^4$
to the charged Higgs boson mass
arise only if there is mixing. One finds
\begin{equation}
\mhpm^2 \simeq \mha^2+\mw^2+{\dmsq_{\pm}\over\sin^2\beta}\,,
\label{CHARGEDMASS}
\end{equation}
with
\begin{equation}
\dmsq_{\pm}={3g^2\over32\pi^2\mw^2}\mt^4\mu^2 f(\mstopi^2,\mstopii^2)\,,
\end{equation}
where
\begin{equation}
f(a,b)\equiv {-1\over a-b}\left[ 1 -{b\over a-b}\ln\left({a\over
b}\right)\right]\,.
\label{fdef}
\end{equation} 
There are also contributions to $\dmsq_{\pm}$ that 
are of order $\mt^2\mb^2\tan^2\beta$ and $\mb^4\tan^4\beta$, 
which could be as important
as the $\mt^4$ terms if $\tanb\sim\mt/\mb$. For small $\mu$, 
Eq.~\ref{CHARGEDMASS}  implies that the
charged Higgs boson has a lower bound of 
\begin{equation}
\mhpm^2>\mw^2\;.
\end{equation}

The scalar Higgs boson mass eigenstates discussed
above are mixtures of the neutral components of the real 
parts of the Higgs-doublet fields.
In the case of the MSSM, 
the mixing angle, $\alpha$, rather than being a free parameter
as in the general two-Higgs-doublet model,
is determined by $\mha$, $\tanb$ and (at one loop) the $\dmsq_{ij}$:
\begin{eqnarray}
\sin 2\alpha &=& {2{\cal M}^2_{12}\over \sqrt{({\cal M}^2_{11}-{\cal
M}^2_{22})^2 + 4 {\cal M}^4_{12}}} \nonumber \\
\cos 2\alpha &=& {{\cal M}^2_{11} - {\cal M}^2_{22} \over
\sqrt{({\cal M}^2_{11}-{\cal M}^2_{22})^2 + 4 {\cal M}^4_{12}}} 
\label{2alphageneral}
\end{eqnarray}
leading to, for example,
\begin{equation}
\tan 2\alpha
 = \frac{(\mha^2+\mz^2)\sin 2\beta +2\dmsq_{12}/\sin^2\beta}
{(\mha^2-\mz^2)\cos 2\beta +(\dmsq_{22}-\dmsq_{11})/\sin^2 \beta}\,.
\label{tan2alpha}
\end{equation}
In our convention $0 < \beta < \pi/2$. The result for $\sin2\alpha$ in
Eq.~\ref{2alphageneral} then implies that $-\pi/2<\alpha<0$ ($+\pi/2>\alpha>0$)
for ${\cal M}^2_{12}<0$ ($>0$). Referring to Eq.~\ref{msqmatrix}, we see that
positive $\alpha$ is only possible if
$\beta$ is near 0 or $\pi/2$ and $\dmsq_{12}<0$.  The ensuing discussion
assumes that $\alpha<0$, as
holds if $\dmsq_{12}$ is small compared to
$(\mha^2+\mz^2)\cos\beta\sin\beta$ and/or positive. Results for the alternative
case are easily worked out.
  
There is an important transition that occurs at
${\cal M}_{11}^2-{\cal M}_{22}^2=0$, \ie\
$\mha=\sqrt{\mz^2-(\dmsq_{22}-\dmsq_{11})/(\cos2\beta\sin^2\beta})$.
For $\mha$ values below this, $\cos2\alpha$ is positive, implying
that $-\pi/4<\alpha< 0$. For larger $\mha$, $-\pi/2<\alpha <-\pi/4$.
In particular, for $\tanb\rta\infty$ (\ie\ $\beta\rta \pi/2$)
and neglecting $\dmsq_{12}$ in ${\cal M}_{12}^2$, 
if $\mha>\sqrt{\mz^2+(\dmsq_{22}-\dmsq_{11})}$
then $\alpha\simeq r(\beta-\pi/2)\to 0$, whereas for
$\mha<\sqrt{\mz^2+(\dmsq_{22}-\dmsq_{11})}$
$\alpha\simeq -\pi/2+r(\pi/2-\beta)\to -\pi/2$ when $\beta\rta\pi/2$,
where 
\begin{equation}
r\equiv{ \mha^2+\mz^2 \over \left|\mha^2-(\mz^2+\dmsq_{22}-\dmsq_{11})\right|}
\, .
\label{rform}
\end{equation}
Note that $r=1$ for
$\mha^2\to\infty$; $r=1$ is also obtained at $\mha^2=0$ if the 
one-loop radiative corrections are small compared to $\mz^2$. 
These different behaviors of $\alpha$
will have important implications for the couplings
of the $\hl$ and $\hh$ in the large $\tanb$ limit, as will be
discussed in more detail shortly. 

Two-loop corrections to the Higgs sector are also significant and can reduce
the one-loop addition to the upper bound on $\mhl$
\cite{habertwoloop,carenatwoloop}. Results
for the leading [${\cal O}(g^2\alpha_s\mt^4)$ and 
${\cal O}(g^4\mt^6)$]
two-loop corrections have been computed in the large $\tanb$
limit \cite{habertwoloop}. 
The most important two-loop effects can be incorporated by
replacing $\mt$ in the formulae above by the running top quark mass
evaluated at $\mt$:  
\begin{equation}
\mt(\mt)= \mt \left[1-{4\over 3\pi}\alpha_s+{1\over 2\pi}\alpha_t\right]
\approx 0.966\mt\,,
\label{mtrun}
\end{equation}
where $\alpha_t=h_t^2/(4\pi)$ and
we maintain the notation $\mt$ for the pole mass.
For $\mt=175\gev$, $\mt(\mt)$ is roughly 169 GeV.
Due to the leading $\mt^4$ behavior of the one-loop
corrections, the replacement of $\mt$ by $\mt(\mt)$ in the one-loop
equations is numerically important, leading, as noted above, to a significant
reduction in the upper bound on $\mhl$.

If the scale of supersymmetry breaking is much larger than $\mz$,
then large logarithmic terms arise in the perturbative
expansion (as seen, for example,
in Eq.~\ref{HBOUND}).  These large logarithms can be re-summed to all orders
in perturbation theory using renormalization
group equations (RGE's). The formula for the full one-loop
radiatively corrected Higgs mass is very complicated \cite{cpr}.
Moreover, the computation of the RGE-improved one-loop corrections
requires numerical integration of a coupled set of RGE's \cite{hhrge}.
(The dominant two-loop next-to-leading logarithmic results
are also known \cite{hemphoang}). Although this program has
been carried out in the literature, the procedure is
unwieldy and not easily amenable to large-scale Monte-Carlo analyses.
Fortunately, very accurate
and simple analytic formulae can be developed that incorporate
the dominant effects of the RGE improvements. In particular,
many of these effects can be included simply by a correct
choice of the scale at which to evaluate the running top-quark
mass in various pieces of the one-loop formulae.
The method can be easily implemented,
and incorporates both the leading one-loop and two-loop effects
and the RGE improvement.  Although the results are conceptually
simple, complications arise when supersymmetric thresholds are fully
taken into account.  The details can be found in Ref.~\cite{habertwoloop},
along with other references to the original literature.
Complementary work can be found in Ref.~\cite{carenatwoloop}.

For the results presented in this review, we employ
a numerical program based on the work
of Ref.~\cite{habertwoloop}, combined with numerical routines
developed for the work of Refs.~\cite{gunionperspectives,gunionerice}.
When referring to these results, we shall use the phrase `\twoloop'
to indicate that the corrections to the Higgs 
boson masses and the mixing angle $\alpha$ have included leading-logarithmic
re-summation as well as the one-loop and two-loop corrections.

At the \twoloop\ level, 
\ie\ after including one- and two-loop radiative corrections
and leading-log re-summation via the RGE's,
the lower bound on $\mhl$ is significantly above $\mz$;
for $m_t=150,175,200$ GeV, if squark mixing
is absent and $\mstopi=\mstopii\equiv\mstop= 1$ TeV, the bound 
(reached for $\tanb\gg 1$ and large $\mha$) is of order
$\mhl < 102,113,127$ GeV, respectively. [These upper bounds
can be compared to the values
$\mhl < 108,122,140$ (106,118,134) GeV obtained from the one-loop 
formulae using the pole (running $\mt(\mt)$) top-quark mass
in Eq.~\ref{SCALARMASS}.]
The lower bounds on $\mhh$ (attained for high $\tanb$ and $\mha\to 0$) are 
essentially the same as the upper bounds on $\mhl$ ---
roughly $102,113,127\gev$ for the above top masses.

The large $\mha$ upper bound on $\mhl$ is
less for smaller values of $\tanb$ and/or smaller $\mstop$.
For example, if $\tanb$ is near 1 then
the mass of the $\hl$ typically arises primarily from loop corrections
to the tree-level mass, with the result that $\mhl\lsim 100\gev$
is quite probable. More specifically, for $\tanb=1.5$, $\mstop=1\tev$
and large $\mha$ 
one finds (at the `\twoloop' level)
$\mhl=60,78,98\gev$ for $\mt=150,175,200\gev$, respectively.

\begin{figure}[htbp]
\let\normalsize=\captsize   
\begin{center}
\centerline{\psfig{file=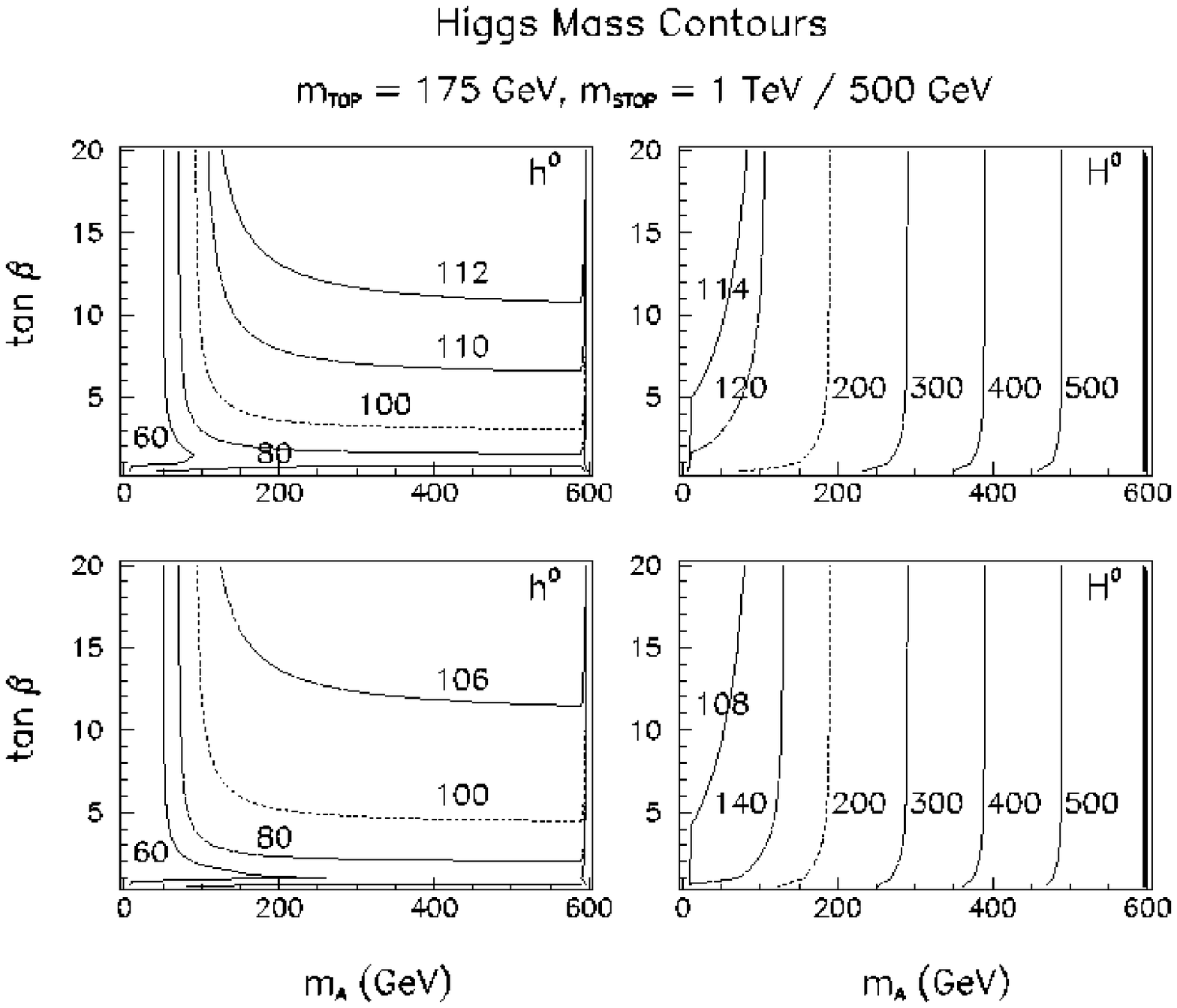,width=12.2cm}}
\begin{minipage}{12.5cm}       
\caption{Contours for the $\hl$ and $\hh$ masses
in $(\mha,\tanb)$ parameter space. Results are given for $\mt=175\gev$,
with $\mstop=1\tev$ (upper plots) and $\mstop=500\gev$ (lower plots).
The masses are computed including \twoloop\ radiative corrections, neglecting
squark mixing.}
\label{rcmasses}
\end{minipage}
\end{center}
\end{figure}

As further illustration,
Fig.~\ref{rcmasses} gives the mass contours in $(\mha,\tanb)$
parameter space for the $\hl$ and
$\hh$ in the case of $\mt=175\gev$ and for degenerate
stop-squark masses of $\mstop=1\tev$ (upper plots) and $500\gev$ (lower plots),
after including \twoloop\ corrections. Note that for the lower
stop mass the radiative corrections are substantially reduced.
For $\mstop\sim 300\gev$, the upper limit of $\mhl$ is $\sim 100\gev$.
Fig.~\ref{rcmasses} not only illustrates the upper
bound on $\mhl$ at large $\mha$ and $\tanb$, but also demonstrates 
the rapid decline of $\mhl$ from its upper bound
and approach of $\mhh$ to its lower bound
in the (as we shall see, unlikely) case that
$\mha<\sqrt{\mz^2+(\dmsq_{22}-\dmsq_{11})}$.
Below, we shall see that when
$\mha>\sqrt{\mz^2+(\dmsq_{22}-\dmsq_{11})}$
the $\hl$ has roughly SM-like couplings while the $\hh$ decouples
from $WW,ZZ$, whereas for lower $\mha$
the $\hh$ has roughly SM-like couplings (squared), and the $\hl$
decouples from $WW,ZZ$.

As we have already seen,
the above results are altered in the presence of significant
squark mixing because of large values for the supersymmetry parameters
$A_t$ and $\mu$, defined earlier.
If $A_t$ and/or $\mu$ are large, then the upper bound on
$\mhl$ increases due to an increase
in the \twoloop\ corrections, which grow
with larger mixing in the stop squark sector.
(At very large $\tanb$, mixing in the sbottom squark sector
can also play a role. We assume that $A_b=A_t$ and denote the
common value by $A$.)
In what follows, in addition to the `no-mixing' scenario
for which we have given specific numerical results above,
we also consider the `typical-mixing' scenario with
$A=-\mu=\msusy$ and `maximal-mixing' scenario with $A=\sqrt 6\msusy$,
$\mu=0$. $\msusy$ is to be identified with the
value of $\mstop$ in the no-mixing scenario. In the maximal-mixing
case, the upper bound on $\mhl$ for $\msusy=1\tev$
at large $\tanb$ and large $\mha$ is $\sim 125\gev$ ($\sim 150\gev$) 
for $\mt=175\gev$ ($200\gev$).

Placing the
MSSM in the context of unified supergravity or string boundary conditions 
leads to substantial prejudice as to what the relevant parameters
are \cite{baerreport,gunionpois}. 
Such models are attractive, not only
in that gauge coupling unification at a high scale $M_X$ is successful, 
but also in that proper electroweak symmetry breaking at scales below a TeV
is more or less automatic. Minimal constraints on the
models from the requirements of  
correct $Z$ mass, uncharged LSP, perturbative Yukawa couplings
and no SUSY particles (such as charginos or sneutrinos) below $\mz/2$
are already sufficient to imply significant constraints on the 
$\mz$-scale Higgs sector parameters.  Typically, there are few or no 
$M_X$-scale parameter choices that yield values of $\mha$ below about $2\mz$.
Additional theoretical prejudices can yield very strong additional
constraints.  For example, unification of the Yukawa couplings
of the $b$ and $\tau$ at $M_X$ is only possible if $\tanb$ is very
near 1 or else very large ($\gsim 30$) \cite{bargerhawaii}.  
As another example,
dilatonic or no-scale supergravity style $M_X$-scale boundary conditions often
lead to modest values for the squark masses (especially for the lightest
stop squark) \cite{bgkp}.

The neutral Higgs sector angles  
$\alpha$ (see Eq.~\ref{tan2alpha}) and 
$\beta$ figure prominently in the couplings of the SUSY Higgs bosons
to weak vector bosons and fermions.  We list these couplings in 
Table~\ref{hcouplings}; these factors
multiply the corresponding standard-model coupling in Fig.~\ref{dpf1}.  The 
neutral scalar Higgs bosons, $\hl$ and $\hh$, couple to the weak bosons 
proportional to $\sin(\beta-\alpha)$ and $\cos(\beta-\alpha)$, 
respectively; they share the squared-coupling of the standard 
Higgs boson to the weak bosons. Similarly, $WW,ZZ$ and $ZA$ share 
the squared-coupling to a given scalar Higgs boson. Note 
that there are no tree-level $W^\mp ZH^\pm$, $AWW$, nor $AZZ$ couplings.

\begin{table}[hbt]
\caption[fake]{Couplings of the neutral Higgs bosons of the minimal 
supersymmetric model.  These multiply the couplings of the standard Higgs boson
given in Fig.~\ref{dpf1}.}
\begin{center}
\begin{tabular}{c|cccc}
& $WW,ZZ$ & $Z\ha$ & $t\anti t$ & $b\anti b$, $\tau^+\tau^-$ \\ \hline
$\hl$& $\sin(\beta-\alpha)$ & $\cos(\beta-\alpha)$ & $\cos\alpha/\sin\beta$ & 
$-\sin\alpha/\cos\beta$ \\
$\hh$ & $\cos(\beta-\alpha)$  & $\sin(\beta-\alpha)$ & $\sin\alpha/\sin\beta$ &
$\cos\alpha/\cos\beta$ \\
$\ha$ & 0 & 0 & $\gamma_5 \cot\beta$ & $\gamma_5 \tan\beta$ \\
\end{tabular}
\end{center}
\label{hcouplings}
\end{table}

\begin{table}[hbt]
\caption[fake]{Couplings of the $\hl$ and $\hh$ 
in the limit of $\alpha\rta r(\beta-\pi/2)$,
applicable for: i) $\mha\rta\infty$, for which $r\to 1$; 
ii) $\tanb\rta\infty$ (\ie\ $\beta\to\pi/2$), 
$\mha>\protect\sqrt{\mz^2+\dmsq_{22}-\dmsq_{11}}$, and $\dmsq_{12}\sim 0$,
with $r$ as defined in Eq.~\ref{rform}.}
\begin{center}
\begin{tabular}{c|cccc}
& $WW,ZZ$ & $Z\ha$ & $t\anti t$ & $b\anti b$, $\tau^+\tau^-$ \\ \hline
$\hl$& $1$ & $0$ & $1$ & $r$ \\
$\hh$ & $0$  & $1$ & $-r\cot\beta$ & $\tan\beta$ \\
\end{tabular}
\end{center}
\label{hcouplingslimi}
\end{table}

\begin{table}[hbt]
\caption[fake]{Couplings of the $\hl$ and $\hh$ 
in the limit of $\alpha\to -\pi/2+r(\pi/2-\beta)$,
applicable for $\tanb\rta\infty$ (\ie\ $\beta\to \pi/2$),
$\mha<\protect\sqrt{\mz^2+\dmsq_{22}-\dmsq_{11}}$, and $\dmsq_{12}\sim 0$,
with $r$ as defined in Eq.~\ref{rform}.}
\begin{center}
\begin{tabular}{c|cccc}
& $WW,ZZ$ & $Z\ha$ & $t\anti t$ & $b\anti b$, $\tau^+\tau^-$ \\ \hline
$\hl$& $0$ & $-1$ & $r\cot\beta$ & $\tan\beta$ \\
$\hh$ & $-1$  & $0$ & $-1$ & $r$ \\
\end{tabular}
\end{center}
\label{hcouplingslimii}
\end{table}

There are three limits which are of particular interest:

\begin{enumerate}

\item $\mha \to \infty$: 
In this limit, the mass of $\hl$ is driven to its upper bound, given
at one-loop by Eq.~\ref{HBOUND}. In addition, $\alpha \to \beta -\pi/2$.
The limits of the couplings are given in Table~\ref{hcouplingslimi},
with $r=1$. The  couplings of $\hl$ become identical 
to those of the standard Higgs 
boson. The $\hh$ decouples from the weak bosons.
The squares of the fermionic couplings of $\hh$ and $\ha$ become identical. 
The masses of the $\hh$, $\ha$, and $\hpm$ become degenerate and
heavy, leaving behind $\hl$, which acts like the minimal standard Higgs boson.

\item $\tanb \to \infty$, $\mha^2>\mz^2+\dmsq_{22}-\dmsq_{11}$: 
In this limit, if squark mixing is small
so that $\dmsq_{12}$ can be neglected, $\alpha\to r(\beta-\pi/2)$
as $\beta\to \pi/2$, where $r$ is defined in Eq.~\ref{rform}
(see discussion associated with Eq.~\ref{tan2alpha}).  
The couplings, given in Table~\ref{hcouplingslimi},
are similar to those obtained in the large $\mha$ limit, 
except for the fact that $r\neq 1$ in general (see Eq.~\ref{rform}).
The $WW,ZZ$ and $t\anti t$ couplings of the $\hl$ become equal to those of
the standard Higgs boson, while the $\hl\to b\anti b$
coupling acquires an extra factor of $r$.
The (enhanced) $\hh\to b\anti b$ coupling becomes equal to
the $\ha\to b\anti b$ coupling, while the (suppressed) $\hh\to t\anti t$
coupling acquires an extra factor of $-r$ relative
to the $\ha\to t\anti t$ coupling. If $\dmsq_{12}\neq 0$, $\alpha$
no longer approaches $0$ exactly, and the couplings are
correspondingly modified.

\item $\tanb\to\infty$, $\mha^2<\mz^2+\dmsq_{22}-\dmsq_{11}$:
If $\dmsq_{12}$ can be neglected,
then $\alpha\to -\pi/2+r(\pi/2-\beta)$ as $\beta\to \pi/2$,
with $r$ given by Eq.~\ref{rform}.
The couplings then have the limits given in Table~\ref{hcouplingslimii}.
The $\hh$ has couplings to $WW,ZZ$ and $t\anti t$ that
are the same as the SM Higgs up to an overall sign, but the $b\anti b$
coupling differs by a factor of $r$ relative to the SM value.
The $\hl$ has the same coupling as the $\ha$ to $b\anti b$, but
its coupling to $t\anti t$ differs by a factor of $r$.
If $\dmsq_{12}\neq0$, $\alpha$ no longer approaches
$-\pi/2$ exactly, and these results are correspondingly modified.

\end{enumerate}

\begin{figure}[htbp]
\let\normalsize=\captsize   
\begin{center}
\centerline{\psfig{file=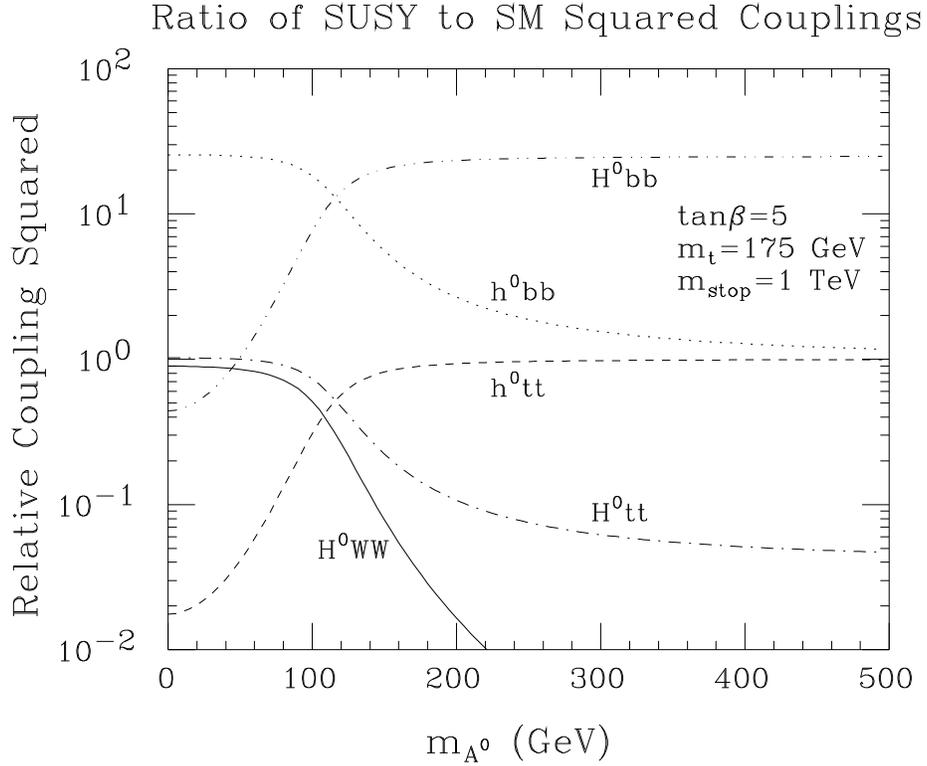,width=12.2cm}}
\begin{minipage}{12.5cm}       
\caption{Ratios of SUSY Higgs squared couplings to 
the corresponding SM values as a function of $\mha$ for the case of
$\tanb=5$, taking $\mt=175\gev$ and $\mstop=1\tev$.
\Twoloop\ radiative corrections to the CP-even Higgs sector
mass matrix are included and squark mixing is neglected.}
\label{couplingstanb5}
\end{minipage}
\end{center}
\end{figure}

These limits, as well as the sometimes rather slow approach to
the $\mha\rta \infty$ limits are illustrated in Fig.~\ref{couplingstanb5}.
There we plot
the ratios of the $\hh WW$, $\hh b\anti b$, $\hh t\anti t$, $\hl b\anti b$
and $\hl t\anti t$ squared couplings to the corresponding $\hsm$ squared 
couplings as a function of $\mha$ for the case of $\tanb=5$, taking
$\mt=175\gev$ and $\mstop=1\tev$, and including \twoloop\ corrections
assuming no squark mixing.
Especially important features to note from this figure are:
\begin{enumerate}
\item the continuing enhancement of the $\hl b\anti b$ squared coupling
out to quite large $\mha$ values;
\item the very rapid fall of the $\hh WW$ squared coupling, $\propto
\cos^2(\beta-\alpha)$, implying a correspondingly
rapid approach of the $\hl WW$ squared coupling to the SM value; and
\item the very rapid rise of the $\hh b\anti b$ squared coupling
as $\mha$ passes above $\mz$.
\end{enumerate}
Each of these features will have crucial phenomenological implications
in what follows.
The lightest SUSY Higgs boson, $\hl$, has particular significance for the 
search for supersymmetry because it is the only particle whose mass is 
bounded from above.  If supersymmetry is relevant to weak-scale physics, 
it may first appear via the discovery of this particle.

\subsection{Higgs decays}

\begin{figure}[htbp]
\let\normalsize=\captsize   
\begin{center}
\centerline{\psfig{file=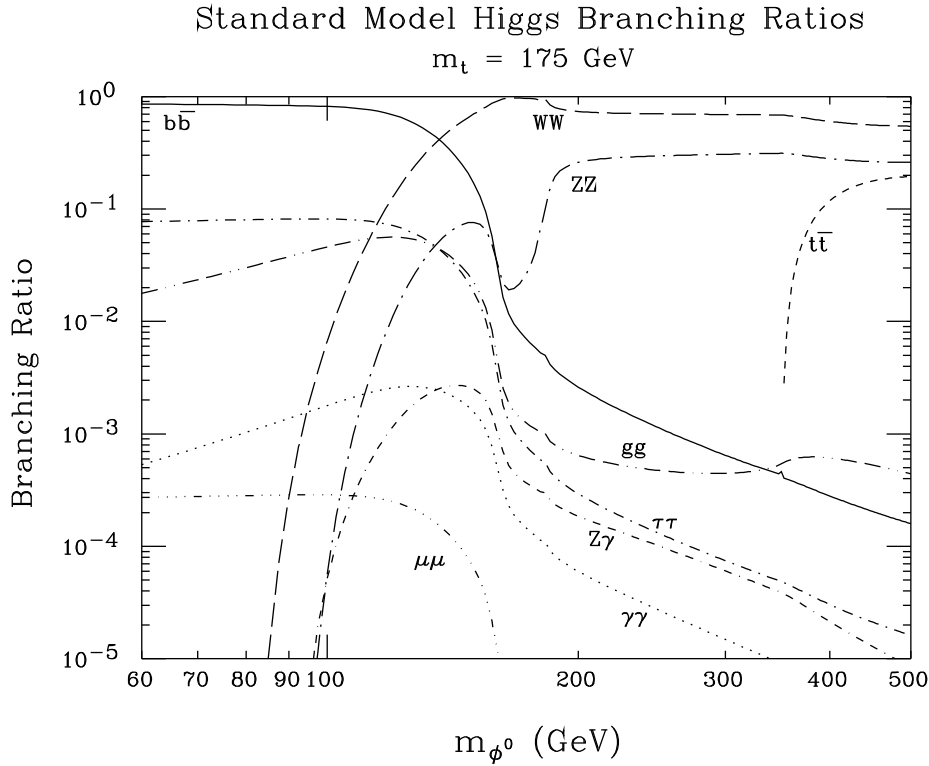,width=12.2cm}}
\begin{minipage}{12.5cm}       
\caption{Branching ratios for the Standard Model Higgs boson.
We have taken $\mt=175\gev$.}
\label{smbr}
\end{minipage}
\end{center}
\end{figure}

\indent\indent The standard Higgs boson couples to the weak bosons 
proportionally to their squared mass, and to fermions proportional to 
their mass, as shown in Fig.~\ref{dpf1}. Thus the Higgs boson tends to decay to 
the heaviest pair of particles which is kinematically allowed \cite{LQT}.  
There are important exceptions to this rule, however.  If the Higgs boson 
is heavier than twice the top-quark mass, it will still decay dominantly
to weak-vector-boson pairs; for $m_t=175$ GeV, the branching ratio to 
$t\anti t$ is at most $20\%$ (for $\mhsm \sim 500$ GeV).  
If the Higgs-boson mass
is not far below the $WW$ threshold, it has a sizable branching ratio to 
one real, one virtual $W$ boson ($W\wstar$), as well as to one real, one virtual 
$Z$ boson ($Z\zstar$) \cite{R}.  The dominant fermionic decay mode in this region
is $b\anti b$, and the suppressed coupling of the Higgs boson to
the bottom quark allows the one real, one virtual weak boson decays to be 
competitive (for $\mhsm \gsim 120$ GeV).  The next largest fermionic branching 
ratio is to 
$\tau^+\tau^-$, which is suppressed relative to $b\anti b$ due to the heavier 
$b$-quark mass and the three $b$-quark colors.
The branching ratio of the standard Higgs boson to various final states is
shown in Fig.~\ref{smbr}. Ref.~\cite{SDGZ} provides an up-to-date
review of the QCD radiative corrections
that must be incorporated in the branching ratio calculations.
The specific results presented here for the SM Higgs and later
for SUSY Higgs bosons are obtained with the following inputs:
i) we take $\lammsbar=290\mev$ for $N_f=4$ flavors, corresponding
to $\alpha_s(\mz)\simeq 0.115$ at two-loops;
ii) we take running masses of $\mc(\mc)=1.23\gev$ and $\mb(\mb)=4.0\gev$;
iii) we use a renormalization scale of $\mu=\mh/2$, where $\mh$
is the mass of the decaying Higgs boson;
iv) partial width contributions arising
from QCD correction diagrams containing one gluon and a $q\anti q$
pair are added to the $q\anti q$ pair partial width, and
not included in the two-gluon partial width;
v) semi-virtual decays of the type $tt^*$, and (for SUSY Higgs)
$Z^*A$ and ${W^\mp}^*\hpm$, are not included;
and, finally,
vi) double virtual $W^*W^*$ and $Z^*Z^*$ decays are 
neglected, but, of course, semi-virtual $WW^*$ and $ZZ^*$
decays are incorporated.

Another phenomenologically-important decay mode 
is to $\gam\gam$, which occurs via a one-loop diagram \cite{GAMGAM}.  
The $W$ boson gives the 
dominant contribution to the loop; the next most important is the top 
quark, which contributes with the opposite sign, but much less than the 
$W$ loop.  The branching ratio peaks at about $2 \times 10^{-3}$, for 
$\mhsm \approx 125$ GeV.  The branching ratio to $Z\gam$, which also occurs at 
one loop, is comparable to 
$\gam\gam$, although it is not phenomenologically useful due to the small
branching ratio of $Z\to \ell\ell$  ($\ell = e,\mu$) \cite{CCF}.
   
The branching ratios (and production cross sections) of the SUSY Higgs 
bosons can be very different from those of the standard Higgs boson.  We 
restrict our discussion to the most important features.  The following
features are evident from Tables~\ref{hcouplings}, \ref{hcouplingslimi}
and \ref{hcouplingslimii}, and are displayed
in Fig.~\ref{couplingstanb5}.
The coupling of the neutral scalar Higgs bosons to 
the weak bosons is at most as large as that of the standard Higgs boson. Thus,
the branching ratio of the SUSY 
Higgs bosons to weak-vector-boson pairs  can be 
suppressed with respect to the standard Higgs.  For example, as $\mha \to 
\infty$, the heavy scalar Higgs decouples from weak-vector-boson pairs.
On the other hand, the coupling of the SUSY Higgs bosons to $b\anti b, 
\taup\taum$ can be increased over that of the standard Higgs for $\tanb > 
1$.  In particular, the $\ha$, and for
$\mha>\sqrt{\mz^2-(\dmsq_{22}-\dmsq_{11})/(\cos2\beta\sin^2\beta)}$ 
the $\hh$, both have enhanced $b\anti b,\taup\taum$ couplings, while
these couplings for the $\hl$ become increasingly SM-like as $\mha$ becomes
large. For $\mha<\sqrt{\mz^2-(\dmsq_{22}-\dmsq_{11})/(\cos2\beta\sin^2\beta)}$ 
the $b\anti b,\taup\taum$
couplings of the $\hl$ are enhanced as $\tanb$
increases, and the squares of the $\hh$ couplings
to $WW,ZZ$ and $t\anti t$ become SM-like; the $\hh\to b\anti b$
coupling is generally somewhat reduced relative to the SM-like value.
(As noted earlier, see Eq.~\ref{tan2alpha} and associated discussion, 
the transition as to whether it is the $\hl$ or $\hh$
that has enhanced couplings is determined by the denominator
of $\tan 2\alpha$, which determines whether $\alpha\to 0$ or $\alpha\to -\pi/2$
when $\beta\to\pi/2$, \ie\ $\tanb\to \infty$.)
Enhanced $b\anti b,\taup\taum$ couplings further
suppress the branching ratio of the SUSY Higgs bosons to weak vector bosons.
In general, even at large $\mha$ the SM-like $\hl$ 
will have slightly enhanced
$b\anti b,\taup\taum$ and slightly suppressed $WW$ couplings
and branching ratios; this is illustrated in Fig.~\ref{couplingstanb5}
and in a later section
in Fig.~\ref{nlcdeviationscontours}. In addition,
this combination implies a suppression of the decay width and branching
ratio of the $\hl$ to the $\gamma\gamma$ channel.

\begin{figure}[htbp]
\let\normalsize=\captsize   
\begin{center}
\centerline{\psfig{file=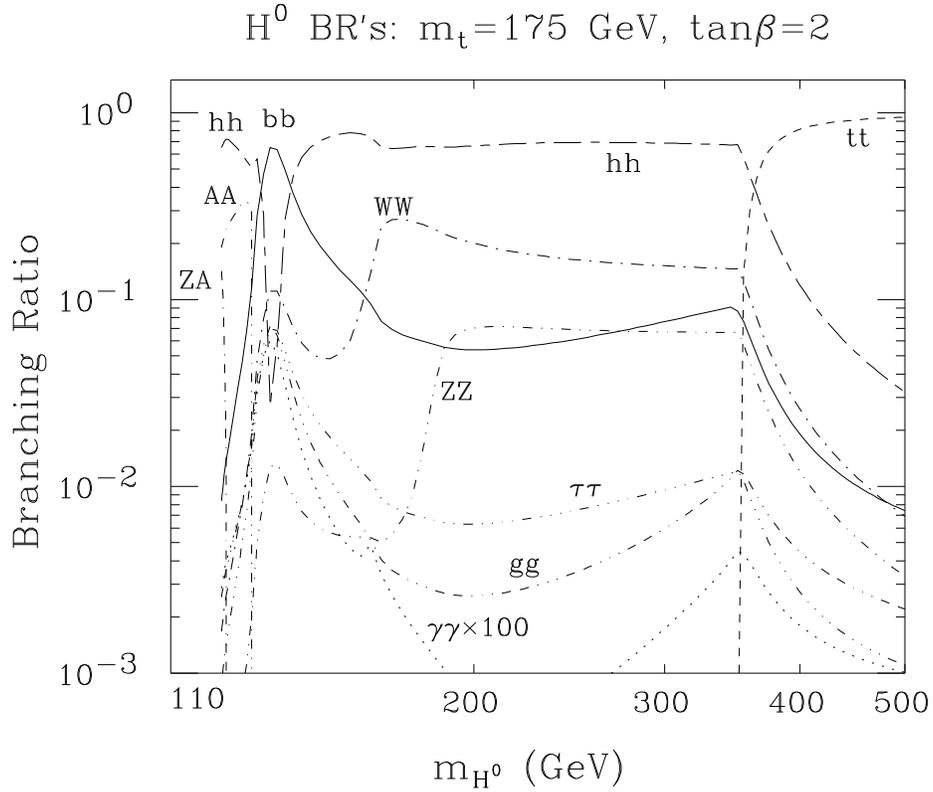,width=12.2cm}}
\begin{minipage}{12.5cm}       
\caption{Selected branching ratios for the MSSM 
$\hh$ for $\tanb=2$ and $\mt=175\gev$. \Twoloop\ radiative
corrections to the CP-even Higgs sector
mass matrix and Higgs self-couplings are included
for $\mstop=1\tev$ and neglecting squark mixing.
SUSY decays are assumed to be absent.}
\label{hhtanb2brs}
\end{minipage}
\end{center}
\end{figure}

MSSM Higgs bosons can also decay to channels that are not present in
the case of the SM $\hsm$.  To determine those additional
channels that are important requires
a rather complete specification of the MSSM parameters.
Aside from the SM decay modes there is the
possibility of decays to channels involving
the other Higgs bosons ($\hl\rta \ha\ha$, 
$\ha\rta Z\hl$, $\hh\rta \hl\hl,\ha\ha$, $\hp\rta \wp \hl,\wp\ha$), 
and all the Higgs bosons can have large branching ratios
for decay to pairs of supersymmetric particles
(for example, $\hl\rta \cnone\cnone$ and
$\hh,\ha\rta \cnone\cnone,\cntwo\cnone,\cpone\cmone,\slep\slep$ ---
here the $\chitil$'s are the charginos
and neutralinos and $\slep$ denotes a slepton).
The decay of $\ha,\hh,\hp$ to channels 
containing the $\hl$ are typically quite important below the 
$\ha,\hh\rta t\anti t$, $\hp\rta t\anti b$ thresholds, 
and in the case of the $\ha$ and $\hh$ can even be dominant over the
normal $b\anti b$ channel unless $\tanb$ is large. For example,
we illustrate the importance of the $\hl\hl$ decay mode for the $\hh$
in Fig.~\ref{hhtanb2brs}, taking $\tanb=2$,
$\mt=175\gev$, $\mstop=1\tev$ and neglecting squark mixing.
At the lowest $\mhh$ values, $\ha\ha$ and, even, $Z\ha$ decays are also
important. At low $\mhh$, the $ZZ^*$ partial width is also rather suppressed
after including the \twoloop\ corrections, which lower the minimum $\mhh$
value.  In combination with the large
partial widths for the $\hl\hl,\ha\ha,Z\ha$ modes,
this implies a rather small $ZZ^*$ branching
ratio at the lowest $\mhh$ values. Similarly,
for small to moderate values of $\tanb$ the supersymmetric-pair channels will
often dominate over all other modes if kinematically allowed.
Of particular note is the $\cnone\cnone$ decay mode, where $\cnone$
is presumed to be the lightest supersymmetric particle and therefore
invisible in an $R$-parity conserving model.
For more graphs of branching ratios in some typical cases (before including
\twoloop\ corrections), 
see Refs.~\cite{gunionperspectives,gunionerice}, especially the
latter for the case of light $\chitil$'s.
In what follows, we will occasionally discuss the complications which these 
two additional classes of decay modes can create.

\subsection{Higgs total widths}

The magnitude of the width of a Higgs boson plays an important role
in determining appropriate techniques for its discovery, and ultimately
is a very important probe of the Higgs boson's interactions and decays.
Higgs boson widths are extremely mass and model dependent.

\begin{figure}[htbp]
\vskip 1in
\let\normalsize=\captsize   
\begin{center}
\centerline{\psfig{file=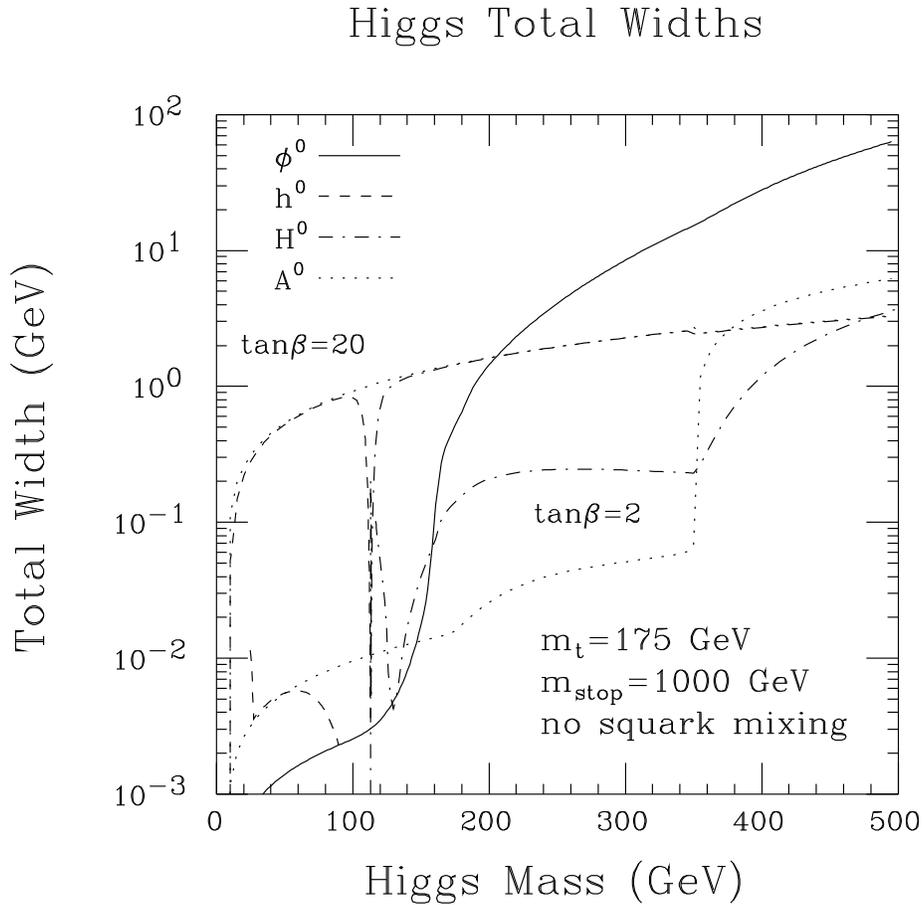,width=12.2cm}}
\smallskip
\begin{minipage}{12.5cm}       
\caption{
Total width versus mass of the SM and MSSM Higgs bosons for $\mt=175\gev$.
In the case of the MSSM, we have plotted results for
$\tan\beta =2$ and 20, taking $\mstop=1\tev$ and including \twoloop\
radiative corrections, 
neglecting squark mixing; SUSY decay channels are assumed to be absent.}
\label{hwidths}
\end{minipage}
\end{center}
\end{figure}

In the case of the SM $\hsm$, the Higgs boson width is very small
for $\mhsm$ below about $2\mw$.  Once $\mhsm\gsim 2\mw$ the Higgs width
increases rapidly as the $WW$ and then $ZZ$ channels open up.
The width as a function of mass is plotted in Fig.~\ref{hwidths}.

The widths of the neutral MSSM Higgs bosons 
are also plotted as a function of mass
in Fig.~\ref{hwidths} for two values of $\tanb$ --- $\tanb=2$ and $\tanb=20$.
We have taken $\mstop=1\tev$ and $\mt=175\gev$
and included \twoloop\ corrections to masses and mixing angles;
stop squark mixing has been neglected and SUSY decay channels
are assumed to be absent. The width of the $\hl$
depends very strongly on its mass.  For masses substantially below
the upper limit (as arise for $\mha\lsim \mz$) and $\tanb$ values
above 1, its width will be much larger than the width of a SM $\hsm$ of 
the same mass.  As $\mhl$ approaches its upper limit, the $\hl$ becomes more
and more SM-like and its width approaches that of the SM $\hsm$.
As far as its couplings to $f\anti f$ and $WW,ZZ$ channels are concerned,
the $\hh$ becomes  roughly (see earlier coupling
discussion) SM-like for $\mhh$ very near its lower limit
({\it i.e.} at small $\mha$). However, 
the $\hh\rta\ha\ha,Z\ha,\hl\hl$
decay channels are open and give large contributions to the total width of the
$\hh$ when $\tanb=2$ (see Fig.~\ref{hhtanb2brs}), 
implying $\Gamma_{tot}(\hh)\gg\Gamma_{tot}(\hsm)$
for $\mhh$ near its lower bound.  The $\hh\rta\hl\hl$ channel is also
responsible for the larger $\hh$ total width as compared to the $\ha$
total width in the $200-300\gev$ mass range when $\tanb=2$.  At $\tanb=20$,
for $\hl,\ha$ when $\mha\lsim \mz$ and for $\hh,\ha$ when $\mha\gsim\mz$
the $b\anti b$ decay channel is dominant due to its very large partial width.
This partial width is large and more or less the same for the two members
of a given pair due to the fact that their $b\anti b$ couplings
are both related to the SM value by a factor of $\tanb$ when $\tanb\gg1$ 
(see Tables~\ref{hcouplings}, \ref{hcouplingslimi}
and \ref{hcouplingslimii}, and associated comments).
Thus, one sees a smooth transition as $\mha$ passes above $\sim\mz$ where 
the $\hl$ and $\hh$ interchange roles.

\subsection{Biases from precision electroweak data}

\indent\indent Sensitivity to the Higgs mass from precision electroweak data
is only logarithmic.  If the analysis is carried out without
ascribing the excess in $R_{b\anti b}$ to new physics (supersymmetry
or whatever), then a not insignificant preference
for light Higgs boson masses emerges.
The current situation has been summarized
in Refs.~\cite{hollikeps,habereps,chankowskieps}. 
A typical example is the recent work of Ref.~\cite{eflhiggs} ---
see also Refs.~\cite{lepewwghiggs,chanporhiggs}. In Ref.~\cite{eflhiggs}
the $\Delta \chi^2=1$ contours yield a Higgs mass of $36
{{+56}\atop {-22}}\gev$ based on precision electroweak data
alone, and a mass of $76{{+152}\atop {-50}}\gev$ 
after including the CDF/D0 $\mt$ measurement.  In the latter case,
the $2\sigma$ upper limit on the mass is $700\gev$, definitely
in the perturbative Higgs sector domain.  The Higgs mass
values in the vicinity of the $\chi^2$ minimum are
remarkably consistent with the MSSM
expectations for the $\hl$. Indeed, Ref.~\cite{eflhiggs} states
that after including theoretical
constraints (such as Higgs mass bounds from stability and experiment)
the $1\sigma$ mass range is found to be slightly more preferred in
the MSSM than in the minimal standard model.  

However, if the $R_{b\anti b}$ excess arises from new physics,
and the precision electroweak data is refit after reducing
$R_{b\anti b}$ to its SM-like value, the above preference
for low values of the Higgs mass essentially disappears \cite{langackerpc}.

\subsection{Goals and organization}

\indent\indent In this report we review the search for the standard Higgs 
boson, and the
Higgs bosons of the minimal supersymmetric standard model, at present and
future colliders.  A short summary of this work appeared in
Ref.~\cite{dpfshort}. The order of presentation is as follows:

\begin{itemize}

\item CERN LEP I - $\sqrt s = 91$ GeV $e^+e^-$ collider.

\item CERN \lepii\ - $\sqrt s = 176-184$ GeV, with an upgrade to 192 GeV.  
We also consider upgrades to 205 and 240 GeV.

\item CERN Large Hadron Collider (LHC) - $\sqrt s = 14$ TeV $pp$ collider.
We also comment on an intermediate-stage LHC with $\sqrt s = 10$ TeV.

\item Fermilab Tevatron and \tevstar\ - $\sqrt s = 2$ TeV $p\anti p$ collider. 
The \tevstar\ (``Tev-star'') is a proposed luminosity upgrade of the Tevatron.

\item Fermilab DiTevatron -  A proposed $\sqrt s = 4$ TeV $p\anti p$ collider 
in the Tevatron tunnel.

\item Next linear $e^+e^-$ collider (NLC) - $\sqrt s = 500 - 1500$ GeV, 
including $\gam\gam$ and $e\gam$ collider modes.

\item A $\mu^+\mu^-$ collider - $\sqrt s = 500 - 4000$ GeV.

\end{itemize}

We attempt to summarize the results of the most current analyses of the
search for Higgs bosons at present and future colliders.  Throughout this 
report, we consider a particle to be discovered when the signal exceeds a five 
sigma fluctuation of the background ($S/\sqrt B > 5$). We also consider the 
measurement of the properties of the Higgs bosons.  

\section{LEP I}

\subsection{Standard Higgs boson}

\indent\indent The standard Higgs boson is searched for in $Z$ decays via 
the process $Z \to \zstar \hsm$, with $\zstar 
\to f\anti f$ ($\zstar$ denotes a virtual 
$Z$ boson), where $f\anti f = \nu\anti \nu, e^+e^-, \mu^+\mu^-$ \cite{BJ}.  
The lower bound on the Higgs mass is given in Table~\ref{hlimits} for all 
four LEP experiments \cite{LEP,RICHARD}. 
Although a combined limit does not exist, one can be 
estimated as $\mhsm >$ 64.5 GeV \cite{RICHARD}.
The most recent summary \cite{grivazeps} does not find a significantly
different result.
Since the branching ratio falls rapidly with increasing Higgs mass, and because
there exist Higgs candidate events at the highest masses, this 
bound will not improve significantly with increased statistics.  Thus the 
ultimate reach of LEP I is likely to be close to the existing bound.

\begin{table}[hbt]
\caption[fake]{Lower limits ($95\%$ C.~L.) on the Higgs-boson mass from the four
LEP experiments \cite{RICHARD}.  The limits include data taken in 1993, and
are slightly higher than the published limits \cite{LEP}. The combined limit
is an estimate \cite{RICHARD} (see also Ref.~\cite{grivazeps}).}
\begin{center}
\begin{tabular}{lc}
& $\mhsm$ (GeV) \\
\\
ALEPH & 60.3 \\
DELPHI & 58.3 \\
L3 & 58.0 \\
OPAL & 56.9 \\ \hline
\\
Combined & 64.5 \\
\end{tabular}
\end{center}
\label{hlimits}
\end{table}

The decay $Z \to \hsm\gam$, which occurs at one loop, does not have a large 
enough branching ratio to be useful \cite{CCF,RICHARD}.

\subsection{SUSY Higgs bosons}

\indent\indent There are two production mechanisms for SUSY Higgs bosons 
at LEP I.  The analogue of the standard Higgs process is $Z \to \zstar\hl$.  
There is also the decay $Z \to \ha\hl$, if both $\hl$ and $\ha$ are light.  
The $Z \to \ha\hl$ decay can yield a variety of final states: $\tau^+\tau^- 
b\anti b$, $\tau^+\tau^- \tau^+\tau^-$, and $b\anti b b\anti b$.  If $\mhl > 
2 \mha$, the decay $\hl \to \ha\ha$ becomes the dominant decay mode of the 
$\hl$, leading to 
$Z \to \ha\hl \to \ha\ha\ha$.  This yields six-body final states with 
$b$'s and $\tau$'s. The contribution of $Z \to \ha\hl$
to the $Z$ width can also be used as a probe of this decay mode.  


\begin{figure}[htbp]
\let\normalsize=\captsize   
\begin{center}
\centerline{\psfig{file=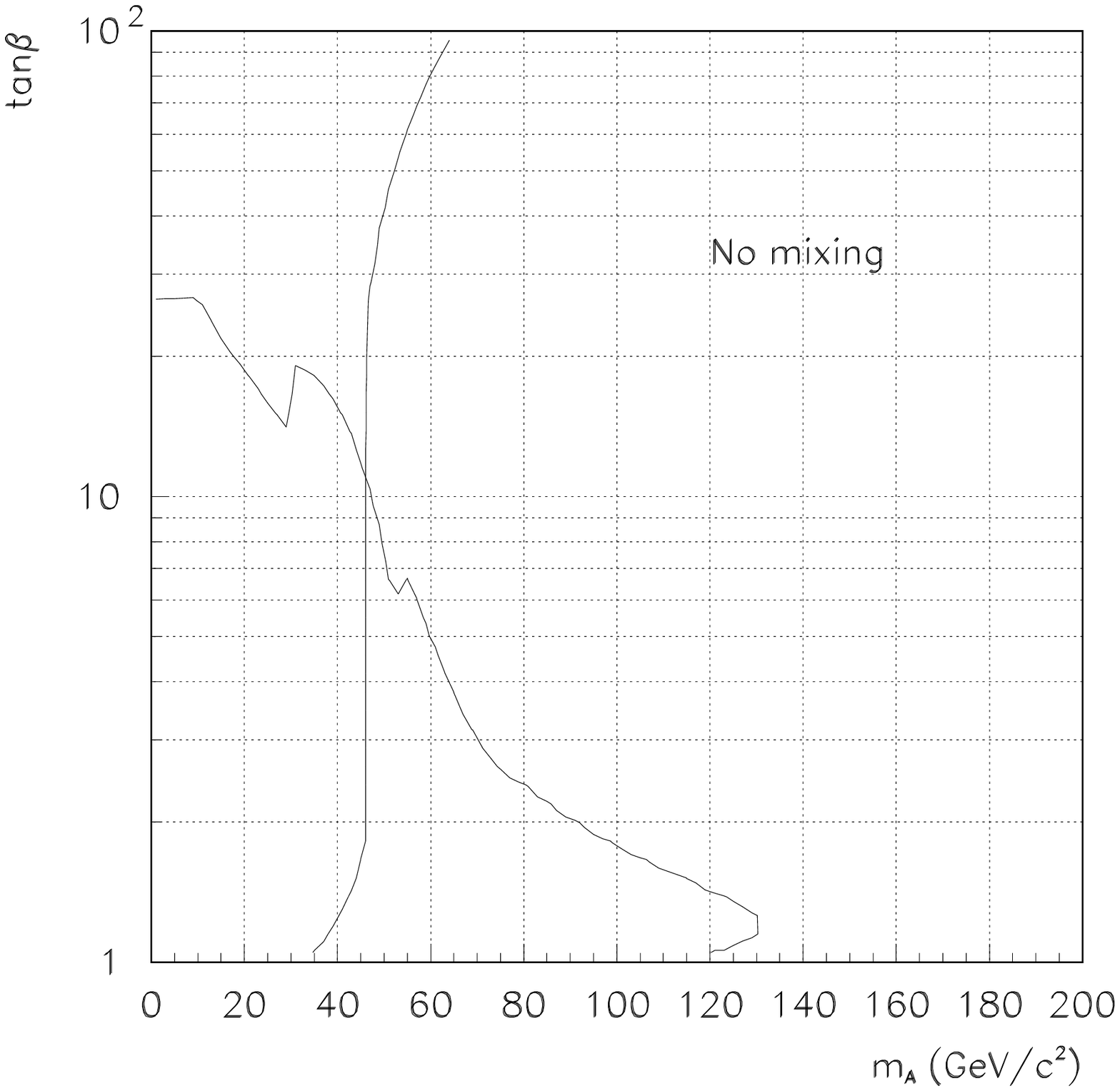,width=12.2cm}}
\begin{minipage}{12.5cm}       
\caption[fake]{Excluded regions in the ($\mha,\tanb$) plane of the minimal 
supersymmetric model from LEP I for the `no-mixing' scenario
with $A=\mu=0$, $\msusy=1\tev$ and $\mt=175\gev$. The regions excluded by
$Z \to \zstar\hl$ and by $Z \to \ha\hl$ (based on the
$Z$ width and the $\tau^+\tau^-b\anti b$,
$\tau^+\tau^-\tau^+\tau^-$, and  $b\anti bb\anti b$ final states)
are indicated separately. The more or less vertical curve
defines the region excluded by $Z\to\ha\hl$. 
Figure from Ref.~\cite{janotnewlep}.}
\label{janot_nomixing}
\end{minipage}
\end{center}
\end{figure}

\begin{figure}[htbp]
\let\normalsize=\captsize   
\begin{center}
\centerline{\psfig{file=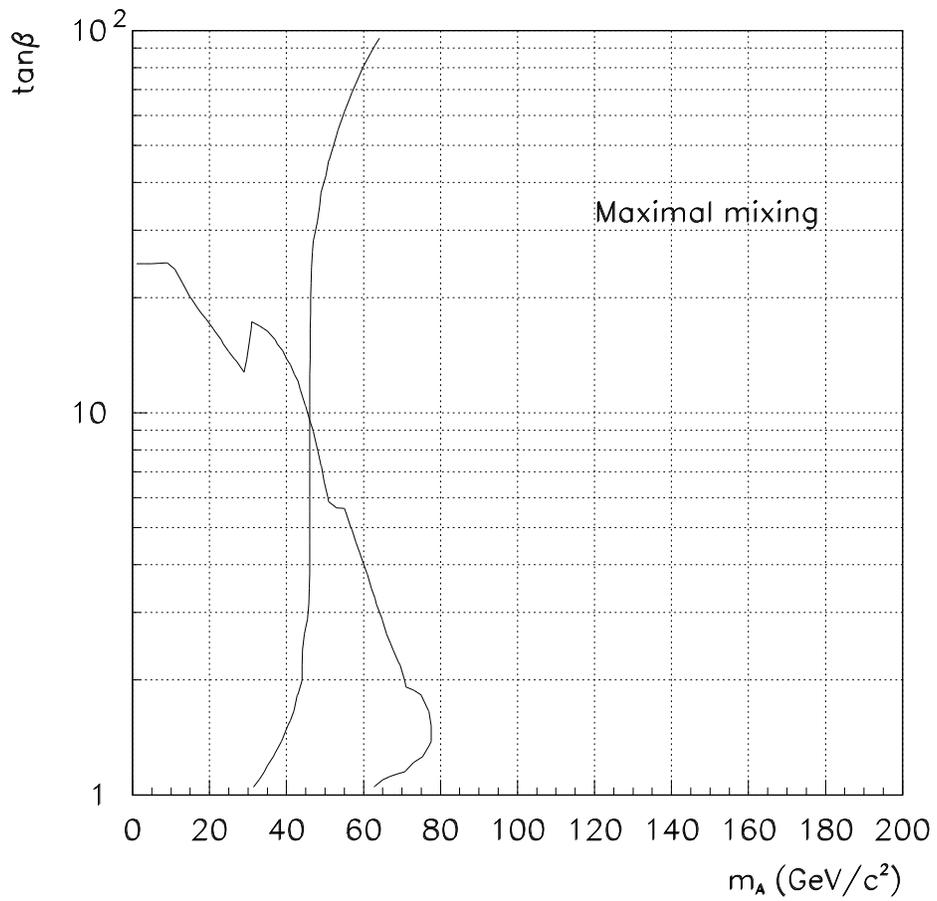,width=12.2cm}}
\begin{minipage}{12.5cm}       
\caption[fake]{Excluded regions in the ($\mha,\tanb$) plane of the minimal 
supersymmetric model from LEP I for the `maximal-mixing' scenario
with $A=\sqrt 6 \msusy$, $\mu=0$, $\msusy=1\tev$ and $\mt=175\gev$.
See caption for Fig.~\ref{janot_nomixing}.}
\label{janot_maximalmixing}
\end{minipage}
\end{center}
\end{figure}

\begin{figure}[htbp]
\let\normalsize=\captsize   
\begin{center}
\centerline{\psfig{file=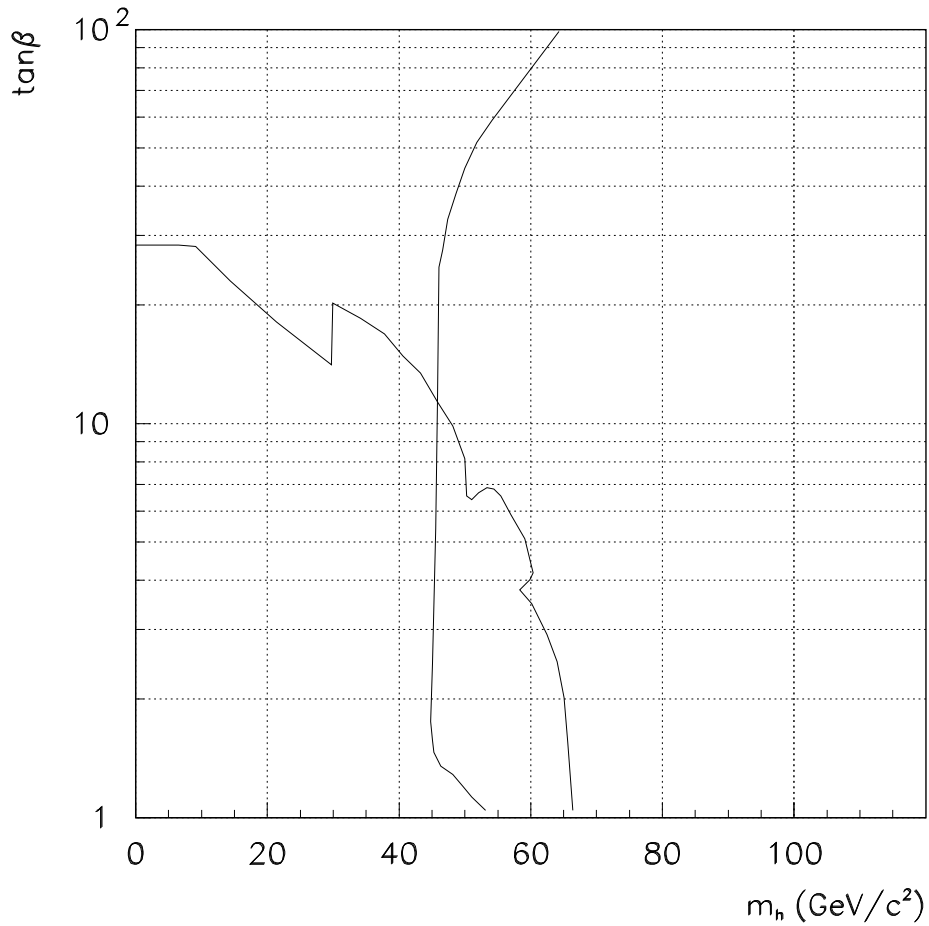,width=12.2cm}}
\begin{minipage}{12.5cm}       
\caption[fake]{Excluded regions in the ($\mhl,\tanb$) plane of the minimal 
supersymmetric model from LEP I for $\msusy=1\tev$ and $\mt=175\gev$;
boundaries for the different mixing scenarios cannot be distinguished
in this parameter plane. See caption for Fig.~\ref{janot_nomixing}.}
\label{janot_nomixing_mhtanb}
\end{minipage}
\end{center}
\end{figure}

The regions of SUSY parameter space that are excluded at LEP I using
$Z\to\ha\hl$ and $Z\to\zstar\hl$ are illustrated
in Figs.~\ref{janot_nomixing}--\ref{janot_nomixing_mhtanb}.
The $Z \to \ha\hl$ decay is crucial in eliminating
large $\tanb$ values at low $\mha$, while 
$Z \to \zstar\hl$ extends the excluded region at lower $\tanb$ values
to somewhat higher $\mha$ values.
Fig.~\ref{janot_nomixing} shows the excluded regions in the 
($\mha,\tanb$) plane from Ref.~\cite{janotnewlep} in the `no-mixing'
scenario with
$\msusy=1\tev$, $\mt=175\gev$ and $A=\mu=0$ (yielding $\mstop\sim\msusy$).
See also Refs.~\cite{SUSYLEP,RICHARD,RS,mfrank,sophp}. 
Excluded regions for the `maximal-mixing' scenario with $\msusy=1\tev$,
$\mt=175\gev$, $A=\sqrt 6\msusy$, and $\mu=0$ are given in
Fig.~\ref{janot_maximalmixing}. Fig.~\ref{janot_nomixing_mhtanb}
gives excluded regions in the $(\mhl,\tanb)$ plane. (The
curves for the different mixing scenarios are very much the same
in this plane.)
From these figures, we see that both $\mha$ and $\mhl$ must
be larger than about $45\gev$.

The heavier scalar Higgs boson, $\hh$, is too heavy to be produced via $Z$ 
decays. The same is true of the charged Higgs boson.

\section{LEP II}

\subsection{Standard Higgs boson}

\begin{figure}[htbp]
\let\normalsize=\captsize   
\begin{center}
\phantom{0}
\vskip -2.25in
\centerline{\psfig{file=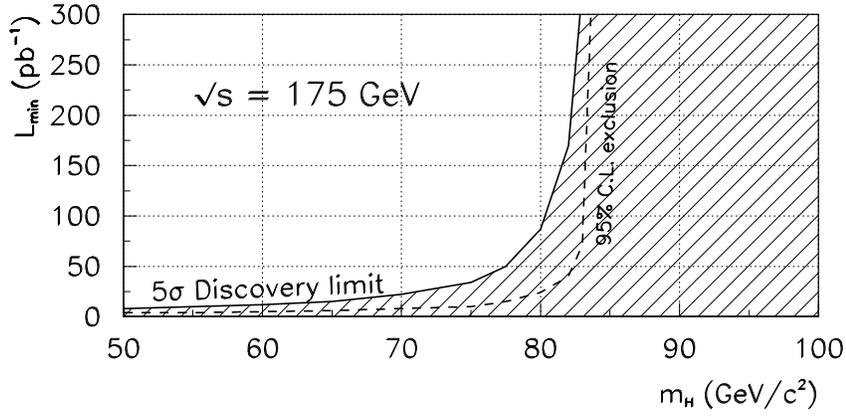,width=12cm}}
\vskip -2.25in
\centerline{\psfig{file=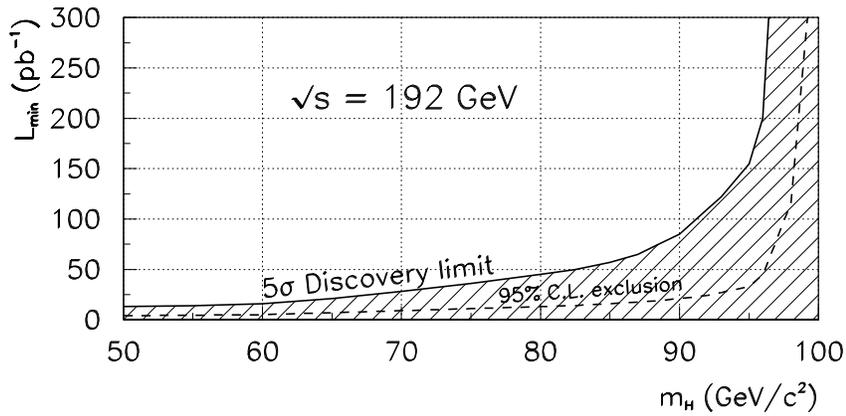,width=12cm}}
\vskip -2.25in
\centerline{\psfig{file=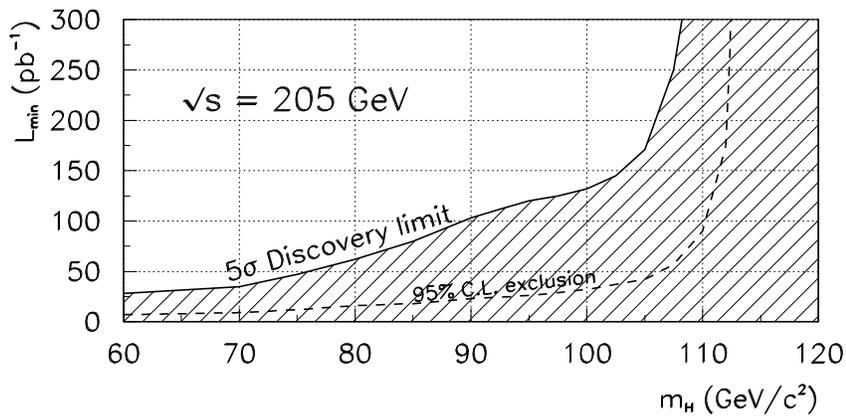,width=12cm}}

\begin{minipage}{12.5cm}       
\caption[fake]{\lepii\ discovery limits for the SM $\hsm$ (denoted by $H$
in this figure) at $\rts=175,192,205\gev$. The luminosity $L_{\rm min}$ is
the minimum required per experiment, assuming that 
data from all four experiments is combined.}                      
\label{lepiiall}
\end{minipage}
\end{center}
\end{figure}

\indent\indent The production of the standard Higgs boson at \lepii\ is via 
$e^+e^- \to Z\hsm$ \cite{JP}.  
The reach in Higgs mass is determined by kinematics and luminosity.
For relevant studies, see Ref.~\cite{LEP2,janotlep,altarellilep}.  
We shall quote the results of the latest study, Ref.~\cite{altarellilep}.
\lepii\ is scheduled to begin operation in 1996 with an energy above the 
$W^+W^-$ threshold.  As illustrated in Fig.~\ref{lepiiall}, 
for the ``standard'' energy of 
$\sqrt s = 175$ GeV a $\gsim 5\sigma$ Higgs signal will be obtained
for masses up to about 82 GeV if an average 
$150 \pbi$ of integrated luminosity is accumulated by each of the four
detectors. However, this energy no longer 
represents the initial \lepii\ goal. The initial energy
of \lepii\ could reach as high as $\sqrt s = 184$ GeV, which would allow 
discovery of the standard Higgs boson up to a mass of about 87 GeV with 
$150 \pbi$ of integrated luminosity per detector.

Larger Higgs masses are accessible with higher energy. The energy can 
potentially be increased to $\sqrt s = 192$ GeV with additional superconducting
cavities, which would allow masses up
to 95 GeV to be probed with $150 \pbi$ of integrated luminosity
per experiment, as illustrated in Fig.~\ref{lepiiall}.
The Higgs mass region $\mhsm \sim \mz$ has a background from $e^+e^- \to
ZZ$ which is comparable to $e^+e^- \to Z\hsm$.  
This can be overcome by tagging $b$-quark jets 
using a silicon vertex detector (SVX) \cite{janotlep}; in the 
mass region of interest, $BR(\hsm\to b\anti b) \approx 85\%$, while $BR(Z \to 
b\anti b) \approx 15\%$.  

An energy of 205 GeV could be achieved by nearly doubling the number of 
superconducting cavities necessary for $\sqrt s = 184$ GeV, and upgrading the
cryogenic system.  
This energy would allow a search for the standard 
Higgs boson up to a mass of 103 GeV with $150 \pbi$ of integrated 
luminosity per experiment, see Fig.~\ref{lepiiall}.

The LEP magnets can keep electrons and positrons in orbit up to 125 GeV.  
However, the energy radiated via synchrotron radiation, already significant at 
$\sqrt s = 175$ GeV, grows as the fourth power of the beam energy. 
A machine energy as high as $\sqrt s = 240$
GeV is conceivable with yet more superconducting cavities and cryogenics 
upgrades.  This would allow the exploration of the standard
Higgs boson up to a mass of roughly 140 GeV with $75 \pbi$ of integrated 
luminosity per detector.

\subsection{SUSY Higgs bosons}

\indent\indent At \lepii, the two main production processes of interest are 
$e^+e^- \to Z\hl$ and $e^+e^- \to \ha\hl$.  Unless $\mstop$ is small,
the heavy scalar Higgs boson, $\hh$, will be too heavy to 
be produced via $\epem\rta Z\hh$, and for the small values
of $\mha$ such that $\ha\hh$ production would be
kinematically allowed, the associated coupling is small.
The processes $\epem\rta b\anti b \hh, b\anti b \ha$ are
kinematically allowed out to larger masses, but generally have very small rates
because of the small $b\anti b\hh$ coupling; the Yukawa coupling would
have to be larger than its perturbative limit for observable rates.

\begin{figure}[htbp]
\let\normalsize=\captsize   
\begin{center}
\centerline{\psfig{file=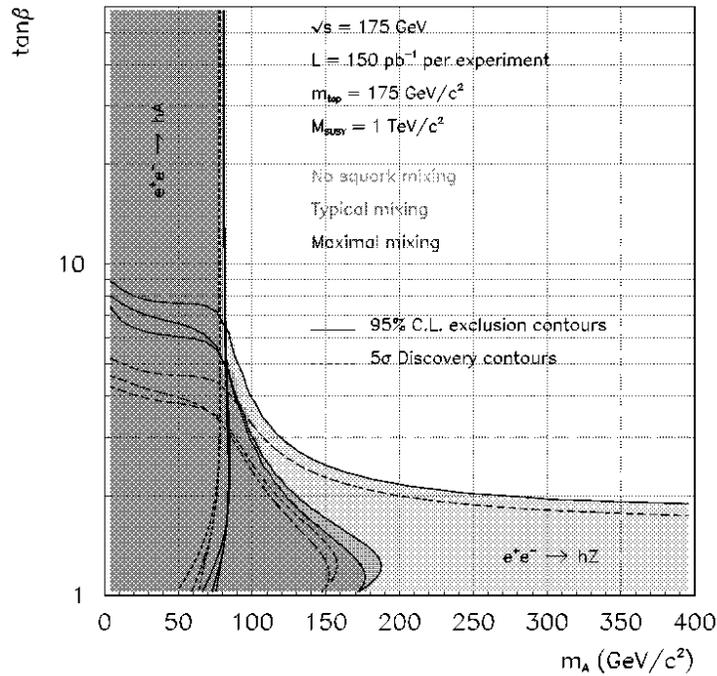,width=11.0cm}}
\begin{minipage}{12.5cm}       
\caption[fake]{Regions in the minimal supersymmetric model parameter space 
for which $Z\hl$ and $\hl\ha$ will be observable
at \lepii\ with $\sqrt s =$ 175 GeV, assuming $\mt=175\gev$
and $\msusy=1\tev$. Results for various
degrees of mixing in the stop-squark sector are shown.  
Figure from Ref.~\cite{altarellilep}.}
\label{175_mix2}
\end{minipage}
\end{center}
\end{figure}

\begin{figure}[htbp]
\let\normalsize=\captsize   
\begin{center}
\centerline{\psfig{file=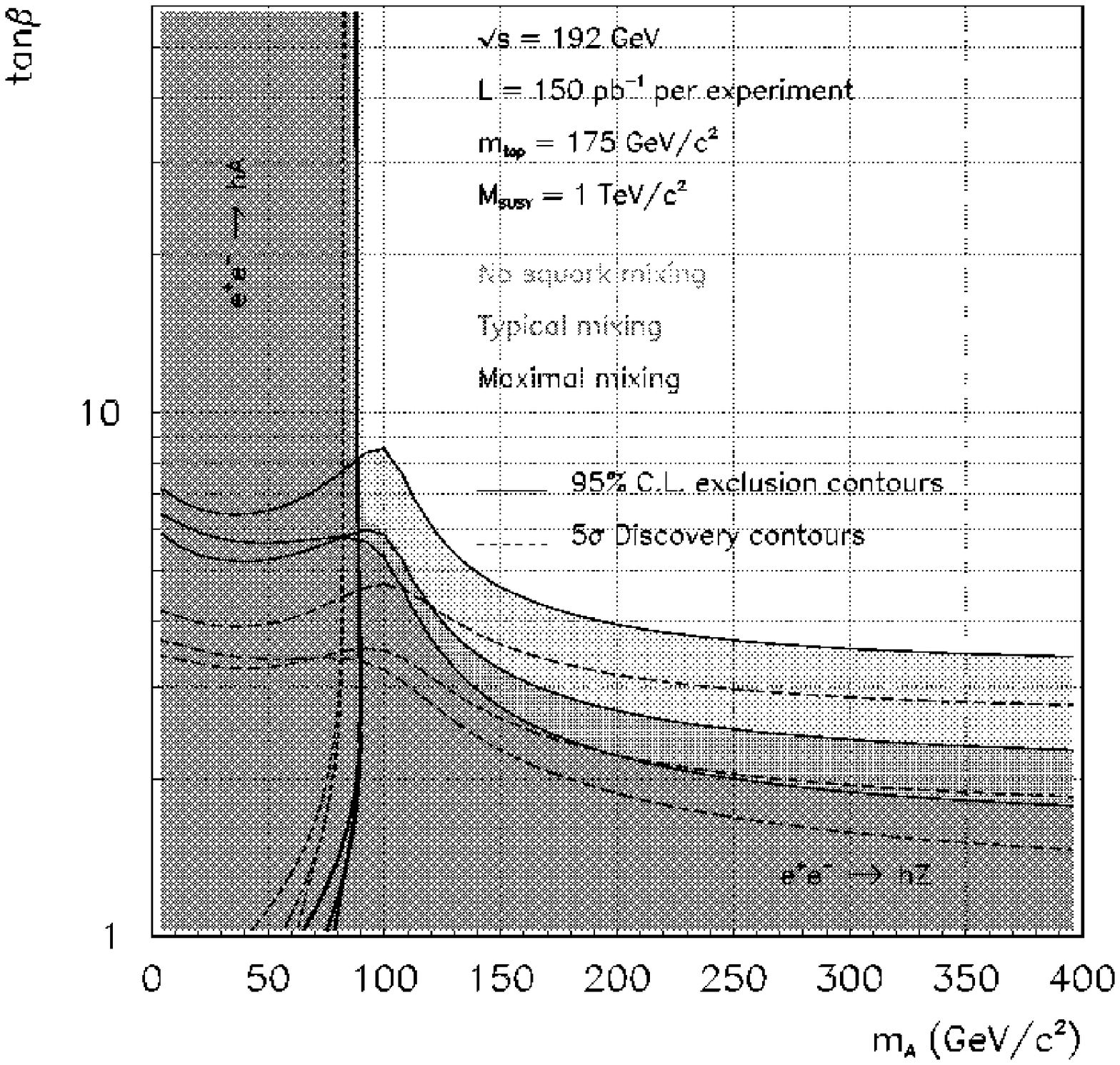,width=11.0cm}}
\begin{minipage}{12.5cm}       
\caption[fake]{Regions in the minimal supersymmetric model parameter space 
for which $Z\hl$ and $\hl\ha$ will be observable
at \lepii\ with $\sqrt s =$ 192 GeV, assuming $\mt=175\gev$
and $\msusy=1\tev$. Results for various
degrees of mixing in the stop-squark sector are shown.  
Figure from Ref.~\cite{altarellilep}.}
\label{192_mix2}
\end{minipage}
\end{center}
\end{figure}

\begin{figure}[htbp]
\let\normalsize=\captsize   
\begin{center}
\centerline{\psfig{file=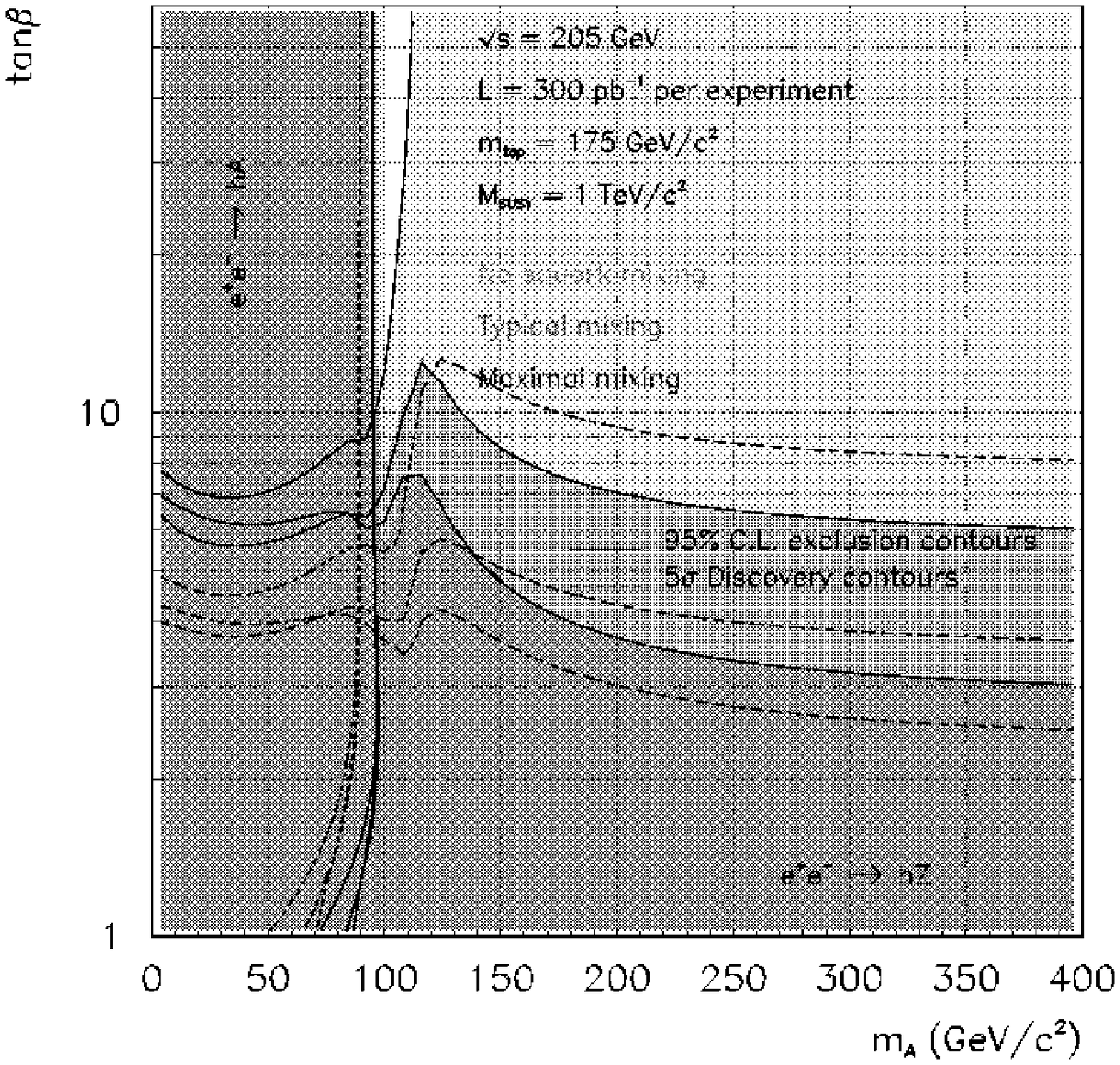,width=11.0cm}}
\begin{minipage}{12.5cm}       
\caption[fake]{Regions in the minimal supersymmetric model parameter space 
for which $Z\hl$ and $\hl\ha$ will be observable
at \lepii\ with $\sqrt s =$ 205 GeV, assuming $\mt=175\gev$
and $\msusy=1\tev$. Results for various
degrees of mixing in the stop-squark sector are shown.  
Figure from Ref.~\cite{altarellilep}.}
\label{205_mix2}
\end{minipage}
\end{center}
\end{figure}

The regions of the $(\mha,\tanb)$ parameter space for which
$Z\hl$ and $\hl\ha$ will be detectable are illustrated
in Figs.~\ref{175_mix2}, \ref{192_mix2}, \ref{205_mix2} 
for $\rts=175,192,205\gev$.
These figures assume $\mt=175\gev$ and $\msusy=1\tev$, and
give contours for the earlier-defined `no-mixing', `typical-mixing',
and `maximal-mixing' scenarios. In the no-mixing
case, $\mstop\sim\msusy$.
Discovery regions increase if $\msusy$ lies below $1\tev$.
For large $\mha$, the $\epem\to\ha\hl$ process is
kinematically forbidden, but $\mhl$ merely approaches its
upper bound, given at one-loop by
Eq.~\ref{HBOUND} (smaller after including \twoloop\ corrections).  
For $\tanb$ near 1 and small mixing this bound is small enough
that the $\hl$ will be produced at an observable rate
via $e^+e^- \to Z\hl$ at large $\mha$,
even for the lowest $\rts=175\gev$ energy. At $\rts=205\gev$,
the bound on $\mhl$ in the absence of mixing implies that $Z\hl$
will be observable for arbitrarily large values of $\tanb$ and $\mha$,
indeed throughout all of parameter space except the $\mha\lsim 100,\tanb\gsim8$
corner where $\hl\ha$ production can instead be detected.

\begin{figure}[htbp]
\let\normalsize=\captsize   
\begin{center}
\centerline{\psfig{file=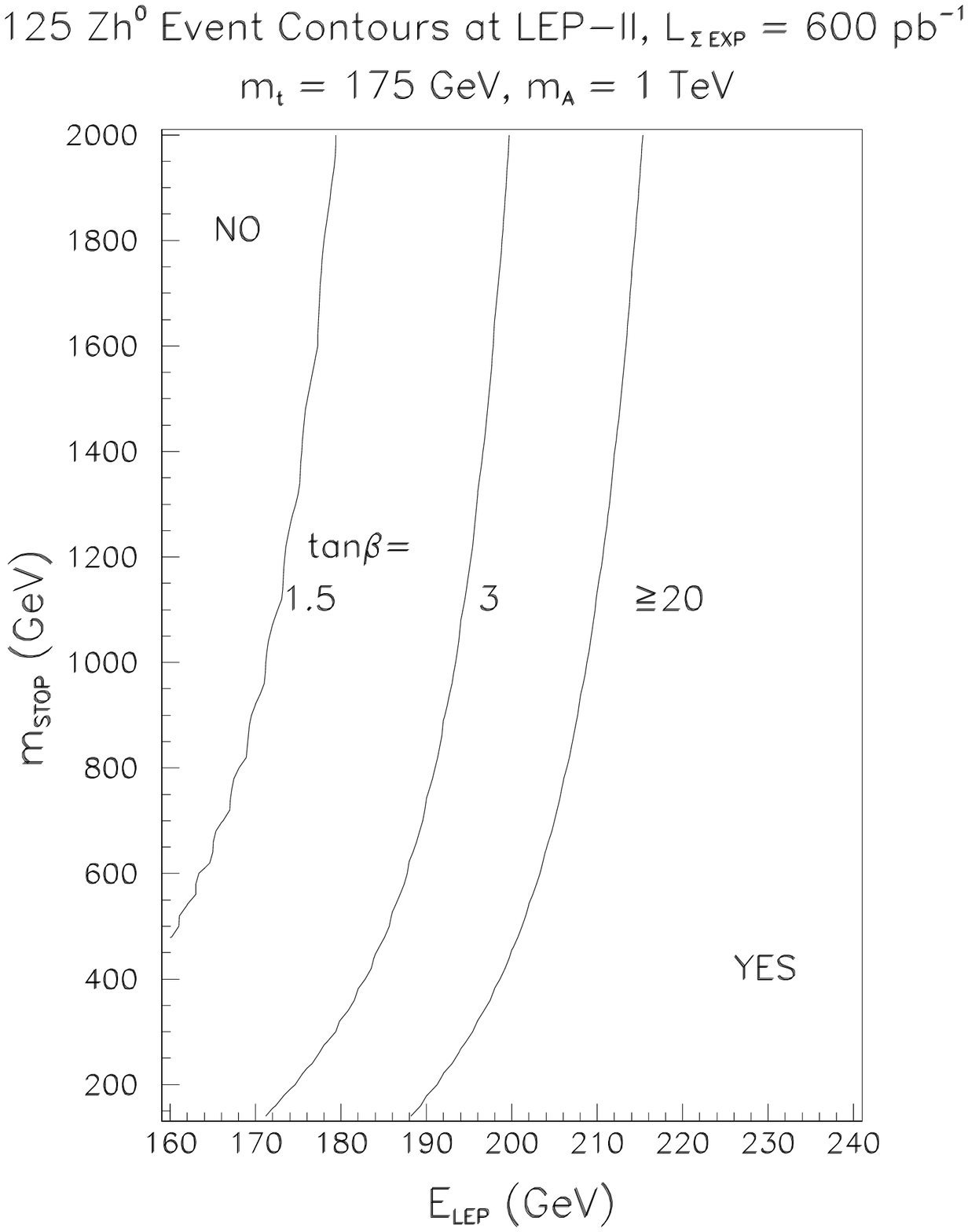,width=12.2cm}}
\begin{minipage}{12.5cm}       
\caption{Fixed $\tanb$ contours in $(\elep,\mstop)$ parameter
space for which 125 $Z\hl$ events are obtained.
We have taken $L=150\pbi$ per experiment (\ie\ $600\pbi$ 
summed over all experiments), $\mt=175\gev$ and $\mha=1\tev$,
and included radiative corrections to the Higgs sector
computed at the \twoloop\ level assuming no squark mixing.}
\label{lepiicontours}
\end{minipage}
\end{center}
\end{figure}

Since large $\mha$ is preferred in many unified models, it is useful 
to further quantify the ultimate
limit on $\hl$ detection at \lepii\ in the case of large $\mha$.
The question is for what portion
of SUSY parameter space is \lepii, running at a certain
energy, guaranteed to find the $\hl$, as a function of the stop squark mass?
(We shall assume negligible mixing, implying
$\msusy=\mstop$; substantial mixing can increase
$\mhl$, implying that higher machine energy would be required.)
A crude approximation to the results of Ref.~\cite{altarellilep}
is that roughly 125 $Z\hl$ events (summed over
all experiments, and before including cuts
and branching ratios) are needed 
in order to achieve a $5\sigma$ discovery signal.
Assuming $\mt=175\gev$ and $L=600\pbi$ (summed over all detectors),
we present in Fig.~\ref{lepiicontours}
contours in $(\elep,\mstop)$ parameter space 
which define the boundary between
the regions where 125 $Z\hl$ events will and will not be obtained
at a given fixed $\tanb$ value 
when $\mha$ is large (and hence $\mhl$ is close to its upper bound
for a given $\tanb$).
Evidently, the machine energy required to guarantee 125 events
for all $\tanb$ values is quite sensitive to $\mstop$.  While
$\elep\sim 205\gev$ is needed for high $\tanb$ and $\mstop\sim 1\tev$
(see also Fig.~\ref{205_mix2}),
$\elep\lsim 200\gev$ is adequate if $\mstop\lsim 500\gev$,
as in many unified scenarios.  At low $\tanb$
(preferred in typical $b-\tau$ Yukawa-unified models) the
machine energy needed to have a good chance of $\hl$
discovery is even more modest. However, 
for the canonical energy of $\elep=192\gev$,
a summed luminosity of $L=600\pbi$ guarantees
$\hl$ discovery only for $\tanb\lsim 3$ if $\mha=\mstop=1\tev$,
as is also apparent from Fig.~\ref{192_mix2}.

Charged Higgs detection at \lepii\ is a possibility
if $Z\to\hp\hm$ is kinematically allowed. For machine energies
$\sqrt s=175$, $190$ and $210\gev$ and $L=500\pbi$, 
more than 50 $\hp\hm$ events
will be present for $\mhpm<77$, $83$, and $88\gev$, respectively.
(Mass reach rises slowly with increased energy at fixed luminosity
due to the slow threshold turn-on of this $P$-wave final state.)
The $\hpm$ decays to a mixture of $cs$ and $\tau\nu$ in the mass
region in question. Sensitivity is present for all final states,
but significantly more than 50 events is generally required unless
$\tanb$ is large enough that the $\tau\nu$ channel is completely dominant.
At $\sqrt s=200\gev$, $L=500\pbi$ allows detection of the $\hpm$ for
masses up to $\sim 67\gev$ at moderate $\tanb$ and for masses as
high as $80\gev$ when $BR(\hp\rta\tau\nu)\sim 1$ (at large $\tanb$)
\cite{sophp}. In the MSSM, $\mhpm$ is bounded from below by $\mw$ 
for most parameter choices, implying that observation
of $\hp\hm$ pair production would almost certainly require LEP II
energies above $200\gev$.

\section{LHC}

\subsection{Standard Higgs boson}

\indent\indent \lepii\ will search for 
the standard Higgs boson up to a mass of at least 80 GeV (for $\sqrt s = 
175$ GeV), and higher if higher energy is attained. 
We are therefore primarily interested in $\mhsm>80$ 
GeV in our deliberations concerning future colliders.

The LHC is a 14 TeV $pp$ collider which will reside in the LEP tunnel.
It is expected to deliver $100 \fbi$ of integrated luminosity per year when 
operating at a luminosity of ${\cal L}=10^{34}/cm^2/s$.  With a bunch 
crossing time of 25 $ns$, this luminosity
yields about 20 interactions per bunch crossing.  For some Higgs searches,
it may be advantageous to run at lower luminosity, to reduce the number of
multiple interactions. It is 
anticipated that the machine will operate at an instantaneous luminosity of 
${\cal L}=10^{33}/cm^2/s$ during the first few years of running. Furthermore, 
the current plan for the construction of the LHC calls for the first stage to 
operate at this low luminosity and with $\sqrt s =$ 10 TeV.  We comment on
the Higgs potential of this first-stage machine at the end of this section.  

\medskip
\begin{figure}[htbp]
\let\normalsize=\captsize   
\begin{center}
\centerline{\psfig{file=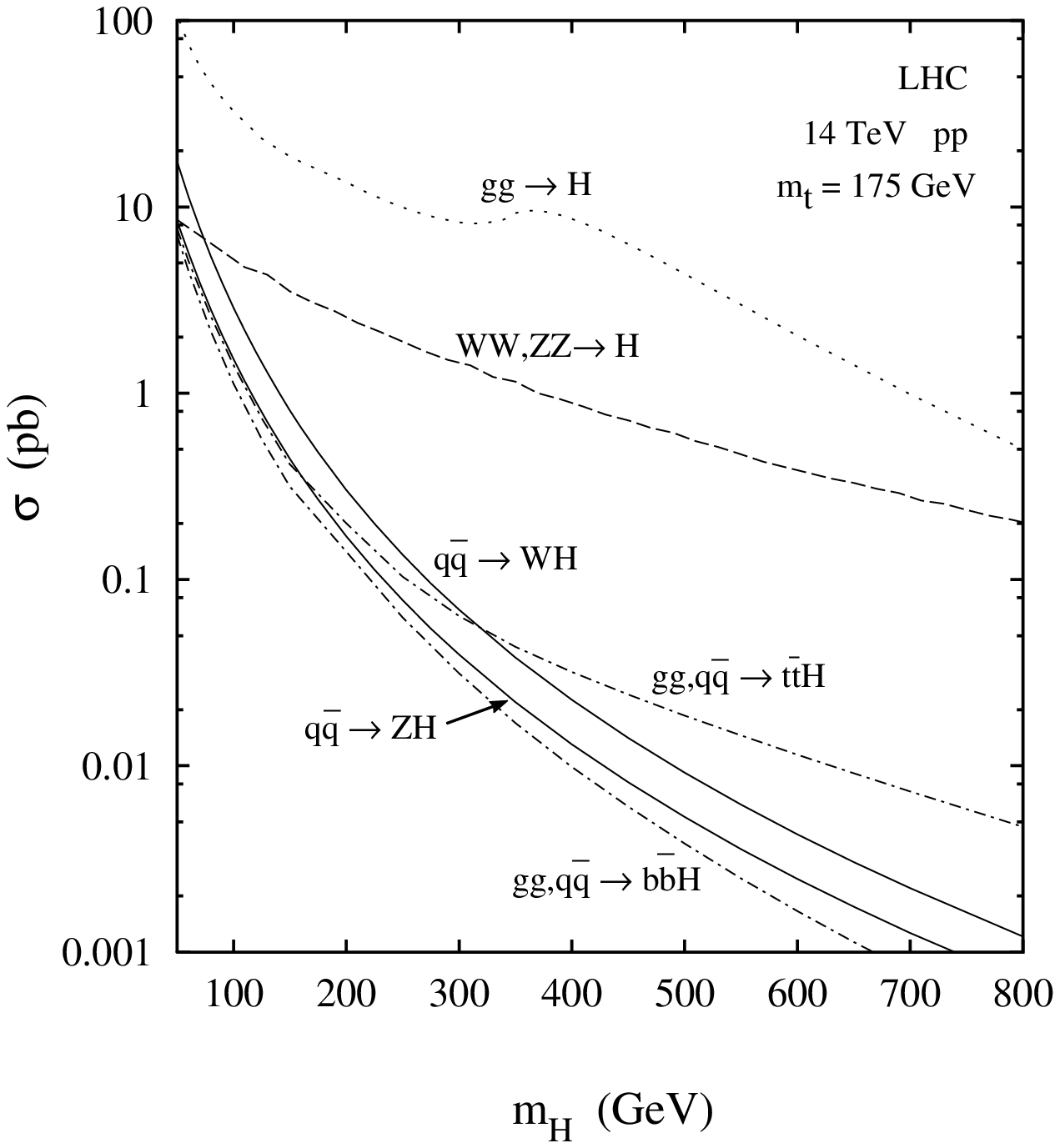,width=12.2cm}}
\begin{minipage}{12.5cm}       
\caption[fake]{Cross sections for the production of the standard Higgs boson 
at the
LHC ($\sqrt s =$ 14 TeV) vs.\ the Higgs-boson mass. The cross sections include
the QCD correction factors listed in Table~\ref{K}. 
In this figure we use $H$ to denote the SM Higgs boson.}
\label{lhc14}
\end{minipage}
\end{center}
\end{figure}

The total cross sections for various Higgs-boson production processes at 
the LHC are 
shown in Fig.~\ref{lhc14} as a function of the Higgs mass.  The dominant cross 
section is $gg\to \hsm$ via a top-quark loop \cite{GGMN}; 
we assume $m_t = 175$ GeV here and throughout.  The next largest cross 
section is from $WW,ZZ \to \hsm$, where the (virtual) $W,Z$ bosons are 
radiated from the incoming quarks and antiquarks \cite{CD}.  For 
relatively light Higgs bosons, two other processes are important.  The 
production of the Higgs boson in association with a $W$ or $Z$ boson is 
the analogue of the \lepii\ process $e^+e^- \to Z\hsm$ \cite{GNY}.  The Higgs 
boson can also be produced in association with a heavy quark-antiquark pair, 
either $t\anti t$ \cite{K} or $b\anti b$ \cite{DW}.  

\begin{table}[hbt]
\caption[fake]{Approximate QCD correction factors ($K$ factors) for various 
Higgs boson
production cross sections at the LHC. The leading-order (LO) cross section
was computed with LO parton distribution functions, while the 
next-to-leading order (NLO) cross section was computed with NLO parton 
distribution functions. The scale $\mu$ is the common 
renormalization and factorization scale. The $K$ factors are from 
Ref.~\cite{GSZ} ($gg\to \hsm$), Ref.~\cite{HVW} ($VV\to \hsm$), 
and Ref.~\cite{HW} ($q\anti q \to V\hsm$).}
\begin{center}
\begin{tabular}{c|cc}
& $\mu$ & $K$ factor \\ \hline
$gg\to \hsm$ & $\mhsm$ & 1.6 \\
$VV\to \hsm$ & $M_V$ & 1.1 \\
$q\anti q \to V\hsm$ & $M_{V\hsm}$ & 1.2 \\
\end{tabular}
\end{center}
\label{K}
\end{table}

The cross sections in Fig.~\ref{lhc14} were calculated at next-to-leading 
order (NLO) in QCD, except for $t\anti t \hsm$ and $b\anti b \hsm$, 
where such a calculation is 
lacking.  The QCD correction may be approximated by an overall 
multiplicative factor, often called a ``$K$ factor''.  The $K$ factor depends 
on both the factorization and renormalization scales; we set these scales 
equal, as is conventional. The most 
meaningful way to report the $K$ factor is as the ratio of the NLO cross 
section convoluted with NLO parton distribution functions, to the LO cross 
section convoluted with LO parton distribution functions; this ratio is 
renormalization- and factorization-scheme independent.  We list the $K$ factors 
in Table~\ref{K}, along with the common renormalization and 
factorization scale used.  The parton distribution functions used are from the 
set CTEQ2 \cite{CTEQ}.

The Higgs mass range can be divided into three separate regions, with 
different search strategies.  A heavy Higgs boson, $2\mz < \mhsm < 700$ GeV,
is searched for principally via the ``gold-plated'' mode, $gg \to \hsm \to ZZ 
\to \ell^+\ell^-\ell^+\ell^-$.  The intermediate-mass region,
80 GeV $< \mhsm < 2\mz$, divides into two segments.  For $120 \gev < \mhsm < 
2\mz$, the principal search mode is the same as the ``gold-plated'' mode,
but with one $Z$ boson off shell \cite{GKWudka}.  The mode $gg \to \hsm \to 
\gam\gam$ is also viable in this mass region.  
The ``light'' intermediate-mass region, 
$80 \gev < \mhsm < 120 \gev$, is the most challenging.  The $\gam\gam$ decay 
mode, using $gg \to \hsm$ as well as $W\hsm$ and $t\anti t\hsm$ production
\cite{GMP}, has long been
thought to be the best signal in this region. However, recent work 
indicates that the decay mode $\hsm \to b \anti b$, where the Higgs is produced
via $W\hsm$ \cite{SMW2,FR,MK,ABC} and $t\anti t\hsm$ \cite{DGV1}, may also 
be viable in part of this same mass range. A 
combination of signals from various modes may be necessary to discover a 
Higgs boson in the ``light'' intermediate-mass region. 

We divide our discussion into the three Higgs mass regions.  Wherever 
possible, we quote the results of the recent studies performed for the
ATLAS \cite{ATLAS} and CMS \cite{CMS} Technical Proposals.  

\subsubsection{{\boldmath$80 \gev < \mhsm < 120 \gev$}}

\indent\indent The branching ratio of $\hsm\to \gam\gam$ exceeds 
$0.5\times 10^{-3}$ 
for $80 \gev < \mhsm < 155 \gev$ (see Fig.~\ref{smbr}), 
and the cross section is sufficiently large in this region that
thousands of $gg\rta\hsm\rta\gam\gam$ 
events are produced in $100 \fbi$.  However, there is 
a large irreducible background from $q\anti q, gg \to \gam\gam$.  The 
background from $jj$ and $\gam j$, where the jet fakes a photon, can be 
reduced to a manageable level via hadronic activity, 
calorimetric isolation, shower shape, and $\gam/ \pi^0$ 
discrimination. The Higgs 
appears as a narrow bump in the $\gam\gam$ invariant-mass spectrum, so the 
statistical significance of the signal depends on the ability to measure 
photon energies and angles, which is detector dependent.  For $100 \fbi$, 
ATLAS claims sensitivity to $110 \gev < \mhsm < 140 \gev$, while 
CMS claims coverage of $85 \gev < \mhsm < 150 \gev$. CMS benefits from an
outstanding electromagnetic calorimeter made from PbWO$_4$ (lead 
tungstate). The low end of the mass range
is the most challenging, due to the small branching ratio and the large 
backgrounds. ATLAS requires five years of running at peak luminosity 
($500 \fbi$) to reach down to a Higgs mass of 80 GeV.  
ATLAS claims a Higgs mass 
resolution of 1.4 GeV for $\mhsm= 100$ GeV, and similar resolution for other 
Higgs masses.  CMS claims a resolution of 0.87 GeV at high luminosity, 
and 0.54 GeV at low luminosity, for $\mhsm = 110$ GeV.  
Thus the Higgs mass will be
measured with very good precision if the Higgs is detected in its 
two-photon decay mode.

The associated-production processes $W\hsm, t\anti t\hsm$, 
followed by $\hsm \to 
\gam\gam$, are also useful \cite{GMP}.  The number of signal events is 
small, due to the relatively small cross sections and the small branching 
ratio.  However, the background is greatly reduced by requiring an isolated
charged lepton from either the $W$ boson or the top quark. The irreducible 
backgrounds are $W\gam\gam$ and $t\anti t\gam\gam$, respectively.  Unlike 
the process $gg \to \hsm \to \gam\gam$, the signal is comparable to the 
background, so the $\gam\gam$ invariant-mass resolution is not as 
important. The charged lepton also helps indicate the primary vertex, 
which aids in the reconstruction of the two-photon invariant mass.
ATLAS finds about 15 signal events in $100 \fbi$ in this mass 
region, yielding a signal with a significance of 4$\sigma$
for $80\lsim\mhsm\lsim 120\gev$.  CMS finds 
a somewhat better signal with their ``shashlik''-type electromagnetic 
calorimeter, which has an invariant-mass resolution of about 1 GeV.  
The statistical significance of the signal is at the $6-7\sigma$
level for $80\lsim\mhsm\lsim 120\gev$.  
The PbWO$_4$ calorimeter improves the invariant-mass 
resolution by about a factor of two, and hence the significance of the 
signal by about a factor of $\sqrt 2$. However, this conclusion
has been questioned in Ref.~\cite{FGR}.

CMS has studied the possibility of detecting the Higgs boson produced in 
association with two (or more) jets, followed by $\hsm\to \gam\gam$.  
The signal 
contains not only $gg \to \hsm$ with jet radiation, but also weak-vector-boson 
fusion, $W\hsm$ and $Z\hsm$, and $t\anti t \hsm$.  Although the number of signal 
events and the significance of the signal is decreased with respect to the 
pure $\hsm\to \gam\gam$ search, the signal to background ratio is greatly 
improved, and is of order unity.  With $100 \fbi$, CMS claims coverage of
the region $70 \gev < \mhsm < 150 \gev$ via this mode, using the PbWO$_4$
calorimeter. 

The CDF collaboration has established high-efficiency 
$b$-tagging via secondary vertices with a silicon vertex detector (SVX) in a 
hadron-collider environment.  This allows the possibility to detect the 
``light'' intermediate-mass
Higgs boson in its dominant decay mode, $\hsm\to b\anti b$ 
\cite{SMW2,FR,MK,ABC,DGV1}.  Below we summarize the results of 
Ref.~\cite{FR}, which is the most complete analysis and is used by the 
ATLAS collaboration.  

\begin{figure}[htbp]
\let\normalsize=\captsize   
\begin{center}
\centerline{\psfig{file=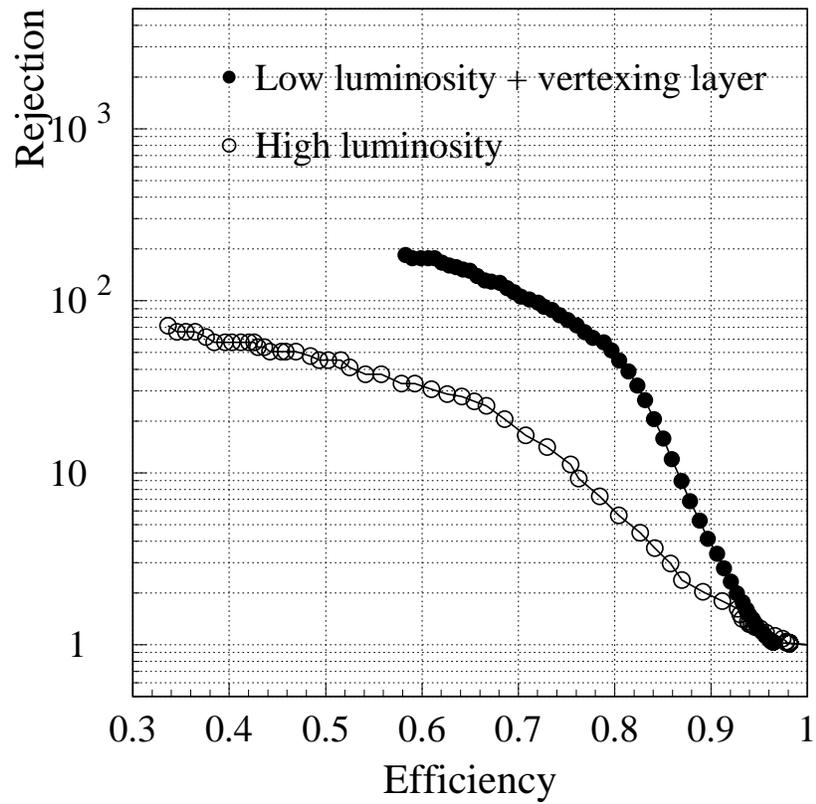,width=12.2cm}}
\begin{minipage}{12.5cm}       
\caption[fake]{Rejection against prompt jets as a function of the
efficiency for tagging $b$-jets in the ATLAS inner detector at low
(closed circles) and high (open circles) luminosity.
}
\label{atlasbtag}
\end{minipage}
\end{center}
\end{figure}

Early on, CDF achieved a $b$-tagging efficiency of $30\%$, with a $1\%$ 
misidentification of light-quark and gluon jets as $b$ jets, using a 
silicon-strip vertex detector \cite{CDF}. The efficiency has recently been 
improved to about $45\%$, maintaining the same misidentification rate 
\cite{TML}. One can anticipate higher efficiency with 
pixel detectors, which are planned for ATLAS.  The results from
Refs.~\cite{ATLAS,fgianotti} are shown in Fig.~\ref{atlasbtag}. 
At low luminosity,
an efficiency of $60\%$ with a fake rate of $1\%$ can be achieved.
At high luminosity $50\%$ efficiency is possible
with $2\%$ mis-tag probability. The decrease in the high luminosity
case arises if the innermost pixel layer, located 4 cm from the beam line,
must be removed because of irradiation problems. 
The additional track density from multiple interactions
at high luminosity is not a problem, since it is
significantly smaller than the total track density inside the $b$ jet 
\cite{ATLAS}.  Soft leptons may also be used to tag $b$ jets which undergo
semileptonic decay, with an efficiency of about $20\%$ and a fake rate of 
$1\%$.  We shall quote results based on $\ebtag=0.6,\emistag=0.01$
and $\ebtag=0.5,\emistag=0.02$ at low and high luminosity, respectively.

\bigskip

The simplest signal is $q\anti q \to W\hsm$, with $W \to \ell\nu$ and $\hsm\to 
b\anti b$, where the charged lepton is used as a trigger.  The best 
significance is attained if one tags both $b$ jets.  This process has 
a large number of backgrounds.  The irreducible backgrounds are $Wb\anti 
b$ and, for $\mhsm \sim \mz$, $WZ$ with $Z\to b\anti b$.  The latter 
background can be normalized via the leptonic decay modes of the weak 
vector bosons.  The irreducible 
background $q\anti q \to t\anti b \to Wb\anti b$ is small at the LHC.  The
dominant reducible backgrounds are $Wjj$, where the jets fake $b$ jets, and
$t\anti t \to W^+W^-b\anti b$, where one $W$ boson is missed.  This last 
background, before rejecting the extra $W$ boson, is much larger than the 
signal.  An event which contains an extra isolated charged lepton, or an
extra jet of $p_T > 15$ GeV, is rejected; this brings the $t\anti t$ 
background down to a manageable level.  Such a low $p_T$ threshold is 
impossible to implement at peak luminosity due to the pile-up of 
minimum-bias events.  A higher threshold allows too much of the $t\anti t$
background to survive.  Thus the $W\hsm$, with $\hsm\to b\anti b$ channel is
likely to be useful only at a luminosity of $10^{33}/cm^2/s$ or below.

The second production process with which to observe the 
decay $\hsm\to b\anti b$ is $gg \to t\anti t \hsm$ \cite{DGV1}. 
The cross section for this process is comparable 
to $q\anti q \to W\hsm$. Including the top-quark decays, the final state is
$W^+W^-b\anti b b\anti b$.  Events are accepted if they contain a lepton from
a $W$ decay plus
three $b$ jets.  The irreducible backgrounds are $t\anti t b\anti b$ and 
$t\anti t Z$, with $Z \to b\anti b$; the latter is small, so the mass region
$\mhsm \sim \mz$ is not particularly difficult.  The reducible backgrounds 
are $t\anti t j$, $Wjjj$, and $Wb\anti b j$, where the jets fake a $b$ jet.
The signal process itself is also a background, as one does not know which
pair of $b$ jets to select to reconstruct the Higgs mass.  This 
combinatoric background is minimized if one chooses the two lowest $p_T$ jets
as the Higgs candidate.

In the latest results, summarized in Fig.~3 of Ref.~\cite{fgianotti},
it is found that at low luminosity ($30 \fbi$ per detector) combining
ATLAS and CMS data will lead to an observable ($\geq 5\sigma$)
signal for the $W\hsm$ with $\hsm\to b\anti b$ 
process for $\mhsm\lsim 105\gev$. By combining ATLAS and CMS
data and the $W\hsm$ ($\hsm\rta b\anti b$) and inclusive $\hsm\to\gam\gam$
signals, $L=60\fbi$ (combined ATLAS+CMS luminosity)
yields a $5\sigma$ signal for $\mhsm\leq 150\gev$.
At $L=100\fbi$, 
$t\anti t \hsm$ (with $\hsm \to b\anti b$), and $W\hsm,t\anti t\hsm$ (with 
$\hsm\to \gam\gam$), ATLAS alone can cover all of the difficult region 
$80 \gev < \mhsm < 120 \gev$ \cite{ATLAS}.

Aside from its discovery potential, the decay mode $\hsm\to b\anti b$ is 
important to establish the coupling of the Higgs to fermions.  Furthermore,
the $W\hsm,t\anti t\hsm$ with 
$\hsm \to b\anti b$ channels can be studied during the initial 
low-luminosity running of the LHC, and may provide the first sighting of 
the Higgs boson.  
  
The process $t\anti t\hsm$ with $\hsm \to b\anti b$ and semileptonic decay 
of both the top and anti-top quarks has been 
suggested to produce an observable signal in the two-jet invariant mass 
distribution without $b$ tagging \cite{GKW}.  The main background is from 
$t\anti tjj$, which is much larger than the signal. 
This signal was also considered in Ref.~\cite{DGV1}, with a negative 
conclusion.  It has not been studied by ATLAS and CMS.  
  
\subsubsection{{\boldmath $120 \gev < \mhsm < 2\mz$}}

\indent\indent The process $gg\to \hsm\to Z\zstar \to \ell^+\ell^-\ell^+\ell^-$ 
is the most 
reliable mode for detection of the ``heavy'' intermediate-mass Higgs boson.
The branching ratio for this decay mode drops rapidly with 
decreasing Higgs mass, to $1\%$ for $\mhsm=120$ GeV. The irreducible 
backgrounds, from $q\anti q, gg  \to Z\zstar,Z\gam^* \to 
\ell^+\ell^-\ell^+\ell^-$, are small.  The dominant 
reducible backgrounds are from $t\anti t \to W^+W^-b\anti b$ and $Zb\anti b$, 
where the weak bosons and $b$ quarks decay to charged leptons. Charged 
leptons also arise from cascade decays of the $b$ quarks. These 
backgrounds can be reduced to a negligible level
via lepton isolation and impact-parameter cuts. 
The lepton isolation cut requires a 
higher threshold at peak luminosity, and the signal efficiency is 
thereby decreased. The significance of the signal is therefore only slightly
improved for one year of running at peak luminosity versus three years at
$10^{33}/cm^2/s$.  CMS concentrates on the four-muon channel at 
peak luminosity. Both ATLAS and CMS find that the smallest Higgs mass that can 
be reached with $100 \fbi$
is about 130 GeV.  To reach as low as 120 GeV requires several years of 
running at peak luminosity. 
The signal significance generally increases 
with increasing Higgs mass, except for a dip near $\mhsm=$ 170 GeV, where the
decay $\hsm\to W^+W^-$ suppresses the branching ratio of $\hsm\to Z\zstar$
(see Fig.~\ref{smbr}).
However, even a Higgs at this mass yields a detectable signal with $100 
\fbi$. ATLAS finds that the Higgs mass resolution in this mode
varies from 1.6 GeV to 2.2 GeV for $\mhsm = 120 - 180$ GeV, so the Higgs mass
measurement is very precise via this decay mode. 

The branching ratio of $\hsm \to \gam\gam$ reaches its maximum at about 125 
GeV. 
This process is discussed in the previous section.  ATLAS claims discovery
in this channel for a Higgs mass up to 140 GeV for $100 \fbi$, with a mass
resolution of 1.7 GeV.  CMS claims discovery up to 150 GeV for $100 
\fbi$, and up to 150 GeV via the $jj\gam\gam$ mode.
By combining ATLAS and CMS data, a $\gsim 7\sigma$ inclusive $\gam\gam$
signal is seen for $\mhsm\lsim 150\gev$ and $L=100\fbi$ per detector
\cite{fgianotti}.

Observation of both $\hsm \to \gam\gam$ and $\hsm \to Z\zstar$ would test the 
relative coupling of the Higgs to $WW$ and $ZZ$, since the $\hsm \to \gam\gam$
amplitude is dominated by a $W$ loop. These couplings are expected to be 
related as in the standard Higgs model, since this follows from custodial 
SU(2).  The $\hsm \to \gam\gam$ amplitude receives 
contributions from all electrically-charged heavy particles which obtain their 
mass via the 
Higgs mechanism, so a large deviation from the expected amplitude would
suggest the presence of such heavy particles. Should 
any of these particles carry color, they would also contribute to the 
production amplitude $gg \to \hsm$. If these heavy particles
are fermions they reduce the $\hsm\rta \gam\gam$ decay width and branching
ratio, but enhance the $gg\to \hsm$ coupling and cross section.
The resulting $gg\to\hsm\to\gam\gam$ production rate can be either
smaller or larger than in their absence; see, for example, 
Ref.~\cite{jggeer}.

The process $\hsm \to W\wstar \to \ell^+\nu\ell^-\nu$ has been suggested as a 
signal for the ``heavy'' intermediate-mass Higgs \cite{GOW}.  The 
Higgs signal manifests itself as an excess of dilepton events. The 
signal-to-background ratio is near unity.  However, the broad signal peak 
lies near the threshold for real $WW$ production, and is difficult to 
extract. This process has not been studied by ATLAS or CMS.

\subsubsection{{\boldmath$2\mz < \mhsm < 700 \gev$}}

\indent\indent The process $gg\to \hsm\to ZZ \to \ell^+\ell^-\ell^+\ell^-$ is 
the so-called ``gold-plated'' mode, due to its striking signal and small  
irreducible background (from $q\anti q, gg \to ZZ$).  ATLAS claims detection 
of this mode up to a Higgs mass of 500 GeV even at low
luminosity ($10 \fbi$); CMS claims detection up to 400 GeV.  Higher masses 
require higher luminosity; with $100 \fbi$, ATLAS claims a  
Higgs mass as high as 800 GeV can be attained, while CMS is confident up 
to 600 GeV and a bit above.  The irreducible 
background is non-negligible at these masses.  
For $\mhsm > 700$ GeV the search strategies become more sophisticated; see 
the report of the working group on ``Strongly-Coupled Electroweak
Symmetry Breaking'' \cite{hanreport}.  

The width of the Higgs boson becomes greater than the  
$\ell^+\ell^-\ell^+\ell^-$ invariant-mass resolution for a sufficiently 
heavy Higgs boson, allowing a measurement of this quantity. 
A 300 GeV Higgs boson has a width of about 8 GeV, and the width grows 
rapidly with increasing Higgs mass (see Fig.~\ref{hwidths}).  
An approximate formula for a heavy 
Higgs boson is $\Gamma_{\hsm} (\tev) =  \frac{1}{2}[\mhsm(\tev)]^3$. 
A direct measurement of the width would allow a determination of the 
$\hsm ZZ$ and $\hsm WW$ couplings 
(assuming they are related via custodial SU(2) symmetry), 
provided $\hsm \to t\anti t$ is kinematically forbidden or has
a relatively small width (as in the standard model). Under
these same assumptions $BR(\hsm\rta ZZ)$ could also be determined.
Since the 
$\ell^+\ell^-\ell^+\ell^-$ rate is proportional to $\Gamma(\hsm\rta
gg)BR(\hsm\rta ZZ)$ one would then be able to compute $\Gamma(\hsm\rta gg)$.
This in turn would allow the 
determination of the $\hsm t\anti t$ coupling, which is responsible for the 
$\hsm gg$ coupling (via a top-quark loop).

There are a variety of other production modes in which to detect the Higgs 
decay to weak-vector-boson pairs.  For example, for high $\mhsm$ values,
$pp\to \wp\wm jj X$ via $\wp\wm\to \hsm\to\wp\wm$ fusion 
(the initial $\wp$ and $\wm$
having been radiated from the two detected jets)
has a cross section that is a significant fraction
of that for $gg\to \hsm\to\wp\wm$.
These production modes are usually regarded as being 
relevant to the Higgs search at masses beyond the reach of the 
gold-plated mode, roughly $\mhsm > 700$ GeV.  Since this is the domain of 
the ``strongly-coupled'' Higgs, we defer a detailed discussion of these 
modes to the ``Strongly-Coupled Electroweak Symmetry Breaking'' subgroup 
\cite{hanreport}.
However, these modes can also be used for lighter Higgs masses
and potentially provide additional sensitivity to the couplings of the $\hsm$
to $\wp\wm$.

The channel $\hsm \to ZZ \to \ell^+\ell^-\nu\anti\nu$ has a six times higher 
rate than the gold-plated mode, but has a less distinctive signal.  The 
Higgs appears as a resonance above the $ZZ \to \ell^+\ell^-\nu\anti\nu$ 
background in the $p_T(\ell^+\ell^-)$ spectrum.  CMS claims a clear signal
with only $10 \fbi$ for $\mhsm=$ 500 GeV.  Since the Higgs width increases 
rapidly with increasing Higgs mass, the signal becomes very broad 
and difficult to recognize above the background at higher masses.  Forward 
jet tagging suppresses the background while maintaining most of the signal
from vector-boson fusion, and yields observable signals up to $\mhsm =$ 800 
GeV with $100 \fbi$.

For $\mhsm > 2 m_t$, the decay mode $\hsm \to t\anti t$ becomes available.  
In the standard model, the branching ratio is at most $20\%$ (for
$\mhsm =$ 500 GeV). The signal is swamped by the
irreducible background from $gg \to t\anti t$ \cite{DSW}.

\subsection{10 TeV LHC}

\indent\indent The LHC was approved by the CERN Council in December 1994.  
One of the 
provisions is that the LHC will be staged, with the first stage being a 
machine of 10 TeV energy and peak luminosity $10^{33}/cm^2/s$.  
Detailed results for the Higgs
discovery potential of such a machine are not yet available. Here we restrict
our comments to very basic observations.

\begin{figure}[htbp]
\let\normalsize=\captsize   
\begin{center}
\centerline{\psfig{file=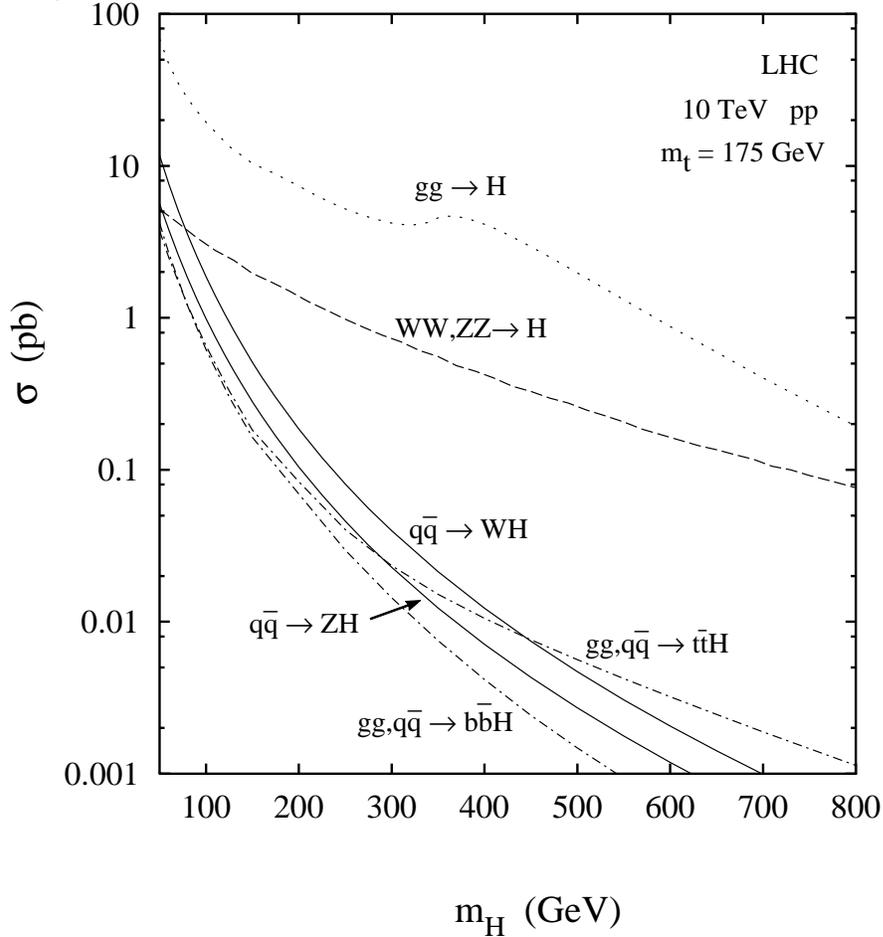,width=12.2cm}}
\begin{minipage}{12.5cm}       
\caption[fake]{Cross sections for the production of the standard Higgs boson 
at the
first-stage LHC ($\sqrt s =$ 10 TeV) vs. the Higgs boson mass. 
The cross sections include the QCD correction factors listed in Table~\ref{K}.
In this figure we use $H$ to denote the SM Higgs boson.}
\label{lhc10}
\end{minipage}
\end{center}
\end{figure}

The various cross sections for production of the standard Higgs are shown
in Fig.~\ref{lhc10}.  The biggest differences from the full 14 TeV collider
occur at the highest Higgs masses.  The dominant cross section, $gg \to \hsm$,
is reduced by about a factor of 2.5 for $\mhsm > 400$ GeV.  The Higgs mass
reach in the gold-plated mode, $\hsm \to ZZ\to \ell^+\ell^-\ell^+\ell^-$,
with $30 \fbi$ should therefore be comparable to the reach with $10 \fbi$ at
14 TeV, about 500 GeV for ATLAS and 400 GeV for CMS.
Although the $gg\rta \hsm$ cross section decreases by only
about 40\% in the intermediate mass region, this decrease must
be combined with the lower luminosity anticipated.  If we assume 
an integrated luminosity of $L=30\fbi$, then
we estimate that, in the $\hsm\rta\gam\gam$
mode, ATLAS would at best achieve a $\sim 3\sigma$ signal
for $\mhsm\sim 120\gev$, whereas with the high
resolution PBWO$_4$ calorimeter CMS would achieve a $5\sigma$ signal
(and not much more) 
for roughly $105\lsim\mhsm\lsim 140\gev$. In the $Z\zstar\rta 4\ell$           
mode, we estimate that for $L=30\fbi$ discovery of the $\hsm$ at 
the $5\sigma$ level would be possible for both ATLAS and CMS for 
$140\gev\lsim \mhsm\lsim 2\mz$, 
except possibly in the vicinity of $\mhsm\sim 170\gev$.
Thus, Higgs discovery would just barely be possible throughout
the upper portion of the intermediate mass region
after three years of running.

\subsection{SUSY Higgs bosons}

\indent\indent Generally speaking, the SUSY Higgs bosons are more elusive than 
the standard Higgs boson at the LHC. 
There is very little of the parameter space in which all four SUSY 
Higgs bosons are observable; rather, one asks if at least one SUSY Higgs 
boson can be detected over the entire parameter space.  This appears to be
the case, using a combination of detection modes. The
early theoretical studies of this issue \cite{KZ,BBKT,GO,BCPS} 
and newer ideas (to be referenced below) have
been confirmed and extended in detailed studies by
the ATLAS and CMS detector groups \cite{ATLAS,CMS}.

\begin{figure}[htbp]
\let\normalsize=\captsize   
\begin{center}
\centerline{\psfig{file=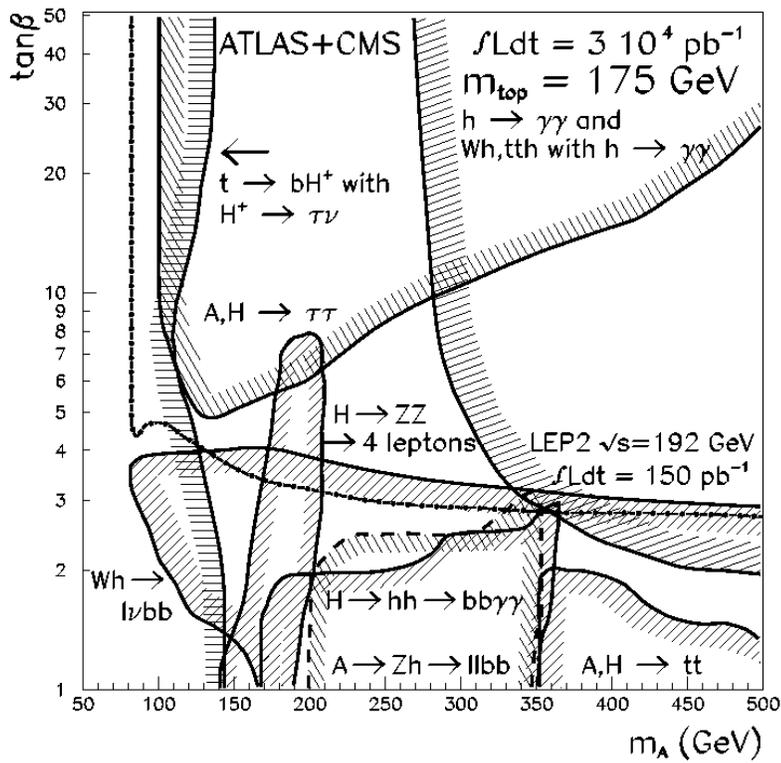,width=12.2cm}}
\begin{minipage}{12.5cm}       
\caption[fake]{Discovery contours ($5\sigma$) in the parameter space of the 
minimal supersymmetric model for ATLAS+CMS at the LHC: $L=30\fbi$.  Figure 
from Ref.~\cite{latestplots}. \Twoloop\ radiative corrections 
to the MSSM Higgs sector are included
assuming $\mstop=1\tev$ and no squark mixing.}
\label{mssmlolum}
\end{minipage}
\end{center}
\end{figure}
\begin{figure}[htbp]
\let\normalsize=\captsize   
\begin{center}
\centerline{\psfig{file=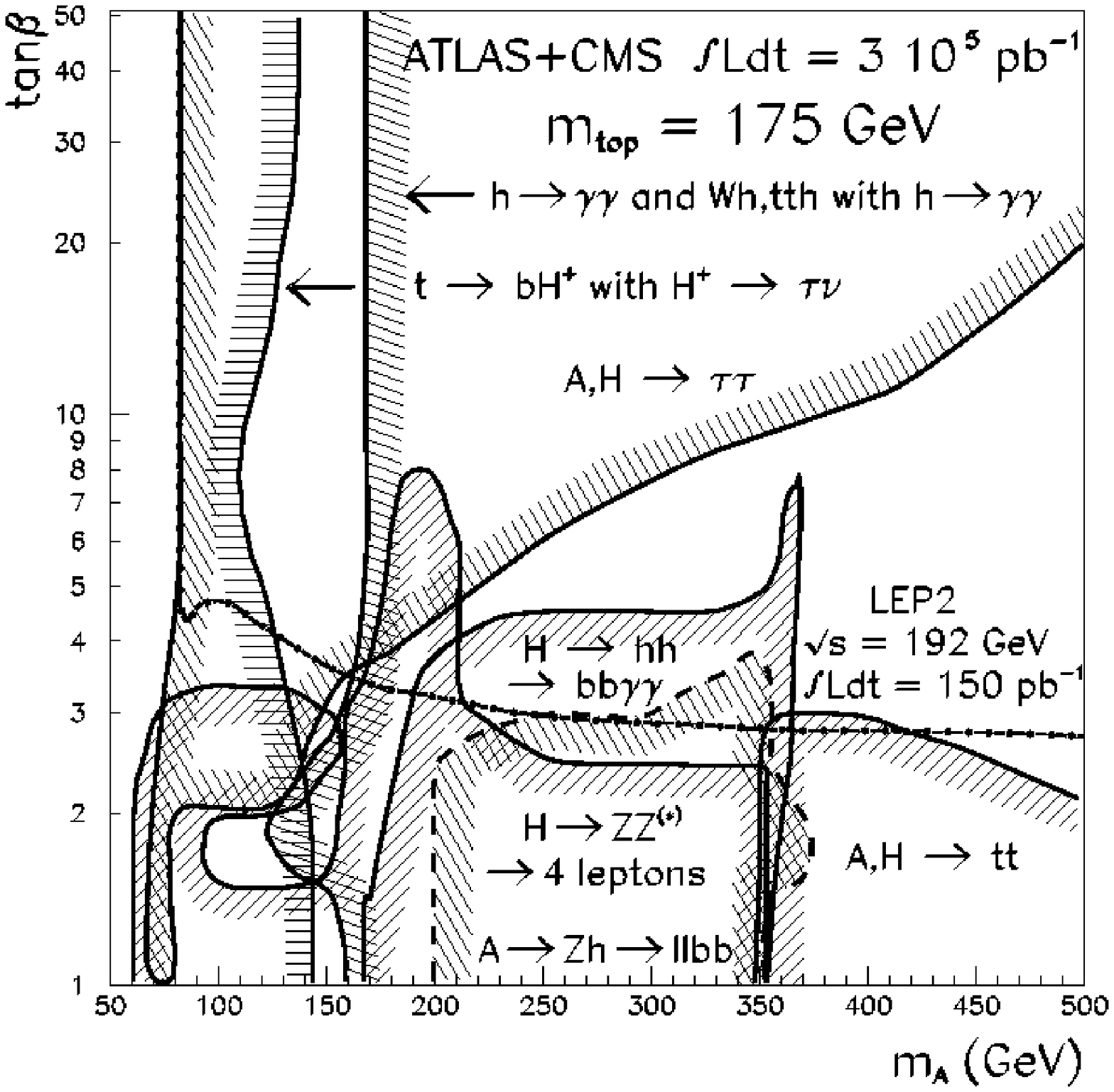,width=12.2cm}}
\begin{minipage}{12.5cm}       
\caption[fake]{Discovery contours ($5\sigma$) in the parameter space of the 
minimal supersymmetric model for ATLAS+CMS at the LHC: $L=300\fbi$.  Figure 
from Ref.~\cite{latestplots}. \Twoloop\ radiative corrections 
to the MSSM Higgs sector are included
assuming $\mstop=1\tev$ and no squark mixing.}
\label{mssmhilum}
\end{minipage}
\end{center}
\end{figure}

The first experimental detector collaboration studies of the
search for SUSY Higgs bosons by ATLAS \cite{ATLAS} and CMS \cite{CMS}
did not include the improved radiative corrections to the $\hl$
and $\hh$ masses. Thus, in the ($\mha,\tanb$) parameter space plots given
in these references, the contours for $\hl$ discovery
must be reinterpreted;  for small stop-squark
mixing the $\hl$ contours apply for a top quark {\it pole} pass $\mt$
of about $190\gev$. Certain aspects of the $\hh$ contours at lower
$\mha$ values are also sensitive to the \twoloop\ corrections.
The discovery contours have also been evolving by virtue of improvements
and alterations in the detector itself. A recent survey
of the experimental studies is contained in 
Refs.~\cite{fgianotti,latestplots}. 
Updated figures for ATLAS+CMS at low ($L=30\fbi$) luminosity and
high ($L=300\fbi$) luminosity from Refs.~\cite{fgianotti,latestplots}
are included below as Figs.~\ref{mssmlolum} and \ref{mssmhilum},
respectively. Note that the ATLAS+CMS notation
means that the signals from the two detectors are combined
in determining the statistical significance of a given signal.
Since not all modes of interest are included
on these plots, we shall also occasionally refer to the original
Technical Proposal figures on MSSM Higgs studies in Refs.~\cite{ATLAS,CMS}.
The CMS figure from Ref.~\cite{CMS} is reproduced below for
the reader's convenience.
We note that all results discussed in the following are
those obtained without including higher order QCD ``$K$'' factors
in the signal and background cross sections.  The $K$ factors
for both signal and background are presumably significant; if
they are similar in size, then statistical significances would
be enhanced by a factor of $\sqrt K$.

\begin{figure}[htbp]
\let\normalsize=\captsize   
\begin{center}
\centerline{\psfig{file=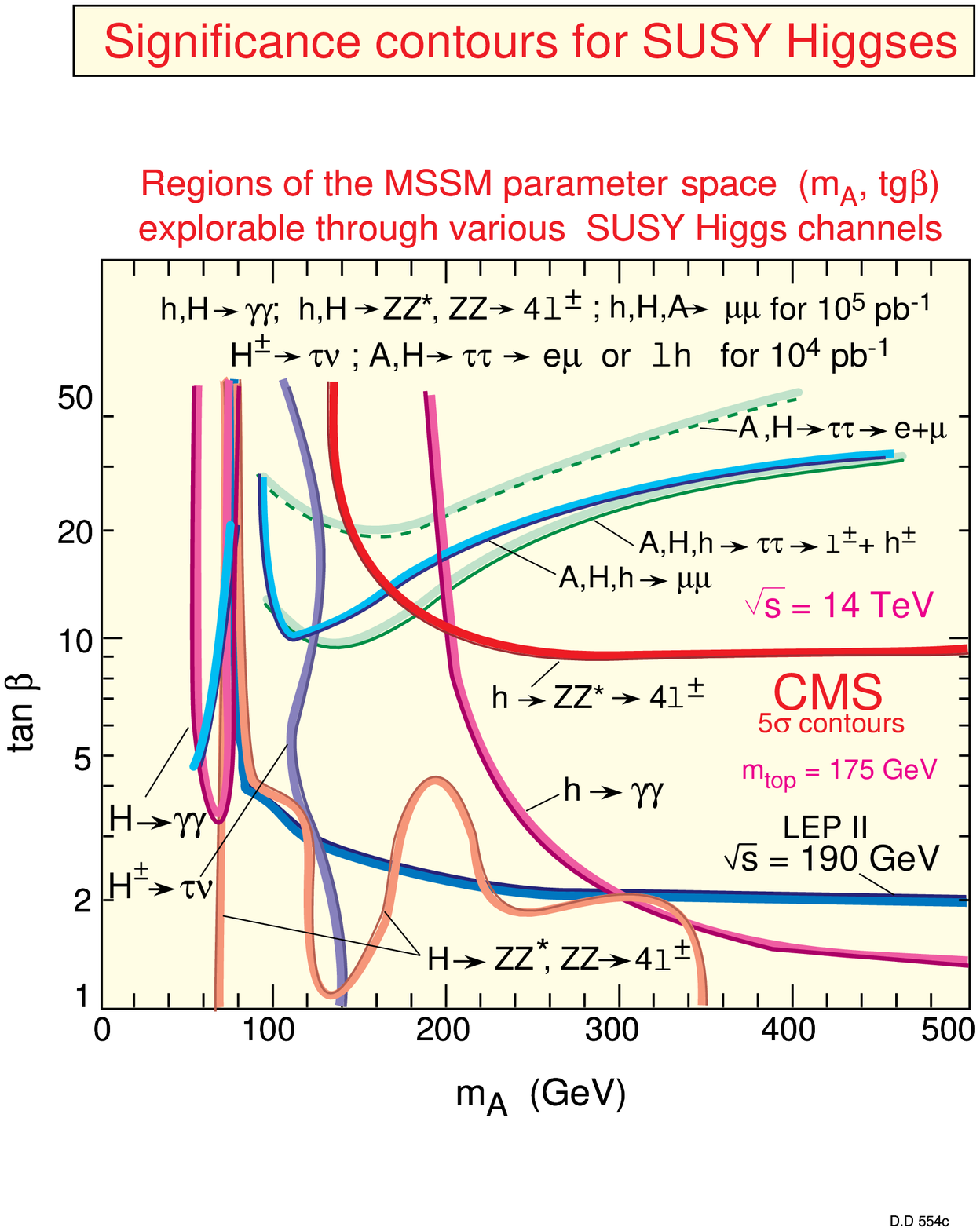,width=12.2cm}}
\begin{minipage}{12.5cm}       
\caption[fake]{Discovery contours ($5\sigma$) in the parameter space of the 
minimal supersymmetric model for the CMS detector at the LHC from
the original Technical Proposal, Ref.~\cite{CMS}.
The contours for $\hl,\hh \to \gam\gam$, $\hl,\hh \to ZZ^{(*)} \to 4\ell$, 
and $\hl,\hh,\ha\rta \mu\mu$
are shown for an integrated luminosity of $100 \fbi$.
The contours for $\hh,\ha \to \taup\taum\rta e\mu,\ell h$ 
($h=$hadron) and $t \to H^+b$ are shown for an
integrated luminosity of $10 \fbi$. Also shown is the contour for the 
discovery region of \lepii. Nominally,
this figure is for $\mt=175\gev$. However, 
after accounting for \twoloop\ Higgs mass corrections, the
$\hl$ and \lepii\ discovery 
contours of this figure should be reinterpreted as those
for a top quark pole mass of $\mt\sim 190\gev$.}
\label{cmsplot}
\end{minipage}
\end{center}
\end{figure}

In the limit $\mha \to \infty$, the $\hh$, $\ha$, and $\hpm$ are all heavy, 
and decouple from the weak bosons. The lightest neutral scalar Higgs boson, 
$\hl$, approaches its upper bound, and behaves like a 
standard Higgs boson. Since this bound (for pole mass $\mt=175\gev$)
is about 113 GeV
(assuming small stop-squark mixing and $\mstop\leq 1\tev$), the primary
channels for $\hl$ detection will be those based on
the $\gam\gam$ decay mode.  The $5\sigma$ contours are
shown in Figs.~\ref{mssmlolum} and \ref{mssmhilum}. At high luminosity
$\hl$ discovery in its $\gam\gam$ decay mode is possible
for $\mha\gsim 170$. For low luminosity the coverage of the $\gam\gam$
mode decreases substantially, reaching only down as far
as $\mha\sim 270\gev$ at high $\tanb$ with no coverage for any $\mha$
if $\tanb\lsim 2$. For top quark masses $\mt\gsim 190\gev$,
the maximum $\mhl$ mass increases to about $122\gev$,
and the $\hl$ will also be observable via $\hl \to Z\zstar \to 4\ell$ over 
an overlapping part of the parameter space (see the $\mt=175\gev$
one-loop contours in the CMS plot, Fig.~\ref{cmsplot}).

For $\tanb \sim 1$, the lightest scalar Higgs is observable at \lepii\ via
$e^+e^- \to \ha\hl,Z\hl$.  
Including \twoloop\ corrections ($\mstop=1\tev$, no squark mixing)
for $\mt=175\gev$ the LEP-192 discovery region asymptotes 
at $\tanb\lsim 3$, assuming $L=150\pbi$ per detector,
as shown in Figs.~\ref{mssmlolum} and \ref{mssmhilum}, as well
as in the earlier Fig.~\ref{175_mix2}.

For $60\lsim \mha \lsim 2\mt$ the heavy scalar Higgs has high enough mass and 
for $\tanb\lsim 3$ maintains enough of a 
coupling to weak vector bosons to allow its discovery via $\hh \to ZZ^{(*)} 
\to 4\ell$ at high luminosity, as shown in Fig.~\ref{mssmhilum}. 
The height in $\tanb$ as a function of $\mha$ 
of the $\hh\rta 4\ell$ discovery region varies significantly for $\mha\lsim
2\mt$ due to large swings in the branching ratio for $\hh\rta\hl\hl$ decays,
rising as high as $\tanb\sim 8$ for $\mha\sim 190\gev$ (where
$BR(\hh\rta\hl\hl)$ actually has a zero).
The importance of the $\hh\rta\hl\hl$ decays makes the $4\ell$
mode of marginal utility at low luminosity except for $\mha\sim 190\gev$,
see Fig.~\ref{mssmlolum}.
At high luminosity, the $\hh\rta 4\ell$ 
contour is cut off for $\mha \approx \mhh > 2 m_t$ due to the dominance
of the the decay $\hh\to t\anti t$.
The $\hh\rta\hl\hl$ and $\hh,\ha\rta t\anti t$ 
channels can also provide Higgs signals. The key ingredient in employing
these channels is efficient and pure $b$-tagging. We will discuss
these modes shortly.

For $\mha \approx 70$ GeV and $\tanb >$ 3 (CMS) or 5 (ATLAS), the heavy 
scalar Higgs has a reasonable $\gam\gam$ branching ratio and
is observable in its two-photon decay mode. This is 
indicated by the narrow vertical strip in Fig.~\ref{cmsplot} from
Ref.~\cite{CMS} (see also the similar plot in 
Ref.~\cite{ATLAS}).
(This region changes little if \twoloop\ corrections to $\mhh$
are included.)

These ``standard'' modes are not enough to cover the entire SUSY 
parameter space, so others must be considered. The uncovered region 
is for large $\tanb$ and moderate $\mha$. Since $\mhl$ is in the ``light''
intermediate mass region, the dominant decays for the light
scalar are $\hl \to b\anti b,\taup\taum$,
The coupling of 
$\hl$ to weak vector bosons is close to full strength, but its coupling to
$b\anti b$ and $\taup\taum$ is enhanced, so its branching ratio to vector boson
pairs and two photons is suppressed.  Thus $\hl$ must be sought in its decay
to $b\anti b$ or $\taup\taum$ in this region. However,
the $\taup\taum$ decay mode of the SUSY Higgs bosons has a large background 
from $Z \to \taup\taum$.  Since $\mhl < 113$ GeV for $\mt=175\gev$
(taking $\mstop=1\tev$ and assuming no squark mixing)
the lightest scalar Higgs
is too close to the $Z$ peak to be observed.  Thus only $\hh$ and $\ha$ are
candidates for observation via the $\taup\taum$ decay mode. For large $\tanb$,
the cross section for the production of these particles in association with
$b\anti b$ is greatly enhanced, and is the dominant production 
cross section \cite{DW}.

The events are triggered by the leptonic decay of one $\tau$, and the other
$\tau$ is allowed to decay hadronically to maximize its branching ratio.
Certain criteria, such as a single charged high-$p_T$ track, are imposed 
to separate the hadronically decaying $\tau$ from an ordinary jet.  This has
an efficiency of about $25\%$ for a jet rejection factor of about 400.
Events are selected with large missing $p_T$ due to the lost $\tau$ neutrinos.
The missing $p_T$ is projected
along the $\tau$ directions, given roughly by the direction of their decay
products since the $\tau$'s are moving relativistically.  This allows the
reconstruction of the $\tau$ momenta and hence the $\taup\taum$ 
invariant mass \cite{EHSV}. The $p_T^{\rm miss}$ resolution
thus directly affects the width of the reconstructed $\tauptaum$
invariant mass. Provided that the calorimetry coverage
extends up to $|\eta|\lsim 5$ (as expected for both CMS and ATLAS),
the expected $p_T^{\rm miss}$ resolution is about 6 GeV for $\ha\to\tauptaum$
events with $\mha\sim 200\gev$ and the reconstructed $\mha$ resolution
is 20 GeV at low luminosity.
Since the $p_T^{\rm miss}$ resolution is deteriorated at full luminosity, 
this search may be best carried out during low-luminosity running.
A comparison of Figs.~\ref{mssmlolum} and \ref{mssmhilum} indicates, however,
some overall improvement in the significance of the $\tauptaum$
signals by going to higher luminosity.
A similar analysis is applied to events in which both $\tau$'s decay 
leptonically, but the lower event rate leads to less statistical
significance.

The region in the ($\mha,\tanb$) plane which is covered by the 
$\hh,\ha \to \taup
\taum$ channel is shown in Figs.~\ref{mssmlolum} and \ref{mssmhilum}. 
For $L=300\fbi$ and $\mt=175\gev$ the region over
which the $\ha,\hh\rta \tau\tau$ discovery channel  is viable
extends all the way down to $\tanb=1$ for $\mha\sim 70\gev$, rising
to $\tanb\sim 20$ by $\mha\sim 500\gev$. (For $\tanb\lsim 2$, the 
$gg\rta \ha\rta \tau\tau$ reaction provides the crucial contribution to
this signal.) This, in particular, means that discovery
of the $\hh,\ha$ will be possible for $80\gev\lsim\mha\lsim 160\gev$ 
and $\tanb\gsim 4$
where the $\hl\to\gam\gam$ modes are not viable and $Z\hl$
production cannot be observed at \lepii.

CMS has explored the decay modes $\hl,\hh,\ha \to \mu^+\mu^-$ for large $\tanb$.
Although the branching ratio is very small, about $3 \times 10^{-4}$, the 
large enhancement of the cross section for $b\anti b\ha$ 
and either $b\anti b \hh$ (high $\mha$) or $b\anti b \hl$ (low $\mha$)
compensates.
The main background is Drell-Yan production of $\mu^+\mu^-$.  The resulting
discovery contours with $100 \fbi$ of integrated luminosity are shown in 
the CMS contour figure, Fig.~\ref{cmsplot}. Very roughly, $\tanb\gsim 10$
is required for $\mha\sim 100\gev$, rising to $\tanb\gsim 30$ by
$\mha\sim500\gev$. The $\mupmum$ contours
are close to the $\tauptaum$ contour obtained with $L=10 \fbi$, but the
$\mu^+\mu^-$ channel yields a cleaner signal identification and better mass 
resolution. Nonetheless, by comparing the above-referenced $\mupmum$
contours to Figs.~\ref{mssmlolum} and \ref{mssmhilum},
we see that even for $L= 30\fbi$, the $\taup\taum$ mode will
probe to lower $\tanb$ values at any given $\mha$.
Both the $\tauptaum$ and $\mupmum$ signals can be enhanced by tagging
the $b$ jets produced in association with the Higgs bosons.
It will be interesting to see how the $\mupmum$ and $\tauptaum$
modes compare once $b$-tagging is required.

The charged Higgs boson of the minimal supersymmetric model is best sought in
top-quark decays, $t\to \hp b$. For $\tanb >1$, the branching ratio 
of $\hp \to \taup\nu_{\tau}$ exceeds $30\%$, and is nearly unity for 
$\tanb > 2$.  CMS and ATLAS have studied the
signal from $t\anti t$ events with one semileptonic top decay and one top
decay to a charged Higgs, followed by $\hp \to \taup\nu_{\tau}$ . The main 
irreducible background is from top decay to a $\tau$ lepton through a $W$ boson.
This can be normalized by measuring the top semileptonic decay rate and using
lepton universality.  Reducible backgrounds are rejected by tagging one or
both $b$ jets in the signal.  CMS and ATLAS find that, with $30 \fbi$ of 
integrated luminosity, a charged Higgs of mass less than about 140 GeV can be
detected for all values of $\tanb$, extending to $\lsim 160\gev$ 
at low or high $\tanb$ values,
in the case of a top-quark pole mass of $175\gev$.  This is
indicated by the approximately vertical contours that begin
at $\mha=140\gev$ at $\tanb=1$ in Figs.~\ref{mssmlolum} and \ref{mssmhilum}.

The results of the Technical Proposals \cite{ATLAS,CMS} indicated that
all of these processes combined were still not enough to guarantee
detection of at least one MSSM Higgs boson throughout
the entire SUSY parameter space. The  
hole that remained in the contour plots
of Refs.~\cite{ATLAS,CMS} (see Fig.~\ref{cmsplot})
at moderate $\mha$ and $\tanb \sim 
3-10$ remains in Fig.~\ref{mssmlolum}, but disappears in Fig.~\ref{mssmhilum}
by virtue of the much more extensive coverage of the $\hl\to\gam\gam$
and $\hh,\ha\to\taup\taum$ modes at high luminosity.
The observability of the $\hh,\ha\to \tauptaum$
modes is such that $L=600\fbi$ (combining ATLAS+CMS) provides more
than adequate coverage of the entire $(\mha,\tanb)$
parameter plane.  At $L=100\fbi$ coverage is already complete.

We shall now turn to a discussion of additional detection modes
that rely on $b$-tagging.  Not only might these modes provide
backup in this `hole' region, they also expand the portions
of parameter space over which more than one of the MSSM Higgs 
bosons can be discovered. Equally important, they allow a direct
probe of the often dominant $b\anti b$ decay channel.
In the early theoretical studies quoted, it was assumed
(following Ref.~\cite{DGV1})
that an efficiency of 25\%-30\% and purity of 99\% for tagging $b$-jets
with $p_T>20\gev$ and central rapidity could be achieved. In obtaining
their most recent results, ATLAS and CMS employ
efficiency (purity) of 60\% (99\%) for $p_T\gsim 15\gev$
for low luminosity running 
and 50\% (98\%) for $p_T\gsim 30\gev$ 
for high luminosity running, obtained solely from
vertex tagging, as outlined earlier; high-$p_T$ lepton tags
could further improve these efficiencies.

The most direct way to take advantage of $b$-tagging is to employ
modes where the Higgs boson decays to $b\anti b$.
We first discuss
Higgs production in association with a $W$ boson \cite{SMW2} or $t\anti t$ 
\cite{DGV2}. Both have the potential to contribute in the hole region
when the $WW$ and $t\anti t$ couplings are of roughly standard
model strength --- the $\hl$ has approximately SM strength couplings
once $\mha\gsim \mz$, while the $\hh$ has roughly SM-like strength couplings
when $\mha\lsim\mz$ and $\mhh$ approaches its lower bound
(a more precise discussion appears in association with
Tables~\ref{hcouplingslimi} and \ref{hcouplingslimii}).  The
$W+$Higgs and $t\anti t+$Higgs processes were discussed above 
in the ``light'' intermediate-mass standard Higgs section. Recall that  
it is uncertain that the $W+$Higgs mode can be used
at high luminosity \cite{ATLAS}, but it can definitely
be employed at low luminosity, yielding a signal for $\mhsm\lsim 105\gev$
if ATLAS and CMS data are combined assuming that
the two detectors have similar capabilities in this channel.
ATLAS finds that the $t\anti t \hsm$ with $\hsm \to b\anti b$ process 
{\it can} be employed at high luminosity.  New results
are not yet available, but the Technical Proposal \cite{ATLAS} claims
coverage up to $\mhsm \sim 100$ GeV with $100 \fbi$ of integrated luminosity.
With $600\fbi$ of luminosity for ATLAS+CMS, this would be
extended to at least $\mhsm\sim 120\gev$.

Coming to the MSSM Higgs, we first note that for $\tanb>1$
the $b\anti b$ coupling of the $\hl$ remains somewhat
enhanced until $\mha$ becomes very large (see Fig.~\ref{couplingstanb5}), 
implying an enhanced value for $BR(\hl\rta b\anti b)$ compared to the $\hsm$.
The $\hl\wp\wm$ and $t\anti t\hl$ couplings reach more-or-less
full strength by $\mha\gsim 100\gev$, implying an enhanced
overall rate for $W\hl$ and $t\anti t\hl$ with $\hl\to b\anti b$ 
once $\mha\gsim 100\gev$ out to fairly large $\mha$.
Thus, based on the results of the Technical Proposals,
the discovery region for the $W\hl$ and $t\anti t\hl$ modes 
should certainly extend to $\mhl$ values at least as large as
the roughly 105 GeV ($L=30\fbi$) and 120 GeV ($L=600\fbi$) limits
for the two modes, respectively, found for $\mhsm$ in the standard model case.
Since $\mhl$ is below $113\gev$ (unless
$\mstop> 1\tev$ and/or squark mixing is large),
see Fig.~\ref{rcmasses}, and since the `hole' region is at moderate $\mha$,
we see that the $W\hl$ and $t\anti t\hl$ modes both are likely to allow $\hl$
detection in the `hole' in the SUSY parameter space.  

The experimental studies of the $t\anti t\hl$ (with $\hl\to b\anti b$)
mode have not been refined
to the point that a corresponding contour has been included
in Figs.~\ref{mssmlolum} and \ref{mssmhilum}. The theoretical results
\cite{DGV2} claim substantial coverage in the hole region
even for the somewhat pessimistic $b$-tagging assumptions employed
in the study. (The radiative corrections were also done there
at one loop, implying larger $\mhl$ values than found 
at the \twoloop\ level.)
Thus, there is considerable cause for optimism.
The impact of the $W\hl$ (with $\hl\rta b\anti b$) mode has been examined
in Refs.~\cite{fgianotti,latestplots}.
The coverage provided by this mode for $L=30\fbi$
after combining the ATLAS signal with a presumably equal
signal from CMS is illustrated in Fig.~\ref{mssmlolum}.
There, the $W\hl$ mode is shown to cover most of the $\mha\gsim 100\gev$,
$\tanb\lsim 4$ region, where $\mhl\lsim 105\gev$ (Fig.~\ref{rcmasses}). 
Unfortunately, it seems that the experimental analysis does
not find enough enhancement for the $\hl$ rate relative
to the $\hsm$ rate in this channel as to
provide backup in the `hole' region of the low-luminosity figure.
It should be noted, however, that the boundary of $\tanb\lsim 4$
is almost certainly a very soft one, depending delicately
on the exact luminosity assumed, precise \twoloop\
radiative corrections employed, and so forth. For instance, as $\mstop$
is lowered below $1\tev$, the upper bound on $\mhl$ decreases rapidly (see
Fig.~\ref{rcmasses}), and the region of viability for this mode
would expand dramatically.

\begin{figure}[htbp]
\let\normalsize=\captsize   
\begin{center}
\centerline{\psfig{file=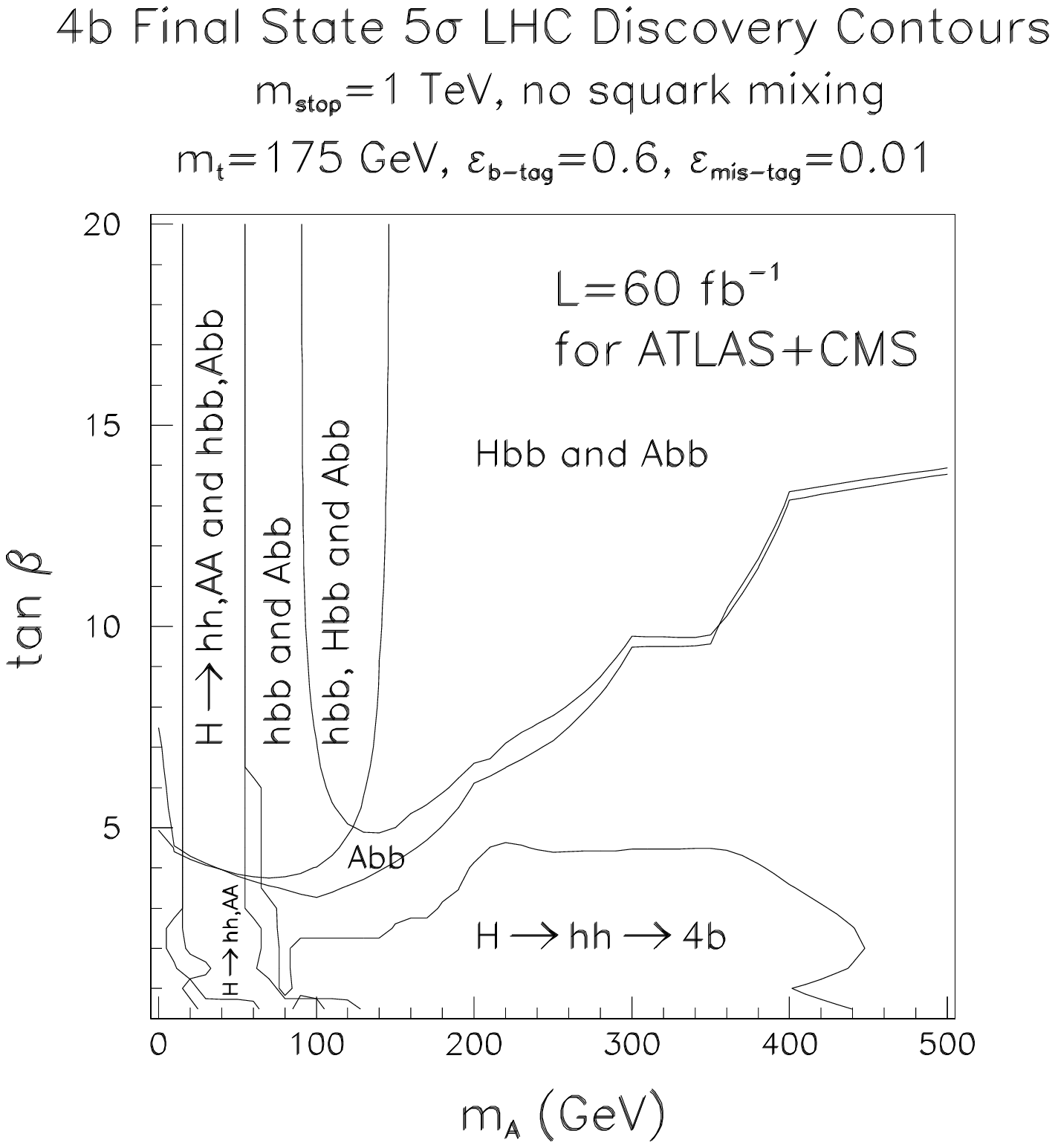,width=12.2cm}}
\begin{minipage}{12.5cm}       
\caption[fake]{$4b$ final state $5\sigma$ discovery regions
for $\hh b\anti b$, $\ha b\anti b$, $\hh\rta\hl\hl$ and $\hh\rta\ha\ha$
in $(\mha,\tanb)$ parameter space at the LHC for 
combined ATLAS+CMS luminosity of $L=60\fbi$, assuming
that an efficient $4b$ trigger can be developed.
\Twoloop\ radiative corrections to the MSSM Higgs sector
are included assuming $\mt=175\gev$,
$\mstop=1\tev$ and no squark mixing.}
\label{4bfigure}
\end{minipage}
\end{center}
\end{figure}

For large $\tanb$, the enhanced cross section 
for associated production of SUSY Higgs bosons with $b\anti b$, 
followed by Higgs decay to $b\anti b$, yields a
four $b$-jet signal \cite{DGV3}. Tagging at least three $b$ jets 
with $p_T>15\gev$ is required
to reduce backgrounds.  It is necessary to establish an efficient
trigger for these events in order to observe this signal;
this is currently being studied by the ATLAS and CMS collaborations.  
The dominant 
backgrounds are $gg \to b\anti bb\anti b$, and $gg\to b\anti bg$ with a 
mis-tag of the gluon jet. Assuming the latest $60\%$ efficiency
and 99\% purity for $b$-tagging at $L=30\fbi$,
the $b\anti b\hl$ reaction yields a viable signal
for $\mha\lsim 125\gev$, $\tanb\gsim 4-5$; $b\anti b \hh$
production probes $\mha\gsim 125\gev$ for $\tanb\gsim 5$, rising
to $\tanb\gsim 15$ for $\mha\gsim 500\gev$; $b\anti b \ha$
production will allow $\ha$ discovery throughout the region defined
by $\tanb\gsim 4-5$ at low $\mha$ rising smoothly 
to $\tanb\gsim 15$ at $\mha\gsim 500\gev$. These results, from
Ref.~\cite{dgv3update} and displayed in Fig.~\ref{4bfigure}, 
are a big improvement over
those obtained for the SDC-like $b$-tagging capabilities assumed
in Ref.~\cite{DGV3}. They imply that the $4b$ final state
could be competitive with the $\taup\taum$ final state modes
for detecting the $\hh$ and $\ha$ if an efficient trigger can be
developed for the former.

For $\mhpm > m_t + m_b$, one can consider searching for the decay of the 
charged Higgs to $t\anti b$.  This signal is most promising when used in 
conjunction with the production process $gg \to t\anti b\hm$, and tagging 
several of the four $b$ jets in the final state \cite{GBPR}. For moderate
$\tanb$, the production cross section is suppressed such that the signal is
not observable above the irreducible $t\anti tb\anti b$ background.  The 
potential of this process is therefore limited to small and large values of
$\tanb$.  With $200 \fbi$, and assuming the now-pessimistic
SDC-like $b$-tagging efficiency and purity,
a signal may be observable for $\tanb< 1.7$ and
$\mhpm < 400$ GeV, and for $\tanb>30$ and $\mhpm<300$ GeV. This
limited region of utility can be expected to expand once
the current ATLAS and CMS $b$-tagging scenarios are utilized.

CMS and ATLAS have 
considered the process $gg \to \ha \to Z\hl \to \ell^+\ell^-b\anti 
b$.  For $\tanb < 2$, the branching ratio of $\ha \to Z\hl$ is about 
$50\%$.  They have demonstrated an observable signal with single and double
$b$ tagging. In Fig.~\ref{mssmhilum} one finds a discovery region
for $200\lsim\mha\lsim 2\mt$ and $\tanb\lsim 3$ for $L=600\fbi$ (\ie\
$L=300\fbi$ for ATLAS and CMS separately),
reduced to $\tanb\lsim 2$ for $L=60\fbi$, Fig.~\ref{mssmlolum}.

Recent results from CMS and ATLAS for the mode
$\hh,\ha\rta t\anti t$ also appear in Figs.~\ref{mssmlolum} and \ref{mssmhilum}.
Even with good $b$-tagging,
the decays $\hh,\ha \to t\anti t$ are challenging to detect at 
the LHC due to the large background from $gg \to t\anti t$ (which can 
interfere destructively with the signal) \cite{DSW}. 
Nonetheless, the preliminary studies indicate that,
for the anticipated $b$-tagging capability, ATLAS and CMS
can detect $\ha,\hh\rta t\anti t$ for 
$\tanb\lsim 2-1.5$ with $L=60\fbi$ and for $\tanb\lsim 3-2.5$ with $L=600\fbi$.
Caution in accepting these preliminary results is perhaps warranted
since they have been obtained by
simply comparing signal and background cross section levels;
excellent knowledge of the magnitude of the 
$t\anti t$ background will then be required since $S/B\sim 0.02-0.1$.

The $\hh\rta\hl\hl$ mode can potentially be employed in the channels
$\hl\hl\rta b\anti b b\anti b$ and $\hl\hl\rta b\anti b \gam\gam$.
The former mode has been explored in Ref.~\cite{DGV4}; using 4 $b$-tagging
(with 50\% efficiency and 98\% purity for $p_T>30\gev$ at $L=600\fbi$)
and requiring that there be two $b\anti b$ pairs of mass $\sim \mhl$
yields a viable signal for $170\lsim\mha\lsim 500\gev$ and $\tanb\lsim 5$.
For $L=60\fbi$, $b$-tagging cuts can be softened and one
can be sensitive to lower masses.  Using 60\% efficiency
and 99\% purity for $p_T\gsim 15\gev$, one finds that $\hh\rta\hl\hl$
and/or $\hh\rta\ha\ha$ can also be detected in the region 
$\mha\lsim 60\gev,\tanb\gsim 1$. This is illustrated in
Fig.~\ref{4bfigure}. Note from this figure that
the ATLAS+CMS $b\anti b\hl$, $b\anti b\hh$, $b\anti b\ha$, $\hh\rta\hl\hl$
and $\hh\rta\ha\ha$ $4b$ final state signals at combined $L=60\fbi$ 
yield a signal for one or more of the MSSM Higgs bosons over a very
substantial portion of parameter space.

Because of uncertainty concerning the ability to trigger
on the $4b$ final state, ATLAS and CMS have examined the 
$\hh\rta\hl\hl\rta b\anti b \gam\gam$
final state.  This is a very clean channel (with $b$ tagging),
but is rate limited. For $L=600\fbi$ (Fig.~\ref{mssmhilum}) a discovery
region for ATLAS+CMS of $175\gev\lsim\mha\lsim 2\mt$, $\tanb\lsim 4-5$ is 
found (using the 50\% efficiency and 98\% purity claimed by ATLAS
at high luminosity); the region is substantially reduced for $L=30\fbi$
(Fig.~\ref{mssmlolum}). It is important to note
that when both $\hh\rta\hl\hl\rta 2b2\gam$ and $4b$ can be observed,
then it will be possible to determine the very important
ratio $BR(\hl\rta b\anti b)/BR(\hh\rta \gam\gam)$.

Putting together all these modes, we can summarize by saying that
for moderate $\mha\lsim 2\mt$ there is an excellent chance
of detecting more than one of the MSSM Higgs bosons.  However, for large
$\mha\gsim 250\gev$ (as preferred in the GUT scenarios)
only the $\hl$ is certain to be found. For $\mha\gsim 250\gev$,
the $\hl$ modes that are guaranteed to be observable are the 
$\hl\rta\gam\gam$ production/decay modes ($gg\rta\hl$,
$t\anti t\hl$, and $W\hl$, all with $\hl\rta\gam\gam$).
Even for $\mha$ values as large as $400-500\gev$, it is also likely
that the production/decay mode $t\anti t\hl$ with $\hl \rta
b\anti b$ can be observed, especially if $\mstop$
is sufficiently below $1\tev$ that $\mhl$ is $\lsim 100\gev$.  
For high enough $\mstop$, $\hl\rta Z\zstar$ might also be detected.
Whether or not it will be possible to
see any other Higgs boson depends on $\tanb$. There are basically
three possibilities when $\mha\gsim 250\gev$.
i) $\tanb\lsim 3-5$, in which case $\ha,\hh\rta t\anti t$  and 
$\hh\rta \hl\hl\rta b\anti b \gam\gam,4b$
will be observable;
ii) $\tanb\gsim 6$ (increasing as $\mha$ increases above $250\gev$),
for which $\ha,\hh\rta \taup\taum$ (and at larger $\tanb$, $\mu^+\mu^-$)
will be observable, supplemented by $b\anti b\ha,b\anti b\hh\rta 4b$
final states;
and iii) $3-5\lsim \tanb\lsim 6$ at $\mha\sim 250\gev$, increasing
to $3-5\lsim \tanb\lsim 13$ by $\mha\sim500\gev$,
which will be devoid of $\ha,\hh$ signals.
Further improvements in $b$-tagging efficiency and purity
would lead to a narrowing of this latter wedge of parameter space.

We must reiterate that the above results have assumed
an absence of SUSY decays of the Higgs bosons. For a light ino sector
it is possible that $\hl\rta \cnone\cnone$ will be the dominant decay.
Detection of the $\hl$ in the standard modes becomes difficult or impossible.
However, it has been demonstrated that detection 
in  $t\anti t\hl$ \cite{guninvis} and $W\hl$ \cite{fjk,cr} production
will be possible after employing cuts requiring large missing energy.
Assuming universal gaugino masses at the GUT scale, our first
warning that we must look in invisible modes would be the observation
of $\cpone\cmone$ production at \lepii.
The $\ha,\hh,\hp$ could all also
have substantial SUSY decays, especially if $\mha$ is large.  
Such decays will not be significant
if $\tanb$ is large since the $b\anti b,\taup\taum,
\mu^+\mu^-$ modes are enhanced, but would generally severely reduce signals
in the standard channels when $\tanb$ is in the small to moderate range
\cite{hhg}.

\subsubsection{Distinguishing the MSSM {\boldmath$\hl$}
from the SM {\boldmath$\hsm$} at the LHC}

Suppose that $\mha$ is moderately large and that $\tanb$ is in the middle
range described above where only the $\hl$ can be detected. 
Then, only detailed measurements of the properties of the $\hl$ could
reveal that it is part of the larger MSSM Higgs
sector, and not just the minimal $\hsm$. In this region of parameter
space, the couplings of the $\hl$ are quite SM-like, and substantial
precision in such measurements would be needed. For a SM-like $\hl$,
we are certain to be able to measure 
\begin{description}
\item[(a)] $\sigma(gg\rta\hl)BR(\hl\rta\gam\gam)$
\item[(b)] $\sigma(W\hl)BR(\hl\rta\gam\gam)$ and 
\item[(c)] $\sigma(t\anti t\hl)BR(\hl\rta\gam\gam)$.
\end{description}
From the latter two we can compute the ratio $\sigma(W\hl)/\sigma(t\anti
t\hl)$. If $\mstop$ is very large, and/or squark mixing
is large,  so that $\mhl$
is above about $120-130\gev$, we shall also be able to measure 
\begin{description}
\item[(d)] $\sigma(gg\rta\hl)BR(\hl\rta Z\zstar)$.
\end{description}
Combining (a) with (d) we obtain the ratio $BR(\hl\rta \gam\gam)/BR(\hl\rta
Z\zstar)$. If $\mstop\lsim 1\tev$ and squark mixing is not large, 
so that $\mhl$ is below about $120\gev$, then we instead observe
\begin{description}
\item[(e)] $\sigma(t\anti t\hl)BR(\hl\rta b\anti b)$ and possibly 
\item[(f)] $\sigma(W\hl)BR(\hl\rta b\anti b)$. 
\end{description}
Using (b)+(f) or, more likely,
(c)+(e), we can compute $BR(\hl\rta\gam\gam)/BR(\hl\rta b\anti b)$.

\begin{figure}[htbp]
\let\normalsize=\captsize   
\begin{center}
\centerline{\psfig{file=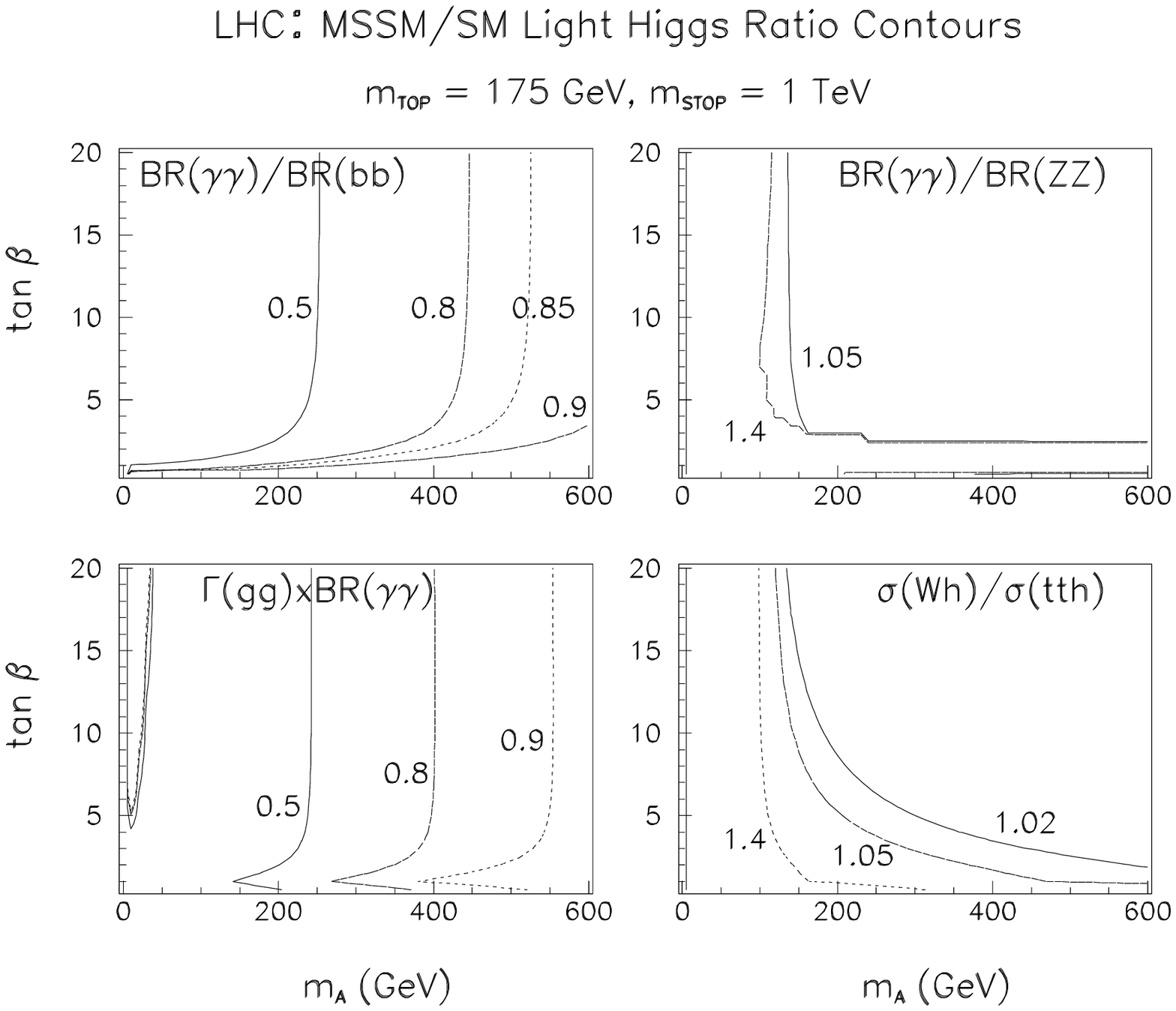,width=12.2cm}}
\begin{minipage}{12.5cm}       
\caption{Contours for ratios of MSSM $\hl$ results to
SM $\hsm$ results for $\mhl=\mhsm$, in $(\mha,\tanb)$ parameter space.
We take $\mt=175\gev$, $\mstop=1\tev$, and neglect squark mixing
in computing the \twoloop\ radiative corrections to the MSSM Higgs sector.}
\label{lhcdeviationscontours}
\end{minipage}
\end{center}
\end{figure}

The three ratios mentioned above as well as the basic rate for (a)
are all sensitive to small differences between the $\hl$ and a standard
$\hsm$ of the same mass.
To quantify this sensitivity, we present \cite{ghhprecision}
in Fig.~\ref{lhcdeviationscontours} contours for
the ratio of the MSSM predictions for these four quantities to those
obtained for a $\hsm$ of exactly the same mass.  For simplicity
we have employed a uniform value of $\mstop=1\tev$ and neglected
squark mixing in computing the MSSM Higgs sector radiative corrections.
From the graphs, it would appear that 
the ratio $BR(\gam\gam)/BR(b\anti b)$ provides the best probe,
since deviations as large as 10\% persist well beyond $\mha=600\gev$.
However, as estimated in a later section (see Table~\ref{nlclhcerrors}), 
determination of
this ratio with such precision from LHC data alone is not likely.
The next most sensitive probe is the cross section times branching ratio 
$\sigma(gg\rta\hl)BR(\hl\rta \gam\gam)$;
deviations as large as 10\% persist
out to values of $\mha$ as large as $\mha\sim 550\gev$. 
In Table~\ref{nlclhcerrors} we estimate that this product can be measured
with roughly $\pm4\%$ accuracy. However, it is
important to keep in mind that we will not be able to distinguish
a deviation due to the difference between the $\hl$ and $\hsm$ from
a deviation due to some new physics contribution to the loops
responsible for the Higgs-$gg$ and/or Higgs-$\gam\gam$ coupling. 
Further, $BR(\h\rta \gam\gam)$ and
$BR(\h\rta b\anti b)$ ($\h=\hl$ or $\hsm$) would both be affected by unexpected
contributions to the total $\h$ decay width --- for example,
an enhanced $\h\rta gg$ partial width and/or
invisible $\hl\rta \cnone\cnone$ decays.
In addition, there is a systematic uncertainty in our ability to
determine $BR(\h\rta b\anti b)$ due to uncertainty in the value of 
the running ($\overline{MS}$) mass $\mb(\mh)$;
\begin{equation}
{\delta BR(b\anti b)\over BR(b\anti b)}
\simeq BR({\rm non}-b){\delta \Gamma(b\anti b)\over \Gamma(b\anti b)}
\sim BR({\rm non}-b) {2\delta \mb\over \mb}\sim 2-3\%\,,
\label{bbbrsystematicerror}
\end{equation}
for a 5\% uncertainty in $\mb$. 
Some current lattice calculations claim that $\mb(\mb)$ (in the $\overline
{MS}$ scheme) can be determined more accurately than this.
Ref.~\cite{latticemb} gives a result $\mb(\mb)=4.0\pm0.1\gev$,
an error of 2.5\%.
Unfortunately, the more model-independent $\hl WW/\hl t\anti t$ coupling 
ratio encoded in the (b)/(c) cross section ratio 
will be very difficult to measure to the 2\% or better
accuracy required to probe out to large values of $\mha$;
in Table~\ref{nlclhcerrors} we estimate an error of roughly $\pm13\%$
for the numerator and denominator individually.

\section{Tevatron and \tevstar}

\indent\indent The Tevatron is currently operating at $\sqrt s = 1.8$ TeV, at 
an instantaneous luminosity of 
${\cal L}\approx 10^{31}/cm^2/s$.  The Main Injector is expected to begin
operation in 1999, increasing the luminosity to ${\cal L}\approx 
2\times 10^{32}/cm^2/s$, 
which provides about $2 \fbi$ of integrated luminosity per year. The machine 
energy will be increased to $\sqrt s = 2$ TeV at that time.

One can consider increasing the luminosity even further.  This idea is
generically referred to as the \tevstar, with an instantaneous
luminosity around $5\times 10^{32}/cm^2/s$. An instantaneous
luminosity of $10^{33}/cm^2/s$ or higher can also be envisioned;
this is referred to as Tev33. (We use \tevstar\ to denote
both \tevstar\ and Tev33 in the following.)
With the 132 $ns$ bunch spacing
of the Tevatron (with the Main Injector), this luminosity results in about
15 interactions per bunch crossing.  The bunch spacing can be reduced
in principle to 19 $ns$ (comparable to the 25 $ns$ timing of the LHC), 
yielding about 2 interactions per crossing.  
The luminosity upgrade and the reduced bunch spacing 
each require detector upgrades, or perhaps a new detector \cite{Tev2000}.

The most promising mode for the standard Higgs at the Tevatron and \tevstar\
is $W\hsm$ production, followed by $\hsm\to b\anti b$ 
\cite{SMW1,SMW2,MK,BBD}.  This process was discussed in the section
on the LHC.  Since the production is via quark-antiquark 
annihilation, the cross section is only a factor of four less at the 
Tevatron than at the LHC (including acceptance).  On the other hand,
the top-quark backgrounds, which are the dominant backgrounds at the LHC,
are much smaller at the Tevatron.
The largest backgrounds at the Tevatron and \tevstar\ are the irreducible 
processes $Wb\anti b$ and $WZ$ (for $\mhsm \approx \mz$), and, for
$\mhsm>100\gev$, the top-quark backgrounds. Each of
these backgrounds is comparable to the signal.  The $WZ$ background 
can be normalized via the leptonic decays of the weak vector bosons. 
However, should the Higgs mass happen to lie near the $Z$ mass, this 
background may be particularly problematic, since the Higgs peak will 
simply add to the $Z$ peak, and observation of the Higgs will rely on an 
excess of events in the peak region. 

The most comprehensive study of the $W\hsm$ with $\hsm\to b\anti b$
process at the Tevatron and the \tevstar\ is Ref.~\cite{Kuhlmann}.
The search at the Tevatron ($2-4\fbi$) will be limited to 
$\mhsm < \mz$, a region that will already have been explored at \lepii.  
Should a Higgs be discovered at \lepii, it would be interesting to search 
for it at the Tevatron to test the relationship between the $\hsm ZZ$ and 
$\hsm W^+W^-$ couplings (which are related by custodial SU(2)).

The most important issue in the analysis of $W\hsm$ with $\hsm\to b\anti b$
at the Tevatron and the \tevstar\ is the $b\anti b$ invariant-mass
resolution.  Ref.~\cite{Kuhlmann} assumes this can be improved
over the current resolution by inventing a jet-clustering algorithm that
takes into account  hard gluon radiation.  With this improved resolution,
a Higgs boson of mass $60-100\gev$ is detectable at the \tevstar\
with $10\fbi$ of integrated luminosity.

To attain Higgs masses much above the $Z$ mass, higher luminosity is needed.
Tev33 can potentially deliver $30 \fbi$ in three years.  Unless the 
bunch spacing is decreased, this will require dealing with about 15 
interactions per bunch crossing.  Fortunately, since the $t\anti t$ background 
is small, one does not have to veto jets with small $p_T$, which is 
difficult to do in this environment. 
Ref.~\cite{Kuhlmann} concludes that a Higgs mass of $120\gev$ requires
approximately $25\fbi$ of integrated luminosity.

Should Higgs masses above $100\gev$ be observable via $W\hsm$ with $\hsm
\to b \anti b$, it would have particular significance for the lightest SUSY
Higgs scalar, $\hl$. For $\tanb > 1$, the
coupling and, hence, partial width for $\hl\rta b\anti b$ are enhanced even
for fairly large $\mha$ values (see Fig.~\ref{couplingstanb5}), 
thereby suppressing
its branching ratio to two photons.  The best chance to observe this
particle may thus lie in its dominant decay mode to $b\anti b$.

The processes $W\hsm,Z\hsm$, with $\hsm \to \taup\taum$ 
and $W,Z\to jj$, may also
provide an opportunity for detection of the intermediate-mass Higgs boson
\cite{MK}. The $\taup\taum$ invariant mass can be reconstructed from the
one-prong decays of the $\tau$'s by projecting the missing $p_T$ (from
neutrinos) along the direction of motion of the charged track from the $\tau$
decay \cite{EHSV,MK}.
The dominant background is $Zjj$ with $Z\to \taup\taum$,
which is much larger than the signal, so this mode is
restricted to Higgs masses somewhat above the $Z$ mass.  A Higgs of mass
$110 - 120$ GeV may be accessible with $30 \fbi$ of integrated luminosity.
This requires high-efficiency $\tau$ identification, a goal worth pursuing.
The signal peak lies on the tail of the rapidly falling $Zjj$ background,
which complicates its extraction.
If $\hsm \to \taup\taum$ can be observed for $\mhsm > 110\gev$, it would 
have particular significance for the lightest Higgs scalar, $\hl$, for 
the same reason as discussed above for the mode $\hl \to b\anti b$.
      
The most comprehensive study of $W\hsm,Z\hsm$, with $\hsm \to \tauptaum$
and $W,Z\to jj$, is Ref.~\cite{Heintz}.  This study concludes
that a Higgs signal (for $\mhsm=120\gev$) is not observable even with
$100\fbi$ of integrated luminosity.  This is largely due to the
fact that this study assumed a missing $p_T$ resolution achievable
in existing detectors, while Ref.~\cite{MK} assumed improved resolution
in upgraded detectors.

The gold-plated mode, $gg\to ZZ^{*} \to \ell^+\ell^-\ell^+\ell^-$, has been
studied in Ref.~\cite{GH}.  Compared with the LHC, the signal cross
section is greatly reduced with respect to the irreducible background
$q\anti q \to ZZ$.  Even with $100 \fbi$ of integrated luminosity,
there is no Higgs mass at which an observable signal can be established,
due both to the background and the small number of signal events.

Finally, we note that, in the case of the SUSY Higgs sector
with $\mstop=1\tev$ and no squark mixing,
the $\hh\rta\hl\hl,\ha\ha\rta 4b$ final
state mode is predicted to yield a $5\sigma$ signal for $\mha\lsim 60\gev$
and $\tanb\gsim 1$ \cite{DGV4} if $\ebtag=0.75$ with $\emistag=0.0025$
could ultimately be achieved.

\section{DiTevatron}

\indent\indent The energy of the Tevatron can be increased by replacing the 
existing magnets with magnets of higher field strength. The
existing magnets have a field strength of 4.4 Tesla; replacing them with
8.8 Tesla magnets (comparable to the 8.36 Tesla LHC magnets)
doubles the energy to $\sqrt s = 4$ TeV.  This is
referred to as the DiTevatron.  A luminosity upgrade, similar to the
\tevstar, can be implemented as well, if desired.  In the following we assume
a luminosity upgrade, and consider the discovery potential with $30 \fbi$
of integrated luminosity.

There is apparently little or no advantage for the process $W\hsm$ with
$\hsm \to b\anti b$ at the DiTevatron as compared with the \tevstar\
\cite{SMW2,GH,FR,MK,BBD}. Although the signal cross section roughly doubles,
the top-quark backgrounds increase much more, and become non-negligible
for all Higgs masses at the DiTevatron.
The statistical significance of the signal is no better at the
DiTevatron than at the \tevstar\ for the same integrated luminosity.

The most comprehensive study of $W\hsm$ with $\hsm \to b\anti b$ at the
DiTevatron is Ref.~\cite{FR}. This study finds that Higgs masses above
$\mz$ cannot be probed with $30 \fbi$. However, several refinements
in the analysis, such as those in Ref.~\cite{Kuhlmann}, would likely
change this conclusion.  Still, there is no reason to believe that
the DiTevatron is superior to the \tevstar\ (for the
same integrated luminosity) for this channel.

The process $W\hsm$ with $\hsm \to \taup\taum$ is less promising at the
DiTevatron compared with the Tevatron due to the increase in the $Zjj$
background relative to the signal \cite{MK}.  The Higgs mass region $110
- 120$ GeV, which may be accessible at the \tevstar, falls below the
level of observability at the DiTevatron, due to the large
increase in the $Zjj$ background.  For the same reason, this process is 
hopeless at the LHC.

A variety of signals involving Higgs-boson decay to vector-boson pairs has been
studied at the DiTevatron \cite{GH}.  The most promising of these is the
gold-plated mode, $gg\to ZZ^{(*)}\to \ell^+\ell^-\ell^+\ell^-$.  In the
intermediate-mass region, it is crucial to be able to efficiently
identify leptons at low $p_T$ in order to have enough events.  A lepton 
$p_T$ threshold of 10 GeV yields about 10 events in $30 \fbi$ at $\mhsm
= 150$ GeV, with no irreducible background.  For $\mhsm > 2 \mz$, the
irreducible $ZZ$ background is comparable to the signal.  With $30 \fbi$,
a Higgs of mass close to 200 GeV may be observable. The extraction of the
signal at this mass is confounded by the fact that it is
close to the threshold for the $ZZ$ background.

The process $gg \to \hsm \to W^+W^- \to \ell\nu jj$ suffers from an enormous
irreducible background, approximately 100 times as large as the signal.
A statistically significant signal is obtained for $\mhsm \approx 170$ GeV.
However, the extraction of the signal is once again confounded by the
fact that it is close to the threshold for the background.

The process $gg \to \hsm \to W^+W^- \to \ell^+\nu \ell^-\anti\nu$ 
has a very large background from $W^+W^-$ production which is about an order
of magnitude above the signal.  The signal is 
broad and lies near the threshold for the background, so is difficult to 
extract.

The process $gg \to \hsm \to ZZ \to \ell^+\ell^- \nu\anti\nu$ yields a
statistically significant signal for $\mhsm = 200 - 250$ GeV. However, the
background from processes such as $Zj$, where the jet transverse momentum is
badly mismeasured or the jet is lost in the forward direction, has not
been included in this analysis, and could render this process unobservable.

The processes involving Higgs decay to vector-boson pairs would be aided by
increased integrated luminosity.  For example, 
if $100\fbi$ can be collected, the gold-plated mode could cover the mass 
regions $\mhsm=200 - 300$ GeV and $\mhsm \sim 150$ GeV.
The actual reach in these modes cannot be established until a more
complete study, including efficiencies and detector simulation, has
been performed.

\section{Determining the properties of the SM Higgs from hadron collider
data}

In this section we summarize the extent to which the expected
observations for a light ($\mhsm\lsim 2\mw$)
SM-like Higgs can be used for a detailed
determination of its properties, including total width
and partial widths for various channels.
After combining all hadron collider information,
let us presume that we measure accurately those cross sections
listed earlier:
\begin{description}
\item[(a)] $\sigma(gg\rta \hsm)BR(\hsm\rta \gam\gam)$,
\item[(b)] $\sigma(W\hsm)BR(\hsm\rta\gam\gam)$, 
\item[(c)] $\sigma(t\anti t\hsm)BR(\hsm\rta\gam\gam)$, and either
\item[(d)] $\sigma(gg\rta\hsm)BR(\hsm\rta Z\zstar)$ (for $\mhsm\gsim 120\gev$) or
\item[(e)] $\sigma(t\anti t\hsm)BR(\hsm\rta b\anti b)$ and
\item[(f)] $\sigma(W\hsm)BR(\hsm\rta b\anti b)$ 
(for, very optimistically, $\mhsm\lsim 120\gev$). 
\end{description}
In order to achieve this list we assume that
if Tev/\tevstar/DiTevatron measurements are employed for (f),
then extrapolation to LHC energy can be done accurately
so that we can analyze the situation as though all data is available
at a single energy. Further, we presume that the Tevatron
$W\hsm$ with $\hsm\rta\taup\taum$ rates are converted to the equivalent
$W\hsm$ with $\hsm\rta b\anti b$ rates assuming the standard relation
between $\hsm b\anti b$ and $\hsm \taup\taum$ couplings.

From (b)+(c) we determine the $\hsm t\anti t$ to $\hsm WW$
coupling ratio.  This allows us to check the SM prediction.
From (a)+(c) we determine the $\hsm gg$ to $\hsm t\anti t$
coupling ratio.
This can then be checked against the assumption of $t$-loop dominance.
The presence of extra (heavy) colored particle loop contributions to the 
$gg\hsm$ coupling would be easily apparent as a large discrepancy.  
For $\mhsm\lsim 120\gev$, from (b)+(f) and also (c)+(e)
we can determine $BR(\hsm\rta\gam\gam)/BR(\hsm\rta b\anti b)$.
Deviations of this ratio from expectations would indicate
extra heavy charged particle loop contributions 
to the $\hsm\gam\gam$ coupling, deviations of the total $\hsm$ width from
expectations (including contributions
from invisible channels), deviations in the $\hsm \rta b\anti b$ coupling,
and/or deviations in the $\hsm WW$ 
coupling that controls the dominate $W$-loop contribution. These effects
would be difficult to disentangle without further information.
For $\mhsm\gsim 120\gev$ (a)+(d) allows determination of $BR(\hsm\rta
\gam\gam)/BR(\hsm\rta Z\zstar)$ which could deviate from
expectations for the same reasons as summarized above.
Of course, one could test the internal consistency of SM choices
for all couplings. The simplest examples of such tests 
for $\mhsm\lsim 120\gev$ would be
to see if the predicted value for $BR(\hsm\rta b\anti b)$ combined
with (e) and (f) yields the expected $\hsm t\anti t$ and $\hsm WW$ 
couplings. For $\mhsm\gsim 120\gev$, the predicted $BR(\hsm\rta Z\zstar)$
+ (d) could be used to extract the $\hsm gg$ coupling which
could in turn be tested against the top-quark loop prediction.

Obviously, it will be important for the experimental collaborations
to assess the errors (both systematic and statistical) that will
accompany the above analysis. We give some very rough estimates
in a later section, see Table~\ref{nlclhcerrors}.
Certainly, though, it is clear that 
a model-independent analysis of couplings
and total width is not possible for $\mhsm\lsim 2\mw$
using hadron collider data alone.
We shall see, however, that by combining with data from a linear
$\epem$ collider (which on its own can also not provide
a complete model-independent determination of the total width
and all couplings) an essentially complete analysis is within reach.
   
\section{A Next Linear {\boldmath$\epem$} Collider}

\indent\indent A Next Linear Collider (NLC) can be expected to play a very 
important
role in unravelling the physics of a light Higgs sector, whether
that of the standard model or that of an extended Higgs sector such 
as a multi-doublet model, including the constrained two-doublet
structure of the minimal supersymmetric standard model. We shall focus
on expectations for the SM and the MSSM, with additional remarks
on a general two-Higgs-doublet model (2HDM) and the inclusion
of extra Higgs singlet fields.  Theories containing
Higgs triplet fields, even if constructed to preserve $\rho\equiv
\mw/(\cos\theta_W\mz)=1$ at tree-level, must be fine-tuned 
in order to maintain $\rho=1$ at one-loop, 
since $\rho$ is infinitely renormalized
\cite{gvw}. Thus, we only briefly remark on Higgs triplets here. 

\subsection{Machine and Detector Considerations}

\indent\indent The probable design of the NLC has progressed
enormously over the last few years.  It has become conventional
to assume that a $\sqrt s=500\gev$ $\epem$ collider can be
built with an instantaneous luminosity of the order $5\times 10^{33}
cm^{-2}s^{-1}$, corresponding to an annual integrated luminosity
in the neighborhood of $L=50\fbi$.  It is also normally
assumed that expansion of the collider to $\sqrt s=1-1.5\tev$ 
with annual $L\sim 200\fbi$ would be feasible.  We shall adopt these
luminosity assumptions in the discussions to follow.

Other characteristics of the machine and detector also play critical roles
in assessing our ability to study light Higgs bosons at the NLC.
First, the collision energy will generally be very well defined; neither simple
initial-state radiation nor, for current designs, beamstrahlung leads
to much spread in the collision energy.  For those relatively rare 
instances in which there is significant radiation, it has been shown
\cite{bck}\ that it is almost always due to the emission
of a photon by only one of the intial-state fermions, not both, leaving
the missing energy in an event equal to the observed longitudinal boost.
Thus, kinematic fitting using center-of-mass energy
constraints can be used to take full advantage of the clean
initial state.
Regarding the detector, the performance level of SLC/LEP-type detectors
is more than adequate for even detailed studies of Higgs boson
branching ratios and the like.  The detector simulations that will
be referenced in what follows typically include:
limited forward calorimeter acceptance, with no detector
elements below $10^{\circ}$ from the beamline;
hermetic calorimetry, which allows accurate energy-flow
determination;
efficient, large acceptance central tracking;
efficient electron, muon, and photon identification; and 
powerful heavy quark flavor tagging via a microvertex detector.

\subsection{Backgrounds at the NLC}

\indent\indent The obvious advantage of $\epem$ colliders over hadron colliders
is that the high-$p_T$ background processes that might obscure a Higgs
signal are relatively small in size and are accurately calculable.
At energies between $\sqrt s=500\gev$ and $1.5\tev$, the dominant backgrounds
are due to hard electroweak and QCD processes, in particular 
$\epem\rta \wp\wm$,
$\epem\rta q\anti q$, and $\epem\rta e\nu W$. Other backgrounds that
have been included in the simulations include $\epem\rta Z\gamma$,
$\epem Z$, $\nu\anti\nu\gamma$, $ZZ$, $\wp\wm\gamma$, $\nu\anti\nu Z$,
$\epem\wp\wm$, $\wp\wm Z$, $e\nu WZ$, $\nu\anti\nu \wp\wm$, $t\anti t Z$,
$\nu\anti\nu ZZ$, $\epem ZZ$, $ZZZ$ and $\nu\anti\nu t\anti t$,
in rough order of descending cross section; see, for example, Ref.~\cite{am}.

\subsection{The SM Higgs boson at the NLC}

\indent\indent The dominant production mechanisms for a SM Higgs boson
are $\epem\rta \nu\anti\nu \hsm$, via $\wp\wm$ fusion,
and $\epem\rta Z\hsm$ via virtual $\zstar$.  For Higgs bosons in the 80 to 200 GeV
range, the cross sections for these two reactions are similar
in size at $\sqrt s=500\gev$ --- for $\mhsm\sim 200\gev$
both yield $\sigma/\sigma_{pt}\sim 0.1$, where $\sigma_{pt}=100\fb/s({\rm
TeV}^2)\sim 400\fb$ \cite{hhg}.  Thus, at an integrated luminosity of $50\fbi$,
for $\mhsm\sim 200\gev$
we have roughly 2000 Higgs boson events of each type with which to work.
As $\mhsm$ decreases, the $\wp\wm$ fusion process rapidly becomes
larger than the $Z\hsm$ process, while for $\mhsm>200\gev$ the
$Z\hsm$ cross section is larger than that from $\wp\wm$ fusion.
Generally speaking, the $Z\hsm$ final state will be the preferred channel
in which to search for a Higgs boson with a $\sqrt s=500\gev$ machine
since observation of the $Z$ allows a direct determination of the
Higgs boson four-momentum.

The branching ratios for a SM Higgs boson are illustrated in Fig.~\ref{smbr}.
At low mass, the important discovery channels are $b\anti b$ and $\taup\taum$.
Various search techniques have been developed for virtually all of the possible
event topologies resulting from the decay of the $\hsm$ when produced
in the $Z\hsm$ channel \cite{desyworkshop,janotlep,janothawaii,jlci}, 
as well as for the
primary modes of the $\wp\wm$-fusion reaction \cite{bbp,desyworkshop}.
For example,  for
$\mhsm$ values such that the $\hsm$ decays primarily
to $b\anti b$ and $\tau\tau$ the following channels have
been analyzed in the case of the $Z\hsm$ production mode: 
$\epem\rta q\anti q \hsm\rta 4j$; $\epem\rta Z\hsm$, with
$\hsm\rta \tau\tau$ (with about 30\% as many events as in the $4j$
topology); $\epem\rta \lplm\hsm$ ($\ell=e,\mu$) by examining
the recoil mass corresponding to
$\mhsm$ using the incoming beams and the lepton pair from the $Z$;
$\epem\rta Z\hsm$, with $Z\rta \nu\anti\nu$ ($b$ tagging needed
for $\mhsm\sim \mz$). The results of these studies, using
all the decay topologies listed above, and the analogous
ones required for other reactions and/or other mass regions,
are summarized below \cite{janothawaii,gunionhawaii}. It is
assumed that a $5\sigma$ excess in events at a particular Higgs mass
is required.

Using the $Z\hsm$ production channel,
a SM Higgs boson with $\mhsm\sim 130\gev$ ($200\gev$) 
can be discovered in one week (one month)
of running at typical design luminosities of $5\times 10^{33}
cm^{-2}s^{-1}$. For heavier masses,
the discovery can be made in either of the the main decay
channels $\hsm\rta \wp\wm$ or $ZZ$. This is possible for any of the decay modes
of the vector bosons, although the hadronic decays are favored due
to their larger branching ratios.  For $\mhsm$ up to $350\gev$, discovery
is possible within one year for $L=50\fbi$ at $\sqrt
s=500\gev$ \cite{desyworkshop}.  
For $50 \fbi$ of data, 
the $\epem\rta \nu\anti\nu \wp\wm\rta \hsm\nu\anti\nu$ 
fusion process is observable up to
$\mhsm = 300$ GeV. For still heavier Higgs bosons, higher machine energies
are required.  The $\epem\rta \nu\anti\nu \hsm$
($WW$-fusion) process provides a clean discovery of any Higgs boson
with $\mhsm\lsim 0.7\sqrt s$ for typical planned luminosities.
For example, the expandability of a linear collider to
$\sqrt s=1\tev$ with yearly luminosity of $200\fbi$ 
would ensure that any Higgs boson with mass below
about $700\gev$ can eventually be discovered at an $\epem$ linear
collider.  Consequently, 
the entire mass range for which the Higgs boson is `weakly
coupled' will be accessible.  The requirements for exploring a
strongly-coupled Higgs sector are delineated in the companion report
of the working group on ``Strongly-Coupled Electroweak
Symmetry Breaking'' \cite{hanreport}.  

Of course, other production mechanisms for the Higgs boson are
also of considerable interest. In particular, 
the $\epem\rta t\anti t \hsm$ process is accessible for 
$\mhsm \lsim 120$ GeV at $\sqrt s=500\gev$ \cite{dkz}.
The $\gamma\gamma$ collider mode of operation, which will be
discussed at more length later, will allow Higgs detection up
to $\mhsm\sim 300-350\gev$, \ie\ it does not extend the reach of
the machine for the standard Higgs \cite{gunionhawaii}.

\subsubsection{Detailed study of Higgs total width, partial widths
 and couplings}

\indent\indent Once the Higgs boson has been found,
a rather precise determination of all of its properties will
be the next order of business.  In particular, 
it would be highly desirable to be able to determine (in a model-independent
fashion) not only its exact mass, but also its total width and its couplings
to all types of particles. We shall see that a reasonably complete
study can be performed, especially if $\epem$ collision data
is combined with hadron collider data and $\gam\gam$ collision data.
The use of measurements beyond $\epem$ collisions is especially crucial when
$\mhsm<2\mw$ (where the total width is too small to be experimentally
measurable from the resonance shape). We first outline the basic
strategies, and then give estimates of the level
of uncertainty that will be encountered due to experimental errors.

From the detection mode in which $Z\hsm\rta
\lplm X$, the Higgs mass peak is observed as a peak in the recoil mass
obtained from the initial
energy and outgoing $Z$ momentum; this peak automatically
includes all possible Higgs decays. 
The value of the total cross section $\sigma(\epem\rta Z\hsm)$
is obtained simply from the production rate and the known $Z\rta \lplm$
branching ratio, thereby allowing an absolute normalization of 
the $\hsm ZZ$ coupling. Given this determination
of $\sigma(Z\hsm)$, measurements of $\sigma(Z\hsm) BR(\hsm\rta X)$
for any channel will yield the absolute value of $BR(\hsm\rta X)$. 
The $\hsm WW$ coupling can be directly determined using the measured
rates for $WW$ fusion reactions
$\epem\rta \nu\anti\nu \hsm\rta \nu\anti\nu X$ for those final
states for which $BR(\hsm\rta X)$ has been determined using the $Z\hsm$
measurements.
The value obtained for the $\hsm WW$ coupling can be cross checked
with that computed from the $\hsm ZZ$ coupling assuming custodial symmetry.
Given the $\hsm WW$ coupling, the partial width for $\hsm\rta WW^{(*)}$
can be computed. 
(Here, $W^{(*)}$ is a real or virtual $W$ depending upon $\mhsm$.)

In principle, the partial width for any channel $X$
(and associated $\hsm\rta X$ coupling)
can be obtained from NLC data alone. After obtaining the partial
width for $\hsm\rta WW^{(*)}$ as outlined above, one would
compute:
\begin{equation}
\Gamma^{\rm tot}_{\hsm}=\Gamma(\hsm\rta WW^{(*)})/BR(\hsm\rta WW^{(*)})\,,
\label{totwidprocedure}
\end{equation}
and then
\begin{equation}
\Gamma(\hsm\rta X)=\Gamma^{\rm tot}_{\hsm}BR(\hsm\rta X)\,.
\label{partialwidprocedure}
\end{equation}
In particular, this procedure would apply for $X=f\anti f$.
Further, by comparing the $Z\hsm\rta \lplm X$ inclusive (recoil-mass) 
channel to the explicit sum over observable
channels, {\it e.g.} the $b\anti b$, $\taup\taum$, $c\anti c+gg$, 
$WW^{(*)}$ and $ZZ^{(*)}$ channels, one can, in principle,
determine if there are additional modes
(including invisible decays) contributing to the total width.  
Unfortunately,  there are difficulties in practice in carrying
out the above program.

For smaller $\mhsm$, roughly $\mhsm\lsim 120-140\gev$, 
the $f\anti f=b\anti b,\taup\taum$ branching ratios will be 
determined with reasonable accuracy ({\it e.g.} $\pm7\%,\pm14\%$,
respectively, for $\mhsm=120\gev$), but the $WW^{(*)}$ and $ZZ^{(*)}$
branching ratios will be very difficult to measure
with reasonable errors (in the case of $WW^{(*)}$, $\pm 50\%,\pm 35\%,\pm 25\%$ 
for $\mhsm\sim 120,130,140\gev$,
respectively). (Typical errors will be reviewed in more detail later).
Thus, neither the total
$\hsm$ width nor the $f\anti f$ absolute partial widths (and couplings) 
could be determined accurately.

At larger $\mhsm$ (roughly beginning about $\mhsm\gsim 145\gev$),
the $WW^{(*)}$ branching ratio should be measured with
good accuracy.
This means that the $\hsm$ total width could be determined
in this region as in Eq.~\ref{totwidprocedure} within
reasonable errors.
(Of course, for $\mhsm$ above about $2\mw$
the total width is larger than $0.2\gev$, large enough
to also allow direct measurement by resonance shape.) 
However, for $\mhsm\sim 140\gev$, errors on the $b\anti b$ and
$\tau\tau$ branching ratios of the $\hsm$ increase to $\gsim 12\%$ and
$\gsim 24\%$, respectively (worsening rapidly as $\mhsm$ increases further).
For $\mhsm\gsim 150\gev$ the $X=f\anti f$ partial widths 
and couplings would then be rather poorly determined through 
Eq.~\ref{partialwidprocedure}.

The $\hsm t\anti t$ coupling might prove more accessible.
In principle, the $t\anti t$ branching ratio measurement
could prove feasible once $\mhsm\gsim 2\mt$, although
for the $\hsm$ this branching ratio is never more than about 20\%.
At low masses, $\mhsm\lsim 120\gev$, we have already noted that
it is probably possible to measure  
$\sigma(t\anti t \hsm)BR(\hsm\rta b\anti b)$ \cite{ATLAS};
this allows a determination of the $\hsm t\anti t$ coupling by employing 
the value of $BR(\hsm\rta b\anti b)$ measured in $Z\hsm$ production. 
Note that were $\sigma(b\anti b\hsm)BR(\hsm\rta b\anti b)$ not
too small to be measurable, one could have used it and the
measured $BR(\hsm\rta b\anti b)$ to determine the total width of the $\hsm$.

Let us now consider what gains can be achieved by 
combining NLC information with LHC/Tev/\tevstar\ measurements.
As summarized in an earlier section, we can presume that
we have determinations of
\begin{description}
\item[(a)] $\sigma(gg\rta \hsm)BR(\hsm\rta \gam\gam)$,
\item[(b)] $\sigma(W\hsm)BR(\hsm\rta\gam\gam)$, 
\item[(c)] $\sigma(t\anti t\hsm)BR(\hsm\rta\gam\gam)$, and either
\item[(d)] $\sigma(gg\rta\hsm)BR(\hsm\rta Z\zstar)$ 
(for $\mhsm\gsim 120\gev$) or
\item[(e)] $\sigma(t\anti t\hsm)BR(\hsm\rta b\anti b)$ and
\item[(f)] $\sigma(W\hsm)BR(\hsm\rta b\anti b)$ 
(for, very optimistically, $\mhsm\lsim 120\gev$). 
\end{description}
Tevatron measurements that contribute are assumed to be
extrapolated to the LHC energy.
Consider first $\mhsm\lsim 120\gev$.
From the NLC we have a reasonable determination of $BR(\hsm\rta b\anti b)$;
(e) and (f) then allow us to determine $\sigma(t\anti t \hsm)$ and
$\sigma(W\hsm)$.
The former (latter) provides us with a cross check on the NLC 
determination of the $\hsm t\anti t$ ($\hsm WW$) coupling.
The value of $\sigma(W\hsm)$ [$\sigma(t\anti t\hsm)$] combined with
(b) [(c)] each allows an independent determination of $BR(\hsm\rta\gam\gam)$,
which can then be combined with (a) to determine the $\hsm gg$ coupling.
This latter can then be checked against the assumption of $t$-loop dominance
given the previous determinations of the $\hsm t\anti t$ coupling.
As noted previously,
the presence of extra (heavy) colored particle loop contributions to the 
$\hsm gg$ coupling would be easily apparent as a large discrepancy.  
Deviations of $BR(\hsm\rta\gam\gam)$ from expectations would
be symptomatic of heavy charged particle loop contributions 
to the $\hsm\gam\gam$ coupling and/or deviations of the total width from
expectations.  For $\mhsm\gsim 120\gev$, the value 
of $BR(\hsm\rta ZZ^{(*)})$ computed and/or measured at the NLC
combined with (d) yields $\sigma(gg\rta\hsm)$
(which determines the $\hsm t\anti t$ coupling if no extra loops are present).
The determination of $\sigma(gg\rta\hsm)$ combined with (a) yields
a determination of $BR(\hsm\rta\gam\gam)$.  An independent determination
of the $\gam\gam$ branching ratio is possible using (b) and the 
value of the $\hsm WW$ coupling as computed from the NLC determination
of the $\hsm ZZ$ coupling and as measured directly in $WW$ fusion production.

Thus, for both $\mhsm\lsim 120\gev$ and $\mhsm\gsim 120\gev$, 
combining NLC with hadron
collider data will provide a series of cross checks and a determination
of $BR(\hsm\rta\gam\gam)$.  An accurate 
determination of the $\hsm\rta b\anti b$
partial width and coupling remains elusive for all $\mhsm$ values,
and a determination of the total $\hsm$ width is not possible
until far enough into the $\mhsm\gsim 120\gev$ region that 
$\sigma(Z\hsm)BR(\hsm\rta W\wstar)$
can be measured with reasonable accuracy at the NLC. This is unfortunate.
For example, for $\mhsm\lsim 120\gev$
an enhanced $b\anti b$ coupling for the Higgs would enhance
both the $b\anti b$ partial width and the Higgs total width without
changing $BR(\hsm\rta b\anti b)$ very much 
(see Eq.~\ref{bbbrsystematicerror}), and thus could easily
escape notice.  And, as noted earlier, unexpected
contributions to the total width could be present and
would lead to deviations in $BR(\hsm\rta b\anti b)$.
However, there is an ace in the hole.  As 
we shall describe in a later section, the $\gam\gam$ collider mode
of operation would allow $\hsm$ discovery and
determination of the partial width $\Gamma(\hsm\rta \gam\gam)$.
This, combined with the hadron collider determination of $BR(\hsm\rta\gam\gam)$
then yields the total $\hsm$ width!  If $BR(\hsm\rta b\anti b)$
is measured with reasonable accuracy (requiring $\mhsm\lsim 145\gev$)
then by combining with the total width determination we obtain
$\Gamma(\hsm\rta b\anti b)$ and a determination of the $\hsm b\anti b$
coupling. In short, for $\mhsm\lsim 2\mw$
{\it all three machines --- LHC, NLC and $\gam\gam$ collider ---
are needed to complete a model-independent study of the $\hsm$}.

Let us now quantify the expected errors in the relevant measurements.
The most optimistic expectations in the `intermediate' $\mhsm< 2\mw$
mass region are reviewed
in Ref.~\cite{kawagoe}. There it is assumed that the machine will
be run at the optimal energy for the $Z\hsm$
cross section, $\sqrt s\sim \mz+\mhsm+ 20/30\gev$, 
that luminosity of $L=30\fbi$ is accumulated, {\it and}
that the detector has the characteristics of the proposed ``super'' performance
JLC detector \cite{jlci}, especially the super momentum resolutions
and the high $b$-tagging efficiency.  With these assumptions, 
$\mhsm$ can be measured to $\sim 0.1\%$
using direct $\mhsm$ reconstruction in
the $Z\hsm\rta \nu\anti\nu jj,\lplm jj,jjjj$ channels. This estimate
is based on the roughly $4\gev$ uncertainty per event
for reconstructing the mass from the detector information, divided
by $\sqrt N$, where $N$ is the total number of events.
Further, the mass resolution for the recoil mass peak in $Z\hsm$ events 
is of order $0.3\gev$ per event for $Z\to\ell^+\ell^-$ decays, leading
to a measurement of $\mhsm$ to $\pm 20\mev$
for $\mhsm\lsim 140\gev$ and $L=50\fbi$.
The total width, $\Gamma_{\hsm}^{\rm tot}$
can be measured down to $\sim 0.2\gev$ using $Z\hsm$ events
with $Z\rta \lplm$ and examining the recoil mass.     
Unfortunately, this latter sensitivity is not adequate for a direct
observation of the SM Higgs boson width until $\mhsm$
approaches the $WW$ decay threshold.  For example, the width is predicted to 
be in the range $\Gamma_{\hsm}\lsim 10 ~{\rm MeV}$ for $\mhsm\lsim 140\gev$;
see Fig.~\ref{hwidths}.
(However, in this mass range the width can be much larger than this 
for some of the Higgs bosons in models such as the MSSM.)  Perhaps
most critical is the determination of the $\epem\rta Z\hsm$ total
cross section, which (as noted above) is the best means for directly determining
the $\hsm ZZ$ coupling.  This determination is best made by
employing the $Z\hsm\rta \lplm X$ mode. Greatest sensitivity
to a Higgs boson is obtained by specifically excluding events in which $X$
is a single $\gamma$ (a channel to which Higgs decay does not contribute)
because of the large radiative contribution to this channel.  The
Higgs boson is then observed as a bump in the recoiling $X$
mass. Ref.~\cite{kawagoe} finds that $\sigma(\epem\rta Z\hsm)$
can be measured with a relative error of $\sim 7\%$, thereby allowing
a $\sim 3.5\%$ determination of the $\hsm ZZ$ coupling. 
Note that this technique automatically sums over all possible
decay modes of the $\hsm$, including invisible channels --- knowledge
of the Higgs decays is not required to extract the $\hsm ZZ$ coupling
from the inclusive recoil mass spectrum.

Turning to the relative (and absolute) branching ratios of the $\hsm$,
for the ``super''-JLC detector
and a Higgs mass in the vicinity of $\sim 110\gev$,
Ref.~\cite{kawagoe} states that $\sigma(\epem\rta Z\hsm) BR(\hsm\rta b\anti b)$
can be measured with a precision of $\sim 2\%$ for $L=80\fbi$
(or $\sim 2.5\%$ for $L=50\fbi$), which yields a roughly
8\% determination of $BR(\hsm\rta b\anti b)$ 
itself, given the 7\% error in $\sigma(Z\hsm)$. 
Further, 
$BR(\hsm\rta \taum\taup)/BR(\hsm\rta b\anti b)$ can be determined 
within $\sim 6\%$, whereas $BR(\hsm\rta c\anti c)/BR(\hsm\rta b\anti b)$
could only be determined within a roughly 100\% error.  However,
even this rough a measurement might be adequate to determine that
the Higgs boson has enhanced $b\anti b$ and suppressed $c\anti c$
couplings as is possible in extended Higgs sector models.

The accuracy of some of these measurements would not be this good if a detector
of the  generic SLC/LEP variety is employed. The expectations for this
case have been examined
with particular care in Ref.~\cite{hbb} (see also
the review of Ref.~\cite{hildrethhawaii}), again for a relatively
light Higgs boson. We shall quote results for $\mhsm\sim 120\gev$
and $\mhsm\sim 140\gev$. The simulations have
assumed that the Higgs mass is known with an accuracy of $\sim 5\gev$
(as would be easily achieved using the various discovery modes),
that the machine energy is set to the optimal value for the $Z\hsm$
mode --- $\sqrt s\sim \mz+\mhsm+ 20/30\gev$, and
that luminosity of $L=50\fbi$ is accumulated.  

To measure $\sigma(Z\hsm) BR(\hsm\rta X)$
for all the relevant channels $X$, Ref.~\cite{hbb}
uses high-impact-parameter track counting (in the precision microvertex
detector) to help distinguish between the $b\anti b$, $\tau\tau$, $WW$,
and $c\anti c +gg$ decay modes.  The $b\anti b$ decay mode is
singled out after $b$-tagging in the available decay topologies,
and $\sigma(Z\hsm) BR(\hsm\rta b\anti b)$ can be measured with a precision 
of about 7\% (12\%) for $\mhsm=120\gev$ ($140\gev$).
The $c\anti c+gg$ modes are dominant if anti-$b$-tagging is implemented,
and the corresponding $\sigma(Z\hsm) BR$ can be measured with a
combined statistical precision of 39\% (116\%), with much larger
errors associated with extracting $c\anti c$ alone.  The $\tau\tau$ mode
can be rather cleanly separated from the others in the $\tau\tau q\anti q$
topology and $\sigma(Z\hsm) BR(\hsm\rta \tau\tau)$ can be measured to
within $\pm14\%$ ($\pm22\%$). 
Finally, measurement of $\sigma(Z\hsm) BR(\hsm\rta W\wstar)$
requires an analysis which selects the $ZW\wstar$ final state. 
This state can be reconstructed in either of the two modes $W\wstar\rta q\anti q
q\anti q$ or $W\wstar\rta q\anti q \ell\nu$ using the $W$, $Z$, and
Higgs boson mass constraints and energy-momentum conservation.
Anti-$b$-tagging removes six-jet events from $\epem\rta t\anti t$,
leaving a relatively clean signal from $\hsm\rta W\wstar$. The product
$\sigma(Z\hsm) BR(\hsm\rta W\wstar)$ can then be determined with a precision, 
which is limited by the relatively small branching
fraction of $\hsm\rta W\wstar$,
of $\pm48\%$ ($\pm24\%$) at Higgs masses of $120\gev$ ($140\gev$).
The same analysis can also be applied to measure the smaller
$\sigma(Z\hsm) BR(\hsm\rta Z\zstar)$, although with rather limited accuracy.

Explicit studies have not been performed by the experimentalists
of the corresponding errors for the various $\sigma(WW\to\hsm)BR(\hsm\to X)$.
($\sigma(WW\to\hsm)$ itself is not directly measurable.)  Nonetheless,
for $100\lsim\hsm\lsim 150\gev$, the rates for $WW\to\hsm$ fusion are
larger than those for $Z\hsm$ associated production,
and backgrounds are well under control after appropriate cuts
and $b$-tagging. For example
Fig.~13b, for $\mhsm=115\gev$, appearing on p. 60 of 
Ref.~\cite{desyworkshop}, shows very little background in the $WW\to\hsm\to
b\anti b$ mode.  The total event rate after cuts would be roughly
of order $S\sim 100$ in the background-free invariant mass region,
implying a $1/\sqrt S= 10\%$ measurement of $\sigma(WW\to\hsm)BR(\hsm\to
b\anti b)$. In general,
it seems reasonable to suppose that errors for $\sigma(WW\to\hsm)BR(\hsm\to X)$
will be comparable to those found
for $\sigma(Z\hsm)BR(\hsm\to X)$ for any given channel, $X$.

Even if the Higgs is too light for direct decays into $t\anti t$,
we have noted that it can be produced at an observable rate 
in $\epem\rta t\anti t\hsm$ for $\mhsm\lsim 120\gev$. By focusing
on the $t\anti t b\anti b$ final state (for example) we obtain
$\sigma(t\anti t\hsm)BR(\hsm\rta b\anti b)$.  
The number of events is a sharply declining function of $\mhsm$ \cite{dkz},
ranging from $\sim 150$ for $\mhsm\sim 80\gev$ to $\sim 20$ for $\mhsm\sim
120\gev$ (assuming $L=50\fbi$, and without cuts {\it etc.}).
Including a factor of 2 reduction for efficiencies and cuts, and assuming
negligible background, we find errors in 
the determination of $\sigma(t\anti t\hsm)BR(\hsm\rta b\anti b)$
ranging from $\sim 10\%$ at $\mhsm=80\gev$ to $\gsim 30\%$ at $\mhsm=120\gev$.
Since $BR(\hsm\rta b\anti b)$
is well determined from the $Z\hsm$ procedures, we can then obtain 
$\sigma (\epem\rta t\anti t\hsm)$ with roughly the same errors;
the cross section is a direct measure of the $\hsm t\anti t$ coupling squared.
For Higgs masses above $2\mt\sim350\gev$, the strategy would be
to measure $\sigma(Z\hsm) BR(\hsm\rta t\anti t)$; a $\sqrt s>500\gev$
machine would be required. But if available, $BR(\hsm\rta t\anti t)$
(predicted to be above 10\% for $\mhsm\gsim 400\gev$) could be obtained
and the $\hsm t\anti t$ coupling determined using knowledge of
the total width.  An estimate of the errors involved is not easily
obtained and requires further study.
For $\mhsm$ values
between $\sim 120\gev$ and $\sim 2\mt$, a determination of
the $\hsm t\anti t $ coupling would require greatly enhanced machine luminosity
and/or energy.

Finally, the rate for $\gam\gam\rta \hsm\rta X$ is proportional to 
the product 
$\Gamma(\hsm\rta \gamma\gamma)BR(\hsm\rta X)$.  The accuracy
with which this product can be measured is estimated in
Ref.~\cite{bordenhawaii} to be $\pm 5\%$ for $\mhsm\lsim 150\gev$
in the channel $X=b\anti b$ and $\pm 10\%$ for 
$180\gev\lsim\mhsm\lsim 300\gev$ 
in the $X=ZZ$ channel. For $\mhsm$ in the $160-170\gev$ range 
the $\hsm\rta WW$ decay is dominant (but has an enormous background)
and the combined $X=b\anti b+ZZ$ error rises to nearly 25\%.  
We have already noted
that for $\mhsm\lsim 120\gev$ the error in $BR(\hsm\rta b\anti b)$
will be $\gsim 7\%$, implying (using quadrature) an
error for $\Gamma(\hsm\rta\gam\gam)$ of $\gsim 9\%$. By $\mhsm\sim 140\gev$
this rises to $\gsim 15\%$ (much worse in the $\mhsm=160-170\gev$ range).

Errors for the LHC measurements have not been quoted by
the experimentalists.  We have made some crude estimates using
the event rate information provided in the Technical Proposals,
Refs.~\cite{ATLAS,CMS}.  In all cases we have used $\sqrt {S+B}/S$
as our fractional $1\sigma$ error estimate; here,
$S$ and $B$ are the signal and background
rates, respectively. For the $gg\to \hsm\to\gam\gam$ mode we have
employed the CMS results appearing in their Figs.~12.3 and 12.5; for
the $W\hsm,t\anti t\hsm$ with $\hsm\to\gam\gam$ modes we employ CMS
Table 12.3; for $gg\to\hsm\to 4\mu$ we employ CMS Table 12.4b;
for the $t\anti t\hsm$ with $\hsm\to b\anti b$ mode we employ
ATLAS Table 11.8. We focus only on high luminosity,
assuming that $L=300\fbi$ is accumulated by each of the
two detectors.  Our errors are obtained by combining the
statistics of the two detectors, corresponding
to a total luminosity of $L=600\fbi$. The results of the figures
and tables mentioned above have been extrapolated to this luminosity.
Note that we have not quoted any error
for the $W\hsm$ with $\hsm\to b\anti b$ channel, since it is not
certain that this channel can be isolated at high luminosity.
However, Tevatron or \tevstar\ data should ultimately provide
errors on this channel that are comparable to the $t\anti t\hsm$
with $\hsm\to b\anti b$ errors quoted for the LHC if $\mhsm\lsim 120\gev$.

The estimated errors for directly measured
quantities are summarized in Table~\ref{nlclhcerrors}.
The particular $\mhsm$ masses for which we have tabulated results
are those for which a significant number of NLC estimates by
experimentalists have appeared, see above.
In the table, question marks (?) indicate that we have had to extrapolate
or make educated guesses. In $Z\hsm$ production,
errors on ratios of branching ratios
are approximately given by adding in quadrature the errors on
particular channel rates; and the error on a particular
$BR(\hsm\to X)$ is given by adding in quadrature
the error on the $\sigma(Z\hsm)$ itself and the error
on $\sigma(Z\hsm)BR(\hsm\to X)$.
In Table~\ref{nlclhcerrors} we have also included
some results for a possible $\mupmum$ collider operating
at $\sqrt s=500\gev$ (the FMC) with a super energy resolution
of $0.01\%$ for the energy of each of the colliding beams.
These results will be discussed in a later section.
Most notable is the great precision with which $\mhsm$ 
and $\Gamma_{\hsm}^{\rm tot}$ can be measured (if $\mhsm\lsim 2\mw$).
It will be important for experimentalists
to refine all these results and obtain errors for a given detector
over the full range of possible Higgs masses.

\begin{table}[htbp]
\caption[fake]{%
NLC ($L=50\fbi$), LHC ($L=300\fbi$ per detector) and FMC ($L=50\fbi$)
Measurement Errors. Question marks indicate
extrapolations of existing studies or crude estimates.}
\begin{center}
\begin{tabular}{|c|ccc|} \hline
NLC ($L=50\fbi$) & & & \\
 Detector ($\mhsm(\gev)$)      & Super-JLC ($110$)     &   
 SLD ($120$) &  SLD ($140$) \\  
\hline
$\mhsm$ (final state) & $\sim 0.1\%$ & $\sim 0.1\%$ & $\sim 0.1\%$ \\
$\mhsm$ (recoil mass) & $\sim 0.02\%$ & $>0.1\%$  & $>0.1\%$  \\
$\Gamma^{\rm tot}_{\hsm}$ & $0.2\gev$ & $0.2\gev$ & $0.2\gev$ \\
$\sigma(Z\hsm)$ & 7\% & 7\% & 7\% \\
$\sigma(Z\hsm)BR(\hsm\rta b\anti b)$ & 2.5\% & 7\% & 12\% \\
$\sigma(Z\hsm)BR(\hsm\rta \tau^+\tau^-)$ & 6\% & 14\% & 22\% \\
$\sigma(Z\hsm)BR(\hsm\rta c\anti c+gg)$ & $100\%$ & $39\%$ & $116\%$ \\
$\sigma(Z\hsm)BR(\hsm\rta WW^*)$ & $\sim 100\%$? & $48\%$ & $24\%$ \\
$\sigma(WW\to\hsm)BR(\hsm\to X)$ & similar to above & for any given 
& final state $X$? \\
$\Gamma(\hsm\rta\gam\gam)BR(\hsm\rta b\anti b)$ & $<5\%$? & 5\% & 5\% \\
$\sigma(t\anti t\hsm)BR(\hsm\rta b\anti b)$ & $\gsim 25\%$? & $\gsim 30\%$?
& unmeasurable? \\
\hline
LHC ($L=600\fbi$) & (110) & (120) & (140) \\
\hline
$\sigma(gg\to\hsm)BR(\hsm\rta\gam\gam)$ & $\sim 4\%$? & $\sim 4\%$? & $\sim
4\%$? \\
$\sigma(W\hsm)BR(\hsm\to\gam\gam)$ & $\sim 13\%$? & $\sim 13\%$? &
unmeasurable? \\
$\sigma(t\anti t\hsm)BR(\hsm\to\gam\gam)$ & $\sim 13\%$? & $\sim 13\%$? &
unmeasurable? \\
$\sigma(gg\to\hsm)BR(\hsm\to Z\zstar)$ & unmeasurable? & large? & $\sim
10\%$? \\
$\sigma(W\hsm)BR(\hsm\to b\anti b)$ & unmeasurable? & unmeasurable? &
unmeasurable? \\
$\sigma(t\anti t\hsm)BR(\hsm\to b\anti b)$ & $\sim20\%$? & $\sim 28\%$? 
& unmeasurable? \\
\hline
FMC ($L=50\fbi$) & (110) & (120) & (140) \\
\hline
$\mhsm$ & $\lsim 0.00015\%$, & $\lsim 0.00015\%$ & $\lsim 0.0003\%$ \\
$\Gamma_{\hsm}^{\rm tot}$ & $\sim 10\%$ & $\sim 10\%$ & $\sim 10\%$ \\
$\Gamma(\hsm\to\mupmum)BR(\hsm\to b\anti b)$ & $\sim 0.3\%$ & $\sim 0.3\%$ &
$\sim 0.5\%$ \\
\hline
\end{tabular}
\end{center}
\label{nlclhcerrors}
\end{table}

\subsubsection{Determination of the Higgs quantum numbers}

\indent\indent A direct verification that the SM Higgs is CP-even would 
be highly desirable.  Of course, if the $Z\hsm$ cross section is such
that the $\hsm ZZ$ coupling is maximal (in agreement
with the SM prediction), then the $\hsm$ must be CP-even.
However, a direct test can be difficult.  Several approaches
have been explored.  These are reviewed in
Refs.~\cite{gunionhawaii,shhawaii,kksz}.
They rely on the fact that a CP-even Higgs boson will 
lead to very different angular production and decay distributions
and/or correlations as compared to a {\it purely} CP-odd Higgs boson. 
The difficulty with these approaches
is that any distribution or correlation that derives
from the coupling to $WW$ or $ZZ$ will only be sensitive to the
CP-even part of the CP-mixed state, unless this CP-even part happens
to be as small relative to the CP-odd part as the CP-odd coupling
to $ZZ,WW$ (arising only at one-loop) is relative to the CP-even
coupling to $ZZ,WW$.  However, in this limit, or in the case of
a purely CP-odd Higgs boson, the production rate will
be so suppressed in production channels relying on the $ZZ,WW$ coupling
that the angular analysis could not be performed with even
marginal statistics anyway.

A few examples will illustrate the potential, but also the difficulties.
In the discussion to follow we generically denote a neutral Higgs boson
by $\hn$, where $\hn=\hsm$ is presumed to be only one of many possibilities.
In $\epem\rta
\zstar\rta Z\hn$ production, consider the distribution $d\sigma/d\cos\theta$,
where $\theta$ is the angle of the produced $Z$ in the center of mass with
respect to the direction of collision of the initial $e^+$ and $e^-$. The
distribution takes the form
${d\sigma\over d\cos\theta}\propto
{8\mz^2\over s}+{\beta^2}\sin^2\theta$ in the CP-even case, as
compared to $1+\cos^2\theta$ for a CP-odd $\hn$,
where $\beta$ is the center-of-mass velocity of the final $Z$.
Thus, in principle one can measure this distribution and determine
the quantum numbers of the $\hn$ being produced. The difficulty with this
conclusion is best illustrated by considering an $\hn$ which is a mixture
of CP-even and CP-odd components, as described above. The crucial point is
that only the CP-even portion of the $\hn$ couples at tree-level to $ZZ$,
whereas the CP-odd component of the $\hn$ couples weakly to $ZZ$
via one-loop diagrams (barring anomalous sources 
for this dimension-5 coupling). Consequently, the $d\sigma/d\cos\theta$
distribution will reflect only the CP-even component of the $\hn$, even
if the $\hn$ has a fairly large CP-odd component.  The $\hn$ would have
to be almost entirely CP-odd in order for the $\cos\theta$ distribution
to deviate significantly towards the CP-odd prediction.  However, in this
case, the $\epem\rta \zstar\rta Z\hn$ production rate would be very small,
and the $\hn$ would probably not be detectable in the $Z\hn$
associated production mode in any case.  In summary, any $\hn$ {\it which is
not difficult to detect 
in the $Z\hn$ mode} will automatically have a $\cos\theta$
distribution that matches the CP-even prediction, even if there is a
significant CP-odd component in the $\hn$. Unfortunately, the $Z\hn$
production rate itself cannot be used as a measure of the CP-even vs.
CP-odd component of the $\hn$; in a general 2HDM, even a purely CP-even
$\hn$ can have a $ZZ$ coupling that is suppressed relative
to SM strength.

Turning to decay distributions, one encounters a similar problem.
In $\hn\rta WW\rta 4~{\rm fermions}$, one can determine the angle $\phi$
between the decay planes of the two $W$'s.  One finds: 
${d\sigma\over d\phi}\propto
1+\alpha \cos\phi+\beta\cos2\phi$ [$\propto 1-(1/4)\cos2\phi$]
for a CP-even [CP-odd] $\hn$,
where $\alpha$ and $\beta$ depend upon the types of fermions observed and 
the kinematics of the final state.
In general these two distributions are distinguishable.  However,
in analogy to the previous case, it is almost entirely the CP-even component
of the $\hn$ which will be responsible for its decays to $WW$, and the
$\phi$ distribution will thus closely match the CP-even prediction, even if
the $\hn$ has a substantial CP-odd component.  
This is explicitly verified in the calculations of Ref.~\cite{sonixu}.
(See also Refs.~\cite{ck}-\cite{skjold}.)
In order for the $\phi$ distribution to deviate significantly towards the
CP-odd prediction, the CP-even component(s) of the $\hn$ must be small.
Consequently, decays of the $\hn$ to $WW$ channels will be substantially
suppressed, and either the $b\anti b$ or $t\anti t$ channel will
dominate. In the $b\anti b$ channel, the CP-even and CP-odd
components of the $\hn$ cannot be separated. The distribution
$d\sigma/d\cos\theta^*$ (where $\theta^*$ is the $b$ angle relative to the
boost direction in the $\hn$ rest frame) is predicted to be flat,
independent of the CP nature of the $\hn$.

Much more promising are $\taum\taup$ and $t\anti t$ decay modes.
The basic ideas and techniques for a pure-CP eigenstate are nicely
reviewed in Ref.~\cite{kksz}.  These techniques are optimized
and extended to the case of a mixed-CP eigenstate in Ref.~\cite{bohguncp}.
There, a fairly realistic evaluation of their potential on a statistics
basis is given.  The techniques rely on measuring the azimuthal
angle ($\phi$) correlation between certain 
`effective-spin' directions associated
with the fermion and anti-fermion in the final state.  These effective
spins can only be defined by reference to the final states
of the $\tau$ and $t$ decays.  Roughly, 
after integrating over all but the angle $\phi$, one obtains
${1\over N}{dN\over d\phi}={1\over 2\pi}(1+\alpha\cos\phi+\beta\sin\phi)$.
For a pure CP state, $\beta=0$ and $\alpha$ has a predicted magnitude
but its sign is $-$ or $+$ depending upon whether the state is
pure CP-even or pure CP-odd, respectively.  In the case
of a mixed-CP state (such as can arise in a general 2HDM),
$\alpha$ is much smaller in magnitude than for a CP-pure state (perhaps
nearly zero), while $\beta$ is quite substantial. (There is
a sum rule of the form $\alpha^2+\beta^2=C^2$, where $C$ is
a kinematically determined constant.)  The statistical
analysis of Ref.~\cite{bohguncp} shows that for the $\hsm$
a statistically reliable determination that $\alpha=-C$ and that $\beta=0$
would be possible at the 10-20\% level in the $\taum\taup$ final
state for $\mhsm\lsim 2\mw$.  For $\mhsm$ between $2\mw$
and $2\mt$, the $WW$ and $ZZ$ decays would be dominant and one
would `automatically' (as described above) observe correlations
between the $WW,ZZ$ decay planes and the like that are typical
of a CP-even state, but that do not guarantee that the state
is purely CP-even.  For $\mhsm>2\mt$, the $t\anti t$ distributions
of the appropriately-defined $\phi$ would again be available.
However, statistics are predicted to be such (at $L=85\fbi$) that
one could at best measure $\alpha$ and $\beta$ at the 50\% level
relative to the maximum possible values of $C$ (\ie\
$\delta\alpha\sim\delta\beta> 0.5 C$).  Substantially higher luminosity
would be called for.

\subsubsection{Summary for the {\boldmath$\hsm$}}

\indent\indent Overall, we find that many of the properties 
of the $\hsm$ can
be determined at a linear $\epem$ collider with enough energy
to produce the $\hsm$ with a significant rate in the first place.
This reflects the clean environment and kinematics of $\epem$
collisions. With regard to couplings and widths,
the most important exceptions are
the {\it total} width and $b\anti b$ partial width (equivalently, coupling).
In particular, the total width
cannot be determined in a model-independent fashion for $\mhsm\lsim 140\gev$,
while the $b\anti b$ partial width is difficult for almost all 
values of $\mhsm$.  However, for $\mhsm\lsim 140\gev$
we have noted that the total width
and $b\anti b$ partial width, as well as $\Gamma(\hsm\rta\gam\gam)$,
{\it can} be determined by combining the NLC
data with data from the LHC and from $\gam\gam$ collisions.
We have also seen that a {\it direct} verification that the $\hsm$
is purely CP even is predicted to be very difficult for $\mhsm\gsim 2\mw$.

\subsection{The General 2HDM}

\indent As noted earlier, in the general 2HDM the neutral Higgs bosons
need not be CP eigenstates. We have already discussed the probes of the 
CP quantum number of a Higgs boson that might be experimentally viable.
Let us for the moment assume that CP is conserved in the 2HDM,
and recall the two important angles that enter into the phenomenology
of the five physical Higgs bosons of the 2HDM: $\tanb=v_2/v_1$
and $\alpha$, the mixing angle in the CP-even neutral Higgs sector.
In the CP-conserving case the $\zstar\rta Z\hl$
[$Z\hh$] cross section is proportional to $\sin^2(\beta-\alpha)$           
[$\cos^2(\beta-\alpha)$], whereas
the $\zstar\rta \hl\ha$ [$\hh\ha$] cross section is proportional to
$\cos^2(\beta-\alpha)$ [$\sin^2(\beta-\alpha)$].                          
In the limit
of a small value for $\cos(\beta-\alpha)$, or $\sin(\beta-\alpha)$, 
either the $Z\hl$ and $\hh\ha$, or
the $Z\hh$ and $\hl\ha$, respectively, cross sections will be large.
That is there is a strong complementarity among these four 
cross sections such that, given sufficient machine energy,
it will always be possible to detect all three neutral Higgs bosons.
However, there are certainly scenarios in which no Higgs boson
would be seen at the NLC.  For example, it is possible that $\zstar\rta Z\hl$
could be kinematically allowed but coupling suppressed, while
$\zstar\rta\hl\ha$ and $\zstar\rta Z\hh$ could be kinematically forbidden.  
No Higgs boson would be seen without raising the machine energy.  
This is not the case in the MSSM, as we shall shortly discuss.

\subsection{MSSM Higgs bosons}

In the MSSM, the important production mechanisms are mainly determined
by the parameters $\mha$ and $\tanb$. In general, we shall
try to phrase our survey in terms of the parameter space
defined by $(\mha,\tanb)$, keeping fixed
the other MSSM parameters. The most important general point
is the previously noted 
complementarity of the $\epem\rta Z\hl$ and $\epem\rta\ha\hh$ 
cross sections, which are proportional to $\sin^2(\beta-\alpha)$,
and the $\epem\rta Z\hh$ and $\epem\rta \ha\hl$ cross sections,
proportional to $\cos^2(\beta-\alpha)$.
Since
$\cos^2(\beta-\alpha)$ and $\sin^2(\beta-\alpha)$ cannot simultaneously
be small, if there is sufficient energy
then there is a large production cross section for all three of the
neutral Higgs bosons.  In the more likely case that $\mha$ is large,
$\sin^2(\beta-\alpha)$ will be large  and the $\hl$ will be
most easily seen in the $Z\hl$ channel, while the $\ha$ and $\hh$
will have a large production rate in the $\ha\hh$ channel if
$\sqrt s$ is adequate. If $\cos^2(\beta-\alpha)$
is large, then $\mha$ must be small and $\ha\hl$ and $Z\hh$ will
have large rates.

Finally, we note that the Higgs sector of the MSSM is automatically
CP-conserving \cite{hhg}.  Thus, if a neutral Higgs boson is found to have
a mixed CP-nature (by employing the techniques discussed earlier,
or those to be reviewed in the $\gam\gam$ collision section)
then supersymmetry requires a more complicated Higgs sector
than the simple two-doublet structure of the MSSM.  Such models
will be discussed later.

\subsubsection{Detection of the {\boldmath$\hl$}}

\begin{figure}[htbp]
\let\normalsize=\captsize   
\begin{center}
\centerline{\psfig{file=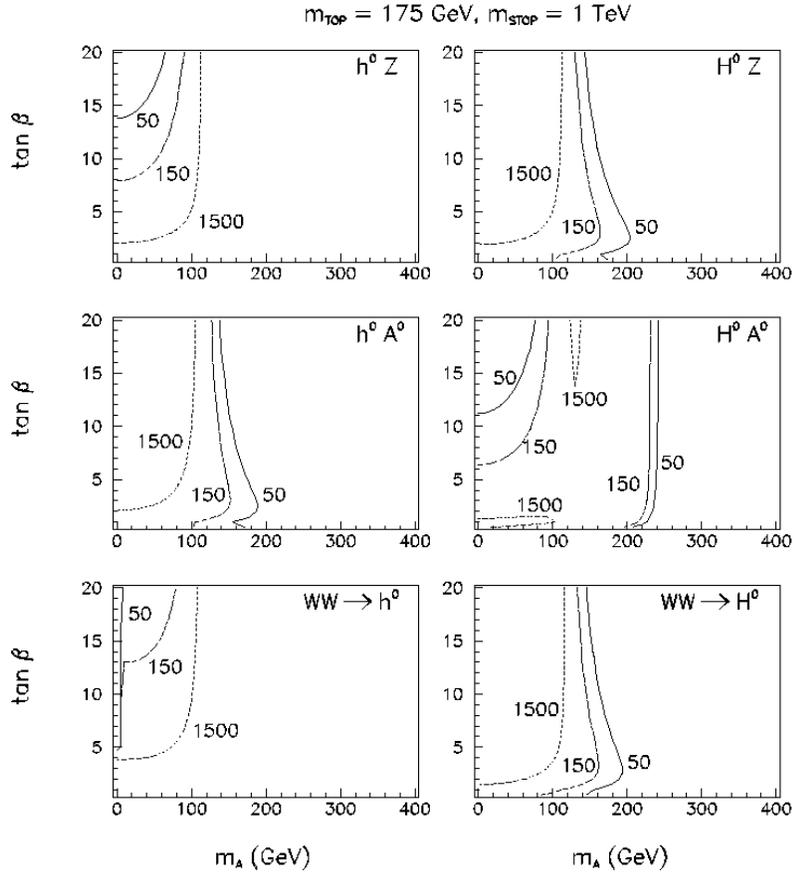,width=12.2cm}}
\begin{minipage}{12.5cm}       
\caption{Event number contours at the NLC for the MSSM Higgs bosons
in $(\mha,\tanb)$ parameter space. \Twoloop\ radiative corrections
to the MSSM Higgs sector have been included, taking $\mstop=1\tev$
and neglecting squark mixing.}
\label{nlccontours}
\end{minipage}
\end{center}
\end{figure}

\indent\indent In the MSSM, the lightest Higgs boson is accessible in the
$Z\hl$ and $WW$-fusion modes for all {\it but} the $\mha\lsim\mz$, 
$\tanb\gsim 7-10$ corner of parameter space.  Outside this
region the $\hl$ rapidly becomes SM-like.
This is illustrated in the first and third windows of Fig.~\ref{nlccontours}, 
where we give
the contours in $(\mha,\tanb)$ parameter space for 50, 150 and 1500
events at $\sqrt s=500\gev$ and $L=50\fbi$, where $\mt=175\gev$
and $\mstop=1\tev$ is assumed.
(This type of figure has appeared in 
Refs.~\cite{desyworkshopsusy,gunionhawaii}. More detailed experimental
discussions can be found in Refs.~\cite{janotlep,janothawaii}.)
After efficiencies, somewhere between 50
and 150 events will be sufficient for detection; 150 events would
be sufficient in the $Z\hl$ mode even if the $\hl$
decays mostly invisibly to a $\cnone\cnone$ pair. (For the
relatively light $\hl$ other SUSY decay modes are unlikely to be present.
For a review of some supergravity/superstring
scenarios in which $\hl\rta \cnone\cnone$
decays are dominant, see Ref.~\cite{gunionzeuthen}.)
If such SUSY decays are not important for the $\hl$,
then to a very good approximation, the entire SM $Z\hsm$ discussion 
can be taken over for $Z\hl$ for large $\mha$.  The interesting
question, to which we shall return below, is for what portion
of moderate-$\mha$ parameter space can
the cross sections, branching ratios and/or couplings be
measured with sufficient precision to distinguish an approximately SM-like
$\hl$ from the $\hsm$ at the NLC. (The results we
obtain should be compared to those summarized earlier
in the case of the LHC.) Presumably, large
numbers of events would be required.  We see from the $Z\hl$ window
of Fig.~\ref{nlccontours}
that at least 1500 events are predicted for all {\it but} the $\mha\lsim 100$,
$\tanb\gsim 2-5$ corner of parameter space.

Although strongly disfavored by model building, there is
still a possibility that $\mha\lsim \mz$. The important production
process will then be $\epem\rta \zstar\rta \hl\ha$, the $\hl ZZ$ and $\hl WW$
couplings being suppressed for such $\mha$ values.
As discussed earlier, if \lepii\ runs at $\sqrt s=200\gev$ 
$\hl\ha$ associated production will be kinematically allowed and detectable for
all but {\it very} large $\mstop$ values.
The NLC could be run at a $\sqrt s$ value optimized for detailed studies
of the $\hl\ha$ final state. At $\sqrt s=500\gev$, at least 1500 events are 
predicted for precisely the $\mha\lsim 100\gev$, $\tanb\gsim 2-5$ section
of parameter space for which fewer events than this would be found
in the $Z\hl$ mode.  Thus, $\epem\rta\hl\ha$ will give
us a large number of $\hl$'s if $Z\hl$ does not.
We will confine ourselves primarily to general remarks regarding
the small-$\mha$ scenario in this report; our focus will
be on large $\mha$.

\subsubsection{Detection of the 
{\boldmath $\hh$, $\ha$ and $\hpm$}}

\indent\indent For $\mha\lsim \mz$, the $\ha$, along with the $\hl$,
can be easily detected in the $\hl\ha$ mode, as discussed above.
If 50 events are adequate, detection of both the $\hl$ and $\ha$
in this mode will even be possible for $\mha$ up to $\sim 120\gev$.
As seen in the fourth and sixth windows of Fig.~\ref{nlccontours},
in this same region
the $\hh$ will be found via $Z\hh$ and $WW$-fusion production
\cite{desyworkshopsusy,janothawaii,gunionhawaii}.
In addition, $\hp\hm$ pair production will be kinematically allowed and easily 
observable \cite{desyworkshopsusy,janothawaii,gunionhawaii}.
In this low to moderate $\mha$ region, the only SUSY decay mode that
has a real possibility of being present is the invisible $\cnone\cnone$
mode for the neutral Higgs bosons.  If this mode were to dominate
the decays of all three neutral Higgs bosons, then only the
$\hh$ could be detected, 
using the recoil mass technique in the $Z\hh$ channel.
However, in the $\hp\hm$ channel the final states would probably
not include SUSY modes and $\hp$ discovery would be straightforward.
If a light $\hpm$ is detected, then one would know that the Higgs
detected in association with the $Z$ was most likely the $\hh$
and not the $\hl$.  A dedicated search for the light $\ha$ and $\hl$
through (rare) non-invisible decays would then be appropriate.

For $\mha\gsim 120\gev$, $\epem\rta\hh\ha$ and $\epem\rta\hp\hm$ must be
employed for detection of the three heavy Higgs bosons.
Assuming that SUSY decays are not dominant, and
using the 50 event criterion, 
the mode $\hh\ha$ is observable up to
$\mhh \sim \mha \sim 240$ GeV, and $\hp\hm$ can be detected up to
$\mhpm =230$ GeV \cite{desyworkshopsusy,janothawaii,gunionhawaii}.
The $\gamma\gamma$ collider mode could potentially
extend the reach for the $\hh,\ha$ bosons up to 400 GeV if $\tanb$
is not large. This is reviewed in Ref.~\cite{gunionerice}, and will
be discussed in more detail later.

The upper limits in the $\hh\ha$ and $\hp\hm$
modes are almost entirely a function of the
machine energy (assuming an appropriately higher integrated luminosity
is available at a higher $\sqrt s$).
At $\sqrt s = 1$ TeV with an integrated luminosity of $200 \fbi$, 
$\hh\ha$ and $\hp\hm$ detection would be extended
to $\mhh\sim\mha\sim\mhpm\sim 450$ GeV 
\cite{desyworkshopsusy,gunionerice,janothawaii,gunionhawaii}. 
As frequently noted, models in 
which the MSSM is implemented in the coupling-constant-unification,
radiative-electroweak-symmetry-breaking context often
predict masses above 200 GeV, suggesting that
this extension in mass reach over that for $\sqrt s =500$ GeV
might be crucial. 
For some review and references, see Refs.~\cite{baerreport,gunionzeuthen}.

\subsubsection{
{\boldmath Distinguishing the MSSM $\hl$ from the SM $\hsm$ at the
NLC}}

\indent\indent We will discuss this issue in the most interesting case where 
a direct discovery of the other MSSM Higgs bosons is not possible,
that is in the limit where $\mha\gsim 230-240\gev$. Although
there are many potentially useful techniques for distinguishing
the $\hsm$ from the $\hl$, only a few could
have the required accuracy for such high $\mha$ values.  
In discussing expected experimental errors below we extrapolate 
the NLC/SLC $\mhsm=120\gev$ results and NLC/JLC $\mhsm=110\gev$ results
(see Table~\ref{nlclhcerrors}) to $\mhsm=113\gev$, the maximum
$\mhl$ value for our canonical choices
of $\mt=175\gev$, $\mstop=1\tev$ and no squark mixing.
Most uncertain is
the error on the $W\wstar$ channel which worsens very rapidly as
$\mhsm$ decreases below $120\gev$.
At $\mhsm=113\gev$ we estimate that 
the $\pm48\%$ that applies at $\mhsm=120\gev$ will worsen to $\sim\pm 70\%$
because of the factor of two decrease in the event rate.

First, there is the simple 
magnitude of the $Z\hsm$ cross section. Deviations as large as 
the best possible error (of order $7\%$, as summarized earlier) are
only possible if $\sin^2(\beta-\alpha)$ deviates from unity
by this amount.  Unfortunately, we shall see below that
deviations of this size are only predicted for $\mha\lsim 160\gev$.

Next there are measurements of $\sigma(Z\hl)BR(\hl\rta X)$
for various channels $X=b\anti b,W\wstar,(c\anti c+gg)$.
As noted earlier, excellent measurement accuracy ($\pm 2.5\%$
for the super-JLC detector and $\pm 7\%$ for SLC/LEP type detectors)
is possible for $\sigma(Z\hl) BR(\hl\rta b\anti b)$ with $L=50\fbi$
of integrated luminosity.
As detailed below (see also
Figs. 17 and 18 of Ref.~\cite{janothawaii}, but keep
in mind that that these figures do not include the full \twoloop\
radiative corrections whereas those presented below do) an excess
in this quantity for the MSSM relative to the SM is predicted to be larger
than $\sim 2.5\%$ for relatively large $\mha$ values; {\it e.g.} 5\% deviations
are predicted for $\mha\sim 300\gev$ at larger $\tanb$ values.
However, the absolute value of $BR(\hl,\hsm\rta b\anti b)$
is quite sensitive to radiative corrections, the `current'
$b$-quark mass (appearing in the $\hsm\to b\anti b$ coupling), 
and `extra' contributions
to the total width. While the radiative corrections 
can eventually be calculated with
the required precision, the value of the running mass ($\overline {MS}$)
$\mb(\mh)$ is likely to remain
somewhat uncertain due to uncertainty in $\mb(\mb)$. This could be critical.
For example, a 5\% uncertainty in $\mb(\mb)$ leads to an
uncertainty in $BR(\h\rta b\anti b)$ ($\h=\hl$ or $\hsm$) of order $3\%$; 
see Eq.~\ref{bbbrsystematicerror}. The value of
$\sigma(Z\hl) BR(\h \rta W\wstar)$ [$\sigma(Z\h) BR(\h\rta c\anti c+gg)$]
can be measured to $\pm48\%$ [$\pm 39\%$]
at $\mh=120\gev$; somewhat larger errors will apply to $\mh=113\gev$
in the $W\wstar$ case --- as noted above, we estimate $\pm70\%$.
The graph presented below will show that such accuracies are adequate 
for distinguishing the $\hl$ from the $\hsm$ only for $\mha\lsim 200\gev$.

The relative branching ratios to different channels
provide additional information. The modes with branching
ratios most sensitive to the small differences between the $\hl$
and $\hsm$ couplings are the $W\wstar$ and $c\anti c+gg$.
The $\hl$ branching ratios to these modes can be much smaller
at large $\tanb$ as compared to the $\hsm$, so long
as $\mha$ is not too large.  Some discussion of such differences
as computed at the one-loop level appears in Refs.~\cite{hbb,hildrethhawaii}.
Below we shall present some results obtained after 
including \twoloop\ corrections
to the $\hl$, taking $\mt=175$, $\mstop=1\tev$ and
neglecting squark mixing. The error in the determination of a ratio $r=x/y$
is computed as
\begin{equation}
{\delta r}=r\sqrt{(\delta x/x)^2+(\delta y/y)^2}
\label{deltar}
\end{equation}
For $\mhsm=113\gev$, we find:
\begin{equation}
r_{W\wstar}\equiv {BR(\hsm\rta W\wstar)\over BR(\hsm\rta b\anti b)}
\sim 0.05\,,\quad
r_{c\anti c+gg}\equiv
{BR(\hsm\rta c\anti c+gg)\over BR(\hsm\rta b\anti b)}\sim 0.125\,,
\label{ratiovalues}
\end{equation}
Using the above-quoted errors for the individual channels 
and the $r$ values given in Eq.~\ref{ratiovalues}, we find from
Eq.~\ref{deltar} the results
$\delta r_{W\wstar}/r_{W\wstar}\sim 0.7$ and $\delta r_{c\anti c+gg}/r_{c\anti
c+gg}\sim 0.4$
(all at the $1\sigma$ level). We shall see below that the former
error renders $r_{W\wstar}$ rather ineffective, while $r_{c\anti c+gg}$
is only capable of probing the difference between the $\hl$
and the $\hsm$ (for the $\mt=175\gev$, $\mstop=1\tev$, no-squark-mixing case
being considered) for $\mha$ values below about $300\gev$.

\begin{figure}[htbp]
\let\normalsize=\captsize   
\begin{center}
\centerline{\psfig{file=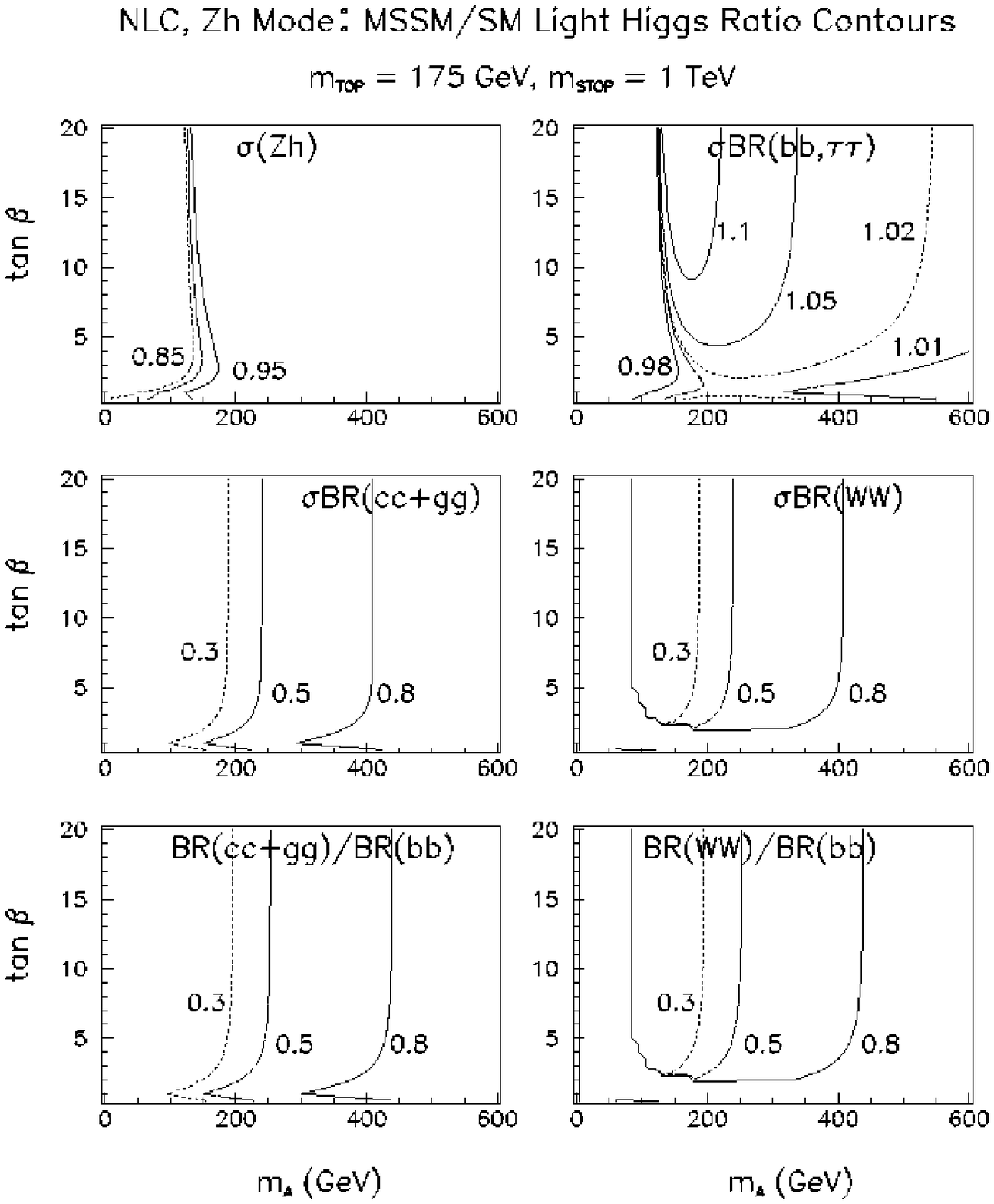,width=12.2cm}}
\begin{minipage}{12.5cm}       
\caption{Contours for ratios of MSSM $\hl$ results to
SM $\hsm$ results for $\mhl=\mhsm$, in $(\mha,\tanb)$ parameter space.
\Twoloop\ radiative corrections are included, taking
$\mstop=1\tev$ and neglecting squark mixing.}
\label{nlcdeviationscontours}
\end{minipage}
\end{center}
\end{figure}

The quantitative predictions \cite{ghhprecision}
for the deviations of $\hl$ predictions
from $\hsm$ predictions for the above quantities are plotted
in Fig.~\ref{nlcdeviationscontours}. There we display
contours in $(\mha,\tanb)$ parameter space
corresponding to fixed ratios of MSSM $\hl$ to SM $\hsm$ predictions
taking $\mhsm=\mhl$. A stop squark mass of $1\tev$ and $\mt=175\gev$
are employed in this figure.
For $\sigma(Z\hl)$, 95\%, 90\% and 85\% contours are shown.
For $\sigma(Z\hl) BR(b\anti b,\tau\tau)$, 98\%, 101\%, 
102\%, 105\%, and 110\% contours
are shown.  In all other cases, 80\%, 50\%, and 30\% contours are shown.
(Our $\sigma(Z\hl) BR(\hl\rta b\anti b)$ results differ somewhat from
those of Ref.~\cite{janothawaii}; we have included \twoloop\ corrections.)
As can be seen, to distinguish the $\hl$ from the $\hsm$ for
$\mha$ values beyond $\sim 200\gev$, a much better than 5\%
measurement of $\sigma(Z\hl)$ must be performed, while
$\sigma(Z\hl) BR(\hl\rta b\anti b~{\rm or}~\tau\tau)$ must be measured
(and reliably computed) to better than 10\%, and $\sigma(Z\hl) BR(\hl\rta
gg+c\anti c)$ or $\sigma(Z\hl) BR(\hl\rta W\wstar)$ to better than 50\%.
As discussed above, such errors are at the limit of what can be achieved
experimentally.  The ratio of branching ratios, $r_{c\anti c+gg}$
provides some sensitivity to $\hl$ vs. $\hsm$ differences for $\mha$ values 
up to about $\mha\sim 300\gev$ so long as $\tanb\gsim 2-3$.
(Values of $\tanb$ near 1 do not
lead to the enhanced $b\anti b$ and $\taup\taum$ partial widths
that cause significant deviations.)

Predicted deviations are smaller at any given $(\mha,\tanb)$ location
if $\mstop$ is significantly below $1\tev$.  For
example, if $\mstop\sim 500\gev$ the 1.01 contour in $\sigma(Z\hl) 
BR(\hl\rta bb,\tau\tau)$
falls in about the same location as the 1.02 contour in the
$\mstop=1\tev$ case illustrated in Fig.~\ref{nlcdeviationscontours},
and the 0.8 contours for the $c\anti c+gg$ ($W\wstar$) deviations
move about $50\gev$ ($10\gev$) lower in $\mha$ at larger $\tanb$.
Experimental errors on the $W\wstar$ channel will
also increase as the large-$\mha$ value of 
$\mhl$ decreases with decreasing $\mstop$.

\subsubsection{The influence of supersymmetric decays} 
 
\indent\indent For some MSSM parameter choices,
the $\hl$ can decay primarily to invisible modes, including
a pair of the lightest supersymmetric neutralinos 
or a pair of invisibly decaying sneutrinos \cite{gunionzeuthen}. 
The $\hl$ would still be easily discovered
at $e^+e^-$ colliders in the $Z\hl$ mode using recoil-mass techniques
\cite{janothawaii,gunionhawaii}.
Of course, if the $\hl$ decays invisibly to $\cnone\cnone$,
this implies that the overall ino mass scale is quite light and that
direct continuum production of $\chitil\chitil$ pairs with visible decays
would also be easily detected at the NLC. These 
additional experimental signals would make it
clear that a supersymmetric model is nature's choice.

The $\hh$, $\ha$, and $\hpm$ decays can be dominated by
chargino and neutralino pair final states and/or slepton pair final states
\cite{hhg,gunionerice,baererice,gunionzeuthen}. Such modes can
decrease the chances of detecting these heavier MSSM
Higgs bosons at a $\epem$ collider;
$\hh$, $\ha$ and $\hpm$ detection up to the earlier
quoted (largely kinematical) limits depends upon the complexity of
their decays. There are MSSM parameter choices such that the decay modes
of these heavier Higgs are extremely diverse
and/or even invisible, in which case their
detection in the normal $\ha\hh$, $\hp\hm$ 
associated production modes could be challenging.
A survey of the possibilities is in progress \cite{gkopair}.

\subsection{Extensions of the MSSM}

\indent\indent As previously discussed, the supersymmetric Higgs sector must
contain at least two-doublets in order to give masses
to both the up and down type quarks.  We have also noted that
gauge unification requires that there be no more than two Higgs
doublets.  Further, Higgs triplet representations create even
more problems than they do in extending the minimal SM Higgs sector.
Thus, the most attractive extensions of the MSSM Higgs sector
are those in which one or more singlet Higgs fields are added.
Indeed, it should be noted that many string-motivated models
contain one or more extra singlet fields.

The minimal such extension is that of a single additional singlet Higgs
field.  This extension was first considered in Ref.~\cite{eghrz}.
The extremely attractive feature of this model is that it provides
a natural source for the $\mu$ term required to give the $\ha$
a non-zero mass.  Recall that in the MSSM, it is necessary to introduce
the superpotential term $W=\mu \hat H_1 \hat H_2$,
where $\hat H_{1,2}$ are the two doublet Higgs superfields. 
The parameter $\mu$
appears in association with the soft-supersymmetry breaking parameter, $B$, 
in the soft mass term $B\mu H_1H_2$ which mixes the two
Higgs scalar doublets, thereby giving mass to the $\mha$.
In many models of supersymmetry breaking, it is most natural that
$\mu$ should have a magnitude of order the unification scale
$M_X$.  There would
be no reason for $B\mu$ (and hence $\mha$) to be of order the
electroweak scale (\ie\ below a TeV) if $\mu$ is of order $M_X$.
Many solutions to this so-called `$\mu$-problem' have been proposed
(see Ref.~\cite{baerreport}), but the most natural is the presence
of an extra Higgs singlet superfield, which we denote by $\hat N$.  
There would then be a superpotential term of the form $W=\lam \hat H_1\hat H_2
\hat N$, and an associated soft-supersymmetry breaking term of the
form $V_{soft}=\lam A_{\lam} H_1H_2 N$, where $A_{\lam}$
is the soft-supersymmetry-breaking parameter associated 
with $\lam$ at the superpotential level.  If $\langle N\rangle\lsim 1
\tev$ and $\lam$ is in the perturbative domain, then at least one of the
 pseudoscalar Higgs bosons (there are two in this model)
would generally have mass below a TeV. Values of 
$\langle N\rangle\lsim 1\tev$ emerge naturally in perturbative RGE/unification
treatments of the model and are technically natural in the sense
that they are protected against large quadratic loop corrections.

In the CP-conserving case, this minimal non-minimal extension of the MSSM
(often denoted MNMSSM) would contain three neutral scalar Higgs bosons,
two neutral pseudoscalar Higgs bosons, and a single charged Higgs pair.
In the CP-violating case, where some of the soft-supersymmetry-breaking
parameters related to the Higgs sector are allowed to be complex
(spontaneous CP violation in the MNMSSM Higgs sector is also possible,
but only if the soft-supersymmetry-breaking potential has
a full complement of terms, and not just the minimal 
$\lam A_{\lam} H_1H_2 N$ and $kA_kN^3$ forms)
there would be five neutral Higgs bosons of mixed-CP character.
Clearly, either version of the MNMSSM would present many new phenomenological
opportunities and issues.  

Consider the CP-conserving model, and label the three CP-even
Higgs bosons as $S_{1,2,3}$ in order of increasing mass.
The first question is whether or not a $\sqrt s=500\gev$ $\epem$
collider would still be guaranteed to discover at least one
of the MNMSSM Higgs bosons.  In the MSSM, there is such a guarantee
because the $\hl$ has an upper mass bound, {\it and} because 
it has near maximal $\hl ZZ$ coupling when $\mhl$ approaches its upper limit.
In the MNMSSM model, it is in principle possible to choose
parameters such that $S_{1,2}$ 
have such suppressed $ZZ$ coupling strength that their $ZS_{1,2}$
and $WW\rta S_{1,2}$ production rates are too low for observation,
while the heavier $S_3$
Higgs is too heavy to be produced in the $Z S_3$ or $WW\rta S_3$ modes.
This issue has been studied recently in Refs.~\cite{kimoh,kot,KW,ETS}.
The result of Ref.~\cite{kot} is that if the model is placed within the
normal unification context, with simple boundary conditions at $M_X$,
{\it and if all couplings are required to remain perturbative
in evolving up to scale $M_X$} (as is conventional), then at least
one of the three neutral scalars will have $\sigma(\epem\rta Z S)\gsim 0.04\pb$
for any $\epem$ collider with $\sqrt s\gsim 300\gev$.
For $L=10\fbi$, this corresponds to roughly 30 events in the incontrovertible
$ZS$ with $Z\rta \lplm$ recoil-mass discovery mode.  To what extent
this result generalizes to models with still more singlets is not
known.  However, the proof of the above statement relies on the observation 
that the more the lighter Higgs bosons decouple from $ZZ$, the lower the upper
bound is on the next heaviest Higgs boson. This could easily generalize.
This result has been confirmed by the recent work of Refs.~\cite{KW,ETS}.
However, Refs.~\cite{kimoh,kot,KW,ETS} all make it clear that
there is no guarantee that \lepii\ will detect a Higgs boson of the MNMSSM.
This is because the Higgs boson with significant $ZZ$-Higgs coupling
can easily have mass beyond the kinematical reach of \lepii.

Once a neutral Higgs bosons is discovered,
it will be crucial to measure all its couplings and to 
determine its CP character, not only to 
try to rule out the possibility that it is the SM $\hsm$, but
also to try to determine whether or not the supersymmetric
model is the MSSM or the MNMSSM (or still further extension).
Measurement of the CP-nature of the Higgs boson
would be especially important.  As previously mentioned, 
in the MSSM CP violation cannot
arise in the Higgs sector \cite{hhg}, whereas we have
noted that Higgs sector CP violation is possible in the MNMSSM
(although spontaneous (explicit) CP violation is not (is) possible 
in the version of the MNMSSM with the most minimal superpotential).
In the CP-violating version of the MNMSSM, the previously discussed
techniques for exploring the CP-nature of any Higgs boson that
is detected in $Z\hn$ production would generally be expected
to be very useful. In particular,
in the unification context many of the neutral Higgs bosons are
most naturally below $2\mw$ in mass, and the $\hn\rta\taup\taum$
mode could well allow a determination of whether or not any
such Higgs boson is a mixed-CP state.

\subsection{The role of a back-scattered laser beam facility}

\indent\indent It is now widely anticipated 
(see, \eg, Ref.~\cite{telnovhawaii,bordenhawaii})
that the NLC can be operated in a mode in which laser beams
are back-scattered off of one or both of the incoming $e^+,e^-$
in such a way as to allow $e\gamma$ or $\gamma\gamma$ collisions
at high luminosity and high energy.  (Operation of the
NLC as an $e^-e^-$ collider
would be even more suitable for the $\gam\gam$ collider mode of operation, and
polarization of both the $e^-$ beams would be possible.)
The luminosity for $\gam\gam$ collisions can even be substantially larger than
that for $\epem$ collisions in the normal mode of operation.
The energy spectrum of the collisions depends upon the
choice of electron and laser beam polarizations, but can be either
broad or peaked.  For instance, in $\gam\gam$ collisions
a spectrum peaked narrowly in the vicinity of $\egamgam\sim 0.8\sqrt s$
is possible. We will focus first on the $\gam\gam$ collision
option, with some remarks on the $e\gam$ collision possibility.

\subsubsection{{\boldmath The SM $\hsm$ at a $\gam\gam$ collider}}

\indent\indent In the last few years the possibility of employing collisions of
back-scattered laser beams to discover the SM Higgs boson at a linear
$\epem$ collider has been explored \cite{ghgamgam,barklow,bbc}.
The event rate is directly proportional to $\Gamma(\hsm\rta\gam\gam)$.
The interest in this mode derives primarily from two facts.
First, observation of the $\hsm$ in this production mode provides
probably the only access to the $\hsm\gam\gam$ coupling at an $\epem$
linear collider.  The $\hsm\rta\gam\gam$ decay channel has (at best) a
branching ratio of order $2\times 10^{-3}$; too few events will
be available in direct $\epem$ collisions to allow detection of such decays.
Second, in principle it might be possible to detect the $\hsm$
for $\mhsm$ somewhat nearer to $\sqrt s$ than the $0.7\sqrt s$ that appears
to be feasible via direct $\epem$ collisions. Indeed, the full $\gam\gam$ 
center-of-mass energy, $W_{\gam\gam}$, goes into creating the $\hsm$, and 
(as noted above) 
the back-scattered laser beam facility can be configured so that
the $W_{\gam\gam}$ spectrum peaks slightly above $0.8\sqrt s$.
The general prospects for $\hsm$ discovery in $\gam\gam$ collisions
are delineated in Refs.~\cite{ghgamgam,barklow,bbc}.  The result
is that by employing
appropriate laser and electron polarizations, a viable signal
in either the $b\anti b$ or $ZZ$ final state decay mode of the $\hsm$
can be extracted for $\mhsm$ up to $\sim 350\gev$; unfortunately, 
as detailed below, beyond this
point the 1-loop $ZZ$ background overwhelms the signal \cite{jikia,berger,DK}.

The importance of determining the $\hsm\gam\gam$ coupling derives from
the fact that it is determined by the sum over all 1-loop diagrams 
containing any charged particle whose mass arises from the Higgs
field vacuum expectation value.  In particular, the 1-loop contribution
of a charged particle with mass $\gsim\mhsm/2$, approaches a constant value
that depends upon whether it is spin-0, spin-1/2, or spin-1. (The
contributions are in the ratio  $-1/3$ : $-4/3$ : 7, respectively.)  
For a light Higgs boson, in the SM the
dominant contribution is the $W$-loop diagram.  The next most important
contribution is that from the top quark loop, which tends to cancel part of
the $W$-loop contribution.  A fourth fermion generation with both a heavy
lepton, $L$, and a heavy $(U,D)$ quark doublet would lead to still further
cancellation. For $\mhsm\gsim 2\mw$, the $W$-loop contribution decreases,
and the heavy family ultimately dominates. Illustrations
are given in Refs.~\cite{jggeer} and \cite{ghgamgam}. For example, the ratio
of $\Gamma(\hsm\rta\gam\gam)$ as computed in the presence of an extra 
generation with $m_L=300\gev$ and $m_U=m_D=500\gev$
to the value computed in the SM rises from below 0.2 at $\mhsm\sim 60\gev$
to above 10 for $\mhsm\gsim 500\gev$. 
Except for $\mhsm$ in the vicinity of $300\gev$,
where the full set (mainly the heavy generation) of contributions
accidentally matches the SM result, 
even a rough measurement (or bound) on
the $\hsm\gam\gam$ coupling would reveal the presence of the otherwise
unobservable heavy generation. Note in particular that a
heavy generation would greatly enhance the event rate (and hence prospects)
for detecting a Higgs boson with mass up near $500\gev$ 
(the probable kinematic limit for a first generation NLC) in $\gam\gam$
collisions. Conversely, for $\mhsm\lsim 2\mw$ such an extra generation
could make it difficult to detect the $\hsm$ in $\gam\gam$ collisions.

Because of the dominance of the $W$ loop contribution in the three family
case, the $\hsm\gam\gam$ coupling is also very sensitive to any deviations
of the $WW\gam$ and $WW\hsm$ couplings from SM values, such as
those considered in Refs.~\cite{derujula,perezt}.
The sensitivity to
anomalies in these couplings can be substantially greater than that
provided by LEP I data, and comparable to that provided by \lepii\ data.

In general, although the $\gam\gam$ mode may not
extend the discovery reach of an $\epem$ collider, it {\it will} allow
a first measurement of the $\hsm\gam\gam$ coupling of any Higgs
boson that is found in direct $\epem$ collisions. The accuracy that can be
expected has been studied in 
Refs.~\cite{bbc,bordenhawaii}. Two final states were considered: the
$\hsm\rta b\anti b$ channel with $b$-tagging, and the $\hsm\rta ZZ$
channel with one $Z$ required to decay to $\lplm$. In the former case, it is
important, as noted in Ref.~\cite{ghgamgam}, 
to polarize the laser beams so that
the colliding photons have $\langle\lam_1\lam_2\rangle$ near 1.  
This suppresses the
$\gam\gam\rta b\anti b$ background which is proportional to
$1-\langle\lam_1\lam_2\rangle$. It was found that
if $35\lsim\mhsm\lsim 150\gev$, then the $b\anti b$ mode will allow a 5-10\%
determination of $\Gamma(\hsm\rta\gam\gam)$, while for
$185\lsim\mhsm\lsim300\gev$ the $ZZ$ mode will provide a 8-11\%
determination.  In the $150-185\gev$ window, the $WW$ and $b\anti b$ decays
are in competition, and the accuracy of the measurement might not be better
than 20-25\%.

Let us discuss in more detail how
high in mass the $\hsm$ can be detected in $\gam\gam$ collisions
in the case of the SM Higgs. For $\sqrt s=500\gev$, the range of
greatest interest is that which cannot be accessed by
direct $\epem$ collisions, \ie\ $\mhsm\gsim 350\gev$. In this mass region,
$\hsm\rta ZZ$ decays provide the best signal. Certainly, the tree-level
$\gam\gam\rta\wp\wm$ continuum background completely overwhelms
the $\hsm\rta\wp\wm$ mode \cite{ghgamgam,mtz}. As summarized in \cite{ghgamgam},
if there were no continuum $ZZ$ background, and if one of the
$Z$'s is required to decay to $\lplm$, the event rate would be adequate
for $\hsm$ detection up to $\mhsm\sim 400\gev$, \ie\ $\mhsm\sim 0.8\sqrt
s$. Unfortunately, even though there is no
tree-level $ZZ$ continuum background, such a background does arise
at one-loop. A full calculation of this background was performed in 
Refs.~\cite{jikia,berger}. The $\wpm$ loop is dominant, and leads to a large
rate for $ZZ$ pairs with large mass, when one or both of the $Z$'s
is transversely polarized. This background is such that $\hsm$
observation in the $ZZ$ mode is probably not possible for $\mhsm\gsim
350\gev$, \ie\ no better than what can be achieved in direct
$\epem$ collisions. 

If the machine energy is significantly increased, then a new possibility
opens up for finding a heavy $\hsm$ in $\gam\gam$ collisions.  The $\hsm$
can be produced in the reaction where each incoming $\gam$
turns into a virtual $WW$ pair, followed by one $W$ from each such pair
fusing to form the $\hsm$; \ie\ a $\gam\gam$ collision version of $WW$-fusion.
This possibility has been evaluated quantitatively in 
Refs.~\cite{brodskyww}-\cite{jikiaww}. The result is that a SM Higgs
boson with $\mhsm$ up to $700\gev$ ($1\tev$) could be found in this
mode at a collider with $\sqrt s=1.5\tev$ ($2\tev$), assuming 
integrated luminosity of $L=200\fbi$ ($300\fbi$).

A number of authors have explored further
backgrounds to detecting the $\hsm$ in the $\gam\gam$ collision mode. 
In general, these additional backgrounds can be kept small,
with the exception of the $\gam\gam\rta ZZ$
continuum background from the $W$-loop graphs just discussed.
First, there is the issue of whether or not the $b\anti b g$
gluon radiation background (which a priori is comparable to the $b\anti b$
background) can be effectively suppressed by the same helicity
choices that were used to suppress the basic $b\anti b$ background.
The speculation \cite{ghgamgam} that this could be accomplished
by vetoing the gluon was quantitatively confirmed in Ref.~\cite{bkso}.

The processes $\gam\gam\rta Z\lplm$ and $Zq\anti q$ yield a reducible
background to the $ZZ$ mode to the extent that the $q\anti q$ or $\lplm$
have mass near $\mz$ \cite{bowserchao}.
The magnitude of this background depends upon the detector
resolution and photon polarizations.
If $\langle\lam_1\lam_2\rangle$ is not near 1, then this background
can be significant (though not as large as the $ZZ$ continuum
background). However, like the $b\anti b$ and
basic $ZZ$ continuum background, these processes are proportional to
$1-\langle\lam_1\lam_2\rangle$ and 
can be suppressed substantially by appropriate
polarization choices for the incoming back-scattered laser beams.

The $b\anti b$ channel receives a background contribution from 
``resolved'' photon processes \cite{ebolietal}.
The most important example is
that where one incoming $\gam$ fragments to a spectator jet and a gluon. 
The subprocess $\gam g\rta b\anti b$ then yields a large $b\anti b$ rate.
However, this background will not be a problem in practice for two reasons.
First, it will be possible to veto against the spectator jet that
accompanies the $g$.  This probably already reduces the background to
a level below the true $\gam\gam\rta b\anti b$ continuum.
Second, for the range of $\mhsm$ such that the $b\anti b$ mode is
appropriate ($\mhsm\lsim 150\gev$), the $\hsm$ will already have been
discovered in direct $\epem$ collisions, \ie\  $\mhsm$ will be known.
To study the $\hsm$ in $\gam\gam$ collisions it will be easy to adjust
the machine energy and laser beam polarizations so that the $\gam\gam$
spectrum is peaked at $W_{\gam\gam}\sim\mhsm\sim 0.8\sqrt s$.
(See, for instance, Ref.~\cite{bbc}.) In this case, the secondary gluon
in the ``resolved''-photon process is quite unlikely to have sufficient
energy to create a $b\anti b$ pair with mass as large as $\mhsm$.

Finally, we note that a variety of other final states 
containing the $\hsm$ can be produced
in $\gam\gam$ collisions.  For instance, the $\gam\gam\rta t\anti t \hsm$
analogue of $\epem\rta t\anti t \hsm$ could provide another measure
of the $t\anti t \hsm$ coupling \cite{boostthsm,cheungtthsm}.
However, because of phase space suppression, the rate for this reaction
is quite small for $\sqrt s=500\gev$, and only becomes competitive
with $\epem\rta t\anti t\hsm$ when $\sqrt s \gsim 1\tev$.
As an aside, radiative corrections to $\gam\gam\rta t\anti t$ and $ZZ$ due to
1-loop Higgs exchange graphs are also sensitive to the $t\anti t\hsm$
coupling. Sufficiently precise measurements of these processes at high
luminosity and energy might allow a determination of the coupling over a
significant range of $\mt$ and $\mhsm$ values, {\it
assuming no other new physics in the 1-loop graphs} \cite{boostthsm}.

Turning now to $e\gam$ collisions, we merely note here that
$e\gam\rta W\hsm\nu\rta jj b\anti b\nu$ may be viable for $\hsm$
searches for $\sqrt s\gsim 1\tev$ \cite{hagwz}-\cite{cheung}.
(The last reference includes some background studies.)
This process is interesting in that it probes the $\gam W\rta \hsm W$
subprocess which is determined by a combination of graphs with
different basic SM couplings.  Should the couplings deviate from SM
predictions, the large cancellations among the graphs might
be reduced and the event rate significantly enhanced. Another mode
of interest is $e\gam\rta e\gam\gam\rta e\hsm$, in which a secondary $\gam$
collides with the primary $\gam$ to create the $\hsm$
\cite{eggn}. The cross section for this process is bigger than that
for $e\gam\rta W\hsm\nu$ and might allow detection of the $\hsm$
at $\sqrt s=500\gev$ in the $b\anti b$ mode. (Resolved photon
backgrounds would have to be suppressed by spectator jet vetoing.)

\subsubsection{The MSSM Higgs bosons at a {\boldmath$\gam\gam$} collider}

\indent\indent The most important limitation of a $\epem$ collider in
detecting the MSSM Higgs bosons is the fact
that the parameter range for which the production process, $\zstar\rta\hh\ha$
has adequate event rate is limited
by the machine energy to $\mha\sim\mhh\lsim \sqrt s/2-20\gev$ (recall that
$\mhh\sim\mha$ at large $\mha$).  At $\sqrt s=500\gev$, this
means $\mha\lsim 230\gev$. Meanwhile, $\epem\rta \hp\hm$ is also limited to
$\mhpm\sim\mha\lsim 220-230\gev$, as we have noted. Thus, it could happen
that only a rather SM-like $\hl$ is detected in $\epem$ collisions at the
linear collider, and none of the other Higgs bosons are observed.

\begin{figure}[htbp]
\let\normalsize=\captsize   
\begin{center}
\centerline{\psfig{file=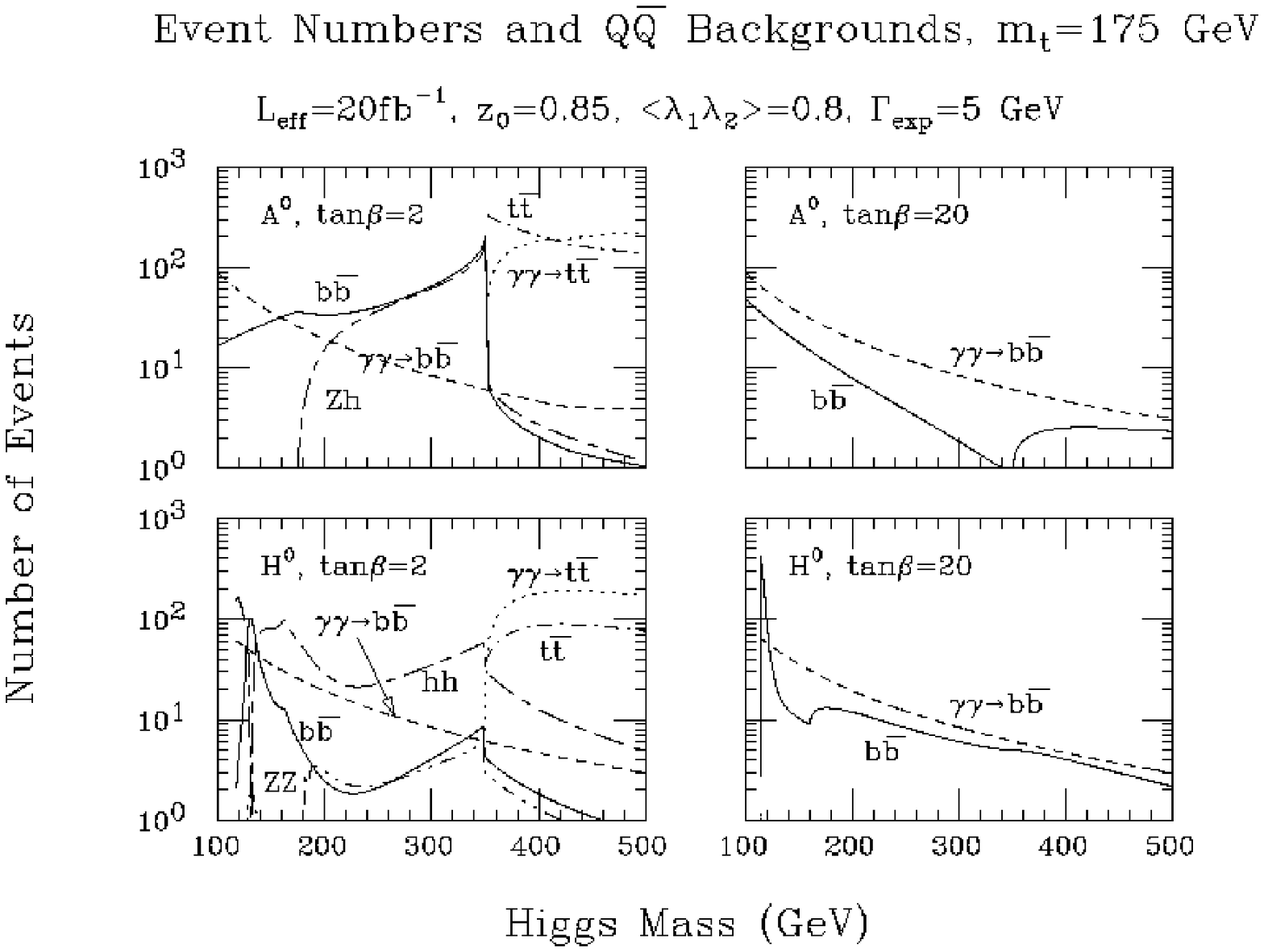,width=12.2cm}}
\begin{minipage}{12.5cm}       
\caption{Number of events as a function of Higgs mass
in various channels for $\gam\gam\rta \hh$ and $\gam\gam\rta \ha$
at $\mt=175\gev$. \Twoloop\ radiative corrections are included,
taking $\mstop=1\tev$ and neglecting squark mixing.
Results for $\tanb=2$ and $\tanb=20$ are shown.}
\label{bslaser}
\end{minipage}
\end{center}
\end{figure}

However, $\gam\gam$ collisions using back-scattered laser beams might allow
discovery of the $\hh$ and/or $\ha$ up to higher masses \cite{ghgamgam}.
Furthermore, detection of the $\hl$ in $\gam\gam$ collisions is
relatively certain to be possible.
Observation of any of the three neutral Higgs bosons
would constitute a measurement of a $\gam\gam$~Higgs coupling, which in
principle is sensitive to loops involving other charged supersymmetric
particles such as squarks and charginos. Fig.~\ref{bslaser} illustrates the
discovery potential for the $\hh$ and $\ha$ at $\mt=175\gev$ in various
final state channels. [Superpartner masses have been taken to be large and
machine energy is assumed to be about 20\% higher than the Higgs mass.
For the $b\anti b$ and $t\anti t$ channels, the continuum $\gam\gam\rta
b\anti b$ and $t\anti t$ background rates are shown for 
$\Delta \mhsm={\rm max}(\Gamma_h,5\gev)$, where $5\gev$
is the most optimistic possible experimental resolution.]
Particularly interesting channels at moderate $\tanb$ and below $t\anti t$
threshold are $\hh\rta\hl\hl$ (leading to a final state containing 4 $b$
quarks) and $\ha\rta Z\hl$.  These channels are virtually background free
unless $\mhl\sim\mw$, in which case the large $\gam\gam\rta \wp\wm$
continuum background would have to be eliminated by $b$-tagging. Above
$t\anti t$ threshold, $\hh,\ha\rta t\anti t$ decays dominate (at moderate
$\tanb$). We see that the event rate is high and that the $\gam\gam\rta
t\anti t$ continuum background is of the same general size as the signal
rate. Discovery of the $\ha$ and $\hh$ up to roughly $0.8\sqrt s$ would be
possible.

For large $\tanb$, it is necessary to look for the $\ha$ and $\hh$ in the
$b\anti b$ final state. For the effective integrated luminosity chosen,
$L=20\fbi$, Fig.~\ref{bslaser} shows that detection will be difficult except at
such low masses that it would also be possible
to observe $\zstar\rta \hh\ha$ 
in $\epem$ collisions. However, it is technically feasible
(although quite power intensive) to run the $\gam\gam$ collider at very
high instantaneous luminosity \cite{bordenpc,telnovhawaii}
such that accumulated
effective luminosities as high as $200\fbi$ can be considered. In this
case, detection of the $\ha$ and $\hh$ in the $b\anti b$ channel
should be possible for masses $\lsim0.8\sqrt s$.

Of course, the above results require re-assessment if the $\ha$
and $\hh$ have supersymmetric decays.  A variety of scenarios of this
type have been examined in Ref.~\cite{gkogamgam}.  The focus
there is on MSSM parameter choices motivated by minimal
supergravity/superstring boundary conditions.  As already noted,
the decays of the Higgs bosons can be very complex, being spread
out over many different modes, with the SUSY modes dominating
unless $\tanb$ is large.  In fact, for a $\gam\gam$
luminosity of $10\fbi$ it is found that 
for most such scenarios the $\ha$
and $\hh$ could not be found for $\mha,\mhh$ in the critical
mass region above $\sim 200\gev$ unless $\tanb$ is large enough
that the $b\anti b$ mode dominates over the many
supersymmetric-pair channels.  
At moderate $\tanb$, an integrated luminosity of $L\gsim 50\fbi$
would be required before statistically significant signals would
be present on a general basis in background-free
modes such as $\hh\rta \hl\hl\rta b\anti b b\anti b$
or $\ha\rta Z\hl \rta Z b\anti b$, and still higher luminosity
would be needed to see the supersymmetric decay mode 
background-free channels, such as $jj+jj+\etmiss$,
$\ell\ell+jj+\etmiss$, or $\ell\ell+\ell\ell+\etmiss$, coming from
$\cntwo\cntwo$ decays of the $\hh$ and $\ha$. Fortunately,
such high integrated luminosities may well lie within the
reach of a properly designed back-scattered-laser-beam facility.

\begin{figure}[htbp]
\let\normalsize=\captsize   
\begin{center}
\centerline{\psfig{file=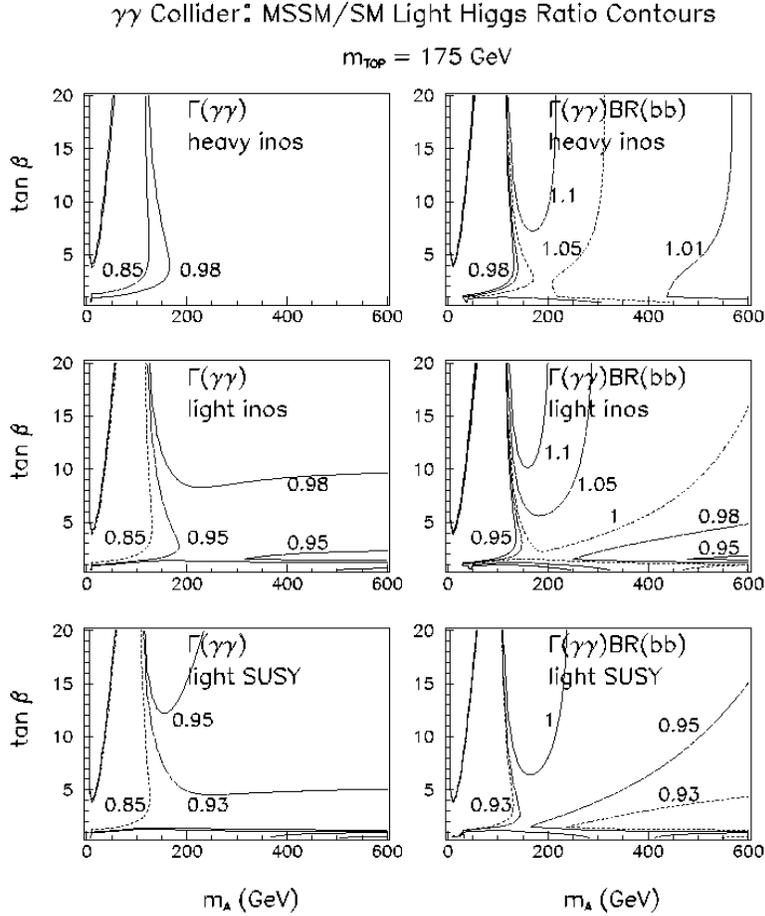,width=12.2cm}}
\begin{minipage}{12.5cm}       
\caption{Ratios of $\Gamma(\gam\gam)$ and $\Gamma(\gam\gam)BR(b\anti b)$
for the $\hl$ relative to the SM $\hsm$ (at $\mhsm=\mhl$)
in the three different cases
described in the text: `heavy inos', where all SUSY particles are heavy;
`light inos', where the neutralinos and charginos are only
just beyond kinematic reach of the $\protect\sqrt s=500\gev$ NLC but
squarks and sleptons are all heavy; and `light SUSY', where 
the inos, sleptons and squarks are all only just beyond reach.
\Twoloop\ radiative corrections are included,
neglecting squark mixing.}
\label{gamgamdeviationscontours}
\end{minipage}
\end{center}
\end{figure}

Assuming that one or more of the MSSM Higgs bosons can be seen
in $\gam\gam$ collisions,
a particularly interesting question is the extent to which the
$\gam\gam$ widths (or simply the production rates in a specific channel)
depend upon the
the SUSY context and/or superpartner masses. Some exploration of this
issue has appeared in Refs.~\cite{ghgamgam,kileng}.
Potentially, these widths are sensitive to  loops containing heavy charged
particles.  However, it must be recalled that supersymmetry decouples when
the SUSY scale is large. (In particular, superpartner masses come primarily
from soft SUSY-breaking terms in the Lagrangian and not from the Higgs
field vacuum expectation value(s).)  We discuss several scenarios.
First, suppose the lightest CP-even Higgs boson has been
discovered, but that no experimental evidence for either the  heavier Higgs
bosons or any supersymmetric particles has been found. Could a measurement
of the $\hl\gam\gam$ coupling provide indirect evidence for physics beyond
the SM? Figure~\ref{gamgamdeviationscontours} \cite{ghhprecision}
illustrates the fact that for our standard scenario 
($\mt=175\gev,\mstop=1\tev$
with all other SUSY particles also heavy --- labelled `heavy inos'
in the figure) the deviations in the width
$\Gamma(\hl\rta\gam\gam)$ [$b\anti b$ event
rate proportional to $\Gamma(\hl\rta\gam\gam)BR(\hl\rta b\anti b)$]
relative to the corresponding results for the SM $\hsm$ are not 
easily observed. If $\mha\gsim 250\gev$, then the
deviations are less than 1\% [8\%]. This is because of decoupling; as the SUSY
breaking scale and the scale of the heavier Higgs bosons become large, all
couplings of the $\hl$ approach their SM values and the squark and
chargino loops become negligible. Ref.~\cite{bordenhawaii}
(see also Ref.~\cite{bbc})
claims measurement accuracies for $\Gamma(\gam\gam)BR(b\anti b)$
that are at best of order $5\%$ (see Table~\ref{nlclhcerrors}
and associated discussion), making $\hl$ {\it vs.} $\hsm$
discrimination very difficult.  Extraction of $\Gamma(\gam\gam)$
would introduce additional errors from the experimental uncertainties
in the measured value of $BR(b\anti b)$ as obtained, for example,
from associated $Z$ plus Higgs production (see Table~\ref{nlclhcerrors}).

Lowering the mass scales of the SUSY particles does not necessarily help.
Also illustrated in Fig.~\ref{gamgamdeviationscontours}
are cases where the MSSM parameters are chosen
such that the neutralinos and charginos
lie only just beyond the kinematic reach of a
$\sqrt s=500\gev$ $\epem$ collider 
(MSSM soft-SUSY-breaking parameter choices are 
$M=-\mu=300\gev,~{\rm all}~\msquark=1\tev$, $A=0$ --- labelled
in the figure by `light inos')                  
and where {\it all} SUSY particles lie only just above the kinematic reach
(MSSM soft-SUSY-breaking parameter choices are 
$M=-\mu=300\gev,~{\rm all}~\msquark=400\gev$, $A=0$ --- labelled
in the figure by `light SUSY'). Deviations larger than $8\%$ do not
occur for $\mha\gsim 250\gev$, again implying
that it will not be easy to distinguish the
$\hl$ from the $\hsm$ using either the 
$\gam\gam$ decay width or the $b\anti b$ final
state event rate. Of course, lowering the SUSY scale still further, so that
some SUSY particles are observable at the NLC can lead to
observable deviations.

\begin{figure}[htbp]
\let\normalsize=\captsize   
\begin{center}
\centerline{\psfig{file=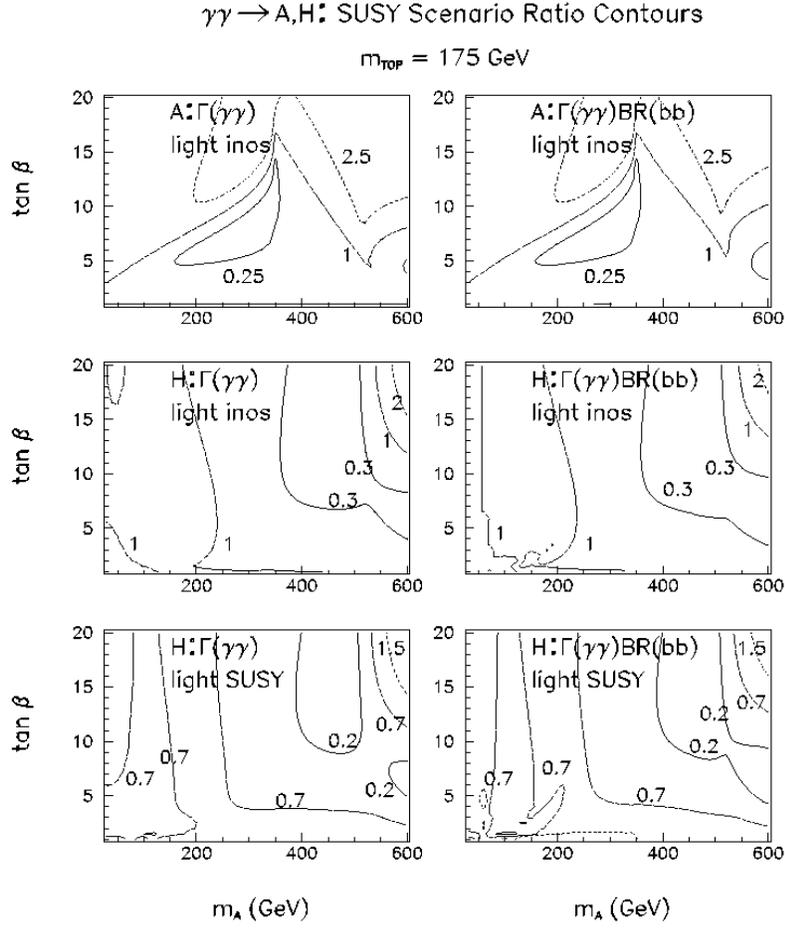,width=12.2cm}}
\begin{minipage}{12.5cm}       
\caption{Ratios of $\Gamma(\gam\gam)$ and $\Gamma(\gam\gam)BR(b\anti b)$
for the $\hh$ and $\ha$: the labelling `light inos' (`light SUSY') refers
to the ratio of the results for the `light inos' (`light SUSY') scenario
to the `heavy inos' scenario --- see text.
\Twoloop\ radiative corrections are included,
neglecting squark mixing.}
\label{gamgamhhharatioscontours}
\end{minipage}
\end{center}
\end{figure}

On the other hand, suppose that the $\hh$ or $\ha$ is light enough to be
seen in $\gam\gam$ collisions ($\lsim 400\gev$).  
In this case, a measurement of its
$\gam\gam$ width or of the rate in $\gam\gam$ collisions
for the $b\anti b$ final state can provide useful information
on the spectrum of charged supersymmetric particles even if the latter are
too heavy to be directly produced. Figure \ref{gamgamhhharatioscontours} 
provides some examples. There we display the ratios of $\Gamma(\gam\gam)$ and
$\Gamma(\gam\gam)BR(b\anti b)$ in the previously described `light inos'
and `light SUSY' scenarios to the results obtained in the `heavy inos'
scenario.  (Results for the `light SUSY' scenario in the case
of the $\ha$ are not plotted; they are
very close to those for the `light inos' scenario
since squark loop contribution to the $\ha\gam\gam$ coupling
are absent in the limit of degenerate $\sq_L$ and $\sq_R$'s.)
Scenario ratios that differ by more than 10\% from unity
are the rule, with factor of 2 or larger differences common.
Thus, observation of the $\hh$ and $\ha$ in the $b\anti b$
final state of $\gam\gam$ collisions could provide insight
as to the mass scale of the SUSY particles in the case that
they cannot be directly produced.

Finally, we note that since charged Higgs
bosons can only be pair produced in $\gam\gam$ collisions, $\epem$
collisions will yield the greatest kinematical reach in $\mhpm$.
For a study of the $\gam\gam\rta \hp\hm$ process see Ref.~\cite{dbcct}.

\subsubsection{Probing a general 2HDM in {\boldmath$\gam\gam$} collisions} 

\indent\indent In general, $\gam\gam$ collisions could play a very vital role 
in sorting out a general 2HDM model. The basic reason is simply that
the $\gam\gam$ coupling receives contributions of 
similar strength from both the CP-even and CP-odd components of 
a typical neutral $\hn$. Further,
any neutral Higgs boson can be produced singly in $\gam\gam$ collisions,
including a pure CP-odd $\ha$, 
whose $\gam\gam$ coupling is determined by fermion
loops (with large mass asymptotic limit of 2, compared to the $-4/3$
quoted earlier for the CP-even $\hsm$). The cross section for any given Higgs
boson depends upon the precise weighting of the different loop diagrams, 
as determined by the appropriate $\beta$- and $\alpha$-dependent coupling 
constant factors \cite{hhg}. Generally speaking the $\gam\gam$ width of all 
the neutral Higgs bosons of the general two-doublet model can be substantial,
and their detection in $\gam\gam$ collisions would be possible over a large
range of parameter space. Thus, simply a determination of the 
cross section(s) for the different Higgs in $\gam\gam$ collisions
would be very useful.

\subsubsection{Determining a Higgs boson's 
CP properties in {\boldmath$\gam\gam$} collisions} 

\indent\indent In the presence of CP violation in the Higgs sector, 
as could be present in the general 2HDM, $\gam\gam$
collisions could play an even more important role. 
Consider a general neutral Higgs, $\hn$, of mixed CP character.
The CP-even component of the $\hn$ will couple to $\gam\gam$ in the fashion of
the $\hsm$, although the relative weights of $W$ and fermion loops can
easily be different. In terms of the polarization vectors $\vec e_{1,2}$ of
the two photons in the photon-photon center of mass, the coupling is
proportional to $\vec e_1\cdot \vec e_2$. The CP-odd component of the $\hn$
will also develop a $\gam\gam$ coupling at one-loop.  As noted earlier,
only fermion loops contribute. The coupling is proportional to $(\vec e_1\times
\vec e_2)_z$ (assuming the photons collide along the $z$ axis). Writing the
net coupling as $\vec e_1\cdot \vec e_2 \cale+(\vec e_1\times \vec e_2)_z
\calo$, one finds that $\cale$ and $\calo$ are naturally of similar size if
the CP-odd and CP-even components of the $\hn$ are comparable.

The most direct probe of a CP-mixed state is provided by
comparing the Higgs boson production rate
in collisions of two back-scattered-laser-beam 
photons of various different polarizations \cite{bgbslaser}.
The difference in rates for photons colliding with $++$ vs. $--$ 
helicities is non-zero only if CP violation is present.
A term in the cross section changes sign when
both $\lam_1$ and $\lam_2$ are simultaneously flipped, and is thus
best measured by taking $\langle\lam_1\lam_2\rangle$ as
near 1 as possible (which suppresses backgrounds anyway), and then
comparing the event rate for $\langle\lam_1\rangle$ and $\langle\lam_2\rangle$ 
both positive, to
that obtained when both are flipped to negative values. Experimentally
this is achieved by 
simultaneously flipping the helicities of both of the initiating
back-scattered laser beams. One finds \cite{bgbslaser} that the asymmetry 
is typically larger than 10\% and is 
observable for a large range of two-doublet parameter space if CP violation
is present in the Higgs potential. 

In the case of a CP-conserving Higgs sector, the earlier outline
makes clear that there is strong dependence
of the $\gamma\gamma\rta \hn$ cross section on the relative orientation
of the transverse polarizations of the two colliding photons. Clearly then,
to extract parallel vs. perpendicular cross section asymmetry
experimentally requires that the colliding photons have substantial transverse
polarization. This is achieved by transversely polarizing the incoming
back-scattered laser beams (while maintaining the ability
to rotate these polarizations relative to one another) and optimizing
the laser beam energy.  This optimization has been discussed in
Refs.~\cite{gk,kksz}, and it is found that $\gam\gam$ collisions
may well allow a determination of whether a given $\hn$ is CP-even or CP-odd.

\subsection{Final NLC remarks}
 
\indent\indent Extension of the energy of the $e^+e^-$ collider beyond 
$\sqrt s =1$ TeV 
is not required for detecting and studying a weakly-coupled SM Higgs boson, 
but could be very important for a strongly-coupled
electroweak-symmetry-breaking scenario. Detection of $\hh\ha$ and $\hp\hm$
production in the MSSM model would be extended to higher masses,
as could easily be required if the MSSM parameters are typical of
those predicted by minimal supergravity/superstring boundary conditions.

\section{A {\boldmath$ \mu^+\mu^-$} collider}

Although design and development are at a very early stage
compared to the NLC, and feasibility is far from proven,
there is now considerable interest in
the possibility of constructing a $\mu^+\mu^-$ collider
\cite{sausi,sausii}.
Two specific muon collider schemes are under consideration. A high energy
machine with 4~TeV center-of-mass energy ($\rts$) and luminosity of order
$10^{35}$~cm$^{-2}$~s$^{-1}$ \cite{palmer} would have an energy reach
appropriate for pair production of $\ha\hh$
up to very large masses and the study of a strongly interacting $WW$ sector. 
A lower energy machine, hereafter called the First Muon Collider (FMC), 
could have c.m.\
energy around 0.5~TeV with a luminosity of order $2\times
10^{33}$~cm$^{-2}$~s$^{-1}$ \cite{palmer} for unpolarized beams. 
Not only would the FMC be able to accomplish everything that the NLC
could (for the same integrated luminosity), but also the FMC
would be extremely valuable for discovery
and precision studies of Higgs bosons produced directly in the $s$-channel.
The most costly component of a muon collider is the muon source 
(decays of pions produced by proton
collisions) and the muon storage rings would comprise a modest fraction of the
overall cost \cite{palmer2}. 
Consequently, full luminosity can be maintained
at all c.m.\ energies where Higgs bosons are either observed or expected
by constructing multiple storage rings optimized 
for c.m.\ energies centered on the observed masses or 
spanning the desired range. We summarize below the results
of Ref.~\cite{bbgh} regarding $s$-channel Higgs discovery and
precision measurements.

For $s$-channel studies of narrow resonances, the energy resolution is an
important consideration. A Gaussian shape for the energy spectrum of each beam
is expected to be a good approximation (beamstrahlung being negligible)
with an rms deviation most naturally in the
range $R = 0.04$\% to 0.08\% \cite{jackson}. \ By additional cooling
this can be decreased to $R = 0.01$\%.
The corresponding rms error $\sigma$ in $\sqrt s$ is given by
\begin{equation}
\sigma = (0.04~{\rm GeV})\left({R\over 0.06\%}\right)\left({\sqrt s\over {\rm
100\ GeV}}\right) \ .
\label{resolution}
\end{equation}
A critical issue is how this resolution compares to the calculated total
widths of Higgs bosons. Widths for the Standard Model Higgs $\hsm$
and the three neutral Higgs bosons $\hl$, $\hh$, $\ha$ 
of the Minimal Supersymmetric Standard Model
(MSSM) were illustrated in Fig.~\ref{hwidths}.
An $s$-channel Higgs resonance
would be found by scanning in $\sqrt s$ using steps 
of size $\sim\sigma$; its mass would be simultaneously determined
with roughly this same accuracy in the initial scan. For sufficiently
small $\sigma$, the Breit-Wigner resonance line-shape would be revealed
and the Higgs width could be deduced.
For $R\alt 0.06\%$, the energy resolution in Eq.~\ref{resolution}
can be smaller than the Higgs widths in many cases;
and for $R\alt 0.01\%$ the energy resolution becomes comparable to even
the very narrow width of an intermediate-mass SM $\hsm$.

In the simplest approximation, the
effective cross section for Higgs production in the $s$-channel
followed by decay to channel $X$, $\sighbar$, 
is obtained by convoluting the standard $s$-channel pole form
with a Gaussian distribution in $\sqrt s$ of rms width $\sigma$.
For $\gamh \gg \sigma$, $\gamh\ll\sigma$, 
$\sighbar$ at $\sqrt s = \mh$ is given by:
\begin{equation}
\sighbar =\left\{ 
\begin{array}{ll} 
{4\pi BR(\h\rta \mm)BR(\h\rta X)\over \mh^2}\,,
&\mbox{$\gamh\gg\sigma$;} \\
{\pi\gamh\over 2\sqrt{2\pi}\sigma}\ 
{4\pi BR(\h\rta \mm)BR(\h\rta X)\over \mh^2}, 
&\mbox{$\gamh\ll\sigma$\, .}
\end{array}
\right.
\label{mmxsec}
\end{equation}
Since the backgrounds vary slowly over the expected energy resolution
interval $\overline\sigma_B=\sigma_B$.
In terms of the integrated luminosity $L$, total
event rates are given by $L\overline\sigma$;
roughly $L=20\fbi$/yr is expected for the FMC. 

The above results do not include the spreading out
of the Gaussian luminosity peak due to photon bremsstrahlung. 
Although this is much reduced as compared to an $\ee$ collider,
it is still important for Higgs bosons that are narrow compared to
the beam resolution.  For such a Higgs boson,
the $s$-channel production rate for $\sqrt s=\mh$
decreases proportionally to the decrease in the peak luminosity.
The amount by which the latter decreases depends upon the resolution $R$
and $\rts$.
For $\sqrt s=\mh=100\gev$, $L_{\rm peak}$ decreases by a factor of 0.61 (0.68)
for $R=0.01\%$ ($R=0.06\%$) when photon bremsstrahlung
is included. By $\rts=500\gev$ these factors have decreased to 0.53 (0.61),
respectively. Meanwhile, the background will decrease by
a smaller factor due to the fact that it varies smoothly with $\sqrt s$
and one must integrate over a final state mass interval of order
the final state mass resolution ($\sim 5\gev$).  As a result $S/\sqrt B$
typically decreases by about 30\% for $\mhsm\sim 100\gev$.
The result is that
for any Higgs boson with width much smaller than the energy resolution
$\sigma$, a factor of $\lsim 3-5$ larger luminosity is required than 
would be computed were bremsstrahlung smearing not included.
For a Higgs boson with width much larger than $\sigma$, the increase
in luminosity required for a given measurement is much less.
The results that follow include the bremsstrahlung smearing.

\subsection{The SM {\boldmath$\hsm$}}

\begin{figure}[htbp]
\let\normalsize=\captsize   
\begin{center}
\centerline{\psfig{file=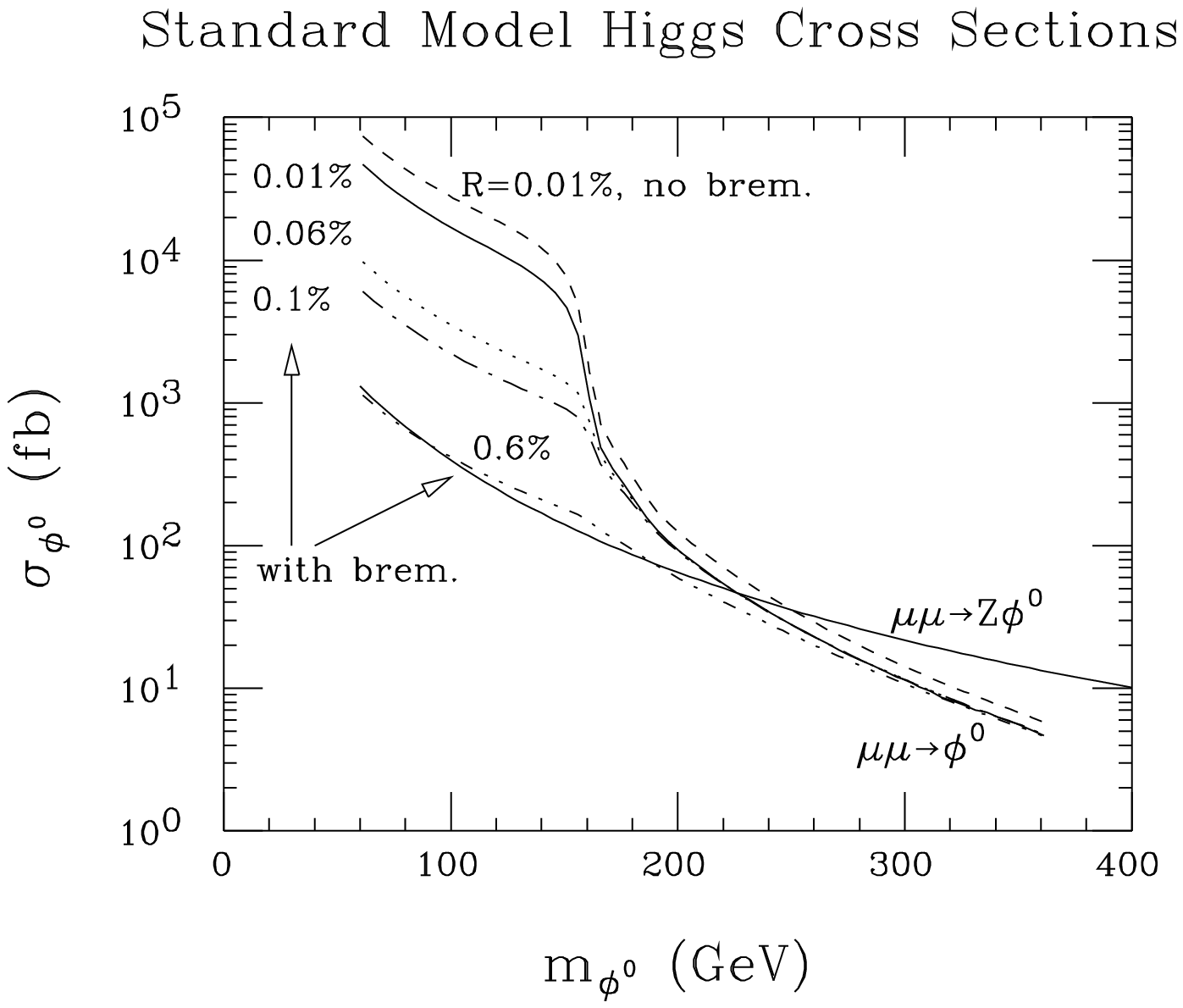,width=12.2cm}}
\begin{minipage}{12.5cm}       
\smallskip
\caption{
Cross sections versus $\mhsm$ for inclusive SM Higgs production: (i) the
$s$-channel $\sighbar$ for $\mm\to \hsm$
with $R = 0.01$\%, 0.06\%, 0.1\% and 0.6\%, 
and (ii) $\sigma(\mu^+\mu^-\to Z\hsm)$
at $ \protect\sqrt s = \mz + \protect\sqrt 2 \mhsm$.}
\label{smxsecmm}
\end{minipage}
\end{center}
\end{figure}

Predictions for $\overline\sigma_{\hsm}$ for inclusive SM
Higgs production are given in Fig.~\protect\ref{smxsecmm} 
versus $\rts=\mhsm$ for resolutions of $R =
0.01$\%, 0.06\%, 0.1\% and 0.6\%.
For comparison, the $\mm\to \zstar\to Z\hsm$ cross section
is also shown, evaluated at the energy  $\sqrt s = \mz + \sqrt 2 \mhsm$ for
which it is a maximum. The $s$-channel $\mm\to \hsm$
cross sections for small $R$ and $\mhsm\lsim 2\mw$ are much larger
than the corresponding $Z\hsm$ cross section.
The increase in the $\mm\to\hsm$ cross section that
results if bremsstrahlung smearing is removed is illustrated in the most
sensitive case ($R=0.01\%$).

The optimal strategy for SM Higgs {\it discovery} at a lepton collider is to
use the $\mm\to Z{\h}$ mode (or $\ee\to Z{\h})$ because no energy scan is
needed. Studies of $\ee$ collider capabilities indicate that the
SM Higgs can be discovered if $\mhsm < 0.7 \sqrt s$. 
If $\mhsm\alt 140\gev$, its mass will be determined to a
precision given by the event-by-event mass resolution
of about 4 GeV in the $\h+Z\rta \tau^+\tau^- +q\anti q$ and $X+\ell^+\ell^-$
channels divided by the square root of the number of events in these channels,
after including efficiencies \cite{barklowresolution,janothawaii}. 
A convenient formula is
$\Delta \mhsm \alt 0.4\ {\rm GeV} \left({10\ {\rm fb}^{-1}\over
L}\right)^{{1\over 2},}$
yielding, for example, $\pm 180$ MeV for $L=50\fbi$ \cite{janothawaii}.
As noted earlier, the super-JLC detector could do even better,
achieving a $\sim\pm 20\mev$ measurement of $\mhsm$ via
the recoil mass spectrum in the $Z\hsm$ mode.
At the LHC the $\hsm \to \gamma\gamma$ mode is
deemed viable for $80 \alt \mhsm \alt 150$~GeV, 
with a better than 1\% mass resolution \cite{ATLAS,CMS}. Once
the $\hsm$ signal is found, precision determination of its mass and 
measurement of its width become the
paramount issues, and $s$-channel resonance production
at a $\mu^+\mu^-$ collider is uniquely suited for this purpose.

\begin{figure}[htbp]
\let\normalsize=\captsize   
\begin{center}
\centerline{\psfig{file=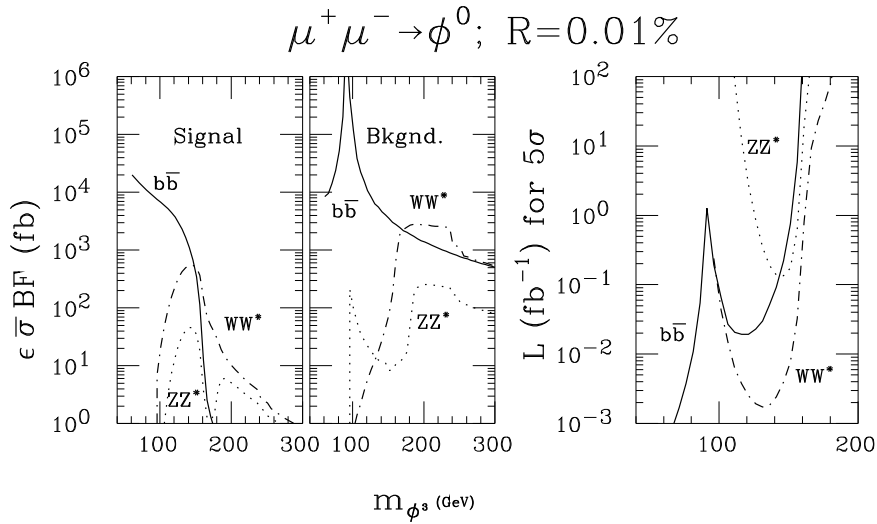,width=11.5cm}}
\bigskip
\begin{minipage}{12.5cm}       
\caption{
The (a) $\hsm$ signal and (b) background
cross sections, $\epsilon \overline\sigma BR(X)$, 
for $X=b\anti b$, and useful $W\wstar$ and $Z\zstar$ final states
(including a channel-isolation efficiency of $\epsilon=0.5$) 
versus $\mhsm$ for SM Higgs $s$-channel production with
resolution $R=0.01\%$. Also shown: (c) the luminosity
required for $S/\protect \sqrt B=5$ in the three channels 
as a function of $\mhsm$ for $R=0.01\%$. Bremsstrahlung
effects are included.
}
\label{smratesmm}
\end{minipage}
\end{center}
\end{figure}

For $\mhsm < 2\mw$ the dominant $\hsm$-decay channels 
are $b\anti b$, $W\wstar$, and
$ZZ^\star$, where the star denotes a virtual weak boson. The light quark
backgrounds to the $b\anti b$ signal can be rejected by $b$-tagging.
For the $W\wstar$ and $Z\zstar$ channels we employ only
the mixed leptonic/hadronic modes ($\ell\nu2j$ for $W\wstar$ and $2\ell2j$,
$2\nu2j$ for $Z\zstar$, where $\ell = e$ or $\mu$ and $j$ denotes a quark jet),
and the visible purely-leptonic $Z\zstar$ modes
($4\ell$ and $2\ell2\nu$), 
taking into account the major electroweak QCD backgrounds.
For all channels we assume a general signal and background
identification efficiency of $\epsilon= 50\%$, after selected acceptance
cuts. In the case of the $b\anti b$ channel, this is to
include the efficiency for tagging at least one $b$.
The signal and background channel cross sections
$\epsilon\overline\sigma BR(X)$ at $\rts=\mhsm$
for $X=b\anti b$, $W\wstar$ and $Z\zstar$ are presented in
Fig.~\ref{smratesmm} versus $\mhsm$ for a resolution $R = 0.01$\%;
$BR(X)$ includes the Higgs decay branching ratios for the signal,
and the branching ratios for the $W,\wstar$ and $Z,\zstar$ 
decays in the  $W\wstar$ and $Z\zstar$ final states for both the signal and the
background. The background level is essentially independent
of $R$, while the signal rate depends strongly on $R$ as illustrated
in Fig.~\ref{smxsecmm}.

The luminosity required to achieve $n_\sigma = S/\sqrt B=5$ (where $S$
and $B$ are the signal and background rates) in the $b\anti b$, $W\wstar$ and 
$Z\zstar$ channels is also shown in Fig.~\ref{smratesmm} --- 
results for $R=0.01\%$ as a function of $\mhsm$ are illustrated.
For $R=0.01\%$, $L=.1\fbi$ would yield
a detectable $s$-channel Higgs signal 
for all $\mhsm$ values between the current LEP\,I limit and $2\mw$
except in the region of the $Z$~peak; a luminosity $L\sim 1\fbi$ at
$\sqrt s = \mhsm$ is needed for $\mhsm\sim \mz$.
For $R=0.06\%$, $n_\sigma=5$ signals typically require about 20 times
the luminosity needed for $R=0.01\%$! Note that a search for the $\hsm$
(or any Higgs with width smaller than the achievable resolution)
by scanning would be most efficient for the smallest possible $R$ due to
the fact that the $L$ required at each scan point
decreases as (roughly) $R^{1.7}$, whereas the number
of scan points would only grow like $1/R$. If the Higgs resonance
is broad, using small $R$ is not harmful since 
the data from a fine scan can be re-binned to test for its presence.

Once the Higgs is observed, the highest priority will be to determine its
precise mass and width.  This can be accomplished by scanning across
the Higgs peak. The luminosity required for this is strongly 
dependent upon $R$ (\ie\ $\sigma$) and the width itself. 
Precision determinations of the Higgs mass and, particularly, its width
are most challenging when the width is
smaller than the $\rts$ resolution, $\sigma$. This is typically
the case for a light SM Higgs boson 
in the intermediate mass range;
{\it e.g.} for $\mhsm=120\gev$, $\Gamma_{\hsm}\sim 0.003\gev$
is significantly smaller than the best possible ($R=0.01\%$) value of
$\sigma\sim .008\gev$.
Nonetheless, unless the beam energy resolution is poor, the 
mass is relatively easily determined by simply measuring the
rates at a few values of $\rts$ near $\mh$.
A model-independent determination of the Higgs width (\ie\ one that does
not rely on knowing normalization of the peak, as fixed by
$BR(\h\to \mupmum)BR(\h\to X)$) is more difficult; it
requires determining accurately
the ratio of the rate at a central $\rts\sim \mh$ position
to the rate at a position on the tail of the peak (as spread out by
the Gaussian smearing).

\begin{figure}[htbp]
\let\normalsize=\captsize   
\begin{center}
\centerline{\psfig{file=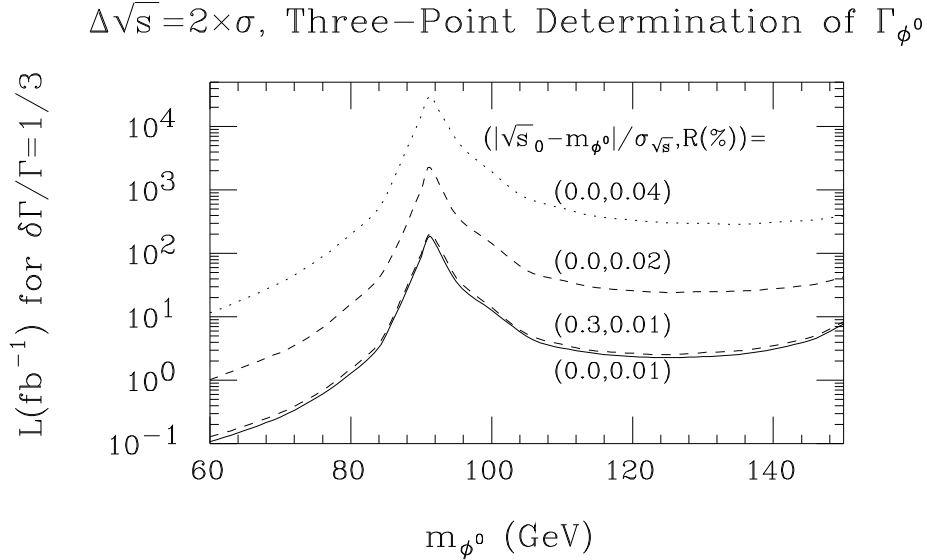,width=12.2cm}}
\begin{minipage}{12.5cm}       
\caption{The luminosity required for a $\delta \Gamma/\Gamma=1/3$
measurement of the $\hsm$ width as a function of $\mhsm$ for various choices of
$(|\protect\rts-\mhsm|/\sigma,R)$.
Bremsstrahlung effects are included.}
\label{widthlummm}
\end{minipage}
\end{center}
\end{figure}

The minimal number of measurements needed to simultaneously 
determine the Higgs mass and width is three.  
Sensitivity is roughly optimized if these three measurements are separated
in $\sqrt s$ by $2\sigma$; the first would be taken at $\sqrt s$
equal to the current best central value of the mass (from
the initial detection scan). The second and third would be at 
$\sqrt s$ values $2\sigma$ below and $2\sigma$ above the first, with
about 2.5 times the integrated luminosity expended on the first measurement
being employed for each of these latter two measurements.
In Fig.~\ref{widthlummm} we plot the total combined luminosity required
for a $\delta \Gamma/\Gamma=1/3$ measurement of the width
in the $b\anti b$ channel as a function of $\mhsm$.  For given $R$,
luminosity requirements vary by up to 25\%, depending
upon luck in placement of the first scan point, as quantified
by the ratio $|\rts -\mhsm|/\sigma$. 
The figure also shows that luminosity requirements increase 
rapidly as $R$ worsens. For the best possible resolution of $R=0.01\%$,
total luminosities of at least $L=3.2$, $2.4$, and $3.1\fbi$ are
needed for a $\delta\Gamma/\Gamma=1/3$ width measurement
at $\mhsm=110$, $120$, and $140\gev$, respectively; and
for $\mhsm\sim \mz$, nearly $L=200\fbi$ is needed. Clearly,
the excellent $R=0.01\%$ resolution is mandatory if
one is to have a good chance of being able to measure the total width 
regardless of what $\mhsm$ turns out to be. 
In the narrow width region ($\mhsm\lsim 150\gev$)
the Higgs mass is simultaneously 
determined using this procedure to the accuracy of 
$\delta\mh\sim0.5\delta \Gamma$. To a first approximation,
$\delta\Gamma$ scales statistically, \ie\  as $1/\sqrt{L}$. 
Thus, $L\gsim 30-35\fbi$ would be required for a 10\% measurement
of the Higgs width for $110\lsim\mhsm\lsim140\gev$.
Allowing for the fact that our `central' position measurement
might be slightly off center, we claim in Table~\ref{nlclhcerrors}
that $L=50\fbi$ is fairly certain to allow a 10\% determination
of the SM Higgs width.  The corresponding errors on the Higgs
mass are also given in the Table.

\begin{figure}[htbp]
\let\normalsize=\captsize   
\begin{center}
\centerline{\psfig{file=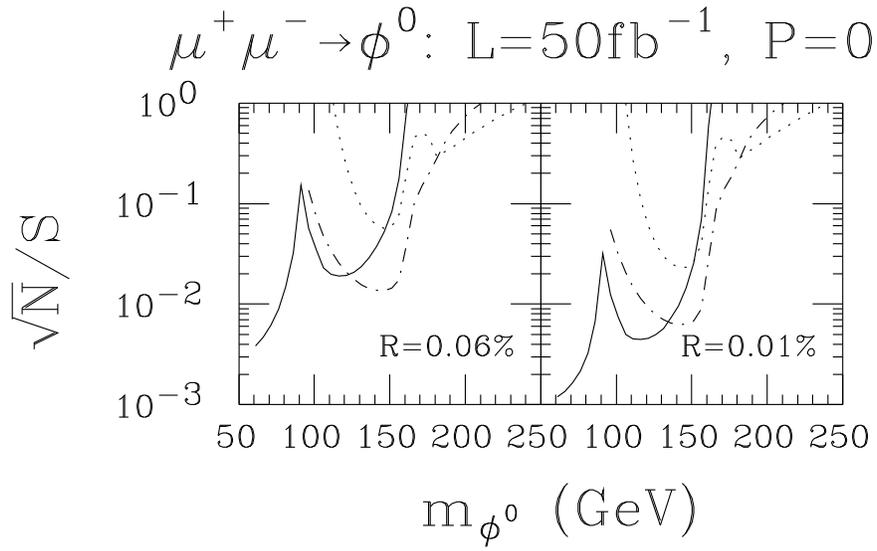,width=11.5cm}}
\bigskip
\begin{minipage}{12.5cm}       
\caption{
The error $\protect\sqrt N/S$ in the measurement of
$\Gamma({\hsm}\to\mu^+\mu^-)\times BR({\hsm}\to X)$ 
for $X=b\anti b$ (solid), and useful $W\wstar$ (dotdash) and 
$Z\zstar$ (dots) final states
(including a channel-isolation efficiency of $\epsilon=0.5$) 
versus $\mhsm$ for SM Higgs $s$-channel production with
resolutions $R=0.06\%,0.01\%$, for an integrated luminosity of $L=50\fbi$. 
Bremsstrahlung effects are included.
}
\label{smerrorsmm}
\end{minipage}
\end{center}
\end{figure}

In addition, the event rate in a given channel measures
$\Gamma({\hsm}\to\mu^+\mu^-)\times BR({\hsm}\to X)$. 
Then, using the branching
fractions (most probably already measured in $Z\hsm$ associated
production), the ${\hsm}\to\mu\mu$ partial width can be determined, 
providing an important test of the Higgs coupling.
The accuracy with which the $X=b\anti b,W\wstar,Z\zstar$
rates can be measured
is plotted in Fig.~\ref{smerrorsmm} as a function of $\mhsm$
for $L=50\fbi$ and resolution choices of $R=0.01\%$ and $0.06\%$.
For $R=0.01\%$, a better than $\pm 1.5\%$ measurement of the $X=b\anti b$
channel rate can be performed for all $\mhsm\lsim 150\gev$.
Thus, in obtaining a direct determination of $\Gamma(\hsm\to\mm)$
we will be limited by the $\sim\pm7\%-\pm 10\%$ measurement of
$BR(\hsm \to b\anti b)$ obtained at the NLC by combining
the $\epem\rta Z\hsm$ inclusive rate with the $\epem\rta Z \hsm\rta Z b\anti b$
partial rate (the uncertainty in the inclusive $Z\hsm$ measurement 
dominates the error).

\subsection{ The MSSM Higgs bosons}

\begin{figure}[htbp]
\let\normalsize=\captsize   
\begin{center}
\centerline{\psfig{file=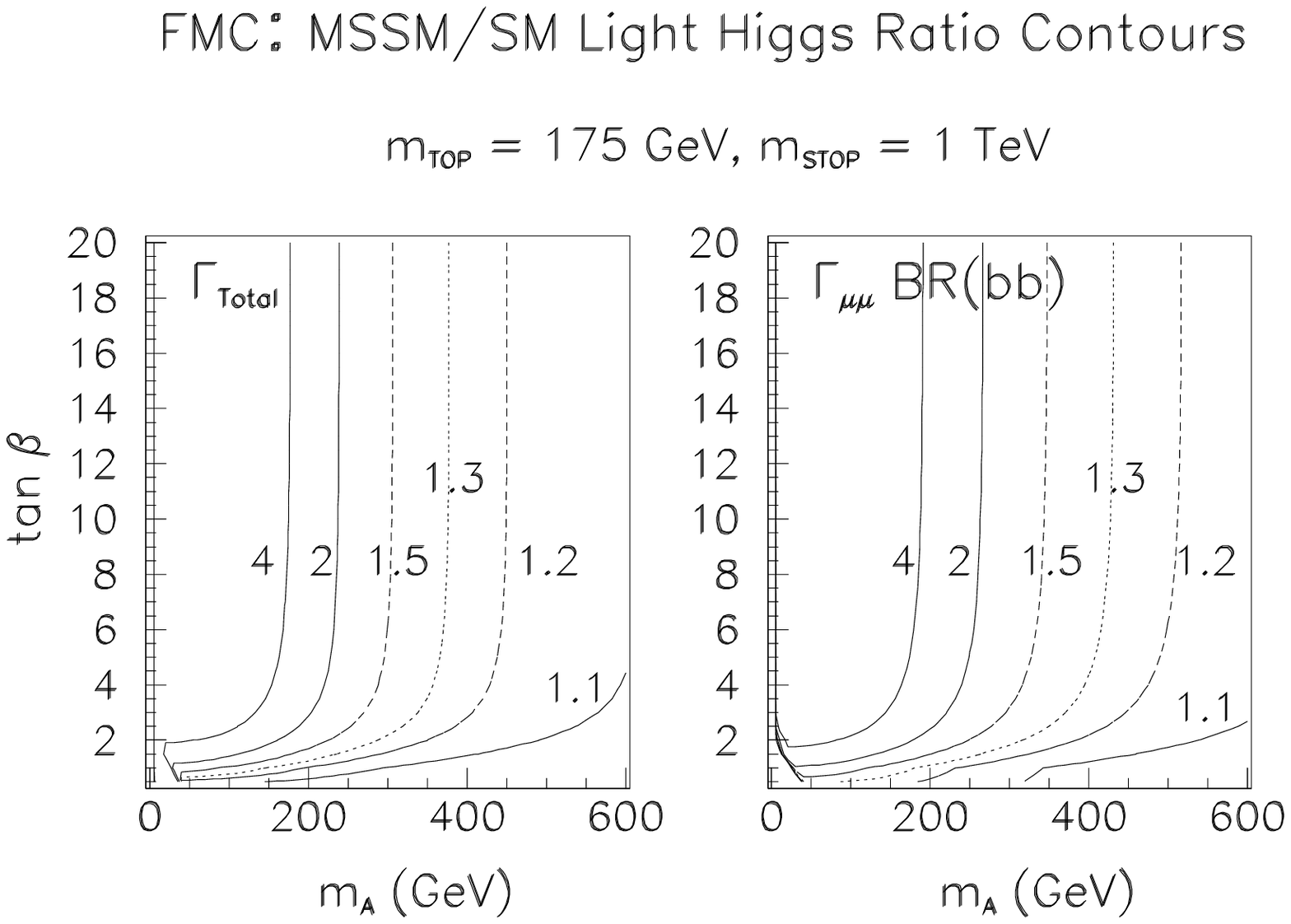,width=12.2cm}}
\begin{minipage}{12.5cm}       
\caption{Ratio of $\hl$ to $\hsm$ predictions at $\mhsm=\mhl$
for the total width $\Gamma_{\h}$ and 
$\Gamma(\h\rta \mm)BR(\h\rta b\anti b)$. We have taken $\mt=175\gev$.
\Twoloop\ radiative corrections are included,
taking $\mstop=1\tev$ and neglecting squark mixing.}
\label{mupmumdeviationscontours}
\end{minipage}
\end{center}
\end{figure}

A $\mm$ collider provides two particularly unique probes of the MSSM
Higgs sector.  First, the couplings of the $\hl$ deviate 
sufficiently from exact SM Higgs couplings that it
may well be distinguishable from the $\hsm$ by measurements of
$\gamh$ and $\Gamma(\h\to\mm)$ at a $\mm$ collider, using the
$s$-channel resonance process (here we use the notation $\h$
for a generic Higgs boson). For instance, in the $b\anti b$ channel
$\gamh$ and $\Gamma(\h\to\mm)\times BR(\h\to b\anti b)$ can both be measured
with good accuracy.
The expected size of the relevant deviations is illustrated in
Fig.~\ref{mupmumdeviationscontours} \cite{ghhprecision}. 
Unless $\mh\sim \mz$, $L=50\fbi$ of luminosity will yield a better
than $\pm 10\%$ determination of $\gamh$ (as noted earlier)
and a better than $\pm 1\%$ determination of 
$\Gamma(\h\to\mm)\times BR(\h\to b\anti b)$ (see Fig.~\ref{smerrorsmm}).  
However, one must also account for systematic uncertainties.
As discussed in the NLC
section, our ability to predict $BR(\h\to b\anti b)$ is limited
by uncertainty in $\mb$, an uncertainty of order $\pm 5\%$ in $\mb$
leading to a $\pm 3\%$ uncertainty in the branching ratio.
Uncertainty in $\mb$ also implies uncertainty in the total width
prediction of order ${\delta \gamh\over\gamh}
\sim 2 {\delta \mb\over \mb} BR(\h\to b\anti b)$ which is of order $8\%-10\%$
for ${\delta \mb\over \mb}\sim \pm 5\%$. 
We see from Fig.~\ref{mupmumdeviationscontours}
that so long as we can keep systematic and statistical errors below the 10\%
level, one can reasonably expect that these quantities will probe the
$\hl$ {\it vs.} $\hsm$ differences for $\mha$ values as large as $400-500\gev$.

The second
dramatic advantage of a $\mm$ collider in MSSM Higgs physics is the ability
to study the non-SM-like Higgs bosons, \eg\ 
for $\mha\agt 2\mz$ the $\hh,\ha$. An $e^+e^-$ collider can only
study these states via $\zstar\to \ha\hh$ production, which could easily be
kinematically disallowed since GUT scenarios typically have $\mha\sim\mhh\agt
200$--250~GeV.  In $s$-channel production the $\hh,\ha$
can be even more easily observable than a SM-like Higgs if $\tanb$ is not near
1. This is because the partial widths 
$\Gamma(\hh,\ha\to \mm)$ grow rapidly with increasing
$\tanb$, implying (see Eq.~\ref{mmxsec})
that $\overline\sigma_{\hh,\ha}$ will become strongly enhanced relative to 
SM-like values. $BR(\hh,\ha\to b\anti b)$ is also enhanced at large $\tanb$, 
implying an increasingly large rate in the $b\anti b$ final state.
Thus, we concentrate here on the $b\anti b$ final states of $\hh,\ha$ 
although the modes $\hh,\ha\to t\anti t$, 
$\hh\to \hl\hl,\ha\ha$ and $\ha\to Z\hl$ can also be useful.

Despite the enhanced $b\anti b$ partial widths, the suppressed (absent)
coupling of the $\hh$ ($\ha$) to $WW$ and $ZZ$ means that,
unlike the SM Higgs boson, the $\hh$ and $\ha$ remain
relatively narrow at high mass, with widths $\Gamma_{\hh},\Gamma_{\ha}
\sim 0.1$ to 6~GeV for $\tanb\lsim 20$ (see Fig.~\ref{hwidths}). 
Since these widths are generally comparable to or
broader than the expected $\sqrt s$ resolution for $R = 0.06$\% and $\rts\agt
200\gev$, 
measurements of these Higgs widths could be straightforward with a scan over
several $\sqrt s$ settings, provided that the signal rates are sufficiently
high. The results of a fine scan can be combined to get a coarse scan
appropriate for broader widths. One subtlety is the fact that the $\hh$
and $\ha$ are sufficiently degenerate in mass at large $\mha\sim\mhh$
that their resonance peaks can overlap substantially.  In this case,
the event rate at say $\rts=\mha$ would include automatically some
$\hh$ events, and vice versa. A detailed scan with small $R$ might
be required to separate the two overlapping peaks and determine
the $\ha$ and $\hh$ widths using a fit employing two Breit-Wigner forms.

\begin{figure}[htbp]
\let\normalsize=\captsize   
\begin{center}
\centerline{\psfig{file=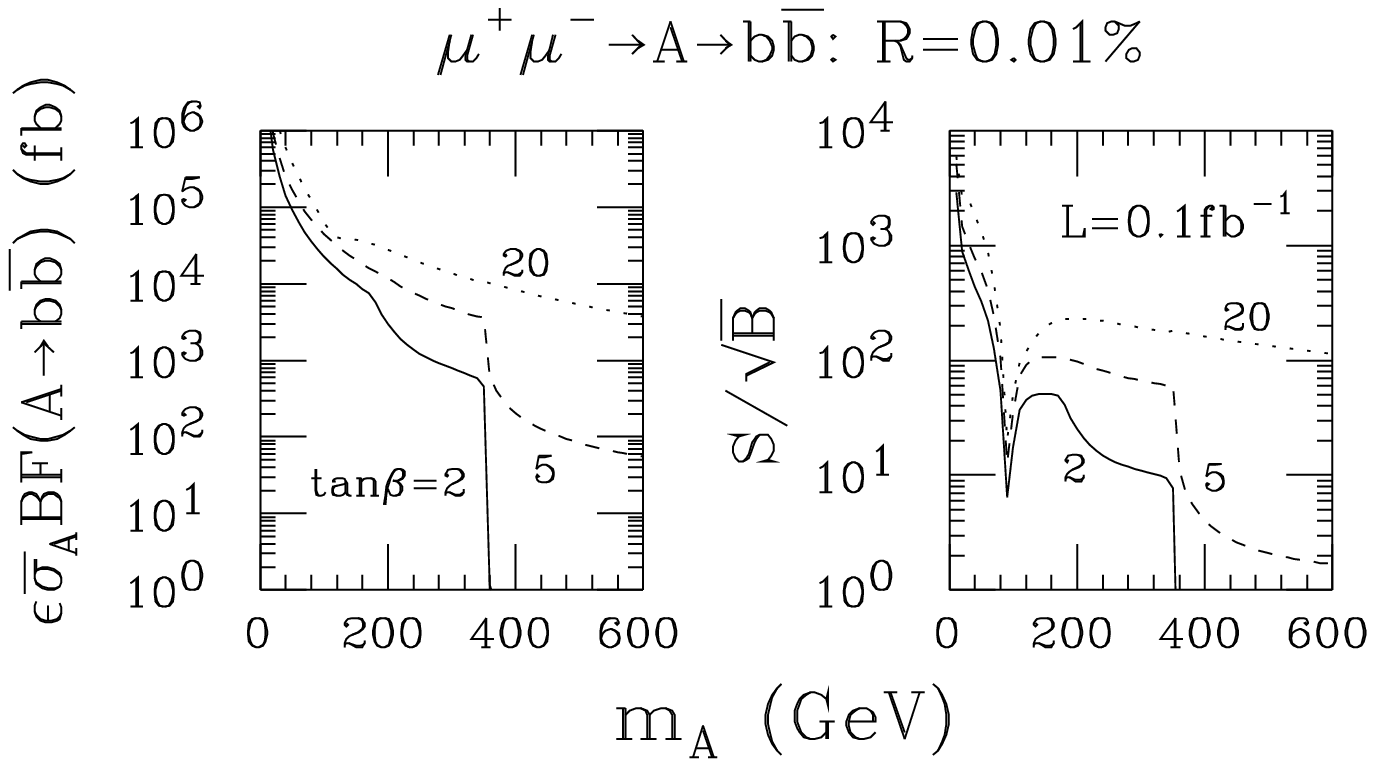,width=11.5cm}}
\begin{minipage}{12.5cm}       
\bigskip
\caption{
(a) The effective $b\anti b$-channel cross section, 
$\epsilon\overline\sigma_{\ha}BR(\ha\to b\anti b)$,
for $s$-channel production of the MSSM Higgs boson 
$\ha$ versus $\protect\sqrt s=\mha$, 
for $\tanb = 2$, 5 and 20, beam resolution $R = 0.01$\% and 
channel isolation efficiency $\epsilon=0.5$; and
(b) corresponding statistical significance of the 
$\ha\to b\anti b$ signal for $L=0.1\fbi$ delivered at 
$\protect\sqrt s=\mha$. Plots are for $\mt=175\gev$.
\Twoloop\ radiative corrections are included,
taking $\mstop=1\tev$ and neglecting squark mixing.}
\label{susymm}
\end{minipage}
\end{center}
\end{figure}

The cross section for $\mm\to \ha\to b\anti b$ 
production with $\tan\beta = 2$, $5$
and 20 (including an approximate cut and $b$-tagging efficiency of 50\%) 
is shown versus $\mha$ in Fig.~\ref{susymm} 
for beam resolution $R = 0.01$\%. Overlapping events from the tail of
the $\hh$ resonance are automatically included. Also shown is
the significance of the $b\anti b$ signal for delivered luminosity $L =
0.1\fbi$ at $\sqrt s = \mha$. Discovery of the $\ha$ and $\hh$ will require an
energy scan if $\zstar\to \hh+\ha$ 
is kinematically forbidden; a luminosity of
20~fb$^{-1}$ would allow a scan over 200~GeV at intervals of 1~GeV with $L =
0.1$~fb$^{-1}$ per point. The $b\anti b$ mode would yield at
least a $10\sigma$ signal at
$\sqrt s = \mha$ for $\tan\beta \agt 2$ for $\mha\lsim 2\mt$
and at least a $5\sigma$ signal
for $\tanb\agt5$ for all $\mha\lsim 500\gev$.
Results for $R=0.06\%$ are displayed in 
the corresponding figure in Ref.~\cite{bbgh}.  The resulting statistical
significance is only noticeably affected in the $\tanb=2$ case, for
which it declines by about a factor of 2.  For $\tanb\gsim 5$,
the $\ha$ is sufficiently broad that very narrow resolution is not helpful.
For $\mha\agt \mz$ ($\mha\alt \mz$), 
the $\hh$ ($\hl$) has very similar couplings to those of the $\ha$ and would
also be observable in the $b\anti b$ mode for $\tanb\gsim 5$.
For $\tanb\sim 2$, $BR(\hh\rta b\anti b)$
is smaller than in the case of the $\ha$ due to
the presence of the $\hh\rta\hl\hl$ decay mode.  For such
$\tanb$ values, detection would be easier in the $\hl\hl$ final state.
Overall, discovery of {\it both} the $\hh$ and $\ha$ MSSM Higgs bosons 
(either separately or as overlapping resonances) would be possible
over a large part of the $\mha\agt\mz$ MSSM parameter space.

\subsection{The advantages of polarized beams}

Polarized beams would allow a
reduction in backgrounds relative to signals in the
discovery and study of any Higgs boson. If longitudinal
polarization $P$ is possible
for {\it both} beams, then, relative to the unpolarized case, the signal
is enhanced by the factor $(1+P^2)$ while the background is
suppressed by $(1-P^2)$. The luminosity required for a signal
of given statistical significance is then proportional to $(1-P^2)/(1+P^2)^2$.
For example, if 85\% polarization
could be achieved with less than a factor of 10 decrease in luminosity,
Higgs studies would benefit.

\subsection{Determining a Higgs boson's CP properties in {\boldmath$\mupmum$ }
Collisions}

A $\mupmum$ collider might well prove to be the best machine
for directly probing the CP properties of a Higgs boson that
can be produced and detected in the $s$-channel mode.
This issue has been explored in Refs.~\cite{bohguncp,atsoncp}
in the case of a general two-Higgs-doublet model.

The first possibility is to measure correlations in the
$\taup\taum$ or $t\anti t$ final states, using the same
techniques as discussed earlier with regard to $Z\h$ production
at an $\epem$ collider. (Note that as in $Z\h$ production
the rest frame of the $\h$ is precisely known for $\mupmum$
collisions.)  The results of the above references imply
that a $\mupmum$ collider is likely to have greater sensitivity
to the Higgs boson CP properties
for $L=20\fbi$ than will the $\epem$ collider for $L=85\fbi$
if $\tanb\gsim 10$ or $2\mw\lsim\mh\lsim 2\mt$.  Indeed, there
is a tendency for the $\mupmum$ CP-sensitivity to be best precisely
for parameter choices such that CP-sensitivity in the $\epem\rta Z\h$
mode is worst. Somewhat higher total luminosity ($L\sim 50\fbi$)
is generally needed in order to use these correlations to distinguish
a pure CP-odd state from a pure CP-even state.

The second possibility arises if it is possible to transversely
polarize the muon beams.  The basic idea is as simple
as that discussed with regard to CP determination using
collisions of two polarized photon beams. Assume that
we can have 100\% transverse polarization and that 
the $\mu^+$ transverse polarization is rotated with respect
to the $\mu^-$ transverse polarization by an angle $\phi$.  The production
cross section for a $\h$ with coupling $a+ib\gamma_5$ then behaves as
\begin{equation}
\sigma(\phi)\propto 1 - {a^2-b^2\over a^2+b^2} \cos\phi 
+ {2ab\over a^2+b^2}\sin\phi\,.
\label{cpmu}
\end{equation}
To prove that the $\h$ is a CP admixture, use the asymmetry
\begin{equation}
A_1\equiv {\sigma(\pi/2)-\sigma(-\pi/2)\over \sigma(\pi/2)+\sigma(-\pi/2)}
= {2ab\over a^2+b^2}\,.
\end{equation}
For a pure CP eigenstate the asymmetry
\begin{equation}
A_2\equiv {\sigma(\pi)-\sigma(-\pi) \over \sigma(\pi)+\sigma(-\pi)}
= {a^2-b^2\over a^2+b^2}
\end{equation}
is $+1$ or $-1$ for a CP-even or CP-odd $\h$, respectively.
Of course, background processes in the final states where
a Higgs boson can be most easily observed ({\it e.g.} $b\anti b$
for the MSSM Higgs bosons) will typically dilute these asymmetries
substantially. Whether or not they will prove useful
depends even more 
upon our very uncertain ability to transversely polarize the muon
beams, especially while maintaining high luminosity.

Note that longitudinally polarized beams are not useful for studying
the CP properties of a Higgs produced in the $s$-channel.
Regardless of the values of $a$ and $b$
in the $\h$ coupling, the cross section is simply proportional
to $1-\lam_{\mu^+}\lam_{\mu^-}$ (the $\lam$'s
being the helicities), and is only non-zero for $L-R$ or $R-L$ transitions,
up to corrections of order $m_\mu^2/\mh^2$.

\subsection{{\boldmath$\mu^+\mu^-$} final remarks and summary}

In summary, $\mm$ colliders offer significant new opportunities for
probing the Higgs sector. The $s$-channel resonance production process is
especially valuable for precision Higgs mass measurements, Higgs width
measurements, and the search for Higgs bosons with negligible $\h ZZ$ couplings,
such as the $\hh,\ha$ Higgs bosons of the MSSM. 
It could also prove valuable for determining the CP properties
of a Higgs boson that is produced at a high rate.
For an extremely narrow Higgs boson, such as a light SM Higgs,
excellent energy resolution is mandatory for precision
measurements of the width and individual channel rates,
and could allow us to distinguish between the SM Higgs and the SM-like
Higgs of the MSSM.
The techniques discussed here in the SM and MSSM theories
are generally applicable to searches for
any Higgs boson or other scalar particle that couples to $\mm$.
If any narrow-width Higgs or scalar particle is observed at either the LHC or
NLC, a $\mm$ collider of appropriate energy would become a priority
simply on the basis of its promise as a Higgs/scalar factory.

\section{Conclusions}

We now present a brief overall summary of the capabilities of present 
and proposed future colliders to
search for the SM Higgs boson and the MSSM Higgs bosons.
The SM Higgs discovery limits for the most useful discovery modes
are summarized in Table \ref{discovery2}.  
The corresponding MSSM Higgs discovery limits
are summarized in Table \ref{discovery3}, where we employ the convenient
MSSM Higgs sector parameterization in terms of $\mha$ and $\tan\beta$.
All values of $\mha\leq 1000$~GeV and
$1\leq\tan\beta\leq 60$ (as preferred in the MSSM) have been surveyed.
Results employ \twoloop\ radiative corrections as computed for $\mt=175\gev$
and $\mstop=1\tev$, neglecting squark mixing.\footnote{The results
of Ref.~\cite{dpfshort} differ slightly since these did not include
\twoloop\ corrections.} Details regarding the claims made in this
summary are to be found in the main body of this review and the
references quoted therein.

\begin{table*}[hp]
\begin{center}
\vskip3pc
\caption{%
Standard Model Higgs boson discovery modes.
All masses are specified in GeV; \lepii\ and LHC
luminosities are detector-summed requirements.}
\label{discovery2}
\vskip1pc
\renewcommand{\arraystretch}{1.3}
\begin{tabular}{|l|c|c|}
\hline
{\bf Machine ($\sqrt{s},\,\int {\cal L}\,dt$)} & {\bf Mode} & 
{\bf Discovery Region}
\\ \hline
LEP   & $\epem\rta \zstar \hsm$ & $\mhsm\lsim 65$
\\ \hline
\lepii, ($175$ GeV, $600$ pb$^{-1}$)  & $\epem\rta Z\hsm$ &
$\mhsm\lsim 82$
\\ \hline
\lepii, ($192$ GeV, $600$ pb$^{-1}$)  & $\epem\rta Z\hsm$ &
$\mhsm\lsim 95 $
\\ \hline
\lepii, ($205$ GeV, $600$ pb$^{-1}$)  & $\epem\rta Z\hsm$ &
$\mhsm\lsim 103 $
\\ \hline
Tevatron, ($2$ TeV, $5$ fb$^{-1}$)  & $p\bar p\rta W\hsm$; $\hsm\rta b\anti
b$ & $\mhsm\lsim 60-80$
\\ \hline
\tevstar, ($2$ TeV, $30$ fb$^{-1}$)  & $p\bar p\rta W\hsm$; $\hsm\rta b\anti
b$ & $\mhsm\lsim 95$ \\
  & $p\bar p\rta W\hsm$; $\hsm\rta \tau^+\tau^-$ & $110\lsim\mhsm\lsim 120$
\\ \hline
DiTeV, ($4$ TeV, $30$ fb$^{-1}$)  & $p\bar p\rta W\hsm$; $\hsm\rta b\anti
b$ & $\mhsm\lsim 95$ \\
  & $p\bar p\rta ZZ\rta 4\ell$ & $\mhsm\sim 200$
\\ \hline
LHC, ($14$ TeV, $600$ fb$^{-1}$)  & $pp\rta \hsm \rta ZZ^{(*)}\rta 4\ell$ &
$120\lsim\mhsm\lsim 800$ \\
  & $pp\rta \hsm \rta \gam\gam$ & $80\lsim\mhsm\lsim 150$ \\
  & $pp\rta t\anti t \hsm,W\hsm$; $\hsm\rta\gam\gam$ & $80\lsim\mhsm\lsim
120-130$ \\
  & $pp\rta t\anti t \hsm$; $\hsm\rta b\anti b$ & $\mhsm\lsim
 120$ \\ \hline
NLC, ($ 500$ GeV, $50$ fb$^{-1}$)  & $\epem\rta Z\hsm$ & $\mhsm\lsim
350$ \\
  & $WW\rta \hsm$ & $\mhsm\lsim 300$ \\
  & $\epem\rta t\anti t \hsm$ & $\mhsm\lsim 120$ \\ \hline
NLC, ($ 1$ TeV, $200$ fb$^{-1}$)  & $\epem\rta Z\hsm$ & $\mhsm\lsim
800$ \\
 &  $WW\rta\hsm$ & $\mhsm\lsim 700$ \\ \hline
FMC, ($ \rts=\mhsm$, $50$ fb$^{-1}$)  & $\mm\rta \hsm$ & $\mhsm\lsim
2\mw$ \\ \hline
\end{tabular}
\end{center}
\vskip3pc
\end{table*}

\begin{table*}[hp]
\begin{center}
\begin{minipage}{17cm}
\caption{%
MSSM Higgs boson discovery modes.  All masses are specified in GeV.
The ordered pair ($\mha$, $\tan\beta$) fixes the MSSM Higgs sector
masses and couplings.  The parameter regime $\mha\leq 1000$~GeV and
$1\leq\tan\beta\leq 60$ is surveyed.
If a range of $\tan\beta$ values is specified below,
then the first (second)
number in the range corresponds to the appropriate minimal (maximal) value of
$\mha$.  Luminosities are those obtained after summing over all
detectors (4 detectors at \lepii\ and ATLAS+CMS at the LHC).
In the case of the LHC we include the $W\h$ with $\h\rta b\anti b$
mode which may not, however, prove viable when running at high luminosity. 
See text for further clarifications.}
\label{discovery3}
\vskip.5pc
\setlength\tabcolsep{4pt}
\renewcommand{\arraystretch}{1.3}
\footnotesize
\begin{tabular}{|l|c|c|}
\hline
{\bf Machine($\sqrt{s},\,\int {\cal L}\,dt$)}& {\bf Mode} &
         \protect\boldmath$(\mha,\tanb)$ {\bf Discovery Region}
\\ \hline
LEP I
 & $\epem\rta \zstar \hl,\hl\ha$ & $(\lsim 45,\gsim 1)$
\\ \hline
\lepii\ ($175$ GeV, $600$ pb$^{-1}$)
 & $\epem\rta Z\hl,\hl\ha$ & $(\lsim 75,\gsim 1)$~or~$(\gsim 75, \lsim 4$--2)
\\ \hline
\lepii\ ($192$ GeV, $600$ pb$^{-1}$)
 & $\epem\rta Z\hl,\hl\ha$ & $(\lsim 80,\gsim 1)~{\rm or}~(\gsim 80,
\lsim 5$--3)
\\ \hline
\lepii\ ($205$ GeV, $600$ pb$^{-1}$)
 & $\epem\rta Z\hl,\hl\ha$ & $(\lsim 90,\gsim 1)~{\rm or}~(\gsim 90,
\lsim 10$--8)
\\ \hline
\tevstar\ ($2$ TeV, $30$ fb$^{-1}$)
 & $p\bar p\rta W\hl$; $\hl\rta b\anti b$; $2b$-tag & $(\gsim 130$--$150,\geq
1)$ \\ \hline
DiTeV ($4$ TeV, $30$ fb$^{-1}$)
 & $p\bar p\rta W\hl$; $\hl\rta b\anti b$; $2b$-tag & $(\gsim 130$--$150,\geq
1)$ \\ \hline
LHC ($14$ TeV, $600$ fb$^{-1}$)
   & $pp\rta \!W\hl\!,t\anti t \hl$; $\hl\!\rta b\anti b$; $2,3b$-tag &
  $(\gsim 130$--$150,\geq 1)$ \\
 & $pp\rta t\anti t$; $t\rta \hp b$; $1b$-tag & $(\lsim 130,\geq 1)$ \\
 & $pp\rta \hh$; $\hh\rta ZZ^{(*)}\rta 4\ell$ & $(\lsim 2\mt,\lsim 3)$ \\
 & $pp\rta \hl,W\hl,t\anti t \hl$; $\hl\rta \gam\gam$ &
    $(\gsim170,\geq 1)$\\
 & $pp\rta b\anti b \ha,\hh$; $\ha,\hh\rta \tau^+\tau^-$ &
   $(\gsim 70,\gsim 1$--40) \\
 & $pp\rta b\anti b \ha,\hh$; $\ha,\hh\rta \mu^+\mu^-$ &
    $(\gsim 100,\gsim 10$--40) \\
 & $pp\rta b\anti b \hl$; $\hl\rta b\anti b$; 3$b$-tag
    & $(\lsim 140,\gsim 4$--6) \\
 & $pp\rta b\anti b \hh$; $\hh\rta b\anti b$; 3$b$-tag
 & $(\gsim 90,\gsim 5$--20) \\
 & $pp\rta b\anti b \ha$; $\ha\rta b\anti b$; 3$b$-tag
    & $(\gsim 45,\gsim 4$--20) \\ 
 & $pp\rta \ha$; $\ha\rta Z\hl$; $Z\hl\rta \ell\ell b\anti b$; 2$b$-tag
    & $(200-2\mt,\lsim 3$) \\ 
 & $pp\rta \hh,\ha$; $\hh,\ha\rta t\anti t$; 2$b$-tag
    & $(\gsim 2\mt,\lsim 3$--1.5) \\ 
 & $pp\rta \hh$; $\hh\rta \hl\hl$; $\hl\hl\rta b\anti b\gam\gam$; 2$b$-tag
    & $(175-2\mt,\lsim 4$--5) \\ 
 & $pp\rta \hh$; $\hh\rta \hl\hl$; $\hl\hl\rta b\anti b b\anti b$; 3$b$-tag
    & $(175-2(\mt+50),\lsim 5$--3) \\ 
\hline
NLC ($500$~GeV, $50$ fb$^{-1}$)
            & $\epem\rta Z\hl$ & visible unless $(\lsim 90,\gsim 8)$ \\
 & $WW\rta \hl$ & visible unless $(\lsim 80,\gsim 13)$ \\
 & $\epem\rta \hl\ha$ & $(\lsim 120, \geq 1)$ \\
 & $\epem\rta Z\hh,WW\rta \hh$ & $(\lsim 140,\geq 1)$ \\
 & $\epem\rta \hh\ha$ & $(\lsim 230,\geq 1)$, unless $(\lsim 90,\gsim 7)$ \\
& $\epem\rta \hp\hm$ & $(\lsim 230,\geq 1)$ \\ \hline
NLC ($ 1$ TeV, $200$ fb$^{-1}$)
 & $\epem\rta \hh\ha$ & $(\lsim 450,\geq 1)$, unless $(\lsim 90,\gsim 7)$ \\
 & $\epem\rta \hp\hm$ & $(\lsim 450,\geq 1)$ \\ \hline
FMC ($\rts=\mh$, $50$ fb$^{-1}$)
 & $\mm\rta \hl$ & $({\rm all},\geq1)$ \\
 & $\mm\rta \hh,\ha$ & $(\lsim \rts_{\rm max},\geq 5)$~{\rm or}
      ~$(\lsim 2\mt,\gsim 2)$ \\ 
 \hline
\end{tabular}
\end{minipage}
\end{center}
\end{table*}

A few examples should help clarify the meaning of the discovery
regions presented in Table \ref{discovery3}.
At \lepii\ with $\sqrt{s}=175$~GeV
and $600~{\rm pb}^{-1}$ integrated luminosity (summing
over all four detectors), either $\hl$ or
$\ha$ (or both) can be discovered via $\epem\rta Z\hl,\hl\ha$ if 
$\mha\lsim 75$~GeV and $\tan\beta\gsim 1$.  A second discovery
region also exists, beginning at $\tan\beta\lsim 4$ (at $\mha=75$~GeV)
and ending at $\mha\lsim 1000$~GeV (at $\tan\beta\sim 2$). 
If $\sqrt{s}$ is increased
to 192~GeV, both discovery regions become larger; in particular, the
latter region now begins at
$\tan\beta\lsim 5$ (at $\mha=80$~GeV) and ends at $\tan\beta\lsim 3$ 
for the maximal value of $\mha$ considered --- see 
the LEP-192 curves in Fig.~\ref{mssmlolum}, for example.
At the NLC-500, $\hl$
can be detected via $\epem\rta Z\hl$ unless $\tan\beta\gsim 8$ and
$\mha\lsim 90$~GeV.  A similar discovery region exception appears for
$WW\rightarrow \hl$. 
The reason such restrictions arise is that in the indicated
region of parameter space, $\hh$ is roughly SM-like
in its couplings, while the $\hl ZZ$ and $\hl W^+W^-$ couplings
are very suppressed.  The $Z\hh\ha$ coupling is likewise
suppressed in the same parameter regime, which explains the other
two discovery region exceptions listed in Table \ref{discovery3}.
Nonetheless, it should be re-stressed that at least one of the MSSM
Higgs bosons is visible throughout all of parameter space. For example,
in the $\tanb\gsim 8$ and $\mha\lsim 90\gev$ region where the $Z\hl$
and $WW\to\hl$ modes lose viability, $\hl\ha$ production occurs
with full strength and would allow discovery of both the $\hl$
and $\ha$.
For a more direct pictorial representation of these $\epem$ regions
see the earlier Fig.~\ref{nlccontours}.

\subsection{A Tour of Higgs Search Techniques at Future Colliders}
 
The goal of the Higgs search at
present and future colliders is to examine the full mass range of
the SM Higgs boson, and the full parameter
space of the MSSM Higgs sector.  The LHC can
cover the entire range of SM Higgs boson masses from the upper limit of
\lepii\ ($\mhsm = 80$--$95$~GeV, depending on machine energy)
to a Higgs mass of 700 GeV (and beyond).  The most difficult region
for high luminosity hadron colliders is the `low' intermediate-mass range
$\mhsm = 80$--$130$~GeV.  The LHC detectors are being designed with the
capability of fully covering the intermediate Higgs mass region.
The \tevstar\ and DiTevatron  may also have some discovery
potential in this mass region.   In contrast, the intermediate mass
Higgs search at the NLC (which makes
use of the same search techniques employed at LEP II) is straightforward.
At design luminosity,
the NLC discovery reach is limited only by the center-of-mass energy of the
machine.   

In the search for the Higgs bosons of the MSSM, two objectives are:
(i) the discovery of $\hl$ and (ii)
the discovery of the non-minimal Higgs states
($\hh$, $\ha$, and $H^\pm$).  Two theoretical results play a key role
in the MSSM Higgs search.  First, the mass of the $\hl$ is bounded. 
For a top-quark pole mass $\mt=175\gev$, $\mstop\sim\msusy\lsim1\tev$, 
and no squark mixing, $\mhl^{\rm max}\simeq 113$~GeV.
For $\mt=175\gev$, $\msusy\sim 1\tev$, and maximal squark mixing,
$\mhl^{\rm max}\simeq 125$~GeV. Second, 
if the properties of $\hl$ deviate significantly from the SM Higgs boson,
then $\mha\lsim{\cal O}(\mz)$ and the $\hh$ and $H^\pm$ masses must
lie in the intermediate Higgs mass region.
As a result, experiments that are sensitive to the intermediate Higgs
mass region have the potential for detecting
at least one of the MSSM Higgs bosons
over the entire MSSM Higgs parameter space  
(parameterized by $\tan\beta$ and $\mha$).
\lepii\ does not have sufficient energy to cover the entire MSSM Higgs
parameter space if $\msusy\sim\mstop\sim 1\tev$ and/or
squark mixing is large,
since then $\mhl^{\rm max}$ lies above the \lepii\ Higgs
mass reach.  The \tevstar, DiTevatron and LHC all possess the
capability of detecting Higgs bosons in the intermediate mass range,
and consequently can probe regions of the MSSM Higgs parameter space
not accessible to \lepii. The most recent results suggest that 
the LHC, when operating at full energy and luminosity,
will guarantee the discovery (or exclusion)
of at least one MSSM Higgs boson for {\it all} values of $\mha$ and $\tan\beta$
not excluded by the \lepii\ search.  A number of specialized
modes have been developed to close the gap in the MSSM parameter space 
that arose when only the most basic $\gam\gam$ and $Z\zstar$
final decays of a neutral Higgs were considered.  However,
these modes do demand efficient and pure
$b$ quark and/or $\tau$ identification in hadron collider events.
Because of the relative simplicity of the NLC
Higgs search in the intermediate mass region, the
NLC is certain to discover at least one MSSM Higgs boson 
(either $\hl$ or $\hh$) if the supersymmetric approach is correct.
If $\hl$ is discovered and proves to be SM-like in its properties, 
then one must be in the region of MSSM Higgs parameter space where
the non-minimal Higgs states are rather heavy and approximately
degenerate in mass.  In this case, the LHC may not be
capable of discovering any Higgs bosons beyond
the $\hl$, and detection of the $\hh$, $\ha$ and
$\hpm$ at the NLC would only be possible for center-of-mass energy 
$\sqrt{s}\gsim 2\mha$.

Below is an outline of the Higgs potential of present and
future colliders. A Higgs boson is deemed observable if a $5\sigma$-excess
of events can be detected in a given search channel. We assume $\mt=175\gev$.
For the MSSM Higgs results we employ $\mstop=1\tev$ and neglect
squark mixing in the radiative mass corrections; further, SUSY decays
are assumed to be absent.
 
\noindent
$\bullet$ LEP --- The current lower bound on the SM Higgs mass is
64.5 GeV, and will increase by at most a few GeV.
The lower bounds on the MSSM Higgs masses (scanning over {\it all}
parameters of the model) are
$\mhl > 45$ GeV, $\mha > 45$ GeV (for $\tan\beta > 1$),
and $m_{H^\pm} > 45$ GeV.
 
\noindent
$\bullet$ \lepii\ --- For $\sqrt s=175$ GeV,
the SM Higgs is accessible via $Z\hsm$ production up
to $82$ GeV for an integrated luminosity of
$600$ pb$^{-1}$ summed over all experiments.
For $\sqrt s=192$~GeV,
$\mhsm$ values as high as $95\gev$ can be probed with $600$ pb$^{-1}$
of data, using
$b$-tagging in the region $\mhsm\sim \mz$. The reach for $\sqrt s=205$~GeV is
about 103 GeV for $L=600\pbi$.  In the MSSM, the
same mass reaches apply to $\hl$ if $\mha\gsim \mz$,
since in this case $Z\hl$ is produced at about the same rate as $Z\hsm$.
If $\mha\lsim \mz$, then the $\hl ZZ$ coupling (which controls the $Z\hl$
cross section) becomes suppressed while the $Z\ha\hl$ coupling becomes
maximal.  In the latter 
parameter region, $e^+e^-\rightarrow\ha\hl$ can be detected 
for all values of $\mha\lsim\sqrt{s}/2-10$~GeV, assuming that $\tan\beta>1$. 
Since $\mhl\lsim 113$ GeV if $\mstop\leq 1\tev$
and squark mixing is negligible, increasing the \lepii\ energy to roughly
$\sqrt s = 220$ GeV while maintaining the luminosity would 
be sufficient to guarantee the detection of at least one Higgs boson 
(via $Z\hl$ and/or $\ha\hl$ final states) over the entire MSSM
Higgs parameter space. Large squark mixing with $\mstop\sim1\tev$ 
would, however, necessitate still larger $\sqrt s$.
 
\noindent
$\bullet$ Tevatron ($\sqrt s = 2$ TeV, ${\cal L} =
10^{32}~{\rm cm}^{-2}~{\rm s}^{-1}$  with the
Main Injector) --- The most promising mode for the SM Higgs
is $W\hsm$ production, followed by $\hsm \to b\anti b$.
With 5~fb$^{-1}$ of integrated luminosity, this mode could potentially
explore a Higgs mass region of 60--80 GeV, a region which will already
have been covered by \lepii\ via the $Z\hsm$ process.
 
\noindent
$\bullet$ \tevstar\ ($\sqrt s = 2$ TeV, ${\cal L}
\ge 10^{33}~{\rm cm}^{-2}~{\rm s}^{-1}$) ---
The Higgs mass reach in the $W\hsm$ mode, with $\hsm \to b\anti b$, is extended
over that of the Tevatron.
With 30~fb$^{-1}$ of integrated luminosity, a Higgs of mass 95 GeV
is potentially accessible, but the peak will
not be separable from the $WZ$, $Z
\to b\bar b$ background.
The mode $(W,Z)\hsm$, followed by $\hsm \to \tau^+ \tau^-$ and $W,Z \to jj$
(where $j$ stands for a hadronic jet), is
potentially viable for Higgs masses sufficiently far above the $Z$ mass.
With 30~fb$^{-1}$ of integrated luminosity, a Higgs in the mass range $110-120$
GeV may be detectable. However, the Higgs peak will be difficult to
recognize on the slope of the much larger $Zjj$, $Z \to \tau^+\tau^-$
background. Both of these modes are of particular significance for $\hl$.
For $\tanb>1$ and $\mha\lsim 1.5 \mz$,
the enhanced coupling of the $\hl$ to $b$'s and $\tau$'s makes it
unobservable at the LHC
via $\hl \to \gamma \gamma$ or $\hl \to Z\zstar \to 4\ell$.
 
\noindent
$\bullet$ DiTevatron ($\sqrt s = 4$ TeV, ${\cal L}
\ge 10^{33}~{\rm cm}^{-2}~{\rm s}^{-1}$) ---
For the SM Higgs, the ``gold-plated'' mode, $\hsm \to ZZ \to 4 \ell$,
requires about 30~fb$^{-1}$
for the optimal Higgs mass in this mode, $\mhsm = 200$ GeV.
If 100~fb$^{-1}$ can be collected, this mode covers the mass region
$\mhsm=200 - 300$ GeV and $\mhsm \sim 150$ GeV.
The $4\ell$ mode is not useful for any of the MSSM Higgs bosons.
The Higgs mass reach in the $W\hsm$, $\hsm
\to b\anti b$ mode is only marginally
better than at the \tevstar, due to the increase in the top-quark backgrounds
relative to the signal.
The mode $(W,Z)\hsm$, with $\hsm \to \tau^+ \tau^-$ and
$W,Z \to jj$, has less promise than at the \tevstar\ 
due to the relative increase in the background.
 
\noindent
$\bullet$ LHC ($\sqrt s = 14$ TeV, ${\cal L} =  10^{33}$%
--$10^{34}~{\rm cm}^{-2}~{\rm s}^{-1}$) ---
For the SM Higgs boson,
the ``gold-plated'' mode, $\hsm \to ZZ^{(*)} \to 4 \ell$,
including the case where
one $Z$ boson is virtual, covers the range of Higgs masses 130--700 GeV and
beyond with 100~fb$^{-1}$.
For $\mhsm > 700$ GeV, the Higgs is no longer ``weakly coupled'', and
search strategies become more complex.
 
The CMS detector is planning an exceptional (PbWO$_4$ crystal)
electromagnetic calorimeter  which will enable
the decay mode $\hsm \to \gamma\gamma$ to cover the Higgs mass range 85--150 GeV
with 100 fb$^{-1}$ of integrated luminosity. This range
overlaps the reach of \lepii\ (with $\sqrt s = 192$ GeV)
and the lower end of the range covered by the gold-plated mode. The ATLAS
detector covers the range 110-140 GeV with this mode.
Both CMS and ATLAS find
that the modes $t\anti t \hsm$ and $W\hsm$, with $\hsm \to \gamma\gamma$, are
viable in the mass range 80--120 GeV with 100~fb$^{-1}$ of integrated
luminosity. Since backgrounds are smaller, these modes do not require
such excellent photon resolutions and jet-photon discrimination
as does the inclusive $\gam\gam$ mode. CMS has also studied the production of
the Higgs in association with two jets, followed by $\hsm\to \gamma\gamma$,
and concludes that this mode covers the Higgs mass range 70--130 GeV.
 
The modes $t\anti t \hsm$ and $W\hsm$, with $\hsm \to b\anti b$, are
useful in the intermediate-mass region.
The reach with 100~fb$^{-1}$ of integrated
luminosity is 100 GeV, reduced to 80 GeV with 30~fb$^{-1}$,
and extended to $120\gev$ for $L=600\fbi$ (summed over experiments).
Overall, ATLAS will cover the Higgs mass region 80--140 GeV with 100
fb$^{-1}$ using a combination of $\hsm \to \gamma\gamma$;
$t\bar t \hsm$, $W\hsm$ with $\hsm\to \gamma\gamma$; and
$t\bar t \hsm$ with $\hsm \to b\bar b$; CMS has not yet completed
their studies of the $b$-tagged modes.

The main search mode for the lightest MSSM Higgs boson is $\hl\to\gamma\gamma$,
which is viable when $\mhl$ is near $\mhl^{\rm max}$. The processes 
$t\anti t \hl$ and $W\hl$, with $\hl \to b\anti b$, can potentially
extend $\hl$ detection to the somewhat lower $\mhl$ values (corresponding to
somewhat lower $\mha$) that would be one way to assure that at least one
MSSM Higgs boson can be detected over all of the MSSM parameter
space. Adapting the ATLAS study for the SM Higgs to $\hl$ suggests
that the required sensitivity could be achieved in the $t\anti t \hl$ mode.
Due to the large top-quark background, the $W\hl$ with $\hl\to b\bar b$
mode at the LHC has little advantage over this mode at the \tevstar\ or the 
DiTevatron for the same integrated luminosity.
Other modes whose inclusion will guarantee discovery of at least one MSSM
Higgs boson appear below.

The other MSSM Higgs bosons are generally more elusive. There are small
regions of parameter space in which $\hh$ can be observed decaying to
$\gamma\gamma$ or $ZZ^{(*)}$.
The possibility exists that the charged Higgs can be discovered
in top decays.
For $\mha>2 \mz$ and moderate to 
large $\tan\beta$, $\hh$ and $\ha$ can have sufficiently
enhanced $b$ quark couplings that
they would be observed when produced in association with
$b\anti b$ and decaying via $\hh, \ha \to \tau^+\tau^-$.
In the most recent studies the region of viability for the $\taup\taum$
modes at $L=600\fbi$ (combined luminosity of ATLAS and CMS)
is bounded from below by 
$\tanb\gsim 1$ at $\mha\sim 70\gev$ rising
to $\tanb\gsim 20$ at $\mha\sim 500\gev$. CMS has demonstrated
that the $\hh$ and $\ha$ can be observed in their $\mupmum$ decay channel,
using enhanced production in association with $b\anti b$,
in a slightly more limited region.

For enhanced couplings of the Higgs bosons to $b$ quarks
(which occurs for large $\tanb$), the modes
$b\anti b (\hl, \hh, \ha)$, with $\hl, \hh, \ha \to b\anti b$,
and $\anti t b H^+$, with $H^+ \to t \anti b$, are potentially viable.
Parton-level Monte Carlo studies suggest that the combined discovery
region for the former processes covers a region that begins 
at $\tanb\gsim 3-4$ at low $\mha$ rising to $\tanb\gsim 14$ at $\mha=500\gev$.

CMS and ATLAS have 
considered the process $gg \to \ha \to Z\hl \to \ell^+\ell^-b\anti 
b$.  They each claim an observable signal with single and double
$b$ tagging in the region $175\lsim\mha\lsim 2\mt$ and $\tanb\lsim 2-3$
for $L=100\fbi$ (in a given detector). 
This region expands somewhat if the data from the two
detectors is combined and a combined $L=600\fbi$ is achieved.
Recent results for the modes
$\hh\rta\hl\hl$ and $\hh,\ha\rta t\anti t$ are also available. 
ATLAS and CMS claim that for $2\mt\lsim\mha\lsim 500\gev$
one can detect $\ha,\hh\rta t\anti t$ for $\tanb\lsim 2$ with $30-100\fbi$
of integrated luminosity; good knowledge of the magnitude of the 
$t\anti t$ background is required. This region expands to $\tanb\lsim 3$
for combined ATLAS+CMS $L=600\fbi$.
The $\hh\rta\hl\hl$ mode can potentially be employed in the channels
$\hl\hl\rta b\anti b b\anti b$ and $\hl\hl\rta b\anti b \gam\gam$.
A parton-level Monte-Carlo study using 3 $b$-tagging
and requiring that there be two $b\anti b$ pairs of mass $\sim \mhl$
yields a viable signal for $2(\mt+50)\gsim\mha\gsim 175$ and $\tanb\lsim 5-3$,
assuming $L=600\fbi$ \cite{DGV4}. 
Because of uncertainty on the ability to trigger
on the $4b$ final state, ATLAS and CMS have examined the 
$\hh\rta\hl\hl\rta b\anti b \gam\gam$
final state.  This is a very clean channel (with $b$ tagging),
but is rate limited. For $L=600\fbi$ of luminosity the discovery
region $175\gev\lsim\mha\lsim 2\mt$, $\tanb\lsim 5$ is claimed.

Putting together all these modes, we can summarize by saying that
for moderate $\mha\lsim 2\mt$ there is an excellent chance
of detecting more than one of the MSSM Higgs bosons.  However, for large
$\mha\gsim 400\gev$ (as often preferred in the GUT scenarios)
only the three $\hl\rta\gam\gam$ production/decay modes ($gg\rta\hl$,
$t\anti t\hl$, and $W\hl$, all with $\hl\rta\gam\gam$)
are guaranteed to be observable if $\tanb$ is neither near 1
nor large.  

\noindent
$\bullet$ $e^+e^-$ linear collider ($\sqrt s = 500$ GeV to 1 TeV,
${\cal L} \ge 5 \times 10^{33}~{\rm cm}^{-2}~{\rm s}^{-1}$) --- At $\sqrt s
 =500$ GeV
with an integrated luminosity of 50~fb$^{-1}$, the SM Higgs boson is observable
via the $Z\hsm$ process up to $\mhsm=350$ GeV.
Employing the same process, a Higgs boson with $\mhsm=130$~GeV
(200~GeV) would be discovered with 1~fb$^{-1}$ (10~fb$^{-1}$) of
data.  Other Higgs production mechanisms are also useful.
With 50 fb$^{-1}$ of data, the $WW$-fusion process is observable up to
$\mhsm = 300$ GeV.
The $t\anti t \hsm$ process is accessible for $\mhsm \lsim 120$ GeV,
thereby allowing a direct determination of the $t\anti t \hsm$ Yukawa
coupling.
The $\gamma\gamma$ collider mode of operation does not extend the reach of
the machine for the SM Higgs, but does allow a measurement of the
$\hsm \to \gamma\gamma$ partial width up to
$\mhsm = 300-350$ GeV.
 
In the MSSM,
for $\mha\lsim \mz$, the $\hl$ and $\ha$ will be detected in the $\hl\ha$ mode,
the $\hh$ will be found via $Z\hh$ and $WW$-fusion production, and $H^+H^-$
pair production will be kinematically allowed and easily
observable. At higher values of the parameter $\mha$,
the lightest Higgs boson is accessible in both the $Z\hl$ and
$WW$-fusion modes.  However, the search for the heavier
Higgs boson states is limited by machine energy.
The mode $\hh\ha$ is observable up to
$\mhh \sim \mha \sim 230-240$ GeV, and $H^+H^-$ can be detected up to
$m_{H^{\pm}} =230$ GeV.
The $\gamma\gamma$ collider mode could potentially
extend the reach for $\hh$ and $\ha$ up to 400 GeV if $\tanb$
is not large.
 
At $\sqrt s = 1$ TeV with an integrated luminosity of 200~fb$^{-1}$,
the SM Higgs boson can be detected via the $WW$-fusion process up to
$\mhsm=700$ GeV. In the MSSM,
$\hh\ha$ and $H^+H^-$ detection would be extended
to $\mhh\sim \mha\sim m_{H^{\pm}}\sim 450$ GeV.
 
\noindent
$\bullet$ Influence of MSSM Higgs decays into supersymmetric particle
final states --- For some MSSM parameter choices,
the $\hl$ can decay primarily to invisible modes, including
a pair of the lightest supersymmetric neutralinos
or a pair of invisibly decaying sneutrinos.
The $\hl$ would still be easily discovered
at $e^+e^-$ colliders in the $Z\hl$ mode using missing-mass techniques.
At hadron colliders, the $W\hl$ and $t\anti t \hl$ modes may
be detectable via large missing energy and lepton plus invisible energy
signals, but determination of $\mhl$ would be difficult.
The $\hh$, $\ha$, and $H^\pm$ decays can be dominated by
chargino and neutralino pair final states and/or slepton pair final states.
Such modes can either decrease or increase the chances of 
detecting these heavier MSSM Higgs bosons
at a hadron collider, depending upon detailed MSSM parameter choices.
At $e^+e^-$ colliders $\hh$, $\ha$ and $H^\pm$ detection up to the earlier
quoted (largely kinematical) limits should in general remain possible.
 
\noindent
$\bullet$ Hadron collider beyond the LHC ---
It could provide increased event rates and more overlap of discovery
modes for the SM Higgs boson.
However, its primary impact would be to extend sensitivity
to high mass signals associated with
strongly-coupled electroweak symmetry breaking scenarios.
In the case of the MSSM Higgs bosons, the increased
event rates would significantly extend the overlap of the various
discovery modes, expanding the parameter regions where several and perhaps
all of the MSSM Higgs bosons could be observed.
 
\noindent
$\bullet$ $e^+e^-$ collider beyond $\sqrt s =1$ TeV --- Such an extension in
energy is not required for detecting and studying a
weakly-coupled SM Higgs boson,
but could be very important for a strongly-coupled
electroweak-symmetry-breaking scenario. Detection of $\hh\ha$ and $H^+H^-$
production in the MSSM model would be extended to higher masses.

\noindent
$\bullet$ $\mm$ collider with $\sqrt s=500\gev$ and 
${\cal L} \ge 2-5 \times 10^{33}~{\rm cm}^{-2}~{\rm s}^{-1}$ --- 
Such a collider would accomplish many of the same tasks as the corresponding
$\ee$ collider, except that the $\gam\gam$ collider option would
not be possible.

\noindent
$\bullet$ $\mm$ collider with $\sqrt s=\mh$ --- 
Such a collider (dubbed the first muon collider or FMC)
can be used to discover a SM Higgs boson with $\mhsm\lsim 2\mw$
or the $\hl$ of the MSSM by scanning. 
The enhanced $b\anti b$ and $\mm$ couplings of the $\hh$ and $\ha$ imply
that they can be discovered by scanning
for $\tanb\gsim 2$ if their masses are below $2\mt$, and for $\tanb\gsim 5$
regardless of mass.

\subsection{Precision Measurements of Higgs Properties}
 
Once discovered, the next priority will be to carry out
detailed studies of the properties of the Higgs bosons.
In the outline below, $\h$ denotes the lightest CP-even
neutral Higgs.  It may be $\hsm$ or $\hl$ (of the MSSM),
or perhaps a scalar arising from a completely different model.
Once discovered, determining the identity of $\h$ 
is the crucial issue.

The largest number of precision measurements
can be performed at an $\ee$ collider (or $\mm$
collider with the same energy and luminosity). The LHC and FMC (run
in the $s$-channel Higgs production mode)
provide more limited information. At an $\ee$ collider 
measurements of cross sections, branching ratios, and angular
distributions of Higgs events will determine most model parameters and
test much of the underlying theory of the scalar sector.
In particular, the $\h ZZ$ coupling-squared can be measured using
inclusive $Z\h$ production with $Z\rta\ell^+\ell^-$ to an accuracy
of better than 10\%. The $\h WW$ coupling can then be computed
assuming custodial symmetry and compared with the value measured
via the $WW\to\h$ fusion production process.  If $\mh\lsim 120\gev$
or $\mh\gsim 2\mt$, there is a good chance that the $\h t\anti t$
coupling can be measured with reasonable accuracy.
Further, for a SM-like Higgs with $\mh\sim 120\gev$,
statistical precisions of 7\%, 14\%, 39\%, and 48\% can
be achieved for
$\sigma(Z\h) \times BR(h \rightarrow b \bar b)$,
$\sigma(Z\h) \times BR(h \rightarrow \tau^+ \tau^-)$,
$\sigma(Z\h) \times BR(h \rightarrow c \bar c+\ gg)$, and
$\sigma(Z\h) \times BR(h \rightarrow W \wstar)$, respectively.

However, in the case of a SM-like Higgs boson, 
the total Higgs width and coupling to $b\anti b$ and $\taup\taum$
would remain poorly determined.  Measurement of the $h\rta\gam\gam$
coupling would require the $\gam\gam$-collider mode of operation.
LHC measurements of $\sigma(t\anti t\h)\times BR(\h\to \gam\gam)$
and $\sigma(W\h)\times BR(\h\to \gam\gam)$ can be combined with
the NLC determinations of the $\h WW$ and (if $\mh\lsim 120\gev$,
$\h t\anti t$) coupling to determine $BR(\h\to\gam\gam)$. This
can be combined with the $\gam\gam$-collider measurement
of the $\h\gam\gam$ coupling itself to obtain a value for the total
$\h$ width.  This would then allow determination of the partial
$\h\rta b\anti b$ width from $BR(\h\rta b\anti b)$ as measured at the NLC,
for example. (An accurate determination of $BR(\h\rta b\anti b)$
requires $\mh\lsim 145\gev$.) However, the accumulation of errors
in reaching the total $\h$ width or partial $\h\to b\anti b$ width
will limit the accuracy with which they can be determined.
For $\mh\lsim 2\mw$, $s$-channel Higgs production 
at a $\mm$ collider with sufficiently good energy
resolution can provide a very accurate
direct scan determination of $\gamh$ with error
of order $\pm10\%$ and a determination of the
$\h\mm$ coupling. The precision of the latter is limited only
by the errors on determinations of $BR(\h\to b\anti b)$ 
(and/or $BR(\h\to W\wstar)$) from other sources.

At an $\ee$ collider with $L=50\fbi$,
the central value of the Higgs mass can be determined 
with a statistical accuracy of $\pm$ 180 MeV for $\mh\lsim 2\mw$
using direct final state mass reconstruction.
A detector with the ultra-excellent resolution of the super-JLC design
could measure $\mh$ to $\pm 20\mev$ using the recoil mass
spectrum in $Z\h$ events where $Z\to \ell^+\ell^-$.
At the LHC, in the intermediate mass region the accuracy
will be of order $\pm 1\gev$ using $\h\to\gam\gam$ modes. 
In this same mass region, the FMC with excellent beam energy
resolution could determine $\mh$ to within $\lsim \pm 0.0003\%$.
Since most Standard Model parameters only have logarithmic
dependence on the Higgs mass, the NLC and FMC accuracies are
far greater than what is needed to check the
consistency of the Standard Model predictions for precision $Z$
electroweak measurements. The LHC accuracy would be adequate. 
Of course, one could use the mass determination to make
accurate predictions for the Higgs
production cross sections and branching fractions.

The expected errors referred to above for all the
different measurements are summarized in Table~\ref{nlclhcerrors}
(appearing earlier) for Higgs masses of $\mh=110$, $120$ and $140\gev$.
It is a matter of some priority for the experimentalists to refine
these estimates and obtain errors for a more complete selection of
Higgs masses. As noted below, a complete model-independent
study of all the properties of a light Higgs boson
can be performed with a useful level of accuracy only if most (if not all)
of the measurements can be made with small errors.
 
For a SM Higgs boson, signals can be selected with small backgrounds and one 
can easily measure the angular
distributions of the Higgs production direction in the process
$e^+e^-\rightarrow Zh$
and the directions of the outgoing
fermions in the $Z^0$ rest frame, and verify the expectations
for a CP-even Higgs boson. For a more general Higgs eigenstate,
since only the CP-even part couples
at tree-level to $ZZ$,  these measurements would still agree with the CP-even
predictions unless the Higgs is primarily
(or purely) CP-odd, in which case the cross section will be much smaller
and the event rate inadequate to easily verify the mixed-CP or CP-odd
distribution forms. Analysis of photon
polarization asymmetries in $\gamma\gamma\rta h$ production rates,
and of angular correlations among secondary
decay products arising from primary
$h\rta\tau\tau$ or $h\rta t\overline t$ final state decays in
$Zh$ production, can both provide much better sensitivity to a CP-odd
Higgs component in many cases. These latter decay correlations
are also very effective in determining the CP character of a Higgs
boson produced directly in the $s$-channel at a $\mm$ collider.
At a $\mm$ collider if polarization of both beams can
be achieved without loss of too much luminosity,
one can also probe the CP parity of a Higgs
using transverse polarization asymmetries.

An important question is whether or not
we can distinguish the SM-like $\hl$ from the 
simple standard model $\hsm$ in the large $\mha$ portion of
MSSM parameter space, $\mha\gsim 400\gev$, where it is more
than likely that only the $\hl$ will be seen directly at the NLC
or LHC. The most promising possibilities are: (i) the NLC
measurement of the rare branching
fraction ratio $r_{c\anti c+gg}\equiv BR(\h\to c\anti c+gg)/BR(\h\to b\anti b)$;
and (ii) the FMC measurements of
the total width (as measured by a careful scan) and
$\Gamma(\h\to \mm)BR(\h\to b\anti b)$ (which
is directly determined from event rate).
The total width can be combined with a direct determination of
$BR(\h\to b \anti b)$ (using $Z\h$ associated production, for example)
to determine in a model-independent way
the $\h\to b\anti b$ partial width and coupling;
similarly $\Gamma(\h\to \mm)BR(\h\to b\anti b)$ 
can be combined with $BR(\h\to b\anti b)$ 
to obtain a model-independent determination of $\Gamma(\h\to \mm)$.
Deviations in the quantities $r_{c\anti c +gg}$, $\Gamma_{\h}^{\rm tot}$, and
$\Gamma(\h\to\mm)$ can potentially distinguish between $\h=\hsm$
and $\h=\hl$ for $\mha$ values above $400\gev$.

Reasonably precise measurements of certain $\sigma \times BR$
combinations will also be possible at hadron colliders.  For example,
at the LHC a good determination of $\sigma(gg\rightarrow
h)\times BR(h\rightarrow \gamma\gamma)$ should be possible for a SM-like
Higgs in the 80 to 150 GeV mass range, and
of $\sigma (gg\rightarrow h)\times BR(h\rightarrow ZZ)$ for $\mh > 130$ GeV;
$\sigma (t\anti t h)\times BR(h\rightarrow b\anti b)$ 
would be roughly determined for $m_h \lsim 100-120$ GeV. However, it would
appear that these cannot probe as far out in $\mha$ as can
the NLC and FMC precision measurements. In addition, deviations
from SM predictions can arise from other types of new physics.

In the absence of a $\mm$ collider, model-independent determinations of
$\Gamma_{\h}^{\rm tot}$ and $\Gamma(\h\to b\anti b)$ require
combining data from the LHC with data from {\it both}
$\epem$ and $\gam\gam$ collisions at the NLC.
Let us repeat in detail
the shortest and probably most accurate procedure outlined earlier.
\begin{description}
\item[(a)] Determine $BR(b\anti b)$ from $Z\h$ events at the \underline{NLC}.
\item[(b)] Combine $BR(b\anti b)$ with $\sigma(WW\to\h)BR(b\anti b)$
at the \underline{NLC} to obtain the $WW\h$ coupling.
\item[(c)] Use the $WW\h$ coupling and a measurement of
$\sigma(W\h)BR(\gam\gam)$ at the \underline{LHC} to determine $BR(\gam\gam)$.
\item[(d)] Combine $BR(b\anti b)$ with the \underline{$\gam\gam$-Collider}
measurement of $\sigma(\gam\gam\to\h)BR(b\anti b)$ to obtain
$\Gamma(\h\to\gam\gam)$.
\item[(e)] Compute $\Gamma_{\h}^{\rm tot}=\Gamma(\h\to\gam\gam)/BR(\gam\gam)$.
\item[(f)] Compute $\Gamma(\h\to b\anti b)=BR(b\anti b)\Gamma_{\h}^{\rm tot}$.
\end{description}
The accumulation of errors will be significant.
However, it is worth re-emphasizing
the basic point that {\it data from all three colliders will be required
in order to complete a {\bf model-independent} determination
of all the properties of a light Higgs boson}.

Overall, if either the minimal
standard model or the minimal supersymmetric standard model,
or something like them, is correct, one or more Higgs bosons
will be discovered at all three types of machines discussed here:
a hadron collider --- perhaps the Tev or \tevstar, but certainly
the LHC; an $\epem$ collider --- perhaps \lepii, but certainly the NLC;
and, assuming target luminosity goals can be met, a $\mm$ collider.
These machines are remarkably complementary, each revealing new
windows on the Higgs boson(s); only the combination of data
from all three types of machines can over constrain the Higgs sector
and provide a fully definitive test of, for example, the minimal standard
model Higgs predictions, including all ($f\anti f$, $VV$, $\gam\gam$, and
$gg$) couplings and the total width.  Higgs physics should provide a rich 
source of vital experimental results far into the future.

\section{Acknowledgements}

We are grateful for advice and assistance from T. Barklow, D. Burke,
A.~Caner, M.~Carena, J.-L.~Contreras, 
A.~Djouadi, D.~Froidevaux, F.~Gianotti, H.~Haber, 
T.~Han, P.~Janot, S.~Keller, T.~Liss,
E.~Richter-Was, A.~Sopczak, M.~Spira, and C.~Wagner.
J.~G. was supported in part by Department of Energy grant DE-FG03-91ER40674
and the Davis Institute for High Energy Physics.
S.~W. was supported in part by Department of Energy grant DE-FG02-91ER40677.

\clearpage
 

\begin{thebibliography}{99}

\bibitem{S} L.~Susskind, \PRD D20 2619 1979 .


\bibitem{W} S.~Weinberg, \PRD D13 974 1976 ; {\bf D19},
1277 (1979).

\bibitem{SSVZ} P.~Sikivie, L.~Susskind, M.~Voloshin, and V.~Zakharov, 
\NPB B173 189 1980 .

\bibitem{V} M.~Veltman, Acta.\ Phys.\ Pol.\ {\bf B8}, 475 (1977).

\bibitem{PT} M.~Peskin and T.~Takeuchi, \PRL 65 964 1990 ; 
B.~Holdom  and J.~Terning, \PLB B247 88 1990 ;
M.~Golden and L.~Randall, \NPB B361 3 1991 .

\bibitem{GW} S.~Glashow and S.~Weinberg, \PRD D15 1958 1977 .

\bibitem{WS} S.~Weinberg, \PRL 19 1264 1967 ; 
A.~Salam, in {\it Elementary Particle Theory: Relativistic Groups and 
Analyticity}, edited by N.~Svartholm, proceedings of the Nobel Symposium 
No.~8 (Almqvist and Wiksell, Stockholm, 1968).

\bibitem{MPP} L.~Maiani, G. Parisi, and R. Petronzio, \NPB B136 115 1978 ; 
N.~Cabbibo,  L.~Maiani, G. Parisi, and R. Petronzio, 
\NPB B158 295 1979 .

\bibitem{DN} R.~Dashen and H.~Neuberger, \PRL 50 1897 1983 .

\bibitem{HKNV} U. Heller, M. Klomfass, H. Neuberger, and P. Vranas, \NPB 405
555 1993 .

\bibitem{AI} G.~Altarelli and G.~Isidori, \PLB B337 141 1994 .

\bibitem{CDF} CDF Collaboration, F.~Abe {\it et al.}, \PRD D50 2966 1994 ; 
\PRL 73 225 1994 ; \PRL 74 2626 1995 .

\bibitem{D0} D0 Collaboration, S.~Abachi {\it et al.}, 
\PRL 74 2632 1995 .

\bibitem{MNPP} G.~Montagna, O.~Nicrosini, G.~Passarino, and F.~Piccinini, 
\PLB B335 484 1994 .

\bibitem{quirosi} J.A.~Casas, J.R.~Espinosa and M.~Quiros,
\PLB B342 171 1995 .

\bibitem{hhg} 
See J.F. Gunion, H.E. Haber, G.L. Kane and S. Dawson,
{\it The Higgs Hunters Guide}, Addison-Wesley Publishing.

\bibitem{HKS}
H.E. Haber, G.L. Kane, T. Sterling \NPB 161 493 1979 .
 
\bibitem{hewett}
For recent summaries, see, for example, J. Hewett, SLAC-PUB-6521 and
C.-H. Chang and C. Lu, AS-ITP-95-23.

\bibitem{haberdecouple}  See, for example, H.E. Haber, SCIPP-94-39,
Workshop on Electro-Weak Symmetry Breaking, Budapest,
Hungary, July 11-15, 1994 and the Joint U.S.-Polish Workshop on Physics
from Planck Scale to Electro-Weak Scale (SUSY 94), Warsaw, Poland,
21-24 September, 1994.

\bibitem{grifolsmendez} J.A. Grifols and A. Mendez, \PRD D22 1725 1980 .

\bibitem{GM} H. Georgi and M. Machacek, \NPB B262 463 1985 .

\bibitem{CG} M. Chanowitz and M. Golden, \PLB B165 105 1985 .

\bibitem{gvw} J.F. Gunion, R. Vega, and J. Wudka, \PRD D42 1673 1990 ;
\PRD D43 2322 1991 .

\bibitem{lrmodels} J.C. Pati and A. Salam, \PRD D10 275 1974 ; R.N.
Mohapatra and J.C. Pati, \PRD D11 566 1975 ;
R.N. Mohapatra and G. Senjanovic, \PRD D12 1502 1975 ; \PRL 44 912 1980 ;
\PRD D23 165 1981 .

\bibitem{leftright} F. Olness and M.E. Ebel, \PRD D32 1769 1985 ;
R.N. Mohapatra and P.B. Pal, \PLB 179B 105 1986 ;
J.F. Gunion, B. Kayser, R.N. Mohapatra, N.G. Deshpande, J.A. Grifols,
A. Mendez, F. Olness, P.B. Pal, Snowmass Summer Study 1986, p. 197;
J.A. Grifols, J.F. Gunion, and A. Mendez, \PLB 197B 266 1987 ;
J.F. Gunion, A. Mendez, and F. Olness, \IJMP A2 1085 1987 ;
J.A. Grifols, A. Mendez, and G.A. Schuler, \MPL A4 1485 1989 ;
J.F. Gunion, J. Grifols, A. Mendez, B. Kayser and F. Olness, 
\PRD D40 1546 1989 ;
L. Roszkowski, \PRD D41 2266 1990 ;
N. G. Deshpande, J.F. Gunion, B. Kayser and F. Olness,
\PRD D44 837 1991 ;
K. Huitu and J. Maalampi, \PLB B344 217 1995 .

\bibitem{lrunification} A sampling of references is:
N.T. Shaban and W.J. Stirling, \PLB B291 281 1992 ;
N.G. Deshpande, E. Keith and T.G. Rizzo, \PRL 70 3189 1993 ; 
M. Bando, T. Sato and T. Takahashi, \PRD D52 3076 1995 ;
B. Brahmachari, \PRD D52 1 1995 ; B. Brahmachari and R.N. Mohapatra,
\PLB B357 566 1995 ; E. Ma, \PRD D51 236 1995 ;
D.-G. Lee and R.N. Mohapatra, \PRD D52 4125 1995 .


\bibitem{ememgunion}
J.F. Gunion {\it Proceedings of the 2nd International Workshop on
``Physics and Experiments with Linear $\epem$ Colliders''},
eds. F. Harris, S. Olsen, S. Pakvasa and X. Tata, Waikoloa, HI (1993), 
World Scientific Publishing, p.~903.



\bibitem{ememtodelmm} J.F. Gunion, UCD-95-36,
Proceedings of the Santa Cruz Workshop on $e^-e^-$ Colliders, 
September 4-5, 1995.

\bibitem{mendezrecent} J.A. Coarasa, A. Mendez and J. Sola,
UAB-FT-378.

\bibitem{GG} H.~Georgi and S.~Glashow, \PRL 32 438 1974 .

\bibitem{QUAD} M.~Veltman, Acta Phys.\ Polon.\ B {\bf12}, 437 (1981);
L.~Maiani, {\it Proceedings of the Summer School on Particle Physics},
Gif-Sur-Yvette, 1979, (Paris, 1980), p.~1; E.~Witten, \NPB B188 513 1981 .

\bibitem{DRW} S.~Dimopoulos, S.~Raby, and F.~Wilczek, \PRD D24 1681 1981 .

\bibitem{TWOHIGGS} S.~Dimopoulos and H.~Georgi, \NPB B193 150 1981 ;
N.~Sakai, \ZP C11 153 1981 ;
L.~Ib\'a\~nez and G.~Ross, \PLB B105 439 1981 ;
M.~Einhorn and D.~R.~T.~Jones, \NPB B196 475 1982 ;
W.~Marciano and G.~Senjanovic, \PRD D25 3092 1982 .

\bibitem{GUT} J.~Ellis, S.~Kelley, and D.~Nanopoulos, \PLB B260 131 1991 ; 
U.~Amaldi, W.~de~Boer, and H.~F\"urstenau, \PLB B260 447 1991 ; 
P.~Langacker and M.~Luo, \PRD D44 817 1991 .

\bibitem{LW} J.~Lykken and S.~Willenbrock, \PRD D49 4902 1994 .

\bibitem{RAD} L.~Ib\'a\~nez and G.~Ross, \PLB B110 215 1982 ;
K.~Inoue, A.~Kakuto, H.~Komatsu, and S.~Takeshita, 
Prog.\ Theor.\ Phys.\ {\bf 68}, 927 (1982);
L.~Alvarez-Gaum\'e, J.~Polchinski, and M.~Wise, \NPB B221 495 1983 ;
J.~Ellis, D.~Nanopoulos, and K.~Tamvakis, \PLB B125 275 1983 .

\bibitem{haberperspectives}
H.E.~Haber and R. Hempfling, \PRL 66 1815 1991 ; Y.~Okada, M.~Yamaguchi
and T. Yanagida, {Prog. Theor. Phys.} {\bf 85} (1991); J.~Ellis,
G. Ridolfi and F. Zwirner, \PLB B257 83 1991 .
For a review and references, see 
H. Haber, in {\it Perspectives on Higgs Physics}, ed. G. Kane, World
Scientific Publishing, p. 79. See also, 
V.~Barger, K.~Cheung, R.~Phillips, and A.~Stange, 
\PRD 46 4914 1992 .

\bibitem{habertwoloop}
H. Haber, R. Hempfling and A. Hoang, CERN-TH/95-216.

\bibitem{erz} J.~Ellis, G.~Ridolfi and F.~Zwirner, 
Ref.~\cite{haberperspectives}.


\bibitem{carenatwoloop}
M. Carena, J.R. Espinosa, M. Quiros and C.E.M. Wagner,
\PLB B355 209 1995 ; J.A. Casas, J.R. Espinosa, M. Quiros and A. Riotto,
\NPB B436 3 1995 .

\bibitem{cpr} P. Chankowski, S.~Pokorski and J.~Rosiek, \PLB B274 191 1992 ;
\NPB B423 437 1994 ; A.~Dabelstein, \ZP C67 495 1995 .

\bibitem{hhrge} H.E.~Haber and R.~Hempfling, \PRD D48 4280 1993 .

\bibitem{hemphoang} R. Hempfling and A.H.~Hoang, \PLB B331 99 1994 .

\bibitem{gunionperspectives}
J.F. Gunion, in {\it Perspectives on Higgs Physics}, ed. G. Kane, World
Scientific Publishing, p. 179.

\bibitem{gunionerice}
J.~Gunion, in {\it Properties of SUSY Particles}, Proceedings of the 23rd
Workshop of the INFN Eloisatron Project, Erice, Italy, Sept. 28 - Oct. 4
1992, eds. L. Cifarelli and V. Khoze, p. 279; and references therein.

\bibitem{baerreport} The general structure and predictions of
unified versions of the MSSM are reviewed in the companion
reports by H. Baer {\it et. al.} and by M. Drees {\it et. al.}.

\bibitem{gunionpois} For a `typical' scenario see J.F. Gunion and H. Pois,
\PLB B329 136 1994 . Further references are given in \cite{baerreport}.

\bibitem{bargerhawaii}
V. Barger \etal,
{\it Proceedings of the 2nd International Workshop on
``Physics and Experiments with Linear $\epem$ Colliders''},
eds. F. Harris, S. Olsen, S. Pakvasa and X. Tata, Waikoloa, HI (1993), 
World Scientific Publishing, p.~840.

\bibitem{bgkp} For a convenient mini-review focusing on these two
models, see H. Baer, J.F. Gunion, C. Kao and H. Pois, \PRD D51 2159 1995 ,
and references therein. See also Ref.~\cite{baerreport}.



\bibitem{LQT} B.~W.~Lee, C.~Quigg, and H.~Thacker, \PRD D16 1519 1977 .

\bibitem{R} T.~Rizzo, \PRD D22 722 1980 ; W.-Y.~Keung 
and W.~Marciano, \PRD D30 248 1984 .

\bibitem{SDGZ} M. Spira, A. Djouadi, D. Graudenz and P.M. Zerwas,
\NPB B453 17 1995 .
                                                                   
\bibitem{GAMGAM} J.~P.~Leveille, \PLB B83 123 1979 ; 
A.~Vainshtein, M.~Voloshin, V.~Zakharov, and M.~Shifman, Yad.\ Fiz.\ {\bf 
30}, 1368 (1979) [Sov.\ J.\ Nucl.\ Phys.\ {\bf 30}, 711 (1979)]. 

\bibitem{CCF} R.~Cahn, M.~Chanowitz, and N.~Fleishon, \PLB B82 113 1979 .

\bibitem{hollikeps} W. Hollik, presented
at the European Physical Society International Europhysics Conference
on High Energy Physics, Brussels, Belgium, July 27 - August 2, 1995.

\bibitem{habereps} H. Haber, presented
at the European Physical Society International Europhysics Conference
on High Energy Physics, Brussels, Belgium, July 27 - August 2, 1995.

\bibitem{chankowskieps} P. Chankowski, presented
at the European Physical Society International Europhysics Conference
on High Energy Physics, Brussels, Belgium, July 27 - August 2, 1995.

\bibitem{eflhiggs} J. Ellis, G.L. Fogli and E. Lisi, CERN-TH/95-202.

\bibitem{langackerpc} P. Langacker, private communication.

\bibitem{lepewwghiggs} The LEP Electroweak Working Group, CERN
Report No. LEPEWWG/95-01.

\bibitem{chanporhiggs} P.H. Chankowski and S. Pokorski, MPI-Ph/95-39.

\bibitem{dpfshort} J.F.~Gunion, A.~Stange, and S.~Willenbrock,  included in
{\it Electroweak Symmetry Breaking and Beyond the Standard Model},
T. Barklow \etal, SLAC-PUB-95-6893.

\bibitem{BJ} J.~Bjorken, in {\it Weak Interactions at High Energy and the 
Production of New Particles}, proceedings of the SLAC Summer Institute on 
Particle Physics, 1976, edited by M.~Zipf (SLAC Report No. 198, 1977).

\bibitem{LEP} ALEPH Collaboration, \PR 216 253 1992 ; 
\PLB B313 299 1993 ;
\\
DELPHI Collaboration, P.~Abreu {\it et al.}, \NPB B421 3 1994 ;
\\
L3 Collaboration, O.~Adriani {\it et al.}, \PLB B303 391 1993 ;
\\ 
OPAL Collaboration, R.~Akers {\it et al.}, \PLB B327 397 1994 .

\bibitem{RICHARD} F.~Richard, presented at the 27$^{th}$
International Conference on High Energy Physics, Glasgow, July 20-27, 1994,
LAL 94-50 (1994).

\bibitem{grivazeps}
J.F.~Grivaz, presented
at the European Physical Society International Europhysics Conference
on High Energy Physics, Brussels, Belgium, July 27 - August 2, 1995.

\bibitem{SUSYLEP} ALEPH Collaboration, D.~Buskulic {\it et al.}, 
\PLB B313 312 1993 ;
\\
DELPHI Collaboration, P.~Abreu {\it et al.}, \NPB B373 3 1992 ; 
\ZP C67 213 1995 ; 
\\ 
L3 Collaboration, O.~Adriani {\it et al.}, \ZP C57 355 1993 ;
\\
OPAL Collaboration, R.~Akers {\it et al.}, \ZP C64 1 1994 .

\bibitem{altarellilep} G. Altarelli {\it et al.}, ``Interim Report
on the Physics Motivations for an Energy Upgrade of LEP2'', 
CERN-TH/95-151 and CERN-PPE/95-78 (1995).

\bibitem{RS} J.~Rosiek and A.~Sopczak, \PLB B341 419 1995 .


\bibitem{mfrank} M. Frank, ALEPH Collaboration, presented
at the European Physical Society International Europhysics Conference
on High Energy Physics, Brussels, Belgium, July 27 - August 2, 1995.

\bibitem{janotnewlep} P. Janot, in preparation.

\bibitem{sophp} A. Sopczak, \IJMP A9 1747 1994 .

\bibitem{JP} D.~R.~T.~Jones and S.~Petcov, \PLB B84 440 1979 .

\bibitem{LEP2} S.~L.~Wu {\it et al.}, in {\it Proceedings of the ECFA Workshop
on LEP 200}, Aachen, 1986, eds. A.~B\"ohm and W.~Hoogland, CERN 87-08, Vol. II,
p.~312.

\bibitem{janotlep} P.~Janot, in {\it `92 Electroweak Interactions and Unified
Theories, Proceedings of the XXVII Rencontre de Moriond}, edited by J.~Tr\^an 
Thanh V\^an (\'Editions Fronti\`eres, Gif-sur-Yvette, 1992), p.~317.

\bibitem{GGMN} H.~Georgi, S.~Glashow, M.~Machacek, and D.~Nanopoulos, 
\PRL 40 692 1978 .

\bibitem{CD} R.~Cahn and S.~Dawson, \PLB B136 196 1984 ;
G.~Kane, W.~Repko, and W.~Rolnick, \PLB B148 367 1984 .

\bibitem{GNY} S.~Glashow, D.~Nanopoulos, and A.~Yildiz, \PRD 18 1724 1978 .

\bibitem{K} Z.~Kunszt, \NPB B247 339 1984 .

\bibitem{DW} D.~Dicus and S.~Willenbrock, \PRD D39 751 1989 .

\bibitem{GSZ} D.~Graudenz, M.~Spira, and P.~Zerwas, \PRL 70 1372 1993 ; 
A.~Djouadi and M.~Spira, private communication.

\bibitem{HVW} T.~Han, G.~Valencia, and S.~Willenbrock, \PRL 69 3274 1992 .

\bibitem{HW} T.~Han and S.~Willenbrock, \PLB B273 167 1990 .

\bibitem{CTEQ} H.~Lai {\it et al.}, MSU-HEP-41024, CTEQ-404 (1994).

\bibitem{GKWudka} J.~Gunion, G.~Kane, and J.~Wudka, \NPB B299 231 1988 .

\bibitem{GMP} J.~Gunion, \PLB B261 510 1991 ;
W.~Marciano and F.~Paige, \PRL 66 2433 1991 .

\bibitem{SMW2} A.~Stange, W.~Marciano, and S.~Willenbrock, \PRD D50 4491 1994 .

\bibitem{FR} D.~Froidevaux and E.~Richter-Was, \ZP C67 213 1995 .

\bibitem{MK} S.~Mrenna and G.~Kane, CALT-68-1938 (1994).

\bibitem{ABC} P.~Agrawal, D.~Bowser-Chao, and K.~Cheung, \PRD D51 6114 1995 .

\bibitem{DGV1} J.~Dai, J.~Gunion, and R.~Vega, \PRL 71 2699 1993 .

\bibitem{ATLAS} ATLAS Technical Proposal, CERN/LHCC/94-43, LHCC/P2 (1994).

\bibitem{CMS} CMS Technical Proposal, CERN/LHCC 94-38, LHCC/P1 (1994).

\bibitem{FGR} D.~Froidevaux, F.~Gianotti, and E.~Richter-Was,
ATLAS Internal Note PHYS-No-64 (1995).

\bibitem{TML} T.~Liss, private communication.

\bibitem{fgianotti} F. Gianotti, to appear in the Proceedings of 
the European Physical Society International Europhysics Conference
on High Energy Physics, Brussels, Belgium, July 27 - August 2, 1995.

\bibitem{latestplots}
D. Froidevaux, F. Gianotti, L. Poggioli,
E. Richter-Was, D. Cavalli, and S. Resconi, 
ATLAS Internal Note, PHYS-No-74 (1995).  

\bibitem{GKW} T.~Garavaglia, W.~Kwong, and D.-D.~Wu, \PRD D48 1899 1993 .


\bibitem{jggeer} These effects were first explored in 
J.F. Gunion and S. Geer, {\it
Proceedings of the ``Workshop on Physics at Current Accelerators and
the Supercollider''}, eds. J. Hewett, A. White, and D. Zeppenfeld,
Argonne National Laboratory, 2-5 June (1993), ANL-HEP-CP-93-92, p.~335.

\bibitem{GOW} E.~W.~N.~Glover, J.~Ohnemus, and S.~Willenbrock, 
\PRD D37 3193 1988 ; V.~Barger, G.~Bhattacharya, T.~Han, and B.~Kniehl, 
\PRD D43 779 1991 .

\bibitem{hanreport} T. Han, M. Golden, and G. Valencia, this volume.

\bibitem{DSW} D.~Dicus, A.~Stange, and S.~Willenbrock, \PLB B333 126 1994 .

\bibitem{KZ} Z.~Kunszt and F.~Zwirner, \NPB B385 3 1992 .

\bibitem{BBKT} H.~Baer, M.~Bisset, C.~Kao, and X.~Tata, \PRD D46 1067 1992 ; 
H.~Baer, M.~Bisset, D.~Dicus, C.~Kao, and X.~Tata, \PRD D47 1062 1993 .

\bibitem{GO} J.~Gunion and L.~Orr, \PRD D46 2052 1992 .

\bibitem{BCPS} V.~Barger, K.~Cheung, R.~Phillips, and A.~Stange, 
\PRD D46 4914 1992 .


\bibitem{EHSV} R.~K.~Ellis, I.~Hinchliffe, M.~Soldate, and J.~van~der~Bij,
\NPB B297 221 1988 .

\bibitem{DGV2} J.~Dai, J.~Gunion, and R.~Vega, \PLB B315 355 1993 .

\bibitem{DGV3} J.~Dai, J.~Gunion, and R.~Vega, \PLB B345 29 1995 .

\bibitem{dgv3update} J.~Dai, J.~Gunion, and R.~Vega, presented
by J. Dai at {\it Tahoe CMS Week},  Granlibakken, Tahoe City, CA,
September 25-27, 1995, and UCD-95-46 (1995).

\bibitem{GBPR} J.~Gunion, \PLB B322 125 1994 ; V.~Barger, R.~Phillips, and 
D.~Roy, \PLB B324 236 1994 .

\bibitem{DGV4} J.~Dai, J.~Gunion, and R.~Vega, preprint UCD-95-25 (1995).

\bibitem{guninvis} J. Gunion, \PRL 72 199 1994 .

\bibitem{fjk} S.G. Frederiksen, N. Johnson, G. Kane, \PRD D50 4244 1994 .

\bibitem{cr} D. Choudhury, D.P. Roy, \PLB B322 368 1994 .

\bibitem{ghhprecision} J.F. Gunion, H.E. Haber, and M. Hildreth,
in preparation.

\bibitem{latticemb} C. Davies, K. Hornbostel, A. Langnau, G. Lepage,
A. Lidsey, C. Morningstar, J. Shigemitsu, and J. Sloan, \PRL 73 2654 1994 .

\bibitem{Tev2000} {\it Report of the Tev2000 Study Group on Future
Electroweak Physics at the Tevatron}, eds. D.~Amidei and R.~Brock,
D0 Note 2589/CDF Note 3177 (1995).

\bibitem{SMW1} A.~Stange, W.~Marciano, and S.~Willenbrock, \PRD D49 1354 1994 .

\bibitem{GH} J.~Gunion and T.~Han, \PRD D51 1051 1995 .

\bibitem{BBD} A.~Belyaev, E.~Boos, and L.~Dudko, \MPL A10 25 1995 .

\bibitem{Kuhlmann} S.~Kuhlmann, in Ref.~\cite{Tev2000}.

\bibitem{Heintz} U.~Heintz, in Ref.~\cite{Tev2000}.

\bibitem{bck} T. Barklow, P. Chen, and W. Kozanecki, in 
{\it $\epem$ Collisions
at 500 GeV: the Physics Potential}, DESY 92-123 (1992), ed.
P.M.~Zerwas, Feb. 4 - Sept. 3 (1991) --- Munich, Annecy, Hamburg, p. 845.

\bibitem{am} A. Miyamoto, {\it Proceedings of the 2nd International Workshop on
``Physics and Experiments with Linear $\epem$ Colliders''},
eds. F. Harris, S. Olsen, S. Pakvasa and X. Tata, Waikoloa, HI (1993), 
World Scientific Publishing, p.~141.


\bibitem{desyworkshop} For recent studies, see P. Grosse-Wiesemann, D. Haidt,
and H.J. Schreiber, in {\it $\epem$ Collisions
at 500 GeV: the Physics Potential}, DESY 92-123A (1992), ed.
P.M.~Zerwas, Feb. 4 - Sept. 3 (1991) --- Munich, Annecy, Hamburg, p. 37;
P. Janot, {\it ibid.}, p. 107; A. Djouadi, D. Haidt, B.A.~Kniehl, B.~Mele,
and P.M.~Zerwas, {\it ibid.}, p. 11; and references therein.

         
\bibitem{janothawaii}
P. Janot, {\it Proceedings of the 2nd International Workshop on
``Physics and Experiments with Linear $\epem$ Colliders''},
eds. F. Harris, S. Olsen, S. Pakvasa and X. Tata, Waikoloa, HI (1993), 
World Scientific Publishing, p.~192; and references therein.

\bibitem{jlci}
See ``JLC-I'', KEK-92-16, December 1992.

\bibitem{bbp} P. Burchat, D. Burke, and A. Petersen, \PRD D38 2735 1988 .


\bibitem{gunionhawaii} For an overview, see
J.F. Gunion, {\it Proceedings of the 2nd International Workshop on
``Physics and Experiments with Linear $\epem$ Colliders''},
eds. F. Harris, S. Olsen, S. Pakvasa and X. Tata, Waikoloa, HI (1993), 
World Scientific Publishing, p.~166.

\bibitem{dkz} A. Djouadi, J. Kalinowski, and P.M. Zerwas, \ZP C54 255 1992 .

\bibitem{kawagoe} K. Kawagoe, 
{\it Proceedings of the 2nd International Workshop on
``Physics and Experiments with Linear $\epem$ Colliders''},
eds. F. Harris, S. Olsen, S. Pakvasa and X. Tata, Waikoloa, HI (1993), 
World Scientific Publishing, p.~660.

\bibitem{hbb} M. Hildreth, T. Barklow, and D. Burke, \PRD D49 3441 1994 .

\bibitem{hildrethhawaii} M. Hildreth,
{\it Proceedings of the 2nd International Workshop on
``Physics and Experiments with Linear $\epem$ Colliders''},
eds. F. Harris, S. Olsen, S. Pakvasa and X. Tata, Waikoloa, HI (1993), 
World Scientific Publishing, p.~635.

\bibitem{bordenhawaii} 
D. Borden,
{\it Proceedings of the 2nd International Workshop on
``Physics and Experiments with Linear $\epem$ Colliders''},
eds. F. Harris, S. Olsen, S. Pakvasa and X. Tata, Waikoloa, HI (1993), 
World Scientific Publishing, p.~323.

\bibitem{shhawaii} 
M.L. Stong and K. Hagiwara, 
{\it Proceedings of the 2nd International Workshop on
``Physics and Experiments with Linear $\epem$ Colliders''},
eds. F. Harris, S. Olsen, S. Pakvasa and X. Tata, Waikoloa, HI (1993), 
World Scientific Publishing, p.~631.

\bibitem{kksz} 
M. Kramer, J. Kuhn, M. Stong, and P. Zerwas, \ZP C64 21 1994 .

\bibitem{sonixu}
A. Soni and R.M. Xu, preprint \PRD D48 5259 1993 .

\bibitem{ck}
D. Chang and W.-Y. Keung, \PLB B305 261 1993 .

\bibitem{ckp}
D. Chang, W.-Y. Keung, and I. Phillips, \PRD D48 3225 1993 .

\bibitem{bargeretal}
V. Barger, K. Cheung, A. Djouadi, B.A. Kniehl and P.M. Zerwas,
\PRD D49 79 1994 .

\bibitem{skjold}
A. Skjold, \PLB B311 261 1993 ; \PLB B329 305 1994 .


\bibitem{bohguncp}
B. Grzadkowski and J.F. Gunion, \PLB B350 218 1995 .

\bibitem{zwirnererice}
F. Zwirner, in {\it Properties of SUSY Particles}, Proceedings of the 23rd
Workshop of the INFN Eloisatron Project, Erice, Italy, Sept. 28 - Oct. 4
1992, eds. L. Cifarelli and V. Khoze, p. 509; and references therein.

\bibitem{desyworkshopsusy}
A. Brignole, J. Ellis, J.F. Gunion, M. Guzzo, F. Olness, G. Ridolfi,
L. Roszkowski and F. Zwirner, in {\it $\epem$ Collisions at 500 GeV:
The Physics Potential}, Munich, Annecy, Hamburg Workshop,
DESY 92-123A, DESY 92-123B, DESY 93-123C, ed. P. Zerwas, p. 613;
A. Djouadi, J. Kalinowski, P.M. Zerwas, {\it ibid.} p. 83 and 
\ZP C57 569 1993 ; and references therein.

\bibitem{gunionzeuthen} For a review and references, 
   see J.~Gunion, in 
   {\it Proceedings of the Zeuthen Workshop on Elementary
   Particle Theory --- ``LEP 200 and Beyond''}, Teupitz/Brandenburg,
   Germany, 10-15 April (1994), eds. T. Riemann and J. Blumlein, p. 253.


\bibitem{baererice}
H. Baer and X. Tata, 
in {\it Properties of SUSY Particles}, Proceedings of the 23rd
Workshop of the INFN Eloisatron Project, Erice, Italy, Sept. 28 - Oct. 4
1992, eds. L. Cifarelli and V. Khoze, p. 244; and references therein.


\bibitem{gkopair} J.F. Gunion, J. Kelly, J. Ohnemus, work in progress.

\bibitem{eghrz}
J. Ellis, J.F. Gunion, H.E. Haber, L. Roszkowski and F. Zwirner,
\PRD D39 844 1989 .

\bibitem{kimoh}
B.R. Kim, S.K. Oh and A. Stephan, 
{\it Proceedings of the 2nd International Workshop on
``Physics and Experiments with Linear $\epem$ Colliders''},
eds. F. Harris, S. Olsen, S. Pakvasa and X. Tata, Waikoloa, HI (1993), 
World Scientific Publishing, p.~860.

\bibitem{kot}
J. Kamoshita, Y. Okada and M. Tanaka, \PLB B328 67 1994 .

\bibitem{KW} S.F. King and P.L. White, preprint SHEP-95-27 (1995).

\bibitem{ETS} U. Ellwanger, M.R. de Traubenberg and C.A. Savoy,
\ZP C67 665 1995 .

\bibitem{telnovhawaii} 
V. Telnov,
{\it Proceedings of the 2nd International Workshop on
``Physics and Experiments with Linear $\epem$ Colliders''},
eds. F. Harris, S. Olsen, S. Pakvasa and X. Tata, Waikoloa, HI (1993), 
World Scientific Publishing, p.~551.


\bibitem{ghgamgam}
J.F. Gunion and H.E. Haber, {\it Proceedings of the 1990 DPF Summer Study on 
High Energy Physics: ``Research Directions for the Decade''}, 
editor E. Berger, Snowmass (1990), p. 206; \PRD D48 5109 1993 .

\bibitem{barklow}
T. Barklow, {\it Proceedings of the 1990 DPF Summer Study on 
High Energy Physics: ``Research Directions for the Decade''}, 
editor E. Berger, Snowmass (1990), p. 440.

\bibitem{bbc}
D. Borden, D. Bauer, and D. Caldwell, \PRD D48 4018 1993 .

\bibitem{jikia}
G.V. Jikia, \PLB B298 224 1993 ; 
{\it Proceedings of the 2nd International Workshop on
``Physics and Experiments with Linear $\epem$ Colliders''},
eds. F. Harris, S. Olsen, S. Pakvasa and X. Tata, Waikoloa, HI (1993), 
World Scientific Publishing, p.~558.

\bibitem{berger}
M.S. Berger, \PRD D48 5121 1993 .

\bibitem{DK}
D.~Dicus and C.~Kao, \PRD D49 1265 1994 .

\bibitem{derujula}
A. de Rujula \etal, \NPB B384 3 1992 .

\bibitem{perezt}
M.A. Perez and J.J. Toscano, \PLB B289 381 1992 .

\bibitem{mtz}
D. Morris, T. Truong, and  D. Zappala, \PLB B323 421 1994 .

\bibitem{brodskyww}
S.J. Brodsky,
{\it Proceedings of the 2nd International Workshop on
``Physics and Experiments with Linear $\epem$ Colliders''},
eds. F. Harris, S. Olsen, S. Pakvasa and X. Tata, Waikoloa, HI (1993), 
World Scientific Publishing, p.~295.

\bibitem{boudjemaww}
M. Baillargeon, G. Belanger, and F. Boudjema, 
ENSLAPP-A-473-94 (1994).

\bibitem{cheungww}
K. Cheung, \PLB B323 85 1994 ; \PRD D50 4290 1994 .

\bibitem{jikiaww}
G. Jikia, preprints hep-ph/9406395; \NPB B347 520 1995 .

\bibitem{bkso}
D.L. Borden, V.A. Khoze, W.J. Stirling, and J. Ohnemus, \PRD D50 4499 1994 .

\bibitem{bowserchao}
D. Bowser-Chao and K. Cheung, \PRD D48 89 1993 .

\bibitem{ebolietal}
O.J.P. Eboli, M.C. Gonzalez-Garcia, F. Halzen and
D. Zeppenfeld, \PRD D48 1430 1993 .

\bibitem{boostthsm}
E. Boos \etal, \ZP C56 487 1992 .

\bibitem{cheungtthsm}
K. Cheung, \PRD D47 3750 1993 .

\bibitem{hagwz}
K. Hagiwara, I. Watanabe, P.M. Zerwas, \PLB B278 187 1992 .

\bibitem{boos}
E. Boos, M. Dubinin, V. Ilyin, A. Pukhov, G. Jikia, S.
Sultanov, \PLB B273 173 1991 .

\bibitem{cheung}
K. Cheung, \PRD D48 1035 1993 .

\bibitem{eggn}
O.J.P. Eboli, M.C. Gonzalez-Garcia, S.F. Novaes, \PRD D49 91 1994 .

\bibitem{bordenpc}
D. Borden, private communication. 

\bibitem{kileng}
B. Kileng, \ZP C63 87 1994 .

\bibitem{gkogamgam} J.F. Gunion, J. Kelly, J. Ohnemus, \PRD D51 2101 1995 .

\bibitem{dbcct}
D. Bowser-Chao, K. Cheung, and S. Thomas,  \PLB B315 399 1993 .

\bibitem{bgbslaser}
B. Grz\c{a}dkowski and J.F. Gunion, \PLB B294 361 1992 .

\bibitem{gk}
J.F. Gunion and J. Kelley, \PLB B333 110 1994 .

\bibitem{sausi} {\it Proceedings of the First Workshop on the Physics Potential
and Development of $\mu^+\mu^-$ Colliders}, Napa, California (1992), Nucl.\
Instru.\ and Meth.\ {\bf A350}, 24 (1994).

\bibitem{sausii} {\it Proceedings of the Second Workshop on the Physics
Potential and Development of $\mu^+\mu^-$ Colliders}, Sausalito, California
(1994), ed.\ by D.~Cline, p. 55.

\bibitem{bbgh} V. Barger, M. Berger, J. Gunion and T. Han, 
\PRL 75 1462 1995 ; UCD-96-1, in preparation.

\bibitem{palmer} R.B.~Palmer and A.~Tollestrup, unpublished report.

\bibitem{palmer2} R.B.~Palmer, private communication.

\bibitem{jackson} G.P.~Jackson and D.~Neuffer, private communications.

\bibitem{barklowresolution} T.~Barklow and D.~Burke, private communication.

\bibitem{atsoncp}
D. Atwood and A. Soni, \PRD D52 6271 1995 .

\end{thebibliography}
\end{document}